\documentclass[twocolumn,amsmath,amssymb,prd]{revtex4}

\usepackage{wasysym}

\def\al{\alpha}
\def\be{\beta}
\def\ga{\gamma}
\def\de{\delta}
\def\ep{\epsilon}

\def\ze{\zeta}
\def\et{\eta}
\def\th{\theta}

\def\ka{\kappa}
\def\la{\lambda}

\def\rh{\rho}

\def\si{\sigma}

\def\ta{\tau}

\def\ph{\phi}

\def\ch{\chi}
\def\ps{\psi}
\def\om{\omega}
\def\Ga{\Gamma}
\def\De{\Delta}

\def\La{\Lambda}
\def\Si{\Sigma}

\def\Ph{\Phi}
\def\Ps{\Psi}
\def\Om{\Omega}

\def\cE{{\cal E}}
\def\cl{{\cal L}}
\def\cL{{\cal L}}
\def\cO{{\cal O}}

\def\mn{{\mu\nu}}

\def\fr#1#2{{{#1} \over {#2}}}
\def\half{{\textstyle{1\over 2}}}
\def\quar{{\textstyle{1\over 4}}}

\def\frac#1#2{{\textstyle{{#1}\over {#2}}}}

\def\vev#1{\langle {#1}\rangle}

\def\abs#1{\left|{#1}\right|}

\def\lsim{\mathrel{\rlap{\lower4pt\hbox{\hskip1pt$\sim$}}
    \raise1pt\hbox{$<$}}}
\def\gsim{\mathrel{\rlap{\lower4pt\hbox{\hskip1pt$\sim$}}
    \raise1pt\hbox{$>$}}}
\def\sqr#1#2{{\vcenter{\vbox{\hrule height.#2pt
         \hbox{\vrule width.#2pt height#1pt \kern#1pt
         \vrule width.#2pt}
         \hrule height.#2pt}}}}

\def\prt{\partial}

\def\etal{{\it et al.}}

\def\pt#1{\phantom{#1}}

\def\ol#1{\overline{#1}}

\def\nsc#1#2#3{\om_{#1}^{{\pt{#1}}#2#3}}
\def\lsc#1#2#3{\om_{#1#2#3}}

\def\vb#1#2{e_{#1}^{{\pt{#1}}#2}}
\def\ivb#1#2{e^{#1}_{{\pt{#1}}#2}}
\def\uvb#1#2{e^{#1#2}}
\def\lvb#1#2{e_{#1#2}}
\def\etul#1#2{\et^{#1}_{{\pt{#1}}#2}}
\def\hul#1#2{h^{#1}_{{\pt{#1}}#2}}

\def\ab{\overline{a}{}}
\def\bb{\overline{b}{}}
\def\cb{\overline{c}{}}
\def\db{\overline{d}{}}
\def\eb{\overline{e}{}}
\def\fb{\overline{f}{}}
\def\gb{\overline{g}{}}
\def\Hb{\overline{H}{}}

\def\kb{\overline{k}{}}
\def\pb{\overline{p}{}}
\def\sb{\overline{s}{}}
\def\kfb{(\overline{k_F})}

\def\dep{\de p}

\def\psb{\overline{\ps}{}}

\def\twiddle{\lower4pt\hbox{\hskip-0pt{$\widetilde{}$}}}
\def\m@th{\mathsurround=0pt}
\def\cmapstochar{\mathrel{\rlap{
  \lower0.1pt\hbox{\hskip-1.75pt{$\mapstochar$}}}
  \raise0pt\hbox{\hskip2.5pt{$\twiddle$}}}}
\def\notsimfill{$\m@th\cmapstochar$}
\def\scroodle#1{\vbox{\ialign{##\crcr\notsimfill\crcr
  \noalign{\kern-4pt\nointerlineskip}
   $\hfil\displaystyle{#1}\hfil$\crcr}}}

\def\cmapstocharbig{\mathrel{\rlap{
  \lower0.1pt\hbox{\hskip0.25pt{$\mapstochar$}}}
  \raise0pt\hbox{\hskip4.5pt{$\twiddle$}}}}
\def\notsimfillbig{$\m@th\cmapstocharbig$}
\def\scroodlebig#1{\vbox{\ialign{##\crcr\notsimfillbig\crcr
  \noalign{\kern-4pt\nointerlineskip}
   $\hfil\displaystyle{#1}\hfil$\crcr}}}

\def\Btw{\scroodlebig{B}}

\def\X{t_{\la \mn \ldots}}
\def\Xupper{t^{\la \mn \ldots}}
\def\Xb{\overline{t}_{\la \mn \ldots}}
\def\Xbupper{\overline{t}^{\la \mn \ldots}}
\def\Xtw{\scroodle{t}_{\la \mn \ldots}}
\def\Btw{\scroodlebig{B}}

\def\Xtildew{\widetilde{t}^{\la \mn \ldots}}
\def\Xtildewl{\widetilde{t}_{\la \mn \ldots}}
\def\Xtwo#1#2{\scroodle{t}{}^{(#1,#2)}_{\la \mn \ldots}}

\def\mt{m^{\rm T}}
\def\ms{m^{\rm S}}
\def\mb{m^{\rm{B}}}
\def\mbp{m^{\prime\rm{B}}}
\def\qt{q^{\rm T}}

\def\af{(a_{\rm{eff}})}
\def\afp{(a^{\prime{\rm B}}_{\rm{eff}})}
\def\afB{(a^{\rm B}_{\rm{eff}})}
\def\cuB{(c^{\rm B})}

\def\afb{(\ab_{\rm{eff}})}
\def\afbb{(\ab^{\rm B}_{\rm{eff}})}
\def\afbp{(\ab^{\prime{\rm B}}_{\rm{eff}})}
\def\afbnp{\ab_{\rm{eff}}}

\def\abt{(\ab^{\rm T}_{\rm{eff}})}
\def\abs{(\ab^{\rm S}_{\rm{eff}})}

\def\afbm{(\ab^\mu_{\rm{eff}})}
\def\cbm{(\cb^\mu)}

\def\afbx#1{(\ab^{#1}_{\rm{eff}})}
\def\cbx#1{(\cb^{#1})}

\def\abw{(\ab^w)}
\def\abe{(\ab^e)}
\def\abp{(\ab^p)}
\def\abn{(\ab^n)}
\def\afbe{\afbx{e}}
\def\afbp{\afbx{p}}
\def\afbn{\afbx{n}}

\def\cbe{(\cb^e)}
\def\cbp{(\cb^p)}
\def\cbn{(\cb^n)}

\def\cbw{\cbx{w}}
\def\afbw{\afbx{w}}
\def\afw{(a^w_{\rm{eff}})}
\def\cw{(c^w)}

\def\cbpr{(c^{\prime{\rm B}})}
\def\cot{(c^{\rm T})}
\def\cs{(c^{\rm S})}

\def\cbt{(\cb^{\rm T})}
\def\cbs{(\cb^{\rm S})}
\def\cbb{(\cb^{\rm B})}

\def\hlv{h^{(1,1)}}
\def\hlnv{h^{(0,1)}}

\def\atw{\scroodle{a}{}}
\def\btw{\scroodle{b}{}}
\def\ctw{\scroodle{c}{}}
\def\ctwb{(\ctw{}^{\rm B})}
\def\dtw{\scroodle{d}{}}
\def\etw{\scroodle{e}{}}
\def\ftw{\scroodle{f}{}}
\def\gtw{\scroodle{g}{}}
\def\Htw{\scroodle{H}{}}

\def\aftw{(\scroodle{a}_{\rm{eff}})}
\def\aftwb{(\scroodle{a}{}^{\rm B}_{\rm{eff}})}
\def\atwt{(\scroodle{a}{}^{\rm T}_{\rm{eff}})}

\def\ctwt{(\ctw{}^{\rm T})}

\def\aloo{\al_{(0,0)}}
\def\alol{\al_{(0,1)}}

\def\zon{\ze_1}
\def\ztw{\ze_2}
\def\zes{\ze_1^{\rm S}}
\def\zet{\ze_1^{\rm T}}
\def\zts{\ze_2^{\rm S}}
\def\ztt{\ze_2^{\rm T}}

\def\xc{\si_1}

\def\vf{j}

\def\noe{N_\oplus}

\def\axw{\afbw_{X}}
\def\ayw{\afbw_{Y}}
\def\azw{\afbw_{Z}}
\def\akw{\afbw_{K}}
\def\cxxw{(\cb^w)_{XX}}
\def\cyyw{(\cb^w)_{YY}}
\def\czzw{(\cb^w)_{ZZ}}
\def\cxyw{(\cb^w)_{(XY)}}
\def\cxzw{(\cb^w)_{(XZ)}}
\def\cyzw{(\cb^w)_{(YZ)}}
\def\ctxw{(\cb^w)_{(TX)}}
\def\ctyw{(\cb^w)_{(TY)}}
\def\ctzw{(\cb^w)_{(TZ)}}
\def\cttw{(\cb^w)_{TT}}
\def\ctjw{(\cb^w)_{(TJ)}}

\def\cqbx#1{(\cb^{#1})_Q}

\def\stt{\sb_{TT}}
\def\sxx{\sb_{XX}}
\def\syy{\sb_{YY}}
\def\szz{\sb_{ZZ}}
\def\sxy{\sb_{(XY)}}
\def\sxz{\sb_{(XZ)}}
\def\syz{\sb_{(YZ)}}
\def\stx{\sb_{(TX)}}
\def\sty{\sb_{(TY)}}
\def\stz{\sb_{(TZ)}}
\def\that{{\hat{t}}}
\def\xhat{{\hat{x}}}
\def\yhat{{\hat{y}}}
\def\zhat{{\hat{z}}}
\def\jhat{{\hat{j}}}
\def\khat{{\hat{k}}}
\def\lhat{{\hat{l}}}
\def\muhat{{\hat{\mu}}}
\def\mnhat{{\hat{\mu}}{\hat{\nu}}}

\def\jearth{{\tilde{j}}}
\def\kearth{{\tilde{k}}}
\def\learth{{\tilde{l}}}
\def\tearth{{\tilde{t}}}
\def\xearth{{\tilde{x}}}
\def\yearth{{\tilde{y}}}
\def\zearth{{\tilde{z}}}
\def\muearth{{\tilde{\mu}}}
\def\mnearth{{\tilde{\mu}}{\tilde{\nu}}}

\def\acc{{\rm a}}

\def\lrpartial{\raise 1pt\hbox{$\stackrel\leftrightarrow\partial$}}

\def\lrDmu{\stackrel{\leftrightarrow}{D_\mu}}

\def\a{$a_\mu$}
\def\b{$b_\mu$}
\def\c{$c_{\mu\nu}$}
\def\d{$d_{\mu\nu}$}
\def\e{$e_\mu$}
\def\f{$f_\mu$}
\def\g{$g_{\la\mu\nu}$}
\def\H{$H_{\mu\nu}$}

\def\G{G_N}

\def\hnr{H_{\rm NR}}
\def\hnrs#1{H_{{\rm NR},#1}}

\def\pb{\overline{p}}

\def\csk{\xi}
\def\cskc{\xi_{\rm clock}}

\def\generala{I}
\def\laba{II}
\def\labb{III}
\def\labc{IV}
\def\labe{V}
\def\labf{VI}
\def\labg{VII}
\def\labh{VIII}
\def\spacea{IX}
\def\spaceb{X}
\def\spacec{XI}
\def\llra{XII}
\def\photona{XIII}
\def\summarya{XIV}
\def\summaryb{XV}

\newcommand{\beq}{\begin{equation}}
\newcommand{\eeq}{\end{equation}}
\newcommand{\bea}{\begin{eqnarray}}
\newcommand{\eea}{\end{eqnarray}}
\newcommand{\bit}{\begin{itemize}}
\newcommand{\eit}{\end{itemize}}
\newcommand{\rf}[1]{(\ref{#1})}

\def\pno#1{PNO(#1)}

\def\np{\nu_P}
\def\ne{\nu_E}

\def\epcombo{e+p}
\def\pen{e+p-n}

\def\nwr#1{n^w_{#1}}

\begin{document}

\title{Matter-gravity couplings and Lorentz violation}

\author{V.\ Alan Kosteleck\'y and Jay D.\ Tasson}

\affiliation{Physics Department, Indiana University,
Bloomington, IN 47405, U.S.A.}

\date{
IUHET 544, June 2010;
version published in Phys.\ Rev.\ D {\bf 83}, 016013 (2011)}

\begin{abstract}

The gravitational couplings of matter 
are studied in the presence of Lorentz and CPT violation.
At leading order in the coefficients for Lorentz violation,
the relativistic quantum hamiltonian is derived from 
the gravitationally coupled minimal Standard-Model Extension.
For spin-independent effects,
the nonrelativistic quantum hamiltonian and 
the classical dynamics for test and source bodies are obtained.
A systematic perturbative method is developed 
to treat small metric and coefficient fluctuations
about a Lorentz-violating and Minkowski background.
The post-newtonian metric and the trajectory 
of a test body freely falling under gravity
in the presence of Lorentz violation are established.
An illustrative example is presented for a bumblebee model.
The general methodology is used to identify 
observable signals of Lorentz and CPT violation
in a variety of gravitational experiments and observations,
including
gravimeter measurements,
laboratory and satellite tests of the weak equivalence principle,
antimatter studies,
solar-system observations,
and investigations of the gravitational properties of light.
Numerous sensitivities to coefficients for Lorentz violation
can be achieved in existing or near-future experiments
at the level of parts in $10^{3}$ down to parts in $10^{15}$.
Certain coefficients are uniquely detectable 
in gravitational searches and remain unmeasured to date.

\end{abstract}

\maketitle

\section{Introduction}
\label{intro}

General Relativity (GR) is known to provide
an accurate description of classical gravitational phenomena
over a wide range of distance scales. 
A foundational property of
the gravitational couplings of matter in GR 
is local Lorentz invariance in freely falling frames.
The realization that a consistent theory of quantum gravity
at the Planck scale $m_P\simeq 10^{19}$ GeV
could induce tiny manifestations of Lorentz violation 
at observable scales
\cite{ksp}
has revived interest in studies of Lorentz symmetry,
with numerous sensitive searches for Lorentz violation 
being undertaken in recent years 
\cite{tables}.

Gravitational signals of Lorentz violation
are more challenging to study
than ones in Minkowski spacetime 
for several reasons,
including the comparative weakness of gravity 
at the microscopic level
and the impossibility of screening gravitational effects
on macroscopic scales.
Both for purely gravitational interactions
and for matter-gravity couplings,
Lorentz violations can be classified and enumerated
in effective field theory
\cite{akgrav}.
Several searches for purely gravitational 
Lorentz violations in this context 
have recently been performed
\cite{gravexpt1,gravexpt2,gravexpt3}
using a treatment in the post-newtonian regime 
\cite{qbakpn}.
These results enlarge and complement
the impressive breadth of tests of GR
performed in the context of the
parametrized post-newtonian (PPN) formalism
\cite{cmw}.

Although the coupling between matter and gravity
has historically been a primary source of insights
into the properties of the gravitational field,
a general study of matter-gravity couplings
allowing for Lorentz violation
in the context of effective field theory
has been lacking to date.
In this work,
we address this gap in the literature
and investigate the prospects 
for searches for Lorentz violation
involving matter-gravity couplings.
Our goal is to elucidate both theoretical and experimental aspects
of the subject.
We seek a post-newtonian expansion for the equation for the trajectory
of a test body moving under gravity in the presence
of Lorentz violation,
allowing also for Lorentz-violating effects 
from the composition of the test and source bodies
and for effects from possible additional long-range modes
associated with Lorentz violation.
We also seek to explore the implications of our analysis 
in a wide variety of experimental and observational scenarios,
identifying prospective measurable signals
and thereby enabling more complete searches 
using matter-gravity couplings.

Despite the current lack 
of a satisfactory quantum theory of gravity,
established gravitational and particle phenomena 
at accessible energy scales
can successfully be analyzed 
using the field-theoretic combination
of GR and the Standard Model (SM).
This combination therefore serves
as a suitable starting point 
for a comprehensive effective field theory
describing observable signals 
of Planck-scale Lorentz and CPT violation
in gravity and particle physics
\cite{kp}.
The present paper adopts this general framework,
known as the gravitational Standard-Model Extension (SME)
\cite{akgrav},
to analyze Lorentz violation in matter-gravity couplings.
Each term violating Lorentz symmetry in the SME Lagrange density
is a scalar density 
under observer general coordinate transformations
and consists of a Lorentz-violating operator
multiplied by a controlling coefficient for Lorentz violation.
Under mild assumptions,
CPT violation in effective field theory 
comes with Lorentz violation
\cite{owg},
so the SME also describes general breaking of CPT symmetry.
This feature plays a crucial role 
for certain signals of Lorentz violation in what follows.

In this work,
our focus is on gravitational Lorentz violation
in matter-gravity couplings,
both with and without CPT violation.
These couplings introduce operator structures
offering sensitivity to coefficients for Lorentz violation
that are intrinsically unobservable in Minkowski spacetime.
In fact,
comparatively large gravitational Lorentz violation in nature 
could have remained undetected in searches to date
because gravity can provide a countershading effect
\cite{akjt},
so this line of investigation has a definite discovery potential.
Several searches for gravitational Lorentz violation
have led to constraints 
on SME coefficients for Lorentz violation 
with sensitivities down to parts in $10^{9}$
\cite{gravexpt1,gravexpt2,gravexpt3,gravexpt4},
and additional constraints can be inferred 
by reanalysis of data from equivalence-principle tests
\cite{akjt}.

The nature of the Lorentz violation plays a crucial role
in determining the physics of matter-gravity couplings.
In Riemann geometry,
externally prescribing the coefficients for Lorentz violation
as fixed background configurations
is generically incompatible with the Bianchi identities 
and hence problematic
\cite{akgrav}.
This issue can be avoided via spontaneous Lorentz breaking
\cite{ksp},
in which a potential term 
drives the dynamical development 
of one or more nonzero vacuum values for a tensor field.
This mechanism implies the underlying 
Lagrange density is Lorentz invariant,
so the coefficients for Lorentz violation
are expressed in terms of vacuum values
and can therefore serve as dynamically consistent backgrounds
satisfying the Bianchi identities.
The presence of a potential driving spontaneous Lorentz violation
implies the emergence of
massless Nambu-Goldstone (NG) modes
\cite{ng}
associated with field fluctuations
along the broken Lorentz generators
\cite{lvng}.
If the potential is smooth,
massive modes can also appear
\cite{lvmm}.
Some features of the NG and massive modes are generic,
while others are specific to details of the model
being considered.
In any case,
the nature of these modes plays a key role
in determining the physical implications 
of spontaneous Lorentz violation.

For the purposes of the present work,
the presence of NG modes is particularly crucial
because they can couple to matter 
and can transmit a long-range force.
A careful treatment of these modes is therefore 
a prerequisite for studies
of Lorentz violation in matter-gravity couplings.
In what follows,
we develop a methodology
to extract the dominant Lorentz-violating effects
in matter-gravity couplings
irrespective of the details of the underlying model
for spontaneous Lorentz violation. 
In effect,
the NG modes are treated via a perturbation scheme
that takes advantage of symmetry properties 
of the underlying Lagrange density 
to eliminate them
in favor of gravitational fluctuations
and background coefficients for Lorentz violation.
This treatment allows leading Lorentz-violating effects
from a large class of plausible models
to be handled in a single analysis.

The portion of this paper developing theoretical issues
spans Secs.\ \ref{theory}-\ref{bumblebee}.
It begins in Sec.\ \ref{theory}
with a review of the SME framework.
We present the field-theoretic action,
describe its linearization,
and discuss observability issues
for the coefficients for Lorentz violation.
We also describe the two notions of perturbative order
used in the subsequent analysis,
one involving Lorentz and gravitational fluctuations
and the other based on a post-newtonian expansion.
Section \ref{quantumthry}
concerns the relativistic and nonrelativistic quantum mechanics
arising from the quantum field theory.
One technical issue is extracting a meaningful quantum theory
in the presence of gravitational fluctuations.
We resolve this issue via a judicious field redefinition,
which yields a hamiltonian that is hermitian
with respect to the usual scalar product for wave functions
and that reduces correctly to known limiting cases.
We construct the relativistic quantum hamiltonian
at leading order in Lorentz violation and gravity fluctuations.
For the spin-independent terms,
we perform a Foldy-Wouthuysen transformation
to obtain the nonrelativistic hamiltonian. 
Section \ref{classicaltheory}
treats the classical dynamics corresponding
to the quantum theory.
The point particle action is presented
and the structure of test and source bodies is discussed.
The equations of motion for a test particle
and the modified Einstein equations are derived.
We describe the methodology for handling
coefficient and metric fluctuations.
Combining the results determines 
the Lorentz-violating trajectory of a test body.
The results are illustrated in 
Sec.\ \ref{bumblebee}
in the context of a special class of models
of spontaneous Lorentz violation
known as bumblebee models.
 
The remaining research sections of the paper,
Secs.\ \ref{experiment}-\ref{photon}, 
concern implications of our theoretical analysis
for experiments and observations.
Section \ref{experiment}
contains basic facts concerning frame choices 
and outlines the canonical Sun-centered frame
used for reporting measurements.
We also consider the sensitivities
to coefficients for Lorentz violation
that can be attained in practical situations.
Section \ref{labtests}
treats laboratory tests near the surface of the Earth
using neutral bulk matter,
neutral atoms,
or neutrons.
The theoretical description of these tests
is presented to third post-newtonian order,
and some generic features of the test-body motion  
are discussed.
A wide variety of gravimeter and equivalence-principle tests
is analyzed for sensitivities
to coefficients for Lorentz violation
that are presently unconstrained.
Satellite-based searches for Lorentz violation
using equivalence-principle experiments 
are studied in Sec.\ \ref{space}.
A generic situation is analyzed,
and the results are applied to major proposed satellite tests. 
Section \ref{exotic}
treats gravitational searches using
charged particles, 
antihydrogen,
and particles from the second and third generations of the SM.
Estimated sensitivities in future tests are provided,
and illustrative toy models for antihydrogen studies
are discussed.
Searches for Lorentz violation using
solar-system observations
are described in Sec.\ \ref{Solar-system tests}.
We consider measurements of coefficients for Lorentz violation
accessible via lunar and satellite ranging
and via studies of perihelion precession.
Section \ref{photon}
addresses various tests involving 
the effects of gravitational Lorentz violation
on the properties of light.
We analyze
the Shapiro time delay,
the gravitational Doppler shift,
and the gravitational redshift,
and we consider the implications 
for a variety of existing and proposed searches of these types.
Finally,
in Sec.\ \ref{summary}
we summarize the paper and tabulate
the various estimated actual and attainable sensitivities
to coefficients for Lorentz and CPT violation
obtained in the body of this work.

Throughout the paper,
we follow the conventions of 
Refs.\ \cite{akgrav} and \cite{qbakpn}. 
In particular,
the Minkowski metric 
is diagonal with entries $(-1, +1, +1, +1)$. 
Greek indices are used for spacetime coordinates,
while Latin indices are used for local Lorentz coordinates.
Appendix A of Ref.\ \cite{akgrav}
provides a summary of most other conventions.
Note that parentheses surrounding index pairs 
in the present work 
denote symmetrization with a factor of one-half.

\section{Framework}
\label{theory}

The focus of this work is 
the study of relativity violations
in realistic matter-gravity interactions.
The basic field theory of relevance  
concerns a single fermion field $\ps$
coupled to dynamical gravity
and incorporating Lorentz and CPT violation.
In this section,
we summarize the action for the model,
describe the linearization procedure,
discuss conditions for the observability of effects, 
and present the perturbation scheme 
developed for the analysis to follow.

\subsection{Action}
\label{Action}

The theory of interest is a special case 
of the gravitationally coupled SME
\cite{akgrav}.
The action can be written as
\beq
S = S_G + S_\ps + S^\prime.
\label{SMEaction}
\eeq 
The first term in this expression 
is the action $S_G$ containing the dynamics
of the gravitational field,
including any coefficients for Lorentz violation
in that sector. 
The geometric framework is a Riemann-Cartan spacetime,
which allows both
the Riemann curvature tensor $R^\ka_{\pt{\ka} \la \mu \nu}$
and the torsion tensor $T^\la_{\pt{\la} \mu \nu}$.
To incorporate fermion-gravity interactions,
the vierbein formalism
\cite{uk} 
is adopted,
with the vierbein $\ivb \mu a$
and the spin connection $\nsc \mu a b$
taken as the fundamental gravitational objects.
In the limit of zero torsion and Lorentz invariance,
$S_G$ reduces to the Einstein-Hilbert action
of General Relativity,
\bea
S_G &\to&
\fr 1 {2\ka}\int d^4 x ~(eR -2e\La),
\label{Ract}
\eea
where 
$\ka \equiv 8\pi G_N$,
$e$ is the vierbein determinant,
and $\La$ is the cosmological constant.

The second term in Eq.\ \rf{SMEaction}
is the action $S_\ps$ for the
fermion sector of the SME.
In this work,
we limit attention to terms in this sector
with no more than one derivative,
which is the gravitationally coupled analogue 
of the minimal SME in Minkowski spacetime. 
In this limit,
the action $S_\ps$ for
a single Dirac fermion $\ps$ of mass $m$
can be written as 
\beq
S_\ps = 
\int d^4 x (\half i e \ivb \mu a \ol \ps \Ga^a \lrDmu \ps 
- e \ol \ps M \ps).
\label{qedxps}
\eeq
In the present context,
the action of the covariant derivative $D_\mu$ on $\ps$ is 
\beq
D_\mu \ps \equiv 
\prt_\mu \ps 
+ \frac 14 i \nsc \mu a b \si_{ab} \ps.
\label{covderivqed}
\eeq
It is convenient to introduce the symbol  
\beq
(\ol \ps \ol D_\mu)
\equiv \prt_\mu \ol \ps 
- \frac 14 i \nsc \mu a b \ol \ps \si_{ab}
\label{conjcovderivqed}
\eeq
for the action of the covariant derivative 
on the Dirac-conjugate field $\ol\ps$.
The action \rf{qedxps}
contains the covariant derivative 
in a combination defined by
\beq
\ol\ch \Ga^a \lrDmu \ps
\equiv
\ol\ch \Ga^a D_\mu \ps
- (\ol\ch \ol D_\mu)\Ga^a\ps .
\label{lrDdef}
\eeq

The symbols $\Ga^a$ and $M$ 
appearing in the action \rf{qedxps} are defined by 
\bea
\Ga^a
&\equiv & 
\ga^a - c_{\mu\nu} \uvb \nu a \ivb \mu b \ga^b
- d_{\mu\nu} \uvb \nu a \ivb \mu b \ga_5 \ga^b
\nonumber\\
&&
- e_\mu \uvb \mu a 
- i f_\mu \uvb \mu a \ga_5 
- \half g_{\la\mu\nu} \uvb \nu a \ivb \la b \ivb \mu c \si^{bc} 
\label{gamdef}
\eea
and
\beq
M
\equiv 
m + a_\mu \ivb \mu a \ga^a 
+ b_\mu \ivb \mu a \ga_5 \ga^a 
+ \half H_{\mu\nu} \ivb \mu a \ivb \nu b \si^{ab} .
\label{mdef}
\eeq
The first term of Eq.\ \rf{gamdef}
leads to the usual Lorentz-invariant kinetic term 
for the Dirac field,
while the first term of Eq.\ \rf{mdef}
leads to a Lorentz-invariant mass.
A term of the form $i m_5 \ga_5$ could also appear in $M$,
but here we suppose it is absorbed into $m$ 
via a chiral field redefinition.
The coefficient fields for Lorentz violation
\a, \b, \c, \d, \e, \f, \g, \H\ 
typically vary with spacetime position.
The coefficient field \H\ is antisymmetric,
while \g\ is antisymmetric in $\la\mu$.
Note the use of an uppercase letter for \H,
which avoids confusion with the metric fluctuation $h_{\mu\nu}$.
The CPT-odd operators for Lorentz violation
are associated with the coefficient fields
\a, \b, \e, \f, and \g.

The form of the action \rf{qedxps}
implies the torsion $T^\la_{\pt{\la} \mu \nu}$
enters the fermion action only via minimal coupling. 
This coupling has the same form as 
that of the coefficient field $b_\mu$,
so the effects of minimal torsion 
can be incorporated into a matter-sector analysis
by replacing $b_\mu$ with the effective coefficient field 
\beq
(b_{\rm eff})_\mu \equiv 
b_\mu + \frac 1 8 T^{\al\be\ga}\ep_{\al\be\ga\mu}.
\label{beff}
\eeq 
Note that nonminimal torsion couplings can be incorporated
into the more general coefficient fields 
appearing in the full SME. 
Nonminimal torsion couplings and their experimental constraints
are discussed in Ref.\ \cite{krt}.

The final term in Eq.\ \rf{SMEaction}
is the action $S^\prime$ containing the dynamics 
associated with the coefficient fields for Lorentz violation.
Addressing possible contributions from this sector
is the subject of Sec.\ \ref{fluct}.

\subsection{Linearization}
\label{Linearization}

For the purposes of this work,
it suffices to consider weak gravitational fields
in a Minkowski-spacetime background.
Under these circumstances,
the Latin local indices 
can be replaced with Greek spacetime indices,
so the weak-field forms of the metric, 
vierbein, 
and spin connection can be written as 
\bea
g_\mn &=& \et_\mn + h_\mn,
\nonumber\\
\lvb \mu \nu 
&=&
\et_{\nu\si} \vb \mu \si
\approx ~
\et_{\mu\nu} + \half h_{\mu\nu} + \ch_{\mu\nu} ,
\nonumber \\
\lsc \la \mu \nu 
&=&
\et_{\mu\rh} \et_{\nu\si} \nsc \la \rh\si 
\nonumber\\
&\approx &
\prt_\la \ch_{\mn}
-\half \prt_\mu h_{\la\nu} 
+\half \prt_\nu h_{\la\mu} 
\nonumber \\
& &
+ \half \left( 
T_{\la \mu \nu} + T_{\mu \la \nu} - T_{\nu \la \mu} 
\right).
\label{weakgrav}
\eea
The quantities $\ch_\mn$ 
contain the six Lorentz degrees of freedom
in the vierbein.

The coefficient fields for Lorentz violation
are expected to acquire vacuum values
through spontaneous Lorentz breaking.
An arbitrary coefficient field $\X$ 
can therefore be expanded about its vacuum value $\Xb$,
\bea
\X &=& \Xb + \Xtw,
\label{texpand}
\eea
where the fluctuation $\Xtw$ 
includes massless NG modes and massive modes
\cite{lvng,lvmm}.
The vacuum value $\Xb\equiv\vev{\X}$ 
is called the coefficient for Lorentz violation.
One can instead choose to expand 
the contravariant coefficient field $\Xupper$,
\bea
\Xupper &=& \Xbupper + \Xtildew,
\eea
where $\Xbupper$ is related to $\Xb$ 
by raising with $\et^{\mu\nu}$.
The reader is cautioned that the relation 
between $\Xtw$ and the index-lowered version 
$\Xtildewl$ of $\Xtildew$ 
involves terms containing contractions of $\Xb$ with $h_\mn$.
This paper uses the expansion \rf{texpand}
and gives expressions in terms of $\Xtw$.
 
To provide a smooth match between our analysis 
and previous work on the matter sector
of the SME in Minkowski spacetime
and on the gravitational sector in
asymptotically Minkowski spacetime, 
we make two assumptions
about the coefficients for Lorentz violation.
First,
we assume they are constant
in asymptotically inertial cartesian coordinates,
\beq
\prt_\al \Xb= 0.
\label{assumption}
\eeq
This preserves translation invariance
and hence energy-momentum conservation
in the asymptotically Minkowski regime.
It also ensures that our barred coefficients 
correspond to the usual coefficients 
for Lorentz and CPT violation
investigated in the minimal SME in Minkowski spacetime
\cite{ck}.
Second,
we assume that the vacuum values $\Xb$
are sufficiently small to be treated perturbatively.
This is standard and plausible,
since any Lorentz violation in nature is expected to be small. 
These two assumptions suffice 
for most of the analysis that follows.
To obtain the leading Lorentz-violating corrections to $h_\mn$ 
without specifying a dynamical model
for the coefficient fields for Lorentz violation,
one further assumption is required,
which is presented in Sec.\ \ref{fluct}.

\subsection{Observability}
\label{Observability}

A given coefficient for Lorentz violation
can lead to observable effects
only if it cannot be eliminated from the Lagrange density
via field redefinitions or coordinate choices 
\cite{akgrav,ck,mbak,km,qbak04,cm,ba-f,rl-b,km-nr}.
In this subsection,
we outline some implications of this fact
relevant to the present work. 

\subsubsection{Field redefinitions}
\label{Field redefinitions}

One result of key interest here is that 
matter-gravity couplings 
can obstruct the removal of some coefficients 
that are unphysical in the Minkowski-spacetime limit.
For example, 
in the single-fermion theory in Minkowski spacetime,
the coefficient $a_\mu\equiv\ab_\mu$ 
for Lorentz and CPT violation 
in Eq.\ \rf{mdef} is unobservable 
because it can be eliminated by the spinor redefinition
\beq
\ps (x) \to {\rm exp} [if(x)] \ps (x)
\label{redefa}
\eeq
with $f(x) = \ab_\mu x^\mu$.
However,
in Riemann or Riemann-Cartan spacetimes
we have $a_\mu \equiv \ab_\mu +\atw_\mu$,
so this redefinition typically leaves
the four components of the fluctuation $\atw_\mu$
in the theory.
Instead,
the redefinition \rf{redefa} with an appropriate $f(x)$
can be used to move one component of 
the coefficient field $a_\mu$
into the other three,
unless $a_\mu$ is constant 
or the total derivative of a scalar 
\cite{akgrav}.
Note that in the presence of gravity
this freedom may be insufficient to eliminate any components
of the coefficient $\ab_\mu$
because the components of the fluctuation $\atw_\mu$
can depend on all four components of $\ab_\mu$
through the equations of motion.
This line of reasoning shows that gravitational couplings 
provide a unique sensitivity to the coefficient $\ab_\mu$,
which can be exploited in various experiments
\cite{akjt}.

Another type of field redefinition can be written 
in the generic form
\cite{akgrav}
\beq
\ps(x) \to [1 + v(x)\cdot\Ga] \ps(x),
\label{redef2}
\eeq
where $\Ga$ represents one of 
$\ga^a$, $\ga_5\ga^a$, $\si^{ab}$
and $v(x)$ is a complex function
carrying the appropriate local Lorentz indices.
This can be viewed as a position-dependent
component mixing in spinor space.
Field redefinitions of this type can be used
to demonstrate the leading-order equivalence 
of observable physical effects
due to certain coefficients for Lorentz violation.
An example relevant in the present context 
is a redefinition involving a real vector $v_a (x)$.
Together with assumption \rf{assumption},
this redefinition can be used to 
show that at leading order in Lorentz violation
the coefficients $a_\mu$ and $e_\mu$
always appear in the combination
\beq
\af_\mu \equiv a_\mu - m e_\mu,
\label{aeff}
\eeq
up to derivatives of fluctuations.
Combining this result with the above discussion
of the redefinition \rf{redefa}
shows that observables involving gravitational couplings 
offer the prospect of measuring $\af_\mu$.

Related ideas can be used to simplify 
the weak-field limit of the theory \rf{qedxps}.
In particular,
the antisymmetric part $\ch_\mn$ of the vierbein
can be removed everywhere
except for possible contributions to 
fluctuations of the coefficient fields, 
by applying the field redefinition
\bea
\ps(x) &\to& 
\exp[ - \quar i \ch_\mn(x) \si^\mn]\ps(x) 
\nonumber\\
&&
\approx
\left( 1 - \quar i \ch_\mn \si^\mn 
- \frac{1}{32} \ch_\mn \ch_{\al \be} \si^{\mn} \si^{\al \be} 
\right) \ps(x).
\nonumber\\
\label{redefch}
\eea
Note that this redefinition takes the form 
of a Lorentz transformation on $\ps$
but that the other fields in the Lagrange density
remain unaffected.
Note also that in the absence of Lorentz violation
$\ch_\mn$ can be removed entirely,
a fact compatible with the interpretation 
in Ref.\ \cite{lvng}
of the role of $\ch_\mn$ in Lorentz-violating theories.
In the remainder of this work,
we assume the redefinition \rf{redefch}
has been performed on the Lagrange density,
so that quantities such as
$\Ga^a$, $M$, and $\ivb{\mu}{a}$
are understood to acquire no contributions from $\ch_\mn$
except possibly through the fluctuations
of the coefficient fields for Lorentz violation.

\subsubsection{Coordinate choices}
\label{Coordinate choices}

The observability
of certain combinations of coefficients for Lorentz violation
is also affected by the freedom to make coordinate choices.
Intuitively,
the key point is that any one sector of the SME
can be used to define the scales of the four coordinates,
to establish the meaning of isotropy, 
and to set the synchronization scheme.
The freedom therefore exists to choose the sector
in which the effective background spacetime metric
takes the form of the Minkowski metric $\et_\mn$.
This implies that in any experimental configuration
there are always 10 unobservable combinations 
of coefficients for Lorentz violation.

As an illustration,
consider the SME restricted
to the single-fermion and photon sectors
\cite{ck,qbak04}.
In the fermion sector,
the 10 relevant coefficient components 
are the vacuum values $\cb_\mn$ 
of the coefficient fields $c_\mn$ in Eq.\ \rf{gamdef}
because these coefficients enter $S_\ps$
in the same way as the metric.
In the photon sector,
the SME Lagrange density contains a term 
\beq
\cL_{\rm photon}\supset
-\frac 14 e (k_{F})_{\ka \la \mn} F^{\ka \la} F^\mn,
\eeq
and the 10 relevant coefficient components
can be shown to be the trace $\kfb^\al_{\pt{\al} \mu \al \nu}$.
At leading order,
the coordinate transformation
\beq
x^\mu \rightarrow x^{\mu^\prime} =
x^\mu - \half \kfb^{\al \mu}_{\pt{\al \mu} \al \nu} x^\nu
\label{ktoctransf}
\eeq
redefines the background metric to take the form $\et_\mn$
in the photon sector.
The effective metric in the fermion sector is then also changed,
with the observable coefficient combination becoming
$\cb_\mn + \kfb^\al_{\pt{\al} \mu \al \nu}/2$.
The orthogonal combination 
$2\cb_\mn - \kfb^\al_{\pt{\al} \mu \al \nu}$
is thus unobservable in any experiments 
involving only these two sectors.
Alternatively,
one could perform the coordinate transformation
\beq
x^\mu \rightarrow x^{\mu^\prime} = 
x^\mu + \cb^\mu_{\pt{\al} \nu} x^\nu,
\label{ctoktransf}
\eeq
which instead redefines the fermion-sector background metric 
to be $\et_\mn$
and at leading order produces 
the effective photon-sector coefficient 
$2\cb_\mn + \kfb^\al_{\pt{\al} \mu \al \nu}$.
This coordinate choice is equally valid for analysis,
and as before the orthogonal combination 
$2\cb_\mn - \kfb^\al_{\pt{\al} \mu \al \nu}$
is unobservable.

Similar results apply for experimental searches
for Lorentz violation involving comparisons 
of different fermion species.
Labeling the species by $w$,
each has a coefficient $\cb^w_\mn$.
Then,
for example,
the effective metric for any one species $X$ 
can be reduced to $\et_\mn$
by a coordinate transformation with $\cb^X_\mn$
of the form \rf{ctoktransf}.
The resulting effective coefficients for the remaining species 
involve the differences $\cb^w_\mn - \cb^X_\mn$,
which in this coordinate scheme 
become the relevant observable combinations of coefficients.

In the gravity sector,
the situation is more involved
because a geometrically consistent treatment of Lorentz violation
generically requires the incorporation of effects from the NG modes
to ensure the Bianchi identities are satisfied
\cite{akgrav}.
It turns out that the 10 relevant coefficient components
in the gravity sector are the vacuum values $\sb_\mn$
of the coefficient fields $s_\mn$
in the gravity-sector SME Lagrange density
\beq
\cL_{\rm gravity}\supset \fr{1}{16\pi G_N} e s^{\mn} R_{\mn}.
\label{sterm}
\eeq
We find that at leading order the transformation
\rf{ktoctransf}
generates an accompanying shift
$\sb_\mn \to \sb_\mn-\kfb^\al_{\pt{\al} \mu \al \nu}$,
while the transformation
\rf{ctoktransf}
produces the shift
$\sb_\mn \to \sb_\mn + 2 \cb_\mn$.

One way to obtain these results 
is to consider the leading-order effect 
of a metric shift on the equations of motion,
and then to match to the known results
\cite{qbakpn} 
for observable effects in a post-newtonian expansion.
Consider,
for example,
the restriction of the SME to the
Einstein-Hilbert action and the single-fermion action $S_\ps$ 
with nonzero $\cb_\mn$ only.
At leading order,
the coordinate transformation
\rf{ctoktransf}
removes $\cb_\mn$ from the fermion action
at the cost of introducing 
a metric shift $g_\mn \to g_\mn - 2 \cb_\mn$ 
in the Einstein-Hilbert term.
The resulting equations of motion
involve the Einstein tensor $G_\mn (g - 2 \cb)$
with shifted argument,
which can be written in terms of 
the Einstein tensor $G_\mn (g)$ for the original metric 
and an effective energy-momentum tensor $\ph^{\cb}_\mn$.
We find
\bea
G_\mn (g - 2 \cb) &=& 
G_\mn (g) - \ph^{\cb}_\mn,
\nonumber\\
\ph^{\cb}_\mn &\approx& 
2 ( 
\et_\mn \cb^{\al\be} R_{\al\be}
- 2 {\cb^\al}_{(\mu} R_{\nu)\al} 
+\half \cb_\mn R
\nonumber\\
&&
+\cb^{\al\be} R_{\al\mu\nu\be}
).
\eea
The trace-reversed form $\Ph^{\cb}_\mn$ of $\ph^{\cb}_\mn$
matches the post-newtonian term $\Ph^{\sb}_\mn$
arising from Eq.\ \rf{sterm}
and given explicitly as Eq.\ (24) of Ref.\ \cite{qbakpn}
with the combination $\sb_\mn$ replaced by $2\cb_\mn$.
The transformation
\rf{ctoktransf}
therefore produces the shift
$\sb_\mn \to \sb_\mn + 2 \cb_\mn$,
as claimed.
A similar line of reasoning verifies
the claimed shift
$\sb_\mn \to \sb_\mn-\kfb^\al_{\pt{\al} \mu \al \nu}$
for the coordinate transformation \rf{ktoctransf}.

To keep expressions compact throughout this work,
we choose to work with coordinates satisfying 
\beq
\kfb^\al_{\pt{\al} \mu \al \nu}=0.
\label{coordchoice}
\eeq
To obtain results valid for arbitrary coordinate choices,
the following substitutions can be applied throughout:
\bea
\cb^w_\mn &\rightarrow&
\cb^w_\mn + \half \kfb^\al_{\pt{\al} \mu \al \nu} ,
\nonumber
\\
\sb_\mn &\rightarrow& \sb_\mn-\kfb^\al_{\pt{\al} \mu \al \nu}.
\label{krename}
\eea
We emphasize that all coordinate choices are equivalent
for theoretical work or for data analysis,
with the choice \rf{coordchoice} adopted here 
being purely one of convenience.

\subsection{Perturbation scheme}
\label{Perturbation scheme}

In this work,
we are interested
in experimental searches for Lorentz violation 
involving gravitational effects on matter.
Many of these searches involve test particles 
moving in background solutions
to the equations of motion 
for gravity and for the coefficient fields.
Since the gravitational fields involved are weak
and since no compelling evidence for Lorentz violation 
exists to date,
any effects are expected to be small.
We can therefore focus attention on perturbative modifications 
to the behavior of test particles.
This subsection describes the scheme we use
to track perturbative orders 
in the construction of the relativistic quantum hamiltonian
and in the subsequent developments.

Perturbative effects on physical observables can arise
through modifications to 
the background coefficient fields $\X$ for Lorentz violation
and to 
the background gravitational field $g_\mn$,
or directly through modifications 
to the equation of motion for the test particle.
The analysis of these effects is simplified 
by introducing an appropriate notion of perturbative order.
Several ordering schemes are possible.
In this work,
we adopt a scheme that tracks the orders 
in the coefficients for Lorentz violation $\Xb$
and in the metric fluctuation $h_\mn$.
The overall perturbative order 
of a given term in an equation is denoted as O($m$,$n$),
where $m$ represents the order in $\Xb$
and $n$ the order in $h_\mn$.
Within this scheme,
the fluctuations $\Xtw$
of the coefficient fields for Lorentz violation
are viewed as secondary quantities
that are determined via their equations of motion
in terms of 
the coefficients for Lorentz violation
and the gravitational field.
There is also a subsidiary notion of order
associated with the usual post-newtonian expansion 
of $h_\mn$ itself.
We denote a $p$th-order term in this latter expansion as \pno{$p$}.
However,
performing an explicit post-newtonian expansion at the initial stage 
would complicate the ensuing analysis,
so in what follows we often write results 
in terms of $h_\mn$
while commenting as needed on the post-newtonian counting.

To preserve a reasonable scope,
this work focuses on dominant perturbative effects 
involving both Lorentz violation and gravity. 
We next discuss the relevant perturbative orders
required to achieve this goal.

Consider first contributions from 
the fluctuation $\Xtw$ of the coefficient fields.
The detailed structure of $\Xtw$ depends on the nature
of the action $S^\prime$ in Eq.\ \rf {SMEaction}.
In the scheme adopted here,
$\Xtw$ can be viewed as a series in $\Xb$ and $h_\mn$ 
of the form
\bea
\Xtw &=& \Xtwo 00 + \Xtwo 01 + \Xtwo 10 
+ \Xtwo 11 + \ldots . \qquad
\label{tscrexp}
\eea
For spontaneous Lorentz breaking,
$\Xtw$ includes massive modes and massless NG modes
\cite{lvmm,adp}.  
In this work,
we suppose the massive modes either are frozen 
or have negligible degree of excitation.
Incorporating their possible effects 
into the analysis of matter-gravity phenomena 
would be of potential interest 
but lies beyond our present scope.
In contrast,
the massless NG modes play a key role in what follows.
Their fate can include 
identification with the photon in Einstein-Maxwell theory
\cite{lvng},
with the graviton in GR
\cite{cardinal,ctw},
or with a new force
\cite{akjt,ahclt,abk},
or they can be absorbed in the torsion 
via the Lorentz-Higgs effect
\cite{lvng}.
In some models the NG modes
can be interpreted as composite photons 
\cite{photons}
or gravitons 
\cite{gravitons}.
We consider below the perturbative orders required 
for the various possibilities.

Suppose first the NG modes correspond to photons,
or more generally to a known force field other than gravity.
The term $\Xtwo 00$ then contains 
conventional Lorentz-invariant terms
describing this field in Minkowski spacetime,
while $\Xtwo 01$ contains conventional 
leading-order gravitational interactions with the field.
Effects from both these terms are therefore
part of the conventional description of the force.
The term $\Xtwo 10$ describes possible Lorentz violations
in Minkowski spacetime involving the known field,
many of which are tightly constrained by experiments
\cite{tables}.
For the purposes of this work,
which focuses on Lorentz violation involving gravity,
we can take this term as experimentally negligible.
The dominant term of interest is therefore
$\Xtwo 11$,
which lies at O($1$,$1$).

If the NG modes correspond to gravitons as,
for example,
in the cardinal model
\cite{cardinal},
then the expansion \rf{tscrexp}
contains no terms at O($m$,0).
The term $\Xtwo 01$ corresponds to 
the gravitational fluctuations $h_\mn$,
and its effects are part of the conventional description
of gravity.
The dominant term of interest is therefore
again $\Xtwo 11$.

If instead the NG modes correspond 
to a presently unobserved force field,
then $\Xtwo 00$ and $\Xtwo 01$ 
describe unobserved Lorentz-invariant effects
in Minkowski spacetime 
and in leading-order gravitational couplings.
These modes must therefore be eliminated from the analysis 
prior to interpretation of observations,
via solving the equations of motion or otherwise.
In what follows,
we suppose this elimination has been performed where needed.
The term $\Xtwo 10$ describes possible Lorentz violation 
in Minkowski-spacetime involving the unknown field.
For present purposes,
we take this term to be experimentally negligible,
although in principle it might offer
a novel way to access certain types of 
presently unobserved interactions at exceptional sensitivities.
The remaining term $\Xtwo 11$
displayed in Eq.\ \rf{tscrexp}
describes the dominant Lorentz-violating gravitational effects 
involving the unknown interaction.
As discussed in Sec.\ \ref{Field redefinitions},
certain Lorentz-violating effects 
are observable only in the presence of gravity,
and so under suitable circumstances
observable experimental signals from $\Xtwo 11$
could arise \cite{akjt}
despite the tight existing experimental constraints 
\cite{scwga}
on the direct observation of additional interactions 
due to the lower-order terms $\Xtwo 00$ and $\Xtwo 01$. 
In scenarios with an unobserved force,
we must therefore also allow for O($1$,$1$) effects
involving $\Xtwo 11$. 

The remaining possibility is that NG modes 
are absorbed into the torsion.
To handle this case,
note that the matter-sector role of minimal torsion
can be treated in parallel with  
the coefficient field $b_\mu$ for Lorentz violation 
according to Eq.\ \rf{beff}.
Existing experimental constraints on minimal torsion components 
are tight, 
lying below $10^{-27}$-$10^{-31}$ GeV
\cite{krt,is,haccss},
so effects involving minimal torsion can be treated
in the same way as those corresponding
to a presently unobserved force field.
We can therefore limit attention to O(1,1) effects as before.
In principle,
any nonminimal torsion components can be treated 
in a similar fashion 
because they play a role analogous 
to coefficient fields in the nonminimal SME,
but effects of this type lie beyond our present scope.

Consider next the metric fluctuation $h_\mn$.
For applications to gravitational tests with matter,
$h_\mn$ can be treated as a background field
obtained by solving the appropriate equation of motion,
which is the modified Einstein equation
in the presence of Lorentz violation.
It can therefore be viewed as the sum
of a Lorentz-invariant piece $h^{(0,1)}_\mn$
with a series of corrections 
of increasing perturbative order in $\Xb$,
\beq
h_\mn= h^{(0,1)}_\mn + \hlv_\mn + h^{(2,1)}_\mn + \dots .
\label{hexp}
\eeq
When we specify the perturbative order 
of an expression containing $h_\mn$
in what follows,
it is understood that the correct terms 
from the above series are included.

For a given expression,
establishing the relevant perturbative order for our analysis
typically involves a combination of experimental restrictions
and theoretical considerations.
As an illustration,
we outline here the reasoning establishing 
the appropriate perturbative orders
in the construction of the relativistic quantum hamiltonian.
First,
note that terms quadratic in $h_\mn$ 
involve \pno{4} and higher.
Since the sensitivity of current laboratory and solar-system tests
lies at the \pno{4} level,
we must keep these terms
but can discard terms cubic in $h_\mn$
and ones involving the product of coefficients for Lorentz violation 
with terms quadratic in $h_\mn$.
The Lorentz-invariant part of the hamiltonian
can therefore be truncated at O(0,2),
while the Lorentz-violating part
can be truncated at O(1,1). 
To maintain consistent post-newtonian counting,
the O(0,2) terms must be limited to \pno{5},
while the O(1,1) terms are limited to \pno{3}.
Next,
note that for laboratory and solar-system tests
the variations in $h_\mn$
over the experimental scale $L$
are small compared to $h_\mn$,
$|\prt_\al h_\mn| \ll |h_\mn/L|$.
For example,
the typical value of the gravitational acceleration
on the surface of the Earth is $g \simeq 10^{-32}$ GeV,
which is tiny compared to the ratio
of the gravitational potential $|h_\mn| \simeq 10^{-9}$
and the size of a typical laboratory experiment
$L \simeq 10^{15}$ GeV$^{-1}$.
Terms in the relativistic hamiltonian
proportional to derivatives of $h_\mn$
can therefore be limited to O(0,1).
Finally,
note that products of Lorentz-violating terms
lead to higher-order effects with operator structures
matching ones already appearing in the fermion sector 
of the Minkowski-spacetime SME.
These are already accessible in nongravitational experiments. 
It therefore suffices for our purpose to restrict
attention to terms at leading order in Lorentz violation.
To summarize,
the construction
of the perturbative relativistic quantum-mechanical hamiltonian
can be limited to terms at perturbative orders 
O(0,1), O(1,0), O(1,1), and O(0,2),
except for terms involving derivatives of the gravitational fields,
which can be limited to O(0,1).

\section{Quantum theory}
\label{quantumthry}

This section studies the quantum mechanics
associated with the fermion action $S_\ps$ in Eq.\ \rf{qedxps}.
We begin in Sec.\ \ref{Time dependence}
by addressing the issue 
of the unconventional time dependence 
arising from the Dirac equation derived from $S_\ps$.
The relativistic quantum-mechanical hamiltonian $H$
is then obtained in Sec.\ \ref{relativham},
and the relevant parts of the nonrelativistic limit $H_{\rm NR}$
are extracted in Sec.\ \ref{quantummech}.

\subsection{Time dependence}
\label{Time dependence}

In the weak-field limit,
the Lagrange density $\cL_\ps$ for the action \rf{qedxps}
takes the schematic form
\beq
\cL = \half i [
\psb(\ga^\mu + C^\mu) \prt_\mu \ps 
- (\prt_\mu \psb)(\ga^\mu + C^\mu) \ps ] 
- \psb D \ps,
\label{lagden}
\eeq
where $C^\mu$ and $D$ represent spacetime-dependent operators
without derivatives acting on $\ps$.
These operators satisfy the conditions
\beq
(\ga^0 C^\mu)^\dag = \ga^0 C^\mu,
\qquad
(\ga^0 D)^\dag = \ga^0 D,
\eeq
and $C^\mu$ is perturbative.

The Euler-Lagrange equations obtained from Eq.\ \rf{lagden}
yield a Dirac equation
with unconventional time dependence,
\beq
i (\ga^0 + C^0) \prt_0 \ps = 
[- i (\ga^j + C^j) \prt_j - \half i \prt_\mu C^\mu + D] \ps.
\label{dirac}
\eeq
This equation differs from the standard Dirac form 
by the presence of $C^0$,
which impedes the interpretation
of the operator acting on $\ps$
on the right-hand side 
as the hamiltonian.
In this subsection,
two approaches addressing this issue 
at first order in $C^\mu$ are discussed.

\subsubsection{Field-redefinition method}
\label{Field-redefinition method}

One method for constructing the hamiltonian 
has been developed in the context of
the SME in Minkowski spacetime
\cite{bkr,kl,kleh}.
It uses an appropriate field redefinition
at the level of the action
to ensure the Dirac equation 
emerges with conventional time dependence.
In typical applications,
the field redefinition is defined perturbatively
at the desired order in $C^\mu$.

For present purposes,
it suffices to work at first order in $C^\mu$.
The appropriate field redefinition is
\bea
\ps &=& A\ch,
\qquad
A \equiv 1 - \half \ga^0 C^0.
\label{chfield}
\eea
The resulting hamiltonian can be written as
\beq
H_\ch = H^{(0)} + H^{(1)}_\ch,
\label{redefham}
\eeq
where $H^{(0)}$ is the hamiltonian
in the absence of $C^\mu$,
given by
\beq
H^{(0)} = - i \ga^0 \ga^j \prt_j + \ga^0 D.
\label{unperth}
\eeq
The correction $H^{(1)}_\ch$
is first order in $C^\mu$
and takes the form
\bea
H^{(1)}_\ch &=& 
- i \ga^0 (C^j - \half C^0 \ga^0 \ga^j 
+ \half \ga^0 \ga^j C^0)\prt_j
\nonumber \\ && 
- \half i (\ga^j \prt_j C^0 + \ga^0 \prt_j C^j)
\nonumber \\ && 
- \half \ga^0 (C^0 \ga^0 D + D \ga^0 C^0).
\eea
The subscript $\ch$ serves as a reminder
that the operator acts on the spinor $\ch$.
Note that the hamiltonian $H_\ch$ is hermitian 
with respect to the usual scalar product
in flat space,
\beq
\langle\ch_1, \ch_2\rangle_{\rm f} = \int d^3 x ~\ch_1^\dag \ch_2.
\eeq
This implies,
for example,
that energies can be calculated in the usual way.

Some physical insight into the field-redefinition method
is obtained by noting that 
the combination $\ga^\mu + C^\mu$ takes the generic form
\beq
\ga^\mu + C^\mu = E^\mu_{\pt{\mu} a} \ga^a,
\eeq
where $E^\mu_{\pt{\mu} a}$ 
can be interpreted as an effective inverse vierbein.
It reduces to the conventional inverse vierbein $\ivb{\mu}{a}$ 
in the purely gravitational case
but includes coefficients for Lorentz violation
when Lorentz symmetry is broken.
This suggests that the field-redefinition method
can be interpreted as transforming 
the problematic situation of a fermion on an effective manifold 
with vierbein components $E^0_{\pt{\mu} a}$
into the physically equivalent but tractable theory 
of a different fermion field
on a manifold with vierbein components 
$E^0_{\pt{\mu} a}=\de^0_{\pt{0} a}$,
in which the hamiltonian is hermitian 
with respect to a conventional scalar product.

\subsubsection{Parker method}
\label{Parker method}

Another method has been presented by Parker
\cite{lp}
in the context of field theory in curved spacetime.
It involves multiplying the Dirac equation
by a suitable factor that removes
the unconventional time dependence
to the desired order in $C^\mu$.
The resulting hamiltonian is hermitian  
with respect to a modified scalar product.

Applying this method at first order in $C^\mu$
requires left-multiplying both sides of Eq.\ \rf{dirac}
with $\ga^0 (1 - C^0 \ga^0)$.
The ensuing hamiltonian can be written as
\beq
\hat{H}_\ps = H^{(0)} + H^{(1)}_\ps,
\eeq
where $H^{(0)}$ is given in Eq.\ \rf{unperth}
and the first-order correction in $C^\mu$ is
\beq
H^{(1)}_\ps = 
- i \ga^0 C^j \prt_j 
+ i \ga^0 C^0 \ga^0 \ga^j \prt_j
- \half i \ga^0 \prt_\mu C^\mu 
- \ga^0 C^0 \ga^0 D.
\eeq
The subscript $\ps$ indicates an operator 
acting on the original spinor $\ps$.

In this method,
the modified Dirac equation
implies a modified continuity equation
and hence requires a modified scalar product.
At first order in $C^\mu$,
the continuity equation is
\beq
\prt_0 [\ps^\dag (1 + \ga^0 C^0) \ps] 
+ \prt_j [\ps^\dag \ga^0 (\ga^j+ C^j) \ps] = 0,
\eeq
and the probability density
can be identified as the combination
$\ps^\dag (1 + \ga^0 C^0) \ps$.
The corresponding scalar product is
\beq
\langle\ps_1,\ps_2\rangle_{\rm P} = 
\int d^3 x ~\ps_1^\dag (1 + \ga^0 C^0) \ps_2.
\label{curvedproduct}
\eeq
Provided $H^{(1)}_\ps$ is time independent,
the hamiltonian $\hat{H}_\ps$ is hermitian
with respect to this modified scalar product
and so quantum-mechanical calculations can proceed.
When $C^0$ is time dependent,
hermiticity with respect to the product \rf{curvedproduct} 
can be restored
by adding an extra term 
\cite{xhlp}.
We thereby obtain the hermitian hamiltonian 
\beq
H_\ps = \half i \ga^0 \prt_0 C^0 + \hat{H}_\ps
\label{pham}
\eeq
at first order in $C^0$.

\subsubsection{Comparison}
\label{Comparison}

The hamiltonians 
$H_\ch$ in Eq.\ \rf{redefham}
and $H_\ps$ in Eq.\ \rf{pham}
typically have different forms.
For our present purposes,
however,
they are physically equivalent
because they give rise
to the same eigenenergies at first order.

To demonstrate this,
first note that the difference $\De H$
between the two hamiltonians
can be written as
\beq
\Delta H =  
H^{(1)}_\ch - H^{(1)}_\ps = 
H^{(0)} A - A H^{(0)},
\eeq
where $A$ is given by Eq.\ \rf{chfield}.
This can be shown 
by manipulation of the field redefinition and the Dirac equation 
or verified by direct calculation.
The physical quantities of interest
are the eigenenergies.
These are obtained as perturbations $E^{(1)}_{\ch,\ps}$
to the unperturbed values $E^{(0)}$,
calculable at first order as 
\beq
E^{(1)}_{\ps,\ch} = 
\langle\ps^{(0)}, H^{(1)}_{\ps,\ch} \ps^{(0)}\rangle_{\rm f},
\eeq
where $\ps^{(0)}$ are solutions
to the unperturbed Schr\"odinger equation
$H^{(0)} \ps^{(0)}= E^{(0)} \ps^{(0)}$. 
However,
the expectation value of $\Delta H$ is zero
in this scalar product,
so the first-order perturbations to the energies 
are identical for both hamiltonians.

The above first-order result suffices for the present work,
although we anticipate equivalence 
also holds at higher orders in a complete analysis.
For our purposes,
the field-redefinition method proves 
technically and conceptually easier
because the hamiltonian is hermitian
with respect to the usual scalar product.
We therefore adopt the field-redefinition method 
in the remainder of this work,
and all references to $H$ implicitly refer to $H_\ch$.

\subsection{Relativistic hamiltonian}
\label{relativham}

The relativistic hamiltonian for the action 
$S_\ps$ in Eq.\ \rf{qedxps}
can be obtained via the field-redefinition method.
At the appropriate perturbative order,
we find the operator $A$ of Eq.\ \rf{chfield} 
takes the form
\bea
A &=& 1 - \half \ga^0 
\Big[ e \ivb{0}{a} \Ga^a
-\ga^0
- \frac{3}{4} e^{(0,1)}  e^{(0,1)} \ga^0
\nonumber \\
&& 
\pt{1 + \half \ga^0(}
- \frac{3}{4} e^{(0,1)} h_{0 \mu} \ga^\mu
- \frac{3}{16} h_{0 \mu} h_{0 \nu} \ga^\mu \ga^0 \ga^\nu
\nonumber \\
&& 
\pt{1 + \half \ga^0(}
- \frac{3}{2} e^{(0,1)} \Ga^0
- \frac{3}{8} h_{0 \mu} 
\left( \Ga^0 \ga^0 \ga^\mu + \ga^\mu \ga^0 \Ga^0 \right)
\Big]. 
\nonumber \\
\eea
Implementing the field redefinition
and varying the transformed Lagrange density
results in a Dirac equation from which 
the hamiltonian $H$ can be identified.

The hamiltonian $H$ can be split into pieces 
according to perturbative order,
\beq
H = H^{(0,0)} + H^{(0,1)} + H^{(1,0)} + H^{(1,1)} + H^{(0,2)}.
\label{relham}
\eeq
The component $H^{(0,0)}$ is the conventional hamiltonian
for the Lorentz-invariant Minkowski-spacetime limit 
of the theory. 
The first-order Lorentz-invariant piece is
\bea
H^{(0,1)} &=& 
\half i (h_{jk} + h_{00} \et_{jk}) \ga^0 \ga^k  \prt^j 
+ ih_{j0}\prt^j 
- \half m h_{00} \ga^0
\nonumber \\ 
&&  
+ \quar \prt^j h^{0k} \ep_{jkl} \ga_5 \ga^0 \ga^l
+ \half i \prt^j h_{j0} 
\nonumber \\
&& 
+ \quar i (\prt_j h_{00} + \prt^k h_{jk}) \ga^0 \ga^j .
\label{Hgravity}
\eea
It represents the first-order correction 
to the conventional hamiltonian $H^{(0,0)}$ 
arising from gravitational and inertial effects.

The first-order correction
to the conventional hamiltonian $H^{(0,0)}$ 
arising from Lorentz violation
can be written 
\bea
H^{(1,0)} &=& \ab_0 
- m\eb_0 
+ 2 i \cb_{(j0)} \prt^j 
- (m\cb_{00}  - i\eb_j\prt^j)\ga^0
\nonumber \\
&&
- \fb_j\prt^j\ga^0\ga_5
+ [\ab_j + i(\cb_{00}\et_{jk} + \cb_{jk})\prt^k] \ga^0 \ga^j 
\nonumber \\
&&
+ (-\bb_0 - 2 i \db_{(j0)} \prt^j)\ga_5
+ (i\Hb_{0j} + 2 \gb_{j(k0)} \prt^k)\ga^j
\nonumber \\
&& 
- [\bb_j + i(\db_{jk}\prt^k + \db_{00}\prt_j) 
- \half m \gb^{kl0} \ep_{jkl}] \ga_5 \ga^0 \ga^j
\nonumber \\
&& 
- (\half \Hb^{kl} \ep_{jkl} + m \db_{j0}) \ga_5 \ga^j 
\nonumber \\
&& 
- i \ep_{jlm} (\gb^{l00} \et^{km} 
+ \half \gb^{lmk})\prt_k \ga_5 \ga^j.
\label{Hlv}
\eea
This result matches the one
previously obtained for the SME in Minkowski spacetime
\cite{kl}
when the change in metric signature
is incorporated.
When minimal torsion is included in the analysis,
its background value enters Eq.\ \rf{Hlv}
through the replacement
$b_\mu \to (b_{\rm eff})_\mu$
specified by Eq.\ \rf{beff}. 
It can be constrained through a reinterpretation
of experiments searching for nonzero $\bb_\mu$ 
\cite{krt,is,haccss}.

An interesting issue is the extent to which
the gravitational and inertial effects 
in Eq.\ \rf{Hgravity}
mimic the Lorentz-violating effects
in Eq.\ \rf{Hlv}.
For example,
in a rotating frame of reference the term 
$\prt^j h^{0k} \ep_{jkl} \ga_5 \ga^0 \ga^l/4$
in Eq.\ \rf{Hgravity}
contains a coupling of the rotation 
to the spin of the particle
with the same operator structure
as the term 
$- \bb_j \ga_5 \ga^0 \ga^j$
in Eq.\ \rf{Hlv}.
At this order,
a frame rotation can therefore mimic potential signals 
arising from a nonzero coefficient $\bb_j$ 
for Lorentz and CPT violation.
This effect has been observed
in tests with a spin-polarized torsion pendulum 
\cite{hccass}.
The same term in Eq.\ \rf{Hgravity}
also contains gravitomagnetic effects
that are in principle observable in tests searching for $\bb_\mu$
if sufficient sensitivity is reached.
Certain Lorentz-violating effects can be
separated from gravitational and inertial effects 
because the former generate time-varying signals
due to the motion of the Earth
and can have flavor dependence,
but a complete separation may be problematic.

The O(1,1) contribution to $H$ can be separated as
\bea
H^{(1,1)} &=& 
H^{(1,1)}_h + H^{(1,1)}_a + H^{(1,1)}_b + H^{(1,1)}_c + H^{(1,1)}_d
\nonumber \\
&&
+ H^{(1,1)}_e + H^{(1,1)}_f + H^{(1,1)}_g + H^{(1,1)}_H. 
\eea
Here,
the term $H^{(1,1)}_h$ arises from Lorentz-violating corrections 
to the metric fluctuation $h_\mn$.
It is given by
\bea
H^{(1,1)}_h &=& 
i\hlv_{j0}\prt^j - \half m \hlv_{00} \ga^0
\nonumber \\
& & + \half i (\hlv_{jk} + \hlv_{00} \et_{jk}) \ga^0 \ga^k  \prt^j. 
\eea
The other terms in $H^{(1,1)}$ are labeled according to 
the type of coefficient for Lorentz violation involved.
The contributions involving the four-component coefficients
\a\ and \b\ of mass dimension one are
\bea
H^{(1,1)}_a &=& 
\atw_0 
- \ab^j h_{j0}
+ \big( \atw_j
- \half \ab_j h_{00}
- \half \ab^k h_{jk} \big) \ga^0 \ga^j
\nonumber \\
\eea
and
\bea
H^{(1,1)}_b &=& 
\big( - \btw_0 + \bb^j h_{j0} \big) \ga_5
\nonumber \\
&&
+ \big( \btw_j - \half \bb_j h_{00} 
- \half \bb^k h_{jk} \big) \ga^0 \ga_5 \ga^j,
\eea
respectively.
In the latter equation,
effects from minimal torsion are included
via the replacement $b_\mu \to (b_{\rm eff})_\mu$
given in Eq.\ \rf{beff}. 

The contributions involving
the dimensionless coefficients \c\ and \d\ are
\bea
H^{(1,1)}_c &=&
i \big[ 
\ctw_{00} \et_{jk}
+ \ctw_{kj}
+ \half \cb_{00} \left( h_{00} \et_{jk} - h_{jk} \right)
\nonumber \\
& & \pt{i \big(}
+ 2 \cb_{(l0)} h^{l0} \et_{jk}
+ \quar \cb_{k0} h_{j0}
- \quar \cb_{j0} h_{k0}
\nonumber \\
& & \pt{i \big(}
- \half \cb_{lj} \left( h_{00} \etul{l}{k} + \hul{l}{k} \right)
- \cb_{kl} \hul{l}{j} \big] \ga^0 \ga^k \prt^j
\nonumber \\
& &
- 2 i \big(
\ctw^{(j0)}
+ \cb_{(k0)} h^{jk}
+ \cb^{jk} h_{k0} \big) \prt_j
\nonumber \\
& &
- m \big(
\ctw_{00}
+ \half \cb_{00} h_{00}
+ 2 \cb_{(j0)} h^{j0} \big) \ga^0
\eea
and
\bea
H^{(1,1)}_d &=& 
2 i \big( 
\dtw^{(j0)} 
+ \db_{(k0)} h^{jk} 
+ \db^{(jk)} h_{k0} \big) \ga_5 \prt_j
\nonumber \\
& &
+ i \big(
\dtw_{00}
+ \half \db_{00} h_{00}
+ 2 \db_{(k0)} h^{k0} \big) \ga^0 \ga_5 \ga^j \prt_j
\nonumber \\
& &
+ i \big[
\dtw_{kj}
+ \quar \db_{k0} h_{j0}
- \quar \db_{j0} h_{k0}
- \half \db_{00} h_{jk}
\nonumber \\
& & \pt{+ i \big(}
- \half \db_{lj} \left( h_{00} \etul{l}{k} + \hul{l}{k} \right)
- \db_{kl} \hul{l}{j} \big] \ga^0 \ga_5 \ga^k \prt^j
\nonumber \\
& &
+ m \big(
\dtw^{j0} \et_{jk}
- \half \db^{j0} h_{jk}
- \db_{kj} h^{j0} \big) \ga_5 \ga^k
\nonumber \\
& &
- \quar i m \db^{j0} h^{k0} \ep_{jkl} \ga^l,
\eea
respectively.
The dimensionless four-component coefficients
\e\ and \f\ generate the expressions 
\bea
H^{(1,1)}_e &=& 
i \big(
\etw_j
+ \quar \eb_0 h_{j0}
- \half \eb_j h_{00}
- \eb^k h_{jk} \big) \ga^0 \prt^j
\nonumber \\
& &
- m \etw_0 
+ m \eb^j h_{j0}
+ \quar m \eb_0 h_{j0} \ga^0 \ga^j
\eea
and 
\bea
H^{(1,1)}_f &=& 
- \big(
\ftw_j
+ \quar \fb_0 h_{j0}
- \half \fb_j h_{00}
- \fb^k h_{jk}  \big) \ga^0 \ga_5 \prt^j,
\nonumber\\
\eea
while the dimensionless coefficient \g\
leads to
\bea
H^{(1,1)}_g &=&
\big( 
2 \gtw_{k(j0)}
- \quar \gb_{l00} h^{l0} \et_{jk}
- \gb_{l(j0)} \hul{l}{k}
- 2 \gb_{k(l0)} \hul{l}{j}
\nonumber \\
& & \pt{\big)}
+ 2 \gb_{k(jl)} h^{l0} \big) \ga^k \prt^j
\nonumber \\
& &
+i \big[ 
\gtw^{l00} \et^{jk}
- \half \gtw^{klj}
- \half \gb^{k00} \left( h_{00} \et^{jl} - h^{jl} \right)
\nonumber \\
& & \pt{+ i \big(}
- \half \gb_{n00} h^{ln} \et^{jk}
- \quar \gb^{jk0} h^{l0}
\nonumber \\
& & \pt{+ i \big(}
+ 2 \gb^{l(n0)} h_{n0} \et^{jk}
- \frac{1}{8} \gb^{kl0} h^{j0}
+ \quar \gb^{klj} h_{00}
\nonumber \\
& & \pt{+ i \big(}
+ \half \gb^{kln} \hul{j}{n}
- \half \gb^{knj} \hul{l}{n} \big] \ep_{klm} \ga_5 \ga^m \prt_j
\nonumber \\
& &
- \half m \big(
\gtw^{jk0}
+ \gb^{km0} \hul{j}{m}
+ \gb^{jkm} h_{m0} \big) \ep_{jkl} \ga^0 \ga_5 \ga^l
\nonumber \\
& &
+ \half m i \gb_{jk0} h^{k0} \ga^0 \ga^j
+ \frac{1}{8} m \gb^{jk0} h^{l0} \ep_{jkl} \ga_5.
\eea
Finally, 
the antisymmetric coefficient \H\ of mass dimension one
contributes
\bea
H^{(1,1)}_H &=& 
- \half \big(
\Htw^{jk}
- \Hb^{jm} \hul{k}{m}
- \half \Hb^{jk} h_{00} \big) \ep_{jkl} \ga_5 \ga^l
\nonumber \\
& &
- i \big( \Htw_{j0} 
+ \half \Hb^{k0} h_{jk} 
+ \Hb_{jk} h^{k0} \big) \ga^j.
\eea

In the above expressions,
the fluctuations of the various coefficient fields
appear in $H$ only at perturbative order O(1,1). 
For most purposes,
it is necessary to find expressions for these fluctuations
prior to using the hamiltonian $H$ in a given analysis. 
This issue is addressed further in Sec.\ \ref{fluct},
where the spin-independent coefficient fluctuations
$\aftw_\mu$ and $\ctw_\mn$ are considered in more detail.

The remaining piece of the hamiltonian $H$
lies at perturbative order O(0,2)
and represents 
the second-order Lorentz-invariant contribution 
from gravitational and inertial effects.
It takes the form
\bea
H^{(0,2)} &=&
i \big( \frac{1}{8} h_{00} h_{00} \et_{jk}
+ \half h_{l0} h^{l0} \et_{jk}
- \quar h_{00} h_{jk}
\nonumber \\
& & \pt{i \big)}
- \frac{3}{8} h_{jl} \hul{l}{k} \big) \ga^0 \ga^k \prt^j
- i h_{k0} h^{jk} \prt_j
\nonumber \\
& &
- m \big( 
\frac{1}{8} h_{00} h_{00} + \half h_{j0} h^{j0} 
\big) \ga^0.
\eea

\subsection{Nonrelativistic hamiltonian}
\label{quantummech}

Most experimental tests of interest in this work 
are nonrelativistic.
In this section,
we use a Foldy-Wouthuysen transformation 
\cite{fw}
to extract from the relativistic hamiltonian $H$ 
the parts of the nonrelativistic hamiltonian $H_{\rm NR}$
relevant for the subsequent analyses.

The Foldy-Wouthuysen transformation is a systematic procedure 
for determining the nonrelativistic content
of certain relativistic quantum-mechanical hamiltonians.
For a massive four-component Dirac fermion,
the transformation generates 
a series expansion in powers of the fermion momentum.
In the present case,
the transformation can be implemented as usual,
but care must be taken to keep track
of both the order in momentum 
and the perturbative order O($m$,$n$)
in coefficients for Lorentz violation
and in gravitational fluctuations. 

Performing the Foldy-Wouthuysen transformation
for the complete hamiltonian $H$ of Eq.\ \rf{relham}
is cumbersome and also unnecessary for our scope
because most attained sensitivities to spin couplings 
are unlikely to be improved by studying
the suppressed effects from gravitational couplings.
However,
only limited sensitivity currently exists
to spin-independent effects 
controlled by the coefficients $\ab_\mu$ and $\eb_\mu$
because these are unobservable 
for baryons and charged leptons in Minkowski spacetime.
In the remainder of this work
we focus on general spin-independent effects,
which are associated with the coefficients 
$\ab_\mu$, $\cb_\mn$, and $\eb_\mu$. 
Since the minimal torsion coupling also involves spin,
this focus implies also disregarding 
nonrelativistic effects due to torsion,
effectively restricting attention
to the limiting Riemann geometry.
Although beyond our current scope,
a Foldy-Wouthuysen analysis incorporating spin-dependent effects 
could lead to additional torsion sensitivities
beyond those obtained 
via searches for $\bb_\mu$
\cite{akgrav,krt,is,gos}.

In the relativistic quantum theory,
the upper two components of the four-component wave function 
describe the particle
while the lower two describe the antiparticle.
The hamiltonian $H$ can be separated 
into an odd part $\cO$ containing terms 
that mix the upper and lower components
and an even part $\cE$ that involves no mixing.
The idea of the Foldy-Wouthuysen method 
is to find a momentum-dependent unitary transformation $S$
in the Hilbert space 
such that the $4\times4$ hamiltonian 
$\widetilde H = e^{iS}H e^{-iS}$ 
is $2\times2$ block diagonal.
The leading $2\times2$ block of $\widetilde H$
then represents the desired nonrelativistic hamiltonian 
$H_{\rm NR}$.
The full transformation $S$ is obtained 
at the desired level of accuracy 
via an iterated series of incremental transformations
reducing the off-diagonal content to the appropriate order.

We proceed by separating the hamiltonian $H$ 
into an odd part $\cO^{(m,n)}_0$ 
and an even part $\cE^{(m,n)}_0$ 
at each perturbative order O($m$,$n$).
A subscript is used to specify the iteration number
of the transformation,
with 0 corresponding to the zeroth iteration. 
The relativistic hamiltonian $H$ can therefore be written as
\bea
H_0 \equiv H &=& 
m\ga^0 
+ \cO^{(0,0)}_0 + \cO^{(0,1)}_0 + \cO^{(1,0)}_0 + \cO^{(1,1)}_0
\nonumber \\
& &
+ \cE^{(0,0)}_0 + \cE^{(0,1)}_0 + \cE^{(1,0)}_0 + \cE^{(1,1)}_0.
\eea
The Foldy-Wouthuysen sequence is then defined
iteratively as
\bea
H_{n+1}&=&
e^{iS}H_n e^{-iS}
\nonumber\\
&=& \sum_{k=0}^\infty \fr{1}{k!} 
\underbrace{ [iS_n,[iS_n, \cdots [iS_n,}_{k\ 
\hbox{\scriptsize commutations with}\ iS_n} H_0]
\cdots ] ] ,
\eea
where
\bea
S_n &=& \fr{-i \ga^0 \cO_n}{2 m}.
\eea
At each stage,
the sum on $k$ is truncated
once the appropriate order in momentum and small quantities 
is reached.
The iteration continues until the hamiltonian is even
at the desired order.
Here,
we proceed to O(1,1) in the small quantities
and to second order in the momentum,
which requires three iterations
and yields a hamiltonian $H_3$.

The desired spin-independent contributions
to the nonrelativistic hamiltonian 
$H_{\rm NR}$
can be separated according to perturbative order
and origin as 
\bea
H_{\rm NR}
&\equiv & H_3 
\nonumber \\
&=& 
\hnr^{(0,0)} + \hnr^{(0,1)} + \hnr^{(1,0)} 
\nonumber \\
&&
+ \hnrs{a_{\rm eff}}^{(1,1)}
+ \hnrs{c}^{(1,1)} 
+ \hnrs{h}^{(1,1)}
+ \hnr^{(0,2)}.
\label{nrh}
\eea
Here,
$\hnr^{(0,0)}$
is the conventional Minkowski-spacetime hamiltonian.
The conventional Lorentz-invariant contributions $\hnr^{(0,1)}$
due to the metric fluctuation can be written as
\bea
\hnr^{(0,1)} &=& 
-\half m \hlnv_{00} - \hlnv_{0k} p^k - \fr{1}{4m} \hlnv_{00}p^2
\nonumber \\ 
&& 
- \fr{1}{2 m} \hlnv_{jk} p^j p^k.
\eea
The leading-order perturbation $\hnr^{(1,0)}$ 
due to Lorentz violation and independent of $h_\mn$
is identical to the Minkowski-spacetime result
given as Eq.\ (4) of Ref.\ \cite{kl}.

The corrections of primary interest for our purposes
lie at perturbative order O(1,1).
The contribution from \a\ and \e\ 
can be written in terms of
the effective coefficient $\af_\mu$ 
introduced in Eq.\ \rf{aeff},
and it takes the form
\bea
\hnrs{a_{\rm eff}}^{(1,1)} &=& 
\aftw_0 
+ \afb_k h^{0k} 
- \fr{1}{m} \afb^j h_{jk} p^k 
\nonumber \\
& & + \fr{1}{m} \left( \aftw_j
- \half \afb_j h_{00} \right) p^j.
\eea
The O(1,1) contribution from \c\
can be written
\bea
\hnrs{c}^{(1,1)} &=& 
- m \left( \ctw_{00}
+ \half \cb_{00} h_{00}
+ 2 \cb_{(k0)} h^{0k} \right)
\nonumber \\
&&
-2 \left( 
\ctw_{(j0)}
+\cb_{(jk)} h^{0k} 
- \cb_{(0k)} h_j^{\pt{j}k} \right) p^j
\nonumber \\
&& 
- \fr{1}{m} \big( \half \ctw_{00} \et_{jk}
+ \ctw_{jk}
+ \quar \cb_{00} h_{00} \et_{jk}
+ \cb_{(l0)} h^{0l} \et_{jk}
\nonumber \\
&& 
\pt{- \fr{1}{m} \Big[}
- \half \cb_{jk} h_{00} 
- \half \cb_{00} h_{jk} 
- 2 \cb_{(jl)} h^l_{\pt{l} k} \big) p^j p^k,
\nonumber \\
\eea
while the O(1,1) contribution
from Lorentz-violating effects on the metric fluctuation is
\bea
\hnrs{h}^{(1,1)} &=& 
-\half m \hlv_{00} - \hlv_{0k} p^k - \fr{1}{4m} \hlv_{00}p^2
\nonumber \\
& & - \fr{1}{2 m} \hlv_{jk} p^j p^k.
\eea

The remaining contribution to $H_{\rm NR}$ 
is the O(0,2) contribution
involving quadratic products of $h_\mn$.
This can be written as 
\bea
\hnr^{(0,2)} & = &
- \half m \left( h_{0j} h^{0j} + \quar h_{00} h_{00} \right)
+ h_{0j} h^{jk} p_k
\nonumber \\
& &
- \fr{1}{m} \big( \frac{1}{16} h_{00} h_{00} \et_{jk}
+ \quar h_{0l} h^{0l} \et_{jk}
\nonumber \\
& &
\pt{- \fr{1}{m} \big(}
- \quar h_{00} h_{jk}
- \half h_{jl} \hul{l}{k} \big) p^j p^k.
\eea

\section{Classical Theory}
\label{classicaltheory}

For many analyses of Lorentz violation
in matter-gravity couplings,
a classical description suffices.
This section considers the classical limit
of the quantum theory discussed above,
focusing on the limit 
involving the coefficient fields $\af_\mu$ and \c.
A suitable classical relativistic action 
for a point particle is presented,
and its application to modeling test and source bodies
is described.
The modified Einstein equation
and the equation for the trajectory of a test body are obtained.
We also discuss the treatment of the coefficient fluctuations 
$\aftw_\mu$, $\ctw_\mn$
and the procedure for determining 
the background gravitational field 
in the presence of Lorentz violation. 

\subsection{Particle action}
\label{action}

The classical action $S_{\rm c}$
corresponding to the action $S$ of Eq.\ \rf{SMEaction} 
can be written as
\beq
S_{\rm c} = S_G + S_u + S^\prime.
\label{classicalaction}
\eeq
As before,
$S_G$ describes the gravitational dynamics,
while $S^\prime$ contains the dynamics
associated with the coefficient fields for Lorentz violation.
The partial action $S_u$ is 
the classical relativistic point-particle limit 
of the action $S_\ps$ for the fermion sector.
In this subsection,
we discuss $S_u$ 
and extend it to describe test and source bodies.

\subsubsection{Point particle}
\label{Point particle}

At leading order in Lorentz violation,
we find
\beq
S_u = \int d\la \left(-m  
\sqrt{-(g_\mn + 2c_\mn) u^\mu u^\nu}
-(a_{\rm eff})_\mu u^\mu \right).
\label{actionsec}
\eeq
In this expression,
the particle path $x^\mu = x^\mu(\la)$
is parametrized by $\la$,
and $u^\mu = dx^\mu/d\la$ 
is the four-velocity of the particle.
As usual, 
a gauge choice for $\la$ is required to
fix the path-reparametrization invariance
and to define the proper time of the particle on shell.
We adopt here the conventional proper-time interval 
\beq
d \ta = \sqrt{-g_\mn d x^\mu d x^\nu}.
\label{propertime}
\eeq
 
The leading-order form \rf{actionsec} of the classical action 
can be deduced in several ways.
At the intuitive level,
the term involving $(a_{\rm eff})_\mu$
has the same structure as the usual
coupling of a classical relativistic particle
to an electromagnetic 4-potential,
and this is consistent with the coupling
of $a_\mu$ in the field-theory action \rf{qedxps}.
Similarly,
the coefficient $c_\mn$ enters Eq.\ \rf{actionsec}
as a shift in the metric,
which is compatible with the way it appears
in the field theory \rf{qedxps}.
In a different vein,
the contributions from $c_\mn$
to the relativistic particle action
have previously been discussed
in the context of the photon sector
\cite{qbak04},
where the appearance of $c_\mn$ as a metric shift
is related to the coordinate choices
discussed in Sec.\ \ref{Coordinate choices}.
The validity of the action \rf{actionsec}
can also be verified by extracting the leading-order terms
from the all-orders expression obtained by construction 
of the exact relativistic dispersion relation
\cite{aknr}.
In the present context,
we can demonstrate explicitly that the action \rf{actionsec}
reproduces the corresponding terms 
in the nonrelativistic hamiltonian $H_{\rm NR}$ 
generated from the Foldy-Wouthuysen transformation
as Eq.\ \rf{nrh}.
This involves expanding the action \rf{actionsec}
to the appropriate orders in velocities and Lorentz violation, 
extracting the conjugate 3-momentum,
constructing the corresponding hamiltonian,
and matching it to $H_{\rm NR}$ in Eq.\ \rf{nrh}.
These methods all confirm that Eq.\ \rf{actionsec}
is the correct leading-order form 
of the relativistic classical action.

The energy-momentum tensor $T_u^\mn$
for the point particle
can be derived from the action \rf{actionsec} 
by variation with respect to the metric,
as usual.
We obtain 
\beq
T_u^{\mn}= 
- \int d \ta \fr
{m u^\mu u^\nu \de^4 (x - x^\prime (\ta))}
{\sqrt{g}\sqrt{1 - 2 c_{\al\be} u^\al u^\be}},
\label{cenergy}
\eeq
where the proper-time interval 
is given by Eq.\ \rf{propertime}.
Note that no contributions from $\af_\mu$
appear in this expression.
This follows from the adoption 
of $(a_{\rm eff})_\mu$ with lower index 
as the coefficient field,
which implies that the contraction $(a_{\rm eff})_\mu u^\mu$
in Eq.\ \rf{actionsec} contains no metric.
Working with $(a_{\rm eff})^\mu$ instead is possible
but less convenient.
It would produce a contribution to $T_u^{\mn}$
along with corresponding changes in the contributions
to the energy-momentum tensor $T^{\prime \mn}$
associated with $S^\prime$,
leading to the same physical results.

\subsubsection{Test and source bodies}
\label{Test and source bodies}

The experiments and observations 
considered in this work
involve bodies B acting as test bodies T or as sources S.
Many of these bodies 
consist of atoms or macroscopic matter
rather than individual particles.
It is therefore useful to extend 
the point-particle action \rf{actionsec}
to an action $S_u^{\rm B}$ for a body B.
This requires consideration of several issues.

One issue arises because
the interactions involved in binding
electrons, protons, and neutrons
into atoms and macroscopic matter
contribute additional Lorentz-violating effects.
This issue appears also 
in the study of fermion-sector SME coefficients 
in the Minkowski-spacetime limit
\cite{kl2}.
However,
for the gravitational tests of interest here,
it is reasonable to assume 
that these interaction effects are small
compared to the propagation effects.

Another issue arises from the spacetime dependence
of the coefficient fields $\af_\mu$ and $c_\mn$,
which implies Lorentz-violating effects
may vary over the region filled by the body.
Most of the test bodies we consider are small,
so it is reasonable to approximate the coefficient fields
as constant across the extent of the body.
This corresponds to the usual approximation
of constant metric fluctuation $h_\mn$ across a test body. 
However,
some of the source bodies we consider are comparatively large,
so some variation of the coefficient fields over the source
is plausible.
This could produce Lorentz-violating effects
of various types,
including possible dependence on the mass moments of the body.
In what follows,
we suppose that the variation of the coefficient fields 
is sufficiently mild and smooth that these effects
can be neglected for the bodies we consider. 
A more comprehensive treatment of this issue 
would be of potential interest 
but lies outside our present scope.

With the above assumptions 
and for most purposes in this work,
a given body B can be modeled 
as a composite particle
with constituents located at a single spacetime point
and having the same 4-velocity,
held together by binding energy.
The body B can then be assigned an effective mass $m^{\rm B}$,
expressed in terms of its constituent particles as 
\beq
\mb = \sum_{w} N^w m^w + \mbp.
\label{bodymass}
\eeq
Here,
$w$ ranges over the particle species forming the body B.
For example,
$w$ can be taken to include 
the electron $e$, the proton $p$, and the neutron $n$
whenever B is an atom or made of ordinary macroscopic matter.
The symbol $N^w$ denotes the number of particles
of type $w$ in the body, 
and $\mbp$ represents the contribution to the mass
from the binding energy.
In practice,
the exact values of $N^w$
are readily obtained for test bodies 
on the atomic or molecular scale,
while estimating $N^w$ for macroscopic test and source bodies 
in the laboratory is straightforward.
When considering the Earth as the source body,
we adopt the estimates
$\noe^e = \noe^p \approx \noe^n = 1.8 \times 10^{51}$
based on recent studies of the bulk Earth composition
\cite{earthcomp}.
The difference $\noe^n-\noe^p \simeq 10^{49}$
is primarily due to the iron core.
The radial variation in neutron content 
is neglected in what follows,
although it might be of interest in more detailed studies.

Similarly,
the Lorentz-violating properties of B 
can be represented via effective coefficient fields 
$\afB_\mu$ and $\cuB_\mn$ for the body.
These can be viewed as the sum of vacuum values
and coefficient fluctuations,
\beq
\afB_\mu = \afbb_\mu + \aftwb_\mu ,
\quad
\cuB_\mn = \cbb_\mn + \ctwb_\mn,
\eeq
in parallel with the point-particle case.
The form of the action \rf{actionsec}
implies that the coefficient field $\afB_\mu$
takes the form
\beq
\afB_\mu = 
\sum_w N^w \afw_\mu  + \afp_\mu,
\label{abody}
\eeq
where $\afp_\mu$ is a possible coefficient field
associated with the binding energy
that contributes to $\afB_\mu$.
Also,
expanding the action $S_u$ for small Lorentz violation
shows that at leading order 
the body coefficient field $\cuB_\mn$ 
can be taken as
\beq
\cuB_\mn = 
\fr{1}{m^{\rm B}} 
\left( \sum_{w} 
N^w m^w \cw_\mn + \mbp \cbpr_\mn \right) ,
\label{cbody}
\eeq
where $\cbpr_\mn$ is associated with the binding energy.

The two contributions 
$\afp_\mu$ and $\cbpr_\mn$ 
describe Lorentz violation arising from the particles
associated with the forces binding together the body B.
These particles are primarily gravitons, gluons, or photons
and are associated with boson fields,
for which the CPT-violating terms 
are expected to be small or zero.
In the minimal SME,
no such terms exist for gravitons,
while for photons and gluons
they can reasonably be assumed to vanish
\cite{akgrav}.
Also,
the relevant photon coefficient $(k_{AF})^\mu$
is constrained well below levels relevant
for this work
\cite{tables}.
Possible CPT-violating contributions 
from other sea particles largely cancel due to 
particle-antiparticle pairings
or are suppressed in loops involving weak interactions.
We therefore approximate
the contributions from $\afp_\mu$ as negligible,
\beq
\afp_\mu \simeq 0,
\label{zeroaprime}
\eeq
in this work.
In contrast,
all the force fields have CPT-even terms
that can be expected to contribute to $\cbpr_\mn$,
so the resulting size of $\cbpr_\mn$
may well be of the same order as $\cw_\mn$
and cannot be neglected.
 
Given the above discussion,
we conclude that the leading-order approximation 
to the classical action $S^{\rm B}_u$ for a body B 
can be written in the simple form
\bea
S^{\rm B}_u &\approx &
\int d\la 
\Big(-m^{\rm B}  
\sqrt{-(g_\mn + 2\cuB_\mn) u^\mu u^\nu}
\nonumber\\
&&
\hskip 40pt
-\afB_\mu u^\mu \Big),
\label{actioneq}
\eea
where $m^{\rm B}$, $\afB_\mu$, and $\cuB_\mn$ 
are given by Eqs.\ \rf{bodymass}, \rf{abody}, and \rf{cbody},
respectively.
In this expression,
$u^\mu$ is the 4-velocity of the body B,
which follows a world line parametrized by $\la$.
This form is convenient for calculational purposes.
Note, however,
that the derivation establishes validity 
of this form of the action 
only at leading order in $\cuB_\mn$.

The model action \rf{actioneq}
for a body B suffices for most situations of interest
in this work.
In a few cases
where the body acts as a gravitational source S,
it is also useful to incorporate dominant effects 
arising from its rotation.
For this purpose,
we treat S as rigid at leading order
and assume that the distribution
of electrons, protons, and neutrons
is approximately uniform throughout it.
For the bodies we consider,
this assumption is good to within an order of magnitude.
The density $\rh$ of S 
can be taken as the mass per unit volume
and approximated as uniform.
For large source bodies such as the Earth,
some results could in principle 
also depend on spherical moments of inertia 
\cite{qbakpn},
but these effects are neglected here. 
The angular velocity $\vec{\om}$ of rotation 
is defined in the frame at rest relative to S
with origin at the center of mass,
which can be identified with the location of S.

\subsection{Equations of motion}
\label{eom}

The primary experimental observables
arising from the classical theory
involve the relative motion of particles.
To investigate the motion of a test particle
in the presence of gravitational sources,
the modified Einstein equation
must be solved for the background metric
and the equation for the particle trajectory
must be found.
In this subsection,
we derive the equations of motion
from the action \rf{classicalaction}
in terms of 
the metric fluctuation $h_\mn$
and 
the coefficient fluctuations $\aftwb_\mu$, $\ctwb_\mn$.
The issue of expressing these fluctuations
in terms of the vacuum values
$\afbb_\mu$, $\cbb_\mn$
for a given distribution of matter
is addressed in the following subsection,
Sec.\ \ref{fluct}.
We conclude the present subsection with comments 
about the implications of Lorentz violation
for the equivalence principle.

\subsubsection{Modified Einstein equation}
\label{einsteineq}

Varying the action \rf{classicalaction}
in Riemann spacetime
with respect to the metric
yields the modified Einstein equation
\beq
G^\mn = T_G^\mn + \ka T_u^\mn + \ka T^{\prime \mn},
\label{smeeinstein}
\eeq
where $G^\mn$ is the Einstein tensor
and the terms on the right-hand side
form the energy-momentum tensor.
The contribution $T_G^\mn$ 
arises from Lorentz violation in the pure-gravity sector.
The energy-momentum tensor $T_u^{\mn}$
for the matter is given in Eq.\ \rf{cenergy}.
The remaining energy-momentum contribution
$T^{\prime\mn}$ arises from the dynamics 
of the coefficient fields for Lorentz violation
and is determined by $S^\prime$.

Taking the covariant divergence of Eq.\ \rf{smeeinstein}
and using the Bianchi identities
shows that the geometry requires
the total energy-momentum tensor to be locally conserved.
The theory can be consistent only if this result 
is compatible with the explicit form
of the energy-momentum tensor.
This requires careful accounting of 
contributions from the massless NG modes
arising from the spontaneous Lorentz breaking 
\cite{akgrav}.
In the general case,
these modes are contained in the fluctuations $\Xtw$.
 
For the pure-gravity sector,
the relevant analysis is given in 
Refs.\ \cite{akgrav,qbakpn}
and can be subsumed as needed in the present context.
For the matter sector,
the NG modes produce no relevant contribution
to the energy-momentum tensor $T_u^{\mn}$
at the post-newtonian order
appropriate for the tests considered here. 
The key point is that
the leading Lorentz-violating effects 
of coefficient fluctuations $\Xtw$
arise at \pno{2} or beyond,
as shown in Sec.\ \ref{Perturbation scheme}.
Since these fluctuations
are accompanied by an additional factor of $G_N$
in the modified Einstein equation \rf{smeeinstein},
they affect the metric only at PNO(4) or beyond.
However,
for the tests considered below
it suffices to work at PNO(3)
for Lorentz-violating terms,
so the coefficient fluctuations $\aftwb_\mu$ and $\ctwb_\mn$
appearing in $T_u^{\mn}$ can be neglected 
in Eq.\ \rf{smeeinstein}.

In contrast,
the contributions to $T^{\prime \mn}$
arising from the dynamics 
of the coefficient fluctuations $\Xtw$
are of potential relevance 
in solving the modified Einstein equation
for the metric.
The specific effects 
associated with the coefficients $\aftwb_\mu$ and $\ctwb_\mn$
are derived in Sec.\ \ref{fluct}.

\subsubsection{Particle trajectory}
\label{Particle trajectory}

The equation of motion for a classical test particle T
is obtained by varying the action \rf{classicalaction}
with respect to the particle position 4-vector $x^\mu$. 
In the absence of Lorentz violation,
this is the geodesic equation.
However,
in the presence of Lorentz violation,
the trajectories of test particles T no longer match 
the geodesics of the spacetime.

Expanding to O(1,1),
the equation of motion can be written as
\bea
\ddot{x}^\mu &=&
- \Ga_{(0,1) \pt{\mu} \al \be}^{\pt{(0,0)} \mu} u^\al u^\be 
\nonumber
\\
&&
- \Ga_{(1,1) \pt{\mu} \al \be}^{\pt{(1,1)} \mu} u^\al u^\be
+ 2 \et^{\mu\ga} \cbt_{(\ga \de)} 
\Ga_{(0,1) \pt{\de} \al \be}^{\pt{(0,1)} \de} u^\al u^\be
\nonumber
\\
&&
+ 2 \cbt_{(\al \be)} 
\Ga_{(0,1) \pt{\al} \ga \de}^{\pt{(0,1)} \al} 
u^\be u^\ga u^\de u^\mu
+ \prt^\mu \ctwt_{\al \be} u^\al u^\be
\nonumber
\\
&&
- 2 \et^{\mu\ga} \prt_\al \ctwt_{(\ga \be)} u^\al u^\be 
- \prt_\ga \ctwt_{(\al \be)} u^\al u^\be u^\ga u^\mu 
\nonumber
\\
&&
- \fr 1 \mt
[\prt^\mu \atwt_\al - \et^{\mu\be} \prt_\al \atwt_\be ] 
u^\al,
\label{oogeo}
\eea
where each dot on $x^\mu$ represents a derivative
using the proper-time interval \rf{propertime}.
The superscript T denotes quantities associated
with the test particle.
The first term on the right-hand side
is the usual geodesic contribution,
where $\Ga_{(0,1) \pt{\mu} \al \be}^{\pt{(0,0)} \mu}$
is the linearized Christoffel symbol.
A Christoffel symbol with subscript (1,1) 
also appears in Eq.\ \rf{oogeo}.
It is defined as the linearized Christoffel symbol
with $h_\mn$ replaced by $\hlv_\mn$.
This introduces matter-sector coefficients
associated with the gravitational source,
along with any gravity-sector coefficients
that may be included in the analysis.

Once the forms of
$h_\mn$, $\aftw_\mu$, and $\ctw_\mn$
are established,
Eq.\ \rf{oogeo} can be used to determine to O(1,1)
the motion of a classical test particle
in a curved but asymptotically flat spacetime
with nonzero coefficients for Lorentz violation
$a_\mu$, $e_\mu$, and $c_\mn$.
Obtaining expressions for 
$h_\mn$, $\aftw_\mu$, and $\ctw_\mn$
is the subject of Sec.\ \ref{fluct}.

Although unnecessary for the present work,
we can comment in passing
about the effects of nongravitational interactions
on the particle trajectory.
Any such interactions can be viewed as introducing
an additional contribution $\al^\mu$
to the right-hand side of Eq.\ \rf{oogeo}.
Using the perturbation scheme of Sec.\ \ref{Perturbation scheme},
this additional acceleration $\al^\mu$
can be expanded as a sum over terms $\al_{(m,n)}^\mu$,
one at each perturbative order O($m,n$).
Notice that,
although the interaction itself is nongravitational,
contributions to $\al_{(m,n)}^\mu$ with $n\neq 0$
can be induced from gravitational couplings
in the interaction sector.
Similarly,
Lorentz-violating contributions to $\al_{(m,n)}^\mu$
can originate from coefficients for Lorentz violation
in the interaction sector.
If we also expand $\ddot{x}^\mu$ as
\beq
\ddot{x}^\mu = 
\ddot{x}^\mu_{(0,0)} + \ddot{x}^\mu_{(0,1)} 
+ \ddot{x}^\mu_{(1,0)} + \ddot{x}^\mu_{(1,1)} + \ldots,
\label{modgeo}
\eeq
then we obtain the following additional terms 
for the particle 4-acceleration $\ddot{x}^\mu$: 
\bea
\ddot{x}^\mu_{(0,0)} &\supset& \al_{(0,0)}^\mu,
\nonumber\\
\ddot{x}^\mu_{(0,1)} &\supset& \al^\mu_{(0,1)},
\nonumber\\
\ddot{x}^\mu_{(1,0)} &\supset& \al^\mu_{(1,0)}
- 2 \et^{\mu \al} \cbt_{(\al \be)} \al^\be_{(0,0)}
\nonumber\\
&&
- 2 \cbt_{(\al \be)} \al^\al_{(0,0)} u^\be u^\mu,
\nonumber\\
\ddot{x}^\mu_{(1,1)} &\supset&
\al^\mu_{(1,1)}
- 2 \et^{\mu \al} \cbt_{(\al \be)} \alol^\be
\nonumber\\
&&
- 2 \cbt_{(\al \be)} \alol^\al u^\be u^\mu
+ 2 h^{\mu \al} \cbt_{(\al \be)} \aloo^\be
\nonumber \\
&&
- 2 \et^{\mu \al} \ctwt_{(\al \be)} \aloo^\be.
\label{oogeo2}
\eea
The trajectory at O(1,$n$) is affected 
both directly by $\al^\mu_{(1,n)}$
and indirectly by combinations of $\al^\mu_{(0,n)}$
with the coefficients for Lorentz violation.
The origin of the indirect terms 
can be traced to the additional factor of $2 c_\mn$
in the action \rf{actionsec}
relative to the conventional proper-time interval \rf{propertime}.

\subsubsection{Implications for the equivalence principle}
\label{Equivalence principle}

The deviations from geodesic motion implied by Eq.\ \rf{oogeo}
can be species dependent
because the couplings 
to the coefficient fields $\af_\mu$ and $c_\mn$
can vary with particle flavor.
This leads to apparent violations
of the weak equivalence principle (WEP),
which stipulates that the motion of uncharged test particles 
is independent of internal structure or composition 
\cite{cmw}.

One implication of this observation
is that experiments designed to test the WEP
are also sensitive to $\afb_\mu$ and $\cb_\mn$.
Since all the WEP violations implied by Eq.\ \rf{oogeo}
are accompanied by effects 
associated with the breaking of rotation and boost symmetries,
the experimental signatures 
associated with $\afb_\mu$ and $\cb_\mn$
typically differ from those
in other scenarios for violations of the WEP.
The latter portion of this work 
discusses in some detail
the role that experiments testing the WEP can play
in searches for Lorentz violation.

The flavor dependence of the coefficient couplings
leads to the philosophical question 
of whether spontaneous Lorentz violation in the matter sector
violates the WEP or merely mimics violations of the WEP.
The issue hinges on the interpretation 
of the term `uncharged test particle.' 
In models with spontaneous Lorentz violation
in the matter sector,
the NG modes couple to test particles 
and so mediate an interaction.
This interaction can be identified
with Einstein-Maxwell electrodynamics 
\cite{lvng},
GR gravity
\cite{cardinal},
an effect on torsion 
\cite{lvng},
or a new force
\cite{akjt,ahclt,abk}.
If the term `uncharged'
is taken in the restrictive sense to mean 
that the test particle is unaffected 
by standard forces such as electrodynamics,
then the trajectory deviations of `uncharged' test particles
caused by nonzero $\afb_\mu$ and $\cb_\mn$ coefficients
represent violations of the WEP.
If instead the term `uncharged'
indicates the test particle cannot have nongravitational couplings
of any kind,
then no violations of the WEP occur.
However,
in this latter case
`uncharged' test particles may be nonexistent 
in the matter sector of the SME,
where generically all particles experience nonzero
$\afb_\mu$ and $\cb_\mn$ coefficients.
We emphasize that the above discussion
is a matter of philosophical classification only,
without impact on the practical issue
of using tests of the WEP to search for Lorentz violation
via the deviations from geodesic motion
described by Eq.\ \rf{oogeo}.

The WEP is subsumed in certain other equivalence principles,
such as the Einstein equivalence principle
or the strong equivalence principle.
These incorporate also aspects 
of local Lorentz invariance and local position invariance.
Since nonzero coefficients $\afb_\mu$, $\cb_\mn$ 
correspond directly to local Lorentz violation,
and since Lorentz violation can be position dependent,
the deviations from geodesic motion
described by Eq.\ \rf{oogeo}
can represent violations of these broader equivalence principles
arising in more than one way.
Several related philosophical issues remain open, 
including 
classifying violations of various equivalence principles
according to properties 
of the coefficients for Lorentz violation
and identifying implications for relations 
such as the Schiff conjecture
\cite{lis,cmw}.
We note also in passing that comments analogous to those above
bear on the philosophical issue of whether 
theories with matter-sector couplings 
to spontaneous Lorentz violation 
constitute metric theories of gravity.

\subsection{Coefficient and metric fluctuations}
\label{fluct}

To solve the equation of motion \rf{oogeo} 
for the trajectory of a test particle,
explicit expressions 
for 
the metric fluctuation $h_\mn$
and 
the coefficient fluctuations $\aftw_\mu$, $\ctw_\mn$ 
are required.
Within a specific model with known action $S^\prime$ 
for the coefficient fields,
these expressions can be obtained by direct calculation.
An illustration of this is provided in Sec.\ \ref{bumblebee}.
However,
in the interest of generality,
it is useful to establish results
valid for a large class of models.
In this subsection,
we outline a procedure to obtain expressions 
for $h_\mn$ and for the generic coefficient fluctuations $\Xtw$
when $S^\prime$ is largely unknown,
and we obtain explicit results 
for $h_\mn$, $\aftw_\mu$, and $\ctw_\mn$ 
applicable to the equation of motion \rf{oogeo}. 
These results are used in later sections of this work
in establishing experimental signatures for Lorentz violation.

\subsubsection{Methodology}
\label{Methodology}

Consider first the metric fluctuation $h_\mn$.
In the perturbation scheme of Sec.\ \ref{Perturbation scheme},
the expansion of $h_\mn$ takes the form \rf{hexp}.
To determine the test-particle trajectory 
at order O(1,1) via Eq.\ \rf{oogeo},
it is necessary to obtain explicit expressions for 
$h^{(0,1)}_\mn$ and $h^{(1,1)}_\mn$.

The Lorentz-invariant contribution $h^{(0,1)}_\mn$
can be obtained in the usual way
as the leading-order solution of the Einstein equation,
taking the Lorentz-invariant part 
of the energy-momentum tensor 
as the source.
To \pno3,
the standard solution can be written in harmonic coordinates as
\bea
h^{(0,1)}_{00} &=& 2 U, 
\quad 
h^{(0,1)}_{0j} = - 4 V^j,
\quad
h^{(0,1)}_{jk} = 2 U \de^{jk}, 
\quad
\label{pnpot}
\eea
where $U$ and $V^j$
are the usual post-newtonian potentials defined as
\bea
U &=& 
\G \int d^3 x^\prime 
\fr{\rh(\vec{x}^\prime, t)}
{|\vec{x}- \vec{x}^\prime|},
\nonumber \\
V^j &=& 
\G \int d^3 x^\prime 
\fr{\rh(\vec{x}^\prime, t) v^j (\vec{x}^\prime, t)}
{|\vec{x} - \vec{x}^\prime|}.
\label{potentialsa}
\eea
In these expressions,
the density $\rh(\vec x^\prime, t)$ 
and the 3-velocity $v^j(\vec x^\prime, t)$
are properties of the source
in the chosen asymptotic inertial frame.
As described in Sec.\ \ref{Test and source bodies},
it suffices in this work 
to use the approximation \rf{actioneq}
for the source-body action $S_u^B$,
so the energy-momentum tensor takes
the generic form \rf{cenergy}.

The Lorentz-violating component $\hlv_\mn$ 
can conveniently be viewed as a sum 
over individual contributions arising from 
each coefficient field for Lorentz violation,
\beq
\hlv_\mn = (\hlv_a)_\mn + (\hlv_b)_\mn + ... + (\hlv_H)_\mn.
\label{lvmetric}
\eeq
This means each coefficient can be treated in turn.
However,
the procedure for determining a particular contribution 
can take different paths depending 
on the type of gravitational coupling of the coefficient field.

A simple case arises for any coefficient field $\X$ 
that is minimally coupled to gravity.
Then,
only the vacuum values $\Xb$
in the expansion \rf{texpand}
contribute to the energy-momentum tensor $T^\mn_u$
of the source at the relevant order.
The key point is that the solution for $\hlv_\mn$ 
at \pno{3}
arises from the combination $\ka T^\mn_u$,
which itself already lies at \pno{2}.
However,
as discussed in Sec.\ \ref{Perturbation scheme},
the coefficient fluctuations of interest here are $\Xtwo 11$
and also lie at \pno{2}.
These fluctuations therefore cannot contribute
to $\hlv_\mn$ below \pno{4}.
As a result,
$\hlv_\mn$ can be found directly 
by solving the modified Einstein equation
with attention limited to the vacuum values $\Xb$.

We note in passing that this procedure 
is consistent with the no-go result
for explicit Lorentz violation in gravity
\cite{akgrav}
even though $\Xb$ can be interpreted 
as a coefficient for explicit breaking.
This is because we are working to O(1,1) and \pno{3},
for which $D_\al \Xb \sim$ O$(h \overline{t})\sim$ O(1,1).
As a result, 
the covariant derivative of $\ka T_u^\mn$
is compatible with the Bianchi identities $D_\mu G^\mn = 0$
at this perturbative order. 
In effect,
the comparatively low perturbative order 
implies that a constant vacuum value
remains consistent with the geometry 
of spontaneous Lorentz breaking.
It is also noteworthy 
that an independent contribution from $T^{\prime\mn}$
may exist that satisfies local conservation
and hence is compatible with the Bianchi identities 
to the relevant perturbative order.
This would also represent a consistent theory, 
albeit a different one.
The two theories involving the coefficient field $\af_\mu$ 
with minimal and with nonminimal gravitational couplings
provide an illustration of this,
as is discussed in the next subsection.

In the simple case with minimal coupling,
once $\hlv_\mn$ has been found,
it remains only to determine 
the direct contributions 
to the equation of motion \rf{oogeo}
arising from the fluctuations $\Xtwo 11$.
For this purpose,
we can apply the requirement 
that the system of the source S 
and the test body T conserves the total 4-momentum $P_\mu$.
For a two-body system,
this implies the force law 
must be antisymmetric upon exchange of S and T. 
Otherwise,
the forces on each body due to the other 
would violate Newton's third law,
and the system would self-accelerate.

At \pno{2} and in the absence of Lorentz violation, 
the relevant force between S and T
can be directly identified as $m \ddot{x}^j$.
At higher order and in the presence of Lorentz violation,
it is simpler to impose conservation 
of the total 4-momentum, $dP_\mu/dt = 0$.
In principle,
$P_\mu$ can be found by adding 
the conjugate momenta of S and T 
obtained from the two-body action.
In practice,
for the perturbative order to which we work,
it suffices to obtain the conjugate momentum for T alone
and require antisymmetry of its time derivative
under the exchange of S and T. 
The constraints fixing the fluctuations $\Xtwo 11$
arise from the $\mu=j$ components of $dP_\mu/dt = 0$,
all at \pno{3} except for terms at \pno{4}
involving the velocities of both S and T.
We remark in passing that extending this treatment 
to higher perturbative orders requires
also incorporating back-reaction effects on the metric,
including gravitational radiation.

The above procedure holds for 
coefficient fields that are minimally coupled to gravity.
If nonminimal curvature couplings also occur in $S^\prime$,
then additional terms involving the coefficient fields $\X$ 
can appear in the energy-momentum tensor
and hence can affect the modified Einstein equation.
The curvature couplings intertwine 
the kinetic contributions from $h_\mn$ and $\Xtw$,
so $\Xtw$ can contribute to the solution for $\hlv_\mn$ at \pno{3}. 
To proceed without specifying $S^\prime$,
we therefore need additional information 
about $\Xtw$. 

In the present work,
the necessary information can be extracted
from the general structure 
of the equation of motion for $\X$
and the symmetries of the theory.
When linearized,
this equation of motion  
can be written as the sum 
of a differential operator acting on $\Xtw$ 
and a source term at most linear in $h_\mn$.
The differential operator can involve arbitrary
powers of $\Xb$ but is independent of $h_\mn$.
The contributions $\Xtwo 11$ of interest are at O(1,1) 
and hence are linear in both $\Xb$ and $h_\mn$.
At this order,
the solution for $\Xtwo 11$ 
can therefore be written as a sum of terms,
each containing up to one power of $h_\mn$
along with some number $n$ of powers of $\Xb$ in the numerator
and $n-1$ powers of $\Xb$ in the denominator.
This expansion of $\Xtwo 11$ 
in terms of $\Xb$ and $h_\mn$
is constrained by two requirements.
One arises from the restriction of $\Xtw$ to NG modes,
which must maintain the extremum of the action.
The solution for $\Xtwo 11$ must therefore obey 
the NG conditions at O(1,1).
The second is the requirement
that $\Xtw$ must transform as expected under diffeomorphisms,
as a consequence of the spontaneous nature of the symmetry breaking
and the requirement of observer general coordinate invariance
\cite{aad}.
It turns out that these two restrictions suffice 
to express $\Xtwo 11$ in terms of $\Xb$ and $h_\mn$
in the cases of interest here.

Once the expression for $\Xtwo 11$ has been found,
$h_\mn$ can be obtained by combining
information from the modified Einstein equation
and the trajectory equation.
The modified Einstein equation yields directly
the piece of $h_\mn$ arising 
from $\Xb$ in the energy-momentum tensor.
Inserting this result and the expression for $\Xtwo 11$
in the trajectory equation
and imposing conservation of the total 4-momentum
of the source and test body as before 
determines the missing piece of $h_\mn$ 
arising from $\Xtwo 11$.

In both minimal and nonminimal cases,
the net result of the above procedure
is a form of the trajectory equation 
in which $h_\mn$ and $\Xtw$
can be replaced with specified gravitational potentials
and the vacuum values $\Xb$.
The solution of the trajectory equation can then proceed.
In the next two subsections,
we apply these methods 
to obtain the relevant contributions to the fluctuations 
from the coefficient fields $\af_\mu$ and $c_\mn$.

\subsubsection{Fluctuations and \c}
\label{cfluct}

The treatment of the coefficient field $c_\mn$
provides an example involving 
the comparatively simple case of minimal coupling to gravity.
Although nonminimal curvature couplings 
to $c_\mn$ could be considered,
these are of lesser interest
in the context of searches for Lorentz violation
because direct signals from $\cb_\mn$ 
already appear for minimal coupling.
We therefore neglect nonminimal couplings to $c_\mn$ here.
In this subsection,
the relevant contributions to the metric fluctuation
$(\hlv_c)_\mn$ are obtained to third post-newtonian order,
and effects from $\ctw_\mn$ at O(1,1) are considered.

Following the procedure outlined in the previous subsection,
we begin with the modified Einstein equation \rf{smeeinstein}.
The relevant energy-momentum tensor for the source S
is given by the expression \rf{cenergy}
expanded to leading order in Lorentz violation
and with $c_\mn$ replaced by $\cs_\mn$.
Solving for the metric fluctuation to \pno{3},
we obtain in harmonic gauge
\bea
(\hlv_c)_{00} &=& 2 \cs_{00} U + 4 \cs_{(j0)} V^j,
\nonumber\\
(\hlv_c)_{0j} &=& - 4 \cs_{00} V^j,
\nonumber\\
(\hlv_c)_{jk} &=& 2 \cs_{00} U \de^{jk}.
\label{metricc}
\eea
A consistent expansion to \pno{3} requires only 
\pno{1} terms in $(\hlv_c)_{0j}$ 
and none in $(\hlv_c)_{jk}$,
but we display 
\pno{3} terms in $(\hlv_c)_{0j}$ 
and \pno{2} terms in $(\hlv_c)_{jk}$
because they are useful in part of the analysis to follow. 

The next step is to examine 
the contributions to the equation of motion \rf{oogeo}
from $\ctw_\mn$ at O(1,1).
The conjugate momentum can be extracted from the action
\rf{actionsec} with $\af_\mu$ set to zero.
Conservation of the total 4-momentum $P_\mu$ of the system
is ensured by the requirement that its time derivative 
be antisymmetric under the exchange of the source S 
and the test body T.
We find that $P_\mu$ is conserved to \pno{3} 
without contributions from $\ctw_\mn$.
This establishes the \pno{3} result 
\beq
\prt_\la \ctw_\mn^{(1,1)} = 0,
\label{ctwres}
\eeq
showing that the NG modes associated with $\ctw_\mn$
play no role at this perturbative order.

The results \rf{metricc} and \rf{ctwres}
complete the determination of the trajectory equation 
for the coefficient field $c_\mn$. 
For a given source S,
the potentials $U$ and $V^j$ can be calculated explicitly.
The effects of $\cs_\mn$ and $\cot_\mn$ 
on the trajectory of the test body T
can therefore be investigated 
in various regimes of experimental interest.
This line of reasoning is pursued beginning 
in Sec.\ \ref{experiment}.

\subsubsection{Fluctuations and $\af_\mu$}
\label{Fluctuations and af}

For the coefficient field $\af_\mu$,
the case of minimal coupling to gravity
is of lesser interest.
The modified Einstein equation is unaffected
because $\af_\mu$ is absent 
from the energy-momentum tensor \rf{cenergy}.
Also, only $\aftw_\mu$ enters the trajectory equation.
Since it is indistinguishable from an electromagnetic field,
it provides no relevant contributions at O(1,1). 
We therefore expand the treatment
to the case of nonminimal curvature couplings,
for which $\afb_\mu$ becomes measurable
\cite{akjt}.
In effect,
the fluctuations $\aftw_\mu$ 
become observable by virtue of their
nonminimal gravitational couplings.

Following the procedure of Sec.\ \ref{Methodology},
the first step is to obtain an expression 
for the fluctuations $\aftw_\mu^{(1,1)}$
originating from the source S 
using the NG condition 
and the requirement of diffeomorphism covariance.
At O(1,1),
the NG condition can be written as
\beq
\aftw_\mu^{(1,1)} \afbb^\mu = \half \afbb_\mu h^\mn \afbb_\nu.
\label{angcond}
\eeq
We find that the contributions to $\aftw_\mu^{(1,1)}$ 
consistent with this equation and with diffeomorphism covariance 
take the form
\beq
\aftw_\mu^{(1,1)} = 
\half \al h_\mn \afbb^\nu 
- \quar \al \afbb_\mu \hul{\nu}{\nu} + \prt_\mu \Ps
\label{aftw}
\eeq
in harmonic coordinates.
Here,
the constant $\al$
is calculable but varies with the specifics of the theory,
typically being determined 
in terms of the coupling constants 
that control the nonminimal couplings. 
The function $\Ps$ contains effects 
proportional to $h_\mn$ and $\abs_\mu$
that are unphysical by virtue of the discussion 
in Sec.\ \ref{Field redefinitions},
so it is disregarded in what follows.

At this stage,
the result \rf{aftw}
can be combined with the modified Einstein equation
and the trajectory equation
to determine the contributions to $\hlv_\mn$
proportional to $\abs_\mu$.
Working at \pno{3},
we find these contributions can be written
in harmonic gauge as
\bea
(\hlv_a)_{00} &=& \fr 2m [2 \al \abs_0 U 
+ \al \abs_j V^j
- \al \abs_j W^j],
\nonumber\\
(\hlv_a)_{0j} &=& \fr 1m [\al \abs_j U
+ \al \abs_k U^{jk}
\nonumber\\
& &
\quad
- \al \abs_0 V^j - \al \abs_0 W^j],
\nonumber\\
(\hlv_a)_{jk} &=& 
\fr 2m [- \al \abs_0 U \de^{jk}
+ \al \abs_0 U^{jk}].
\label{metrica}
\eea
Paralleling the case of the coefficient field $c_\mn$,
we have kept here
\pno{3} terms in $(\hlv_a)_{0j}$
and \pno{2} terms in $(\hlv_a)_{jk}$
as a convenience for the analysis to follow. 
In Eq.\ \rf{metrica},
$U$ and $V^j$ are the post-newtonian potentials
defined in Eq.\ (\ref{potentialsa}).
Additional potentials $U^{jk}$ and $W^j$ also appear,
defined by
\bea
U^{jk} &=& 
\G \int d^3 x^\prime 
\fr{\rh(\vec{x}^\prime, t) 
(\vec{x}- \vec{x}^\prime)^j 
(\vec{x}- \vec{x}^\prime)^k}
{|\vec{x}- \vec{x}^\prime|^3} ,
\nonumber\\
W^j &=& 
\G \int d^3 x^\prime 
\fr{\rh(\vec{x}^\prime, t) 
v_k (\vec{x}^\prime, t) 
(\vec{x}- \vec{x}^\prime)^j 
(\vec{x}- \vec{x}^\prime)^k} 
{|\vec{x} - \vec{x}^\prime|^3}.
\nonumber\\
\label{potentialsb}
\eea

The results \rf{aftw} and \rf{metrica}
fix the form of the contributions 
involving the coefficient field $\af_\mu$
to the equation of motion \rf{oogeo}.
Modifications of the trajectory of a test body T
arising from nonzero values 
of the coefficients $\abs_\mu$ and $\abt_\mu$
can therefore be studied at third post-newtonian order.
The resulting experimental signals
are discussed starting in Sec.\ \ref{experiment}.

\section{Example: bumblebee model}
\label{bumblebee}

In this section,
we examine a specific model
and demonstrate how it fits 
into the general theory developed above.
This discussion is included solely for illustrative purposes
and is inessential to the development of the paper.
In particular,
the analyses of experimental signals in subsequent sections 
are independent of this specific model,
so the reader can proceed directly 
to Sec.\ \ref{experiment} if desired.

\subsection{Bumblebee model}
\label{Bumblebee model}

Bumblebee models are theories
in which spontaneous Lorentz violation
is induced by a potential $V(B^\mu)$
for a vector field $B^\mu$ 
\cite{ksbb}.
As an illustration
of the general theoretical treatment 
of Sec.\ \ref{classicaltheory},
we consider here 
a specific and comparatively simple bumblebee model
and study its matter-gravity couplings.
A discussion of generic models 
of vacuum-valued vectors coupled to gravity
including references to the substantial early literature 
can be found in Sec.\ III A of Ref.\ \cite{lvmm},
while some more recent papers are listed in Ref.\ \cite{bumblebee}.
Discussions of various stability issues 
with these models are given in Ref.\ \cite{stability}.

The action $S_B$ for the specific bumblebee model 
of interest here can be written as 
\bea
S_B &=& S_G + S_{Bu} + S_B^\prime
\nonumber\\
&=&
\int d^4 x ~e\cl_G
+ \int d\ta ~\cl_{Bu}
+ \int d^4 x ~e\cl_B^\prime.
\nonumber\\
\label{SBaction}
\eea
The form of this action corresponds 
to that of the general action \rf{classicalaction}.
The term $S_G$ is the usual Einstein-Hilbert action \rf{Ract},
with cosmological constant chosen as $\La=0$
for this illustrative case.
The term $S_{Bu}$ represents the matter-bumblebee coupling,
while $S_B^\prime$ contains the bumblebee dynamics,
including the potential $V$ 
triggering spontaneous Lorentz violation.

For the classical lagrangian $\cl_{Bu}$
describing the matter-bumblebee coupling, 
we choose the expression
\bea
\cl_{Bu} &=&
- m \sqrt{-(g_\mn + 2\ztw B_\mu B_\nu) u^\mu u^\nu}
+ \zon B_\mu u^\mu .
\nonumber\\
\label{bbpp}
\eea
Here,
$\zon$ and $\ztw$ are coupling constants
that can vary with the particle species.
Where needed in what follows,
we distinguish the coupling constants 
for a source body S and a test particle T
by superscripts:
$\zes$, $\zet$, $\zts$, $\ztt$.
Note that the lagrangian \rf{bbpp} could be viewed
as the point-particle limit of a quantum field theory,
in parallel with the derivation 
for the general theory \rf{actionsec}.

For the Lagrange density $e\cl_B^\prime$
determining the dynamics of the bumblebee field,
we take 
\bea
e\cl_B^\prime &=& 
- \quar eB^\mn B_\mn - eV
+ \xc eB^\mu B^\nu R_\mn,
\label{bbmodel}
\eea
where
the field strength is
$B_\mn = \prt_\mu B_\nu - \prt_\nu B_\mu$.
The coupling constant $\xc$ 
is sometimes written $\xc = \xi/2\ka$ in the literature
\cite{akgrav,lvmm}.
The potential $V$ has the form
\beq
V = V(B^\mu B_\mu \pm b^2),
\eeq
where $b^2$ is a real number.
Where a definite form is needed 
in the calculations to follow,
we adopt for simplicity
the smooth quadratic potential 
\beq
V = \la (B^\mu B_\mu \pm b^2)^2/2.
\label{smooth}
\eeq
In any event,
the potential is assumed to induce 
a nonzero vacuum expectation value for the bumblebee field,
which we denote by
$b_\mu \equiv \vev{B_\mu}$
following standard usage,
where $b^\mu b_\mu= \mp b^2$.
Denoting the bumblebee fluctuation about the vacuum value
by $\Btw_\mu$,
we can expand
\beq 
B_\mu = b_\mu + \Btw_\mu
\label{Bexp}
\eeq
in parallel with Eq.\ \rf{texpand}.

A match can be made between the bumblebee action \rf{SBaction}
and the general action \rf{classicalaction}
by identifying the various coefficient fields for Lorentz violation 
with specific combinations of the bumblebee field.
The term $\cl_{Bu}$ corresponds to nonzero
coefficient fields $\af_\mu$ and $c_\mn$,
given by  
\bea
\af_\mu &=& \zon B_\mu,
\nonumber \\
c_\mn &=& \ztw (B_\mu B_\nu - \quar g_\mn B_\al B^\al).
\label{acmatch}
\eea
It is also necessary to introduce
an additional scalar field $k$,
defined as 
\beq
k = \half \ztw B_\al B^\al,
\label{kmatch}
\eeq
which normally can be disregarded in the SME context
because it is Lorentz invariant.
In the presence of $k$,
the general action \rf{actionsec}
is slightly modified,
with lagrangian now given by the expression 
\bea
\cl_u &=& - 
m \sqrt{-(g_\mn + k g_\mn + 2 c_\mn) u^\mu u^\nu}
+ \af_\mu u^\mu .
\nonumber\\
\label{bbppu}
\eea
The term $\cl_B^\prime$ yields 
nonzero coefficient fields $s^\mn$ and $u$
in the pure-gravity sector,
given by
\cite{qbakpn}
\bea
s^\mn &=& \xi B^\mu B^\nu 
- \frac 1 4 \xi B_\al B^\al g^{\mn},
\nonumber\\ 
u &=& \frac 1 4 \xi B_\al B^\al.
\label{sumatch}
\eea

Note that in this model only a single field $B_\mu$
underlies all the coefficient fields
$\af_\mu$, $c_\mn$, $k$, $s_\mn$, and $u$.
It follows that searches for Lorentz violation
that are sensitive to any one of these coefficient fields
could provide information constraining the others,
at least in part. 
This special feature of the bumblebee model
may not extend to models with more complicated field structure.
Note also that the coefficient field $k$ in the matter sector
is a species-dependent analogue
of the coefficient field $u$ in the gravity sector.
A nonzero value of $k$ can introduce apparent WEP violations.
Since these are Lorentz invariant,
the resulting phenomenology 
lacks the various time dependences that characterize 
WEP violations resulting from Lorentz breaking.

\subsection{Solving the Model}
\label{Solving the Model}

Given the action \rf{SBaction},
we can illustrate by direct calculation
the correspondence between results from the bumblebee model
and ones from the general SME-based approach
developed in Sec.\ \ref{classicaltheory}.
For this purpose,
it suffices to work at lowest nontrivial order
in the couplings $\zon$, $\ztw$, and $\xc$.
We focus here on
observable effects arising from the identifications
of $\af_\mu$, $c_\mn$, and $k$
in Eqs.\ \rf{acmatch} and \rf{kmatch}.
Observable effects involving $s^\mn$ and $u$
as defined in Eq.\ \rf{sumatch}
are studied in Ref.\ \cite{qbakpn}.

The basic goal is to predict effects
such as trajectory deviations
for given values of $\afb_\mu$ and $\cb_\mn$.
As discussed in Sec.\ \ref{Fluctuations and af},
observability of $\afb_\mu$ involves nonminimal couplings,
so in the present context 
we can expect dominant effects from $\afb_\mu$
to be proportional to the product $\zon \xc$.
In contrast, 
dominant observable effects from $\cb_\mn$ 
are generated directly from $\ztw$.

We remark in passing that
the special bumblebee model considered here
is experimentally viable
provided the sizes of $\zon$ and $\ztw$
are compatible with existing constraints
on long-range spin-independent forces
\cite{scwga}.
The proportionality of $\afb_\mu$ to $\zon \xc$
implies the model can yield Lorentz-violating effects 
involving large $b_\mu$ 
that are detectable only in gravitational experiments 
\cite{akjt}. 

In this subsection,
working at the appropriate perturbative order
and taking the newtonian limit where useful,
we obtain and solve 
the bumblebee equation of motion
and the modified Einstein equation.
These results suffice to determine
the trajectory equation for a test particle 
in terms of the vacuum value $b_\mu$ of the bumblebee model.
Comparison to the general SME-based approach
developed in Sec.\ \ref{classicaltheory}
yields an explicit match 
for $\afb_\mu$, $\cb_\mn$, $\overline{k}$
in terms of the couplings $\zon$, $\ztw$, $\xc$
and the vacuum value $b_\mu$.

\subsubsection{Bumblebee equation}
\label{Bumblebee equation}

Consider first the equation of motion for the bumblebee field,
which follows from varying the action $S_B$.
At the perturbative order of interest,
this equation takes the form
\bea
\prt^\mu B_\mn &=& 
2 V^\prime b_\nu - 2\xc b^\mu R_\mn 
+ \zes j_\nu + \ldots ,
\label{bbeqmot}
\eea
where
\beq
j_\nu = \int d \ta ~u_\nu \de^4 (x- x^\prime )
\eeq
is the source 4-current at the relevant order.
In Eq.\ \rf{bbeqmot},
the prime on $V$ denotes a derivative with respect to the argument,
while the ellipsis indicates that source terms 
proportional to $\zts$ exist 
but provide no observable contributions 
to the order at which we work.
Adopting the smooth quadratic potential \rf{smooth}
and the expansion \rf{Bexp},
the bumblebee equation can be written 
\bea
(\et_\mn\Box  - \prt_\mu \prt_\nu 
- 4 \la b_\mu b_\nu ) \Btw^\mu 
=&& 
\nonumber \\
&&
\hskip -100pt
- 2 \la b_\nu b^\al b^\be h_{\al\be}
- 2\xc b^\al R_{\al\nu}
+ \zes j_\nu .
\hskip 30pt
\label{bbfe2}
\eea
The idea is to solve this expression 
for the fluctuations $\Btw^\mu$
so they can be eliminated from the analysis
as needed.

The solution can be obtained in momentum space
with the propagator chosen as a suitable Green function 
\cite{qbakpn}.
The appropriate bumblebee propagator is
\bea
K^\mn (p) &=& -\fr {\et^\mn}{p^2} 
+ \fr {(b^\mu p^\nu + b^\nu p^\mu)}{p^2 b_\al p^\al}
- \fr {(4 \la b^\al b_\al + p^2) p^\nu p^\mu}
{4 \la p^2 (b_\al p^\al)^2 }
\nonumber\\
\label{prop}
\eea
in momentum space.
Note that the additional poles 
in this and following expressions
can be understood as a consequence of 
residual gauge freedom
to the order at which we work
\cite{gl}.
Using this propagator,
we find the solution
\beq
\Btw_\mu (p) = 
\fr {p_\mu b^\al b^\be h_{\al\be}}{2 b^\al p_\al}
+ (\Btw_{\xc})_\mu (p)
+ (\Btw_{\zon})_\mu (p)
+ (\Btw_{\ztw})_\mu (p),
\label{btwp}
\eeq
where 
$(\Btw_{\xc})_\mu (p)$, 
$(\Btw_{\zon})_\mu (p)$, and 
$(\Btw_{\ztw})_\mu (p)$
are contributions to 
$\Btw_\mu (p)$
proportional to 
$\xc$, $\zes$, and $\zts$
respectively.
In the limit of vanishing $\zes$ and $\zts$,
the solution \rf{btwp} 
reduces to the known result 
\cite{qbakpn}
once the conversion from $\Btw_\mu$ to $\widetilde B^\mu$
described in Sec.\ \ref{Linearization}
is implemented.

Explicitly,
the quantity $(\Btw_{\xc})_\mu$ is given in momentum space by
\cite{qbakpn}
\bea
(\Btw_{\xc})_\mu (p) &=&
- \fr {\xc b_\mu R}{p^2} 
+ \fr {\xc p_\mu R}{4 \la b^\al p_\al} 
+ \fr {\xc p_\mu b^\al b_\al R}{p^2 b^\al p_\al}
\nonumber\\  
&&
+ \fr {2 \xc b^\al R_{\al\mu}}{p^2} 
- \fr {2\xc p_\mu b^\al b^\be R_{\al\be}}{p^2 b^\al p_\al}.
\label{bxi}
\eea
For the piece proportional to $\zon$,
we obtain
\bea
(\Btw_{\zon})_\mu (p) &=&
- \fr {\zes \vf_\mu}{p^2}
+ \fr {\zes b_\mu p^\nu \vf_\nu}{p^2 p_\al b^\al}
+ \fr {\zes p_\mu b^\nu \vf_\nu}{p^2 p_\al b^\al}
\nonumber\\
& & 
- \fr {\zes b_\al b^\al p_\mu p^\nu \vf_\nu}{p^2 (p_\be b^\be)^2}
- \fr {\zes p_\mu p^\nu \vf_\nu}{4 \la (p_\al b^\al)^2}.
\label{btwze}
\eea
The remaining term in Eq.\ \rf{btwp},
which contains contributions proportional to $\zts$,
is irrelevant to the order at which we work.

We emphasize that the explicit solution \rf{btwp}
for the bumblebee fluctuation $\Btw_\mu$
is obtained by direct calculation from the action $S_B$.
This calculation depends 
on knowledge of the bumblebee dynamics 
as described by the Lagrange density \rf{bbmodel}.
In contrast,
the general SME-based method 
presented in Sec.\ \ref{classicaltheory}
to obtain an arbitrary coefficient fluctuation $\Xtw$
replaces the need for complete knowledge of $S_B$ 
with the judicious use
of perturbation theory, 
the NG constraint,
diffeomorphism invariance,
and Newton's third law.

\subsubsection{Modified Einstein equation}
\label{Modified Einstein equation}

Varying the action with respect to the metric
yields the modified Einstein equation.
At leading order in $\Btw_\mu$ and lowest order in $h_\mn$,
this equation takes the form
\bea
G_\mn &=& 2\xc \ka [b^\al \prt_\al \prt_{(\mu} B_{\nu)} 
+ b_{(\mu} \prt_\al \prt_{\nu)} B^\al
- b_{(\mu} \Box B_{\nu)}
\nonumber \\
& & 
\hskip 20pt
- \eta_\mn b^\al \prt_\al \prt_\be B^\be]
+ 2 \ka V^\prime b_\mu b_\nu
+ \ka (T_{Bu})_\mn ,
\nonumber\\
\label{bbein}
\eea
where contributions from $B_\mu$ 
are understood to be limited 
to the appropriate perturbative order.
Note that $V^\prime$ contributes at most through massive modes
at this order,
so it plays no role in the present context. 

The matter-sector contribution $(T_{Bu})_\mn$ 
to the energy-momentum tensor can be written
\beq
(T_{Bu})_{\mn}= 
- \int d \ta 
\fr{m u_\mu u_\nu \de^4 (x - x^\prime (\ta))}
{\sqrt{1 - 2 \ztw^S b_\al b_\be u^\al u^\be}}.
\label{bbenergy}
\eeq
This explicit expression is the bumblebee analogue
of the general form \rf{cenergy}.
As expected,
terms proportional to $\zes$
are absent from Eq.\ \rf{bbenergy},
confirming that minimal couplings 
cannot generate Lorentz violation 
of the $\afb_\mu$ type
in the modified Einstein equation.
In this model,
the only nonzero observables proportional to $\zes$ 
arise through the bumblebee fluctuations \rf{btwze}.

Inserting the solution for $B_\mu$
at the appropriate order,
the modified Einstein equation \rf{bbein} can be solved.
To match the analysis in the general SME-based method 
of Sec.\ \ref{classicaltheory}
the harmonic gauge must be used.
At zeroth order in Lorentz violation,
the conventional metric is reproduced.
For simplicity in this illustrative model,
we limit consideration at the next order
to the newtonian limit for $\Btw_\mu$.
This avoids possible complications
from the residual gauge invariance,
while permitting a complete match 
to the results of the SME method.

The solution for the metric
can be constructed directly 
from the trace-reversed form
of the modified Einstein equation \rf{bbein}.
Using the NG condition
\beq
b_\mu \Btw^\mu = \half b^\mu b^\nu h_\mn
\eeq
and the bumblebee equation \rf{bbfe2},
we can write 
\beq
R_\mn = 
2\xc \ka
[b^\al \prt_\al \prt_{(\mu} B_{\nu)}
-\zes b_{(\mu} \vf_{\nu)}] 
+ \ka (S_{Bu})_\mn,
\eeq
where $(S_{Bu})_\mn$ is the trace-reversed version
of the energy-momentum tensor \rf{bbenergy}.
In the newtonian limit,
the first term on the right-hand side
is higher-order in time derivatives
and so is negligible.

Expanding $(S_{Bu})_\mn$ to the appropriate order,
we find that the O(1,1) modifications 
of the metric fluctuation $h_{00}$
are given in terms of the bumblebee vacuum value $b_\mu$ by
\beq
\hlv_{00} = 
(4 \xc \zes + \zts m b_0)
\fr{2 \G b_0}{r} .
\label{bbmetric}
\eeq
The first term arises from the bumblebee fluctuations
via the nonminimal couplings
and is the bumblebee analogue of Eq.\ \rf{metrica}.
The second term arises directly
from the energy-momentum tensor $(T_{Bu})_\mn$ 
of the source S
and corresponds to Eq.\ \rf{metricc}.
To complete the match to the general SME analysis
of Sec.\ \ref{classicaltheory},
it remains to apply this result to determine
the deviations from geodesic motion
of a test particle T.

\subsubsection{Particle trajectory}
\label{Particle trajectory bumblebee}

The equation of motion for a test point particle T
in the presence of the bumblebee field $B_\mu$ 
and the metric $h_\mn$
can be obtained by varying the action $S_{Bu}$
with respect to $x^\mu$.
At leading order in the fluctuations,
this yields
\beq
\ddot{x}^\mu = 
- \Ga_{(0,1) \pt{\mu} \al \be}^{ \pt{(0,1)} \mu} u^\al u^\be 
+ \ddot{x}_{\zon}^\mu
+ \ddot{x}_{\ztw}^\mu, 
\label{bbgeo}
\eeq
where $\Ga_{(0,1) \pt{\mu} \al \be}^{\pt{(0,0)} \mu}$
is the linearized Christoffel symbol.
The terms
$\ddot{x}_{\zon}^\mu$ and $\ddot{x}_{\ztw}^\mu$
represent contributions to $\ddot{x}^\mu$ 
proportional to $\zet$ and $\ztt$,
respectively. 
The above equation is the bumblebee analogue
of the equation of motion \rf{oogeo}
obtained for the general SME analysis 
in Sec.\ \ref{classicaltheory}.

The explicit form of the quantity $\ddot{x}_{\zon}^\mu$ 
takes the form
\beq
\ddot{x}_{\zon}^\mu = 
- (\Ga_{\zon})_{\pt{\mu} \al \be}^{\mu} u^\al u^\be 
- \fr{\zet}{\mt} g^\mn
(\prt_\nu \Btw_\al - \prt_\al \Btw_\nu) u^\al,
\label{zegeo}
\eeq
where $(\Ga_{\zon})_{\pt{\mu} \al \be}^{\mu}$
contains terms proportional to $\zes$
that enter via Lorentz-violating corrections to the metric.
The piece of $\Btw_\mu$
contributing to this equation 
at the relevant order can be written as 
\beq
(\Btw_{\xc})_\mu = 
\xc h_\mn b^\nu
- \half \xc b_\mu h_\al^{\pt{\al} \al} + \ldots
\label{bbxi}
\eeq
in harmonic coordinates,
where the ellipsis represents terms
that play no role within our approximations.
The latter equation is the bumblebee analogue
of the SME result \rf{aftw}
for the coefficient fluctuation $\aftw_\mu$,
and as expected it yields contributions 
to the trajectory equation
proportional to the product $\xc \zet$.

The quantity $\ddot{x}_{\ztw}^\mu$ can be written
\bea
\ddot{x}_{\ztw}^\mu &=& 
- (\Ga_{\ztw})_{\pt{\mu} \al \be}^{\mu} u^\al u^\be
+ 2 \ztt \et^\mn b_\nu b_\la 
\Ga_{(0,n) \pt{\la} \al \be}^{\pt{(0,n)} \la} u^\al u^\be
\nonumber \\
& &
+ 2 \ztt b_\al b_\be 
\Ga_{(0,1) \pt{\al} \nu \la}^{\pt{(0,1)} \al} 
u^\be u^\mu u^\nu u^\la
+ 2 \ztt b_\al \prt^\mu \Btw_\be u^\al u^\be
\nonumber \\
& &
- 2 \ztt \et^\mn (b_\nu \prt_\al \Btw_\be 
+ b_\be \prt_\al \Btw_\nu) u^\al u^\be
\nonumber \\
& &
- 2 \ztt b_\al \prt_\nu \Btw_\be u^\al u^\be u^\mu u^\nu,
\label{etgeo}
\eea
where $(\Ga_{\ztw})_{\pt{\mu} \al \be}^{\mu}$
contains terms proportional to $\zts$
that enter via Lorentz-violating corrections to the metric.
In this equation,
only the first term in Eq.\ \rf{btwp}
produces a relevant contribution
to the fluctuation $\Btw_\mu$
in the present context,
which matches the SME result \rf{ctwres}
for the coefficient fluctuation $\ctw_\mn$.

At this stage,
we can verify the conservation
of total 4-momentum of the system
of the source S and test body T,
as described in Sec.\ \ref{Methodology}.
Substituting for $\Btw_\mu$ and $\hlv_{00}$
in the trajectory equation
reveals the antisymmetry under interchange
of S and T required to satisfy Newton's third law. 
We can also complete
the correspondence between the bumblebee model
and the general SME-based analysis  
of Sec.\ \ref{classicaltheory}
by making the identifications
\bea
\al &=& 2\xc,
\qquad
\afb_\mu = \zon b_\mu ,
\nonumber\\
\cb_\mn &=& \ztw b_\mu b_\nu + \quar \ztw \et_\mn b^2,
\qquad
\kb = \half \ztw b^2,
\label{match}
\eea
which can be obtained 
by matching Eqs.\ \rf{bbmetric} and \rf{bbxi}
to the SME results 
\rf{metricc}, 
\rf{metrica}, 
and \rf{aftw}.

\section{Experimental basics}
\label{experiment}

In the remainder of this paper,
we apply the theoretical framework developed above 
to explore some experimental prospects 
for detecting Lorentz violation
through matter-gravity couplings.
As before, 
we adopt coordinates satisfying the condition \rf{coordchoice},
which produces simplified expressions
without the photon-sector coefficients 
$\kfb^\al_{\pt{\al} \mu \al \nu}$.
The primary focus is on signals involving the coefficients 
$\afbw_\mu$ and $\cbw_\mn$. 
Certain effects associated with the coefficient $\sb_\mn$
in the pure-gravity sector are also considered.

In the present section,
we provide some basic information broadly applicable 
to searches for Lorentz violation,
including an outline of frame conventions
and a discussion of sensitivities
to coefficient combinations.
Each subsequent section
addresses a particular class of experimental searches.
Section \ref{labtests} 
examines tests with ordinary neutral matter 
in Earth-based laboratories,
while Sec.\ \ref{space} 
studies satellite-based searches 
with ordinary matter.
Section \ref{exotic}
considers more exotic laboratory and satellite-based tests,
including ones using 
charged particles, 
antimatter, 
and particles beyond the first generation.
Section \ref{Solar-system tests}
addresses solar-system observations,
including lunar and satellite ranging
and measurements of perihelion precession. 
Finally,
Sec.\ \ref{photon}
considers signals from photon-gravity couplings.

\subsection{Frames}
\label{Frames}

A substantial advantage of the SME framework 
is the ability to compare signals for Lorentz violation
across a wide variety of experiments and observations.
To facilitate these comparisons,
it is useful to report search results 
in a canonical inertial frame.

In Minkowski spacetime,
the canonical frame 
is a Sun-centered celestial-equatorial frame 
\cite{km},
which is approximately inertial
over the time scales of most searches.
In this frame,
the $Z$ axis is aligned with the rotation axis of the Earth,
while the $X$ axis points from the Earth to the Sun
at the vernal equinox.
The origin of the time coordinate $T$ 
is the time when the Earth crosses the Sun-centered $X$ axis
at the vernal equinox. 

For post-newtonian investigations 
involving gravitational effects in the solar system,
the canonical frame 
is identified with an asymptotically Minkowski frame
that is comoving with the rest frame of the solar system
and that coincides with the canonical Sun-centered frame
\cite{qbakpn}.
In this Sun-centered frame,
cartesian coordinates are denoted by
\beq
x^\Xi = (T,X^J) = (T,X,Y,Z)
\eeq
and are labeled with capital Greek indices.
Also,
we write  
\beq
{\bf e}_{\Xi} = ({\bf e}_{T}, {\bf e}_{J})
\eeq
for the corresponding coordinate basis vectors.

Various types of observers appear in the analyses below,
including ones at rest 
in an Earth-centered frame,
in a laboratory frame,
in a satellite frame, 
and others. 
The corresponding frames 
are specified as needed in the sections that follow.
Among the sets of basis vectors having generic applicability
are one for an observer
at rest in the Sun-centered frame
and a related one for an observer 
in uniform motion relative to the Sun-centered frame.
We summarize these two sets briefly here.

For an observer at rest 
at the point $(T,\vec X)$
in the Sun-centered frame,
$dX^J/dT=0$.
Suitable basis vectors are denoted as ${\bf e}_{\mu}$
with $\mu = (t, j)$,
and they can be written as
\cite{qbakpn}
\bea
{\bf e}_{t} &=& 
\de_{\pt{T}t}^{T} [ 1 + \frac 12 h_{TT} (T, \vec X) 
+ {\rm{\pno{4}}}]
{\bf e}_{T} ,
\nonumber\\
{\bf e}_{j} &=& 
\de^J_{\pt{J}j} [ {\bf e}_{J}
-\frac 12 h_J^{\pt{J}K} (T, \vec X) {\bf e}_{K} ] 
+ \de^J_{\pt{J}j} h_{TJ} (T, \vec X) {\bf e}_{T} .
\nonumber\\
\label{restobserver}
\eea
This basis is orthonormal.

If the observer is in motion
with four-velocity $u^\Xi$ in the Sun-centered frame,
then an appropriate set of basis vectors
can be taken as ${\bf e}_{\hat \mu}$
with $\hat\mu = (\hat t, \hat j)$,
where the components $({\bf e}_{\hat t})^\Xi$
are identified with the four-velocity,
$({\bf e}_{\hat t})^\Xi = u^\Xi$.
This basis is given by
\cite{qbakpn}
\bea
{\bf e}_{\hat t} &=& 
\de^{t}_{\pt{t}\hat t} (1 + \frac 12 v^2) {\bf e}_t 
+ v^j {\bf e}_j,
\nonumber\\
{\bf e}_{\hat j} &=& 
\de^j_{\pt{j}\hat j} v^k R_{kj} {\bf e}_t 
+ \de^j_{\pt{j}\hat j} 
(\de^{kl} + \frac 12 v^k v^l ) R_{lj} {\bf e}_k,
\label{moveobserver}  
\eea
where $v^j$ is the coordinate velocity 
of the observer in the frame \rf{restobserver},
and $R_{jk}$ implements the appropriate rotation.
This basis is also orthonormal.

In applying the above equations,
the relevant contributions
to the metric fluctuation
and the observer velocity
can be obtained 
from the modified Einstein equations
and from the equation of motion of the observer.
Note that the results typically depend on 
coefficients for Lorentz violation.
Also,
some simplifying assumptions can usually be adopted
without loss of generality.
For example,
in certain laboratory experiments
the contributions to the metric fluctuation $h_{\Xi \La}$
sourced by the energy-momentum tensor of the Sun 
can safely be neglected.

We remark in passing that the above definition
of the Sun-centered frame
could be sharpened in various ways,
such as
allowing for the precession and nutation of the Earth,
establishing the vernal equinox via the centroids of bodies,
and incorporating the motion of the Sun 
with respect to the center of the solar system.
Some of these effects may 
allow additional sensitivities to Lorentz violation
via the resulting time dependence of the standard frame.
Note also that 
the notion of parallelism
used in the Minkowski-spacetime definition 
of the Sun-centered frame
is inapplicable in the context of curved spacetime.  
One way to address this latter issue
is to define the $Z$ axis
so that ${\bf e}_\zhat$ aligns with the spin axis of the Earth
after Eqs.\ \rf{restobserver} and \rf{moveobserver}
with $R_{jk} =0$ and the appropriate velocity are applied.
For the various searches considered in this work,
the standard definition of the Sun-centered frame suffices.
A more complete investigation of these issues 
is of potential interest but lies beyond our present scope.

\subsection{Sensitivities}
\label{Sensitivities}

In the following sections,
we consider the observational effects 
of the coefficients 
$\afbw_\mu$, 
$\cbw_\mn$
in the matter sector
and $\sb_\mn$
in the pure-gravity sector.
This subsection offers some comments 
about attainable sensitivities
to these coefficients.

Measurement of the coefficients $\afbw_\mu$ 
is of particular interest
because they are virtually unexplored to date.
The existence of the field redefinitions
described in Sec.\ \ref{Field redefinitions}
means that observation of effects from $\afbw_\mu$ 
requires either flavor-changing physics
or gravitational couplings.
At the level of quarks,
the flavor-changing weak interactions
have been used to provide access to observables involving 
differences of two coefficients $\abw_\mu$ 
with $w$ including second- and third-generation quarks
\cite{mesons,akmesons}.
Flavor oscillations can also be used to constrain 
the coefficients $\ab_\mu$ in the neutrino sector,
where they form $3\times 3$ matrices in flavor space
\cite{neutrinos}.
However,
to date gravitational couplings have been used to obtain
sensitivity only to limited combinations
of the 12 independent components of the SME coefficients 
$\abe_\mu$, 
$\abp_\mu$, 
$\abn_\mu$ 
for electrons, protons, and neutrons
\cite{akjt,gravexpt4}.
These coefficients are otherwise unconstrained
and could be comparatively large,
so they offer interesting prospects for further investigation
in gravitational tests. 

In the present context,
we can extend the single bound on $\abw_\mu$ 
given in Ref.\ \cite{akjt}
by taking advantage of the result 
of Sec.\ \ref{Field redefinitions}
that $\abw_\mu$ always appears at leading order 
with $(\eb^w)_\mu$ in the combination $\afbw_\mu$ 
given by Eq.\ \rf{aeff}.
Using this result 
immediately yields a constraint 
on three of the independent components of $\afbw_\mu$
for electrons, protons, and neutrons,
given as
\beq
|\al \afbe_T + \al \afbp_T - 0.8 \al \afbn_T|
< 1 \times 10^{-11} \ {\rm GeV}
\label{aeffprl}
\eeq
at the 90\% confidence level.

In contrast,
many of the coefficients $\cbw_\mn$ are readily observable
in nongravitational experiments.
Nonetheless,
gravitational tests offer additional opportunities
to achieve sensitivities to $\cbw_\mn$,
including some components 
that are unmeasured to date. 
For electrons, protons, and neutrons,
there are 27 independent observable symmetric coefficients 
$\cbw_\mn$.
A compilation of existing limits 
on $\cbw_\mn$ for different flavors $w$
is given in Ref.\ \cite{tables}. 

The coefficients $\sb_\mn$ 
lie in the pure-gravity sector of the minimal SME
and therefore can be measured only 
in the gravitational context.
The corresponding post-newtonian corrections 
to the gravitational field are known
\cite{qbakpn,qbgem}.
Constraints on most of the nine independent components 
of $\sb_\mn$ 
have been obtained using a variety of techniques,
including among others 
perihelion-precession studies, 
lunar laser ranging,
atom-interferometer gravimetry,
and laboratory and space-based experiments
\cite{qbakpn,gravexpt1,gravexpt2,gravexpt3}.
All these analyses disregard matter effects.
In this work,
we show that Lorentz violation in the matter sector 
can contribute in different ways to signals 
involving the coefficients $\sb_\mn$.

For all the coefficients
$\afbw_\mu$, $\cbw_\mn$, $\sb_\mn$,
the effects of interest here involve 
gravitational couplings to matter.
It is therefore reasonable to expect
that the best sensitivities to Lorentz violation
are associated with couplings to dominant gravitational effects.
This suggests that tests
with high sensitivity to Newton gravity
are of particular interest.
As described in Sec.\ \ref {Equivalence principle},
the flavor dependence of the coefficients for Lorentz violation
implies that WEP tests also lie in this category. 

Many of the signals sought in gravity tests
require ancillary measurements of time and distance.
These typically involve matter in some form,
and they may introduce additional Lorentz-violating effects
beyond those comprising the direct signal of interest.
However,
most of these additional effects are negligible
in the present context
because the corresponding coefficients
are tightly constrained
via tests in Minkowski spacetime
\cite{tables},
whereas sensitivities in gravitational tests
are typically substantially reduced 
by the weak gravitational field.
Among the coefficients of interest in the present work,
this issue is relevant only to $\cbw_\mn$
because $\afbw_\mu$ and $\sb_\mn$
are unobservable in Minkowski-spacetime tests
and because we adopt the coordinate choice \rf{coordchoice}
making unobservable the photon-sector coefficients 
$\kfb^\al_{\pt{\al} \mu \al \nu}$.
Among the coefficients $\cbw_\mn$ for ordinary matter,
the neutron-sector coefficients $\cbn_\mn$
are the least well constrained at present.
Their effects may therefore be important 
for certain tests,
in which case a detailed analysis of the measurement method
may be necessary.

Another consideration relevant for identifying sensitivities
in tests with atoms or bulk matter
is the role of the contributions from binding energy.
In some cases,
accounting for these contributions
can disentangle effects from different coefficients,
thereby producing additional independent sensitivities.
This can occur when coefficients from two or more 
sectors are involved,
either directly within a WEP test 
or indirectly via comparison of results 
obtained for different bodies.
In the remainder of this subsection,
we discuss this possibility for the coefficients
$\afbw_\mu$ and $\cbw_\mn$ in turn.

Consider first combinations of the coefficients $\afbw_\mu$.
Following the discussion in Sec.\ \ref{Test and source bodies},
a body B has an effective coefficient $\afB_\mu$
given by Eqs.\ \rf{abody} and \rf{zeroaprime}.
The dimensionless quantity relevant for a test
is $\afbb_\mu/m^B$,
and comparisons involving two bodies
therefore appear as the difference
of two quantities of this form.
For two neutral bodies 
involving bound electrons, protons, and neutrons,
this difference can be expanded as follows:
\bea
\sum_w \left( \fr{N^w_1}{m_1} - \fr{N^w_2}{m_2} \right) \afbw_\mu 
&=& 
\nonumber \\
&& 
\hskip-40pt
\fr{N_1^p N_2^n - N_1^n N_2^p}{m_1 m_2}  m^n \afbx{\pen}_\mu
\nonumber \\
&& 
\hskip-40pt
+\fr{N_1^p m_2^\prime - N_2^p m_1^\prime}{m_1 m_2} 
\afbx{\epcombo}_\mu 
\nonumber \\
&& 
\hskip-40pt
+\fr{N_1^n m_2^\prime - N_2^n m_1^\prime}{m_1 m_2} \afbn_\mu.
\label{expbinding}
\eea
Here,
the numbers of particles of species $w$
for the two bodies are $N^w_1$, $N^w_2$,
and $m_1^\prime$, $m_2^\prime$ 
are the binding-energy contributions 
to the masses $m_1$, $m_2$ of the two bodies,
as defined in Eq.\ \rf{bodymass}.
Also, we define 
\bea
\afbx{\epcombo}_\mu &=& 
\afbe_\mu + \afbp_\mu,
\nonumber\\
\afbx{\pen}_\mu &=& 
\afbx{\epcombo}_\mu 
- \fr{m^e + m^p}{m^n} \afbn_\mu .
\label{afbpen}
\eea
When the contributions from binding energy 
are neglected in Eq.\ \rf{expbinding},
the linear combination $\afbx{\pen}_\mu$ 
of coefficients becomes the sole observable
involving $\afbw_\mu$ 
in a comparison of two bodies,
with the effect scaled by their difference in species content.
However,
incorporating the binding energy in the analysis
introduces the last two terms in Eq.\ \rf{expbinding},
revealing that the effects of 
$\afbe_\mu + \afbp_\mu$ and $\afbn_\mu$ 
vary differently with the content of the bodies.
This allows the possibility of independent measurements 
of $\afbe_\mu + \afbp_\mu$ and $\afbn_\mu$.
Note that the sensitivity of such measurements is typically
an order of magnitude less than that
of measurements of $\afbx{\pen}_\mu$ 
due to the appearance of ratios of the form $m^\prime/m$.

Next,
consider combinations of the coefficients $\cbw_\mn$.
For a body B,
the effective coefficient $\cuB_\mn$ 
is a dimensionless quantity given by Eq.\ \rf{cbody}.
As discussed in Sec.\ \ref{Test and source bodies},
nonzero Lorentz-violating contributions from the binding energy
given by the coefficients $\cbpr_\mn$
are expected to exist, 
along with the usual binding-energy contributions $m^\prime$ 
to the body mass.
It turns out that these Lorentz-violating contributions
impede the use of binding energy 
to extract additional independent sensitivities
to combinations of the coefficients $\cbw_\mn$.
To see this,
consider two neutral bodies as before,
and expand the analogue of Eq.\ \rf{expbinding}
to get
\bea
\sum_w \left( \fr{N^w_1}{m_1} - \fr{N^w_2}{m_2} \right) 
m^w \cbw_\mn 
&=& 
\nonumber \\
&&
\hskip-80pt
\fr{N_1^p N_2^n - N_1^n N_2^p}{m_1 m_2} m^n m^p \cbx{\pen}_\mn
\nonumber \\
&&
\hskip-80pt
+\fr{N_1^p m_2^\prime - N_2^p m_1^\prime}{m_1 m_2} 
m^p \cbx{\epcombo}_\mn
\nonumber \\
&&
\hskip-80pt
+\fr{N_1^n m_2^\prime - N_2^n m_1^\prime}{m_1 m_2} m^n \cbn_\mn
\nonumber \\
&&
\hskip-80pt
+ (m^e + m^p)
\fr{N_2^p m_1^\prime \cbx{\prime 1}_\mn
- N_1^p m_2^\prime \cbx{\prime 2}_\mn}{m_1 m_2} 
\nonumber \\
&&
\hskip-80pt
+ m^n 
\fr{N_2^n m_1^\prime \cbx{\prime 1}_\mn
- N_1^n m_2^\prime \cbx{\prime 2}_\mn}{m_1 m_2},
\label{expbindingc}
\eea
where we introduce
\bea
\cbx{\epcombo}_\mn &=&
\fr{m^e}{m^p}\cbe_\mn + \cbp_\mn ,
\nonumber\\
\cbx{\pen}_\mn &=& 
\cbx{\epcombo}_\mn 
- \fr{m^e+ m^p}{m^p} \cbn_\mn .
\label{cbpen}
\eea
When binding-energy effects are neglected,
$\cbx{\pen}_\mn$ becomes the only observable 
combination of the coefficients $\cbw_\mn$
in gravitational tests comparing two bodies.
Including the binding-energy terms as in Eq.\ \rf{expbindingc}
shows that bodies with different species content
can exhibit distinct effects.
Although it seems unlikely that nonzero effects
at order $m^\prime/m$ in a variety of bodies 
would cancel sufficiently well to evade detection altogether,
the appearance of the unknown coefficients $\cbpr_\mn$
makes it infeasible at present
to extract unambiguous independent measurements 
on combinations of the coefficients $\cbw_\mn$
using binding-energy effects.

\section{Laboratory tests}
\label{labtests}

This section considers some sensitive laboratory tests
with ordinary neutral bulk matter, neutral atoms,
and neutrons performed on or near the surface of the Earth.
The basic theory for these tests
is developed in Sec.\ \ref{labtheory},
while Secs.\ \ref{ffgravimeter}, \ref{fcgravimeters},
\ref{epfreefall}, and \ref{tpend}
consider signals and sensitivities
attainable in a variety of terrestrial searches. 
More exotic laboratory tests with charged particles,
antimatter, and particles beyond the first generation
are considered in Sec.\ \ref{exotic}.

Terrestrial experiments 
seeking gravitational Lorentz violation using ordinary matter
can be classified either as gravimeter tests or as WEP tests.
In gravimeter tests,
the basic idea is to seek variations 
either in the gravitational force on a test body 
or in its gravitational acceleration.
The corresponding signals
originate in the time dependence 
of laboratory coefficients for Lorentz violation 
induced by the rotation of the apparatus
and the rotation and revolution of the Earth.
These signals can be interpreted as an effective time variation
of the Newton gravitational constant $\G$.
In WEP tests,
the idea is to compare 
either the gravitational force between two different bodies
or their relative gravitational acceleration.
The corresponding signals,
which can be instantaneous or time-varying,
are sensitive to differences between the coefficients
associated with different species of matter.

Lorentz violation can introduce deviations
from Newton's second law,
so the distinction between force and acceleration
can be important.
This distinction implies
the two classes of gravimeter and WEP tests
can each be further subdivided into two categories,
force-comparison tests and free-fall tests.
The basic idea of a free-fall test
is to search for a time or composition dependence
in the gravitational acceleration 
of a freely falling test body by monitoring its motion.
The idea of a force-comparison test
is to balance the gravitational force experienced by a test body
with a second force,
investigating changes in the equilibrium
arising from the time or species dependence
of the laboratory coefficients for Lorentz violation.
The force comparison can be achieved 
either by using a seesaw arrangement to
balance the gravitational forces
on test bodies of different composition,
which constitutes a force-comparison WEP test,
or by using a nongravitational force 
to counter the gravitational force on the test body,
which represents a force-comparison gravimeter test.

We thus have four categories of possible laboratory tests
with ordinary matter.
In what follows,
each is considered in a separate subsection.
Free-fall gravimeter tests,
including searches with freely falling corner cubes
and with atom interferometers, 
are considered in Sec.\ \ref{ffgravimeter}.
Force-comparison gravimeter tests
using mechanical and superconducting gravimeters
are studied in \ref{fcgravimeters}.
Free-fall WEP tests,
which come in a wide variety of forms,
are considered in Sec.\ \ref{epfreefall}.
Finally,
force-comparison WEP tests 
are discussed in Sec.\ \ref{tpend},
with focus on a torsion-balance configuration.

Table \generala\ 
provides a list of some conventions adopted in this section
for the analyses of laboratory tests.
Many of the quantities are self-explanatory.
The laboratory speed $V_L$ 
is due to the rotation of the Earth
and depends on the laboratory colatitude $\ch$.
The relative time $T_\oplus$ 
involves a convenient choice of origin,
measured from any instant 
when the $\hat y$ axis in the laboratory frame
and the $Y$ axis of the Sun-centered frame 
coincide.
Using $T_\oplus$ instead of the canonical time $T$ 
in the Sun-centered frame 
introduces a phase $\ps$ in the analysis.
The angle $\ze$ is defined in terms of
the component accelerations 
$\acc_\xhat$, $\acc_\zhat$ of a test body 
along $\hat x$, $\hat z$ in the laboratory frame,
\beq
\ze = \tan^{-1}({\acc_\xhat}/{\acc_\zhat}).
\label{zetadef}
\eeq
At leading order,
$\ze$ is approximated by 
the ratio of the usual Newton 
centripetal and gravitational accelerations,
$\ze\simeq 10^{-3}$.
It represents the angular deviation from the vertical 
at the location of the laboratory
of a plumb line or of a test body in free fall.

\begin{center}
\begin{tabular}{ll} 
\multicolumn{2}{c}
{Table \generala.\ Notation for laboratory tests.}\\
\hline
\hline
Quantity \pt{aaaaaaaaaa} & Definition \\
\hline
$R$                     & mean Earth-Sun distance\\
$R_{\oplus}$            & mean Earth radius \\
$\Om$                   & mean Earth orbital frequency \\
$\om$                   & mean sidereal frequency \\
$\om_e$                 & apparatus rotation frequency \\
$V_{\oplus} = \Om R$    & mean Earth orbital speed \\
$V_L$                   & laboratory rotational speed \\
$T_{\oplus}$            & relative time \\
$\et$                   & inclination of Earth orbit \\
$\ch$                   & laboratory colatitude \\
$\ps = \om(T_\oplus-T)$ & phase induced by $T_\oplus$\\
$\ze \approx \om^2 R_\oplus /\sin (2\ch) g$   & deviation angle\\ 
\hline
\hline
\end{tabular}
\end{center}

\subsection{Theory}
\label{labtheory}

The relevant observables for laboratory tests
of Lorentz symmetry in gravity 
are the motions of test bodies relative to the Earth
and relative to each other.
These observables can be obtained
from the action for a test body,
evaluated at the appropriate post-newtonian order
and expressed in laboratory coordinates.

Consider the action $S^{\rm B}_u$ for a test body 
given in Eq.\ \rf{actioneq},
with the gravitational field of the Earth acting as the source S.
The corresponding lagrangian $L^{(3)}_{a,c}$ 
describing the motion of the test body T 
at \pno3 
can be constructed by expanding $S^{\rm B}_u$
with B$\equiv$T.
The solution for the metric fluctuation $h_\mn$ at this order
is obtained from the general expressions \rf{pnpot},
\rf{metricc}, and \rf{metrica},
with the Earth treated as a rigid rotating source S 
as described in Sec.\ \ref{Test and source bodies}.
In what follows,
we neglect the gravitational fields of other bodies
such as the Sun,
although in a more detailed treatment
these could be incorporated using similar methods.

For laboratory searches,
it is convenient to begin calculations in an Earth-centered frame
with coordinates denoted by
$x^\muearth = (\tearth,\xearth,\yearth,\zearth)$.
At leading order,
the spatial components of the Earth-centered basis 
are taken to coincide with those of the Sun-centered frame,
and $\tearth = T$.
In the Earth-centered frame,
we find
\bea
L^{(3)}_{a,c} &=& \half \mt \left(1 
+ \cbt_{\tearth \tearth} 
+ 2 \cbt_{\tearth \jearth} v_\jearth \right) v_\kearth v_\kearth
\nonumber \\
&& 
+ \mt \cbt_{\jearth \kearth} v_\jearth v_\kearth
\nonumber \\
&& 
+ \fr{\G \ms \mt}{r} \Big[ 1 
+ \fr{2 \al}{\mt} \abt_\tearth 
+ \fr{2 \al}{\ms} \abs_\tearth
\nonumber \\
&&
\hskip 60pt
+ \fr{\al}{\ms} \abs_\jearth v_\jearth 
\nonumber \\
&&
\hskip 60pt
+ \cbt_{\tearth \tearth} 
+ \cbs_{\tearth \tearth} 
+ 2 \cbt_{(\tearth \jearth)} v_\jearth
\Big]
\nonumber \\
&&
+ \fr{\G \mt}{r^3} \al \abs_\jearth x_\jearth x_\kearth v_\kearth
\nonumber
\\
& & 
+ \fr{\G \ms \mt}{5 r^3} R_\oplus^2
\ep_{\jearth \kearth \learth} \om_\kearth x_\learth
\Big[ \fr{2 \al}{\mt} \abt_\jearth+ \fr{\al}{\ms} \abs_\jearth
\nonumber
\\
& &
\pt{+ \G \ms \mt R_\oplus^2
\ep_{\jearth \kearth \learth} \om_\kearth x_\learth ]}
+ 2 \cbs_{(\tearth \jearth)} \Big],
\label{nonrellagrangiano3}
\eea
where $r = \sqrt{x_\jearth x_\jearth}$.
This expression contains
the conventional Newton kinetic and potential terms
for a test body T moving in the gravitational field of S,
along with a series of corrections
that depend on the coefficients 
$\afbb_{\muearth}$ and $\cbb_{\mnearth}$.
Some of these additional terms are motional,
analogous to centrifugal effects,
and some are gravitational,
including ones analogous to gravitomagnetic effects.
Effects from the Earth's motion about the Sun
are implicitly included  
via the dependence of $\afbb_{\muearth}$ and $\cbb_{\mnearth}$
on the orbital speed $V_\oplus$.
This dependence can be made explicit by expressing the
coefficients in Sun-centered coordinates
instead of Earth-centered ones.

To obtain results applicable to laboratory tests,
the result \rf{nonrellagrangiano3}
must be transformed from the Earth-centered frame
to the laboratory frame.
We denote the laboratory coordinates by $x^\muhat$,
where the spatial coordinates $x^\jhat$
are taken to coincide with the standard SME conventions
for a laboratory on the surface of the Earth
\cite{km}.
In the laboratory,
the $\xhat$ axis points South,
the $\yhat$ axis points East,
and the $\zhat$ axis points towards the local zenith.
To the required post-newtonian order,
$\tearth = \that$ and
the coordinate location of the laboratory
in the Earth-centered frame can be written 
\cite{qbakpn}
\beq
\vec{\xi} = 
R_\oplus (\sin \ch \cos(\om_\oplus T + \ph), 
\sin \ch \sin(\om_\oplus T + \ph), \cos \ch).
\label{vecxi}
\eeq
The transformation between the two sets of spatial coordinates
can therefore be written
\beq
x_\jearth = \xi_\jearth + R_{\jearth \jhat} x_\jhat,
\label{earthlab}
\eeq
where $R_{\jearth \jhat}$
is the relevant rotation
between the bases of the laboratory 
and the Earth-centered frames.
Note that Eq.\ \rf{earthlab} implies the coefficients
$\afbb_{\muhat}$ and $\cbb_{\mnhat}$
in the laboratory frame
acquire implicit dependences on the laboratory speed $V_L$
and on the sidereal frequency $\om$,
which arise from the rotation of the Earth.

The inclusion of the Earth's rotation in the analysis
implies the laboratory frame is noninertial.
The structure of the first few terms 
in Eq.\ \rf{nonrellagrangiano3}
reveals that inertial forces in the laboratory
couple to $\cbt_{\mnhat}$,
which can result in nongravitational Lorentz-violating effects 
comparable in size to the gravitational ones of interest.
We therefore incorporate 
these nongravitational effects in our subsequent analyses. 
In practice,
this means effects proportional to the centrifugal acceleration
$\om^2 R_\oplus \approx 10^{-3} g$ must be considered.

In what follows,
we consider effects up to and including \pno3.
The leading \pno3 effects 
are proportional to the speed $V_\oplus$ of the Earth
as it revolves about the Sun
and are of order $gV_\oplus \approx 10^{-4}g$,
where $g = \G \ms/R_\oplus^2$
for a laboratory on the surface of the Earth.
This yields sensitivity to various components 
of the coefficients $\afbw_{\mu}$ and $\cbw_{\mn}$.
For some laboratory tests,
it is advantageous to consider also \pno3 effects
proportional to the smaller speed $V_L$ of the laboratory 
due to the rotation of the Earth,
which are of order $gV_L \approx 10^{-6}g$.
The benefit arises in two ways. 
First,
inclusion of the boost $V_L$ introduces
effects proportional to $\afbw_\mu$
that vary sidereally instead of annually.
This offers access to $\afbw_\mu$ for measurements 
conducted on comparatively short time scales,
albeit at a sensitivity reduced by about two orders of magnitude.
Second,
certain laboratory tests have greater sensitivity
to forces in the $\xhat$ and $\yhat$ directions
than to ones in the $\zhat$ direction.
The inclusion of effects from $\afbw_\mu$ and $\cbw_\mn$
that are proportional to $V_L$
can then introduce new sensitivities
or improve existing ones. 
 
So far, 
modifications to the trajectory of the test body 
arising from the coefficients $\sb_\mn$
have been disregarded.
However,
it is straightforward to incorporate these 
in the lagrangian at \pno2
because the coordinate choices made here
are consistent with those of Ref.\ \cite{qbakpn}
at this perturbative order.
In the laboratory frame,
we find the \pno2 contribution from $\sb_\mn$
to the lagrangian of the test body can be written 
\bea
L^{(2)}_s &=& 
\mt g (
\sb_{\zhat \xhat} \xhat + \sb_{\zhat \yhat} \yhat
- \half \sb_{\zhat \zhat} \zhat 
- \frac{3}{2} \sb_{\that \that} \zhat ).
\label{snonrellagrangian}
\eea
It turns out that $L^{(2)}_s$ suffices
to achieve sensitivity to $\sb_\mn$ at \pno3.
The point is that the leading \pno3 effects 
are proportional to $V_\oplus$,
while inclusion of effects proportional to $V_L$ 
offers no additional benefit in this case
for the tests we consider.
The coefficients $\sb_\mn$ are species independent,
so they are unobservable in WEP tests. 
Moreover,
inspection of $L^{(2)}_s$
reveals that the coefficients $\sb_{\mnhat}$ 
already vary at the sidereal frequency
through the transformation to the Sun frame.

In the laboratory frame,
the \pno3 lagrangian $L^{(3)}_{a,c,s}$ 
obtained from Eq.\ \rf{nonrellagrangiano3} 
and incorporating effects from $\sb_{\mnhat}$ 
via Eq.\ \rf{snonrellagrangian}
is somewhat lengthy in form.
As an illustration of its structure and implications,
we can restrict attention to its \pno2 limit $L^{(2)}_{a,c,s}$.
We find
\bea
L^{(2)}_{a,c,s} &=& 
\half \mt (1 + \cbt_{\that \that}) \dot{x}_\jhat \dot{x}_\jhat
+\mt \cbt_{\jhat \khat} \dot{x}_\jhat \dot{x}_\khat
\nonumber
\\
& & 
- \mt g \Big[ 1 
+ \fr{2 \al}{\mt} \abt_\that 
+ \fr{2 \al}{\ms} \abs_\that 
\nonumber
\\
& &
\pt{- \mt g z \Big[}
+ \cbt_{\that\that} + \cbs_{\that\that} 
+ \frac{3}{2} \sb_{\that \that} + \half \sb_{\zhat \zhat} \Big]
z
\nonumber
\\
& & 
+ \mt g (\sb_{\zhat \xhat} x + \sb_{\zhat \yhat} y).
\label{nonrellagrangian}
\eea
Varying this result yields the Euler-Lagrange equations of motion,
which we can express in the form of the modified force law
\beq
F_\jhat = m_{\jhat\khat} \ddot{x}_\khat.
\label{labeom}
\eeq
At this perturbative order,
the inertial and gravitational forces
acting on the test particle are given by
\bea
F_\xhat &=& \mt g \sb_{\zhat \xhat},
\nonumber
\\
F_\yhat &=& \mt g \sb_{\zhat \yhat},
\nonumber
\\
F_\zhat &=& 
- \mt g \Big[1 
+ \fr{2 \al}{\mt} \abt_\that 
+ \fr{2 \al}{\ms} \abs_\that 
\nonumber
\\
& &
\pt{- \mt g \big[}
+ \cbt_{\that\that} + \cbs_{\that\that} 
+ \frac{3}{2} \sb_{\that \that} + \half \sb_{\zhat \zhat}
\Big],
\nonumber\\
\label{lvforcez}
\eea
while 
\beq
m_{\jhat\khat} = 
\mt \left(1 + \cbt_{\that\that} \right) \de_{\jhat\khat} 
+ 2 \mt \cbt_{(\jhat\khat)}
\label{inertialm}
\eeq
is the effective inertial mass.

These results reveal the generic feature
that the gravitational force $F_\jhat$
acquires tiny corrections both along the $\hat z$ direction
and perpendicular to it.
Also,
the response of the test body 
deviates slightly from the direction 
of the applied force
because the effective inertial mass $m_{\jhat\khat}$
depends on the coefficients $\cbt_{\mnhat}$.
In principle,
some of these effects are detectable in sensitive laboratory tests,
and the corresponding signals are discussed 
using \pno3 results in the following subsections.

Some coefficients appear in combinations
that are challenging to separate in laboratory tests.
This is true,
for example,
of the coefficients
$\al \abt_T$ and $\cbt_{TT}$.
Consider for simplicity the scenario with only
isotropic Lorentz violation in the Sun-centered frame,
where the nonzero coefficients are
$\al \abt_T$ and
$\cbt_{TT} = 3\cbt_{XX} = 3\cbt_{YY} = 3\cbt_{ZZ}$.
In the laboratory frame,
$\abt_\that\approx \al \abt_T$ 
and 
$\cbt_{\that\that} \approx \cbt_{TT}$
up to boost factors.
These coefficients therefore cannot be readily separated
in gravimeter tests,
which depend on time variations
from anisotropic effects.
Moreover,
inspection of the \pno2 lagrangian \rf{nonrellagrangian}
reveals that 
if $3 \al \abt_\that = \mt \cbt_{\that\that}$
then the contributions of  
$\al \abt_\that$ and $\cbt_{\that\that}$
to the effective inertial and gravitational masses are identical.
The combination $\al \abt_T - \mt \cbt_{TT}/3$
therefore cannot be readily separated 
in conventional WEP tests either.
Note that WEP tests comparing a particle and its antiparticle
can in principle evade this difficulty 
because the sign of $\al \abt_T$ differs between the two. 
Another possibility would be to compare matter with light,
an option considered further in Sec.\ \ref{photon}.

\subsection{Free-fall gravimeter tests}
\label{ffgravimeter}

In this subsection,
we consider laboratory tests
that monitor the motion of a test body in free fall
near the surface of the Earth.
The equation of motion for the test body
can be obtained from the \pno3 lagrangian $L^{(3)}_{a,c,s}$ 
described in Sec.\ \ref{labtheory}.
Its explicit form is lengthy.
However, 
all information relevant for present purposes 
is contained in its solution 
expressed to the desired perturbative order.
This solution can be written in the form
\beq
x_\jhat = (x_o)_\jhat + (v_o)_\jhat t + \half \acc_\jhat t^2,
\label{labmotion}
\eeq
where the test body has initial position $\vec x_o$ 
and initial velocity $\vec v_o$.
The quantities of interest 
in searches for gravitational Lorentz violation 
are the components $\acc_\jhat$
of the acceleration of the test body 
in laboratory coordinates.

For purposes of data analysis and
reporting sensitivities to coefficients for Lorentz violation,
it is useful to express the acceleration components $\acc_\jhat$
in a form that displays explicitly 
the time variation and the dependence on particle species.
In free-fall gravimeter tests,
the time variation appears at frequencies
$0$, $\om$, $2\om$, $\om\pm\Om$, $2\om\pm\Om$, and $\Om$,
which are collectively labeled as $n$ in what follows.
The dependence on particle species
arises from the composition of the test and source bodies.
It is characterized by the label $w$,
which ranges over $e$, $p$, $n$ for ordinary matter.

\begin{center}
\begin{tabular}{lc}
\multicolumn{2}{c} 
{Table \laba.\ 
Amplitudes for the acceleration $\acc_\xhat$.} \\
\hline
\hline
Amplitude & Phase\\ 
\hline
&\\
$A^w_0 = m^w \sin \ch \cos \ch [\cxxw + \cyyw - 2 \czzw ]$ &
0 \\
$A^w_{\om} = 
2 m^w \cxzw \cos 2 \ch + \fr{2}{5} V_L \al \afbw_Y \cos \ch$ & 
$\ps$ \\
$A^{\prime w}_{\om} = 
\fr{1}{5} V_L \big[ \al \afbw_Y + 2 m^w \ctyw \big]\cos \ch$ &
$\ps$ \\
$B^w_{\om} = 
2 m^w \cyzw \cos 2 \ch - \fr{2}{5} V_L \al \afbw_X \cos \ch$ &
$\ps$ \\
$B^{\prime w}_{\om} = 
- \fr{1}{5} V_L \big[ \al \afbw_X + 2 m^w \ctxw \big] \cos \ch$ &
$\ps$ \\
$A^w_{2 \om} = \half m^w (\cxxw - \cyyw) \sin 2 \ch $ &
$2 \ps $ \\
$B^w_{2 \om} = m^w \cxyw \sin 2 \ch $  &
$2 \ps$ \\
$A^w_{\om + \Om} = - m^w V_\oplus \ctxw \sin \et \cos 2 \ch$ &
$\ps $ \\
$B^w_{\om + \Om} = - m^w V_\oplus 
\big[ \ctyw \sin \et $ & \\
$\pt{B^w_{\om + \Om} = - m^w V_\oplus \big[}
- \ctzw (1 - \cos \et) \big] \cos 2 \ch$ & 
$\ps$ \\
$A^w_{\om - \Om} = -m^w V_\oplus \ctxw \sin \et \cos 2 \ch$ &
$\ps$ \\
$B^w_{\om - \Om} = -m^w V_\oplus \big[\ctyw \sin \eta$ & \\
$\pt{B^w_{\om - \Om} = - m^w V_\oplus \big[}
+ \ctzw (1 + \cos \et) \big] \cos 2 \ch $ & 
$\ps$ \\
$A^w_{2 \om + \Om} = - \half m^w V_\oplus \ctyw (1 - \cos \et) \sin 2 \ch$ &
$2 \ps$ \\
$B^w_{2 \om + \Om} = \half m^w V_\oplus \ctxw (1 - \cos \et) \sin 2 \ch $ &
$2 \ps$ \\
$A^w_{2 \om - \Om} = \half m^w V_\oplus \ctyw (1 + \cos \et) \sin 2 \ch$ &
$2 \ps $ \\
$B^w_{2 \om - \Om} = - \half m^w V_\oplus \ctxw (1 + \cos \et) \sin 2 \ch $ &
$2 \ps$ \\
$A^w_{\Om} = - m^w V_\oplus \big[ \ctyw \cos \et$ & \\
\pt{ $A^w_{\Om} = - m^w V_\oplus \big[ $} 
$- 2 \ctzw \sin \et \big] \sin 2 \ch$ &
$0$ \\
$B^w_{\Om} = m^w V_\oplus \ctxw \sin 2 \ch $ &
$0$ \\
&\\
\hline
\hline
\end{tabular}
\end{center}

\begin{center}
\begin{tabular}{lc}
\multicolumn{2}{c} 
{Table \labb.\ 
Amplitudes for the acceleration $\acc_\yhat$.} \\
\hline
\hline
Amplitude & Phase\\ 
\hline
&\\
$C^w_{0} = m^w V_L \ctzw \sin 2 \ch$ & 
$0$ \\
$C^w_{\om} = 2 m^w \cyzw \cos \ch $ & \\
$ \pt{C^w_{\om} =}
- \frac{2}{5} V_L \al \afbw_X
+ 2 m^w V_L \ctxw \sin^2 \ch $ & 
$\ps$ \\
$C^{\prime w}_{\om} = - \frac{1}{5} V_L \big[ \al \afbw_X
+ 2 \ctxw \big]$ & 
$\ps$ \\
$D^w_{\om} = -2 m^w \cxzw \cos \ch $ & \\
$ \pt{D^w_{\om} =}
+ 2 m^w V_L \ctyw \sin^2 \ch
- \frac{2}{5} V_L \al \afbw_Y$ &
$\ps$ \\
$D^{\prime w}_{\om} = - \frac{1}{5} V_L \big[ 2 m^w \ctyw
+ \al \afbw_Y \big]$ &
$\ps$ \\
$C^w_{2 \om} = 2 m^w \cxyw \sin \ch $ &
$2 \ps $ \\
$D^w_{2 \om} = - m^w (\cxxw - \cyyw) \sin \ch $ &
$2 \ps$ \\
$C^w_{\om + \Om} = m^w V_\oplus \big[ \ctzw (1-\cos \et)$ & \\
$ \pt{C^w_{\om + \Om} = m^w V_\oplus \big[} 
- \ctyw \sin \et \big] \cos \ch $ &
$\ps $ \\
$D^w_{\om + \Om} = m^w V_\oplus \ctxw \sin \et \cos \ch $ &
$\ps $ \\
$C^w_{\om - \Om} = - m^w V_\oplus \big[ \ctzw (1 + \cos \et) $ & \\
$ \pt{C^w_{\om - \Om} = - m^w V_\oplus \big[}
+ \ctyw \sin \et \big]\cos \ch $ &
$\ps$ \\
$D^w_{\om - \Om} = m^w V_\oplus \ctxw \sin \et \cos \ch $ &
$\ps$ \\
$C^w_{2 \om + \Om} = m^w V_\oplus \ctxw (1 - \cos \et) \sin \ch $ &
$2 \ps$ \\
$D^w_{2 \om + \Om} = m^w  V_\oplus \ctyw (1 - \cos \et) \sin \ch $ &
$2 \ps$ \\
$C^w_{2 \om - \Om} = - m^w V_\oplus \ctxw (1 + \cos \et) \sin \ch $ &
$2 \ps $ \\
$D^w_{2 \om - \Om} = - m^w V_\oplus \ctyw (1 + \cos \et) \sin \ch $ &
$2 \ps$ \\
&\\
\hline
\hline
\end{tabular}
\end{center}

For the $\hat x$ component of the acceleration,
some calculation yields an expression of the form
\bea
\acc_\xhat &=& 
\om^2 R_\oplus \sin \ch \cos \ch 
\nonumber \\
& &
\hskip -15pt
+g \sum_{n,w} \Big[ 
\left(\fr{N^w}{\mt} A_n^w 
+ \fr{N^w_\oplus}{\ms} A_n^{\prime w} 
+ \frac{1}{3} A_n \right) \cos (\om_n T + \ps_n) 
\nonumber \\
& & 
\hskip -15pt
\pt{\sum_n} 
+ \left(\fr{N^w}{\mt} B_n^w + 
\fr{N^w_\oplus}{\ms} B_n^{\prime w} 
+ \frac{1}{3} B_n \right) \sin (\om_n T + \ps_n)\Big].
\nonumber \\
\label{xacc}
\eea
In this equation, 
the amplitudes 
$A_n^w$, $A^{\prime w}_n$, $B_n^w$, $B^{\prime w}_n$
contain the coefficients for Lorentz violation
$\afbw_\mu$, $\cbw_\mn$ and hence depend on particle species.
These amplitudes and their associated phases
are listed in Table \laba.
The remaining amplitudes $A_n$, $B_n$
contain the coefficients $\sb_\mn$ 
from the gravitational sector,
which are independent of the composition of the test body.
These amplitudes can be obtained 
from the amplitudes $A_n^w$, $B_n^w$
by the substitutions 
$m^w \rightarrow 1$,
$\cbw_{\Si \Xi} \rightarrow \half \sb_{\Si \Xi}$,
and $\afbw_\Xi \rightarrow 0$,
disregarding contributions proportional to $V_L$.

The $\yhat$ component of the acceleration
can be decomposed similarly.
We find
\bea
\acc_\yhat &=&
\sum_{n,w} g \Big[ 
\left(\fr{N^w}{\mt} C_n^w 
+ \fr{N^w_\oplus}{\ms} C_n^{\prime w} 
+ \frac{1}{3} C_n \right) \cos (\om_n T + \ps_n) 
\nonumber \\
& & 
\hskip -10pt
\pt{\sum_n} 
+ \left(\fr{N^w}{\mt} D_n^w 
+ \fr{N^w_\oplus}{\ms} D_n^{\prime w} 
+ \frac{1}{3} D_n \right) \sin (\om_n T + \ps_n) \Big]. 
\nonumber \\
\label{yacc}
\eea
The amplitudes $C_n^w$, $C^{\prime w}_n$, 
$D_n^w$, $D^{\prime w}_n$
depend on particle species through the coefficients
$\afbw_\mu$, $\cbw_\mn$
and are listed in Table \labb,
along with the corresponding phases.
The remaining amplitudes $C_n$, $D_n$ are obtained
from $C_n^w$, $D_n^w$
using the substitutions
$m^w \rightarrow 1$,
$\cbw_{\Si \Xi} \rightarrow \half \sb_{\Si \Xi}$,
and $\afbw_\Xi \rightarrow 0$,
disregarding contributions proportional to $V_L$ as before.

For the $\zhat$ component of the acceleration,
we obtain
\bea
\acc_\zhat &=&
-g + \om^2 R_\oplus \sin^2 \ch 
\nonumber \\
& &
\hskip-15pt
+ \sum_{n,w} g \Big[ \left(\fr{N^w}{\mt} E_n^w 
+ \fr{N^w_\oplus}{\ms} E_n^{\prime w} 
+ \frac{1}{3} E_n \right) \cos (\om_n T + \ps_n) 
\nonumber \\
& & 
\hskip-15pt
\pt{\sum_{n,w}} 
+ \left(\fr{N^w}{\mt} F_n^w 
+ \fr{N^w_\oplus}{\ms} F_n^{\prime w} 
+ \frac{1}{3} F_n \right) \sin (\om_n T + \ps_n)  \Big].
\nonumber
\\
\label{zacc}
\eea
The amplitudes $E_n^w$, $E^{\prime w}_n$, $F_n^w$, $F^{\prime w}_n$
depend on particle species
via the coefficients $\afbw_\mu$, $\cbw_\mn$.
The amplitudes $E_n$, $F_n$
are independent of species
and given in terms of the coefficients $\sb_\mn$.
All these amplitudes 
and their phases are provided in Table \labc.

In principle,
the results of a free-fall laboratory test 
using any gravimeter can be analyzed with the above equations.
The dominant effects appear at different frequencies
for different coefficients,
so the time scale of data taking in a given experiment 
affects the breadth of its reach in coefficient space.
Also, 
each signal frequency can be expected to have distinct systematics.
For example,
dominant effects from the coefficients $\afb_J$
occur at the annual frequency $\Om$,
for which seasonal systematics are relevant.
Note that all the Lorentz-violating effects
can be accessed at or near the sidereal frequency $\om$,
although in some cases at reduced sensitivity.

At least two kinds of devices can be classified
as free-fall gravimeters:
falling corner cubes,
and matter interferometers.
Falling corner cubes,
which typically are sensitive only to the direction 
of the free-fall motion,
are used to monitor 
time variations of the gravitational field for geodesy
and other geophysical purposes
\cite{mf}.
In principle,
they are of interest for 
free-fall gravimeter tests of Lorentz violation.
However,
matter interferometers presently
carry several advantages over falling corner cubes
in this context.
They are slightly more sensitive,
some types can sense accelerations 
in more than one direction,
and the composition of the test body 
can be determined more readily.
We therefore focus on matter interferometers in this subsection,
revisiting the use of both falling corner cubes 
and interferometers
in the context of free-fall WEP tests in Sec.\ \ref{epfreefall}.

\begin{center}
\begin{tabular}{lc}
\multicolumn{2}{c} 
{Table \labc.\ 
Amplitudes for the acceleration $\acc_\zhat$. } \\
\hline
\hline
Amplitude & Phase\\ 
\hline
&\\
$E^w_{0} = -2 \al \afbw_T 
+ 2 m^w \czzw \cos^2 \ch$ & \\
$ \pt{E^w_{\om} =}
+ m^w \left( \cxxw + \cyyw \right) \sin^2 \ch$ &  
$0$ \\
$E^{\prime w}_{0} = -2 \al \afbw_T - m^w \cttw$ &  
$0$ \\
$E^w_{\om} = 2 m^w \cxzw \sin 2 \ch
- \fr45 V_L \al \afbw_Y \sin \ch$ &  
$\ps$ \\
$E^{\prime w}_{\om} = - \fr45 V_L \left( 3 \al \afbw_Y
+ m^w \ctyw \right) \sin \ch$ &  
$\ps$ \\
$F^w_{\om} = 2 m^w \cyzw \sin 2 \ch
+ \fr{4}{5} V_L \al \afbw_X \sin \ch$ & 
$\ps$ \\
$F^{\prime w}_{\om} = \fr{4}{5} V_L \left( 3 \al \afbw_X
+ m^w \ctxw \right) \sin \ch$ & 
$\ps$ \\
$E^w_{2 \om} = m^w (\cxxw - \cyyw) \sin^2 \ch$ & 
$2 \ps $ \\
$F^w_{2 \om} =  2 m^w \cxyw \sin^2 \ch$ &
$2 \ps$ \\
$E^w_{\om + \Om} = - m^w V_\oplus \ctxw \sin \et \sin 2 \ch$ &
$\ps $ \\
$F^w_{\om + \Om} = - m^w V_\oplus \big[ \ctyw \sin \et $ & \\
$ \pt{F^w_{\om + \Om} = - m^w V_\oplus \big[}
- \ctzw (1 - \cos \et) \big] \sin 2 \ch $ & 
$\ps$ \\
$E^w_{\om - \Om} = - m^w V_\oplus \ctxw \sin \et \sin 2 \ch$ &
$\ps$ \\
$F^w_{\om - \Om} = - m^w V_\oplus \big[ \ctyw \sin \eta $ & \\
$ \pt{F^w_{\om - \Om} = m^w V_\oplus \big[}
+ \ctzw (1 + \cos \et) \big] \sin 2 \ch $ & 
$\ps$ \\
$E^w_{2 \om + \Om} = - m^w V_\oplus \ctyw (1 - \cos \et) \sin^2 \ch$ &
$2 \ps$ \\
$F^w_{2 \om + \Om} = m^w V_\oplus \ctxw (1 - \cos \et) \sin^2 \ch$ &
$2 \ps$ \\
$E^w_{2 \om - \Om} = m^w V_\oplus \ctyw (1 + \cos \et) \sin^2 \ch$ &
$2 \ps $ \\
$F^w_{2 \om - \Om} = - m^w V_\oplus \ctxw (1 + \cos \et) \sin^2 \ch$  &
$2 \ps$ \\
$E^w_{\Om} = 2 V_\oplus \al (\ayw \cos \et + \azw \sin \et)$ & \\
$ \pt{E^w_{\Om} =}
- 2 m^w V_\oplus \big[\ctyw \cos \et \sin^2 \ch $ & \\
$ \pt{E^w_{\Om} = + 2 m^w V_\oplus \big[}
+ 2 \ctzw \sin \et \cos^2 \ch \big] $ &
$0$ \\
$E^{\prime w}_{\Om} = 2 V_\oplus \al (\ayw \cos \et + \azw \sin \et)$ & \\
$ \pt{E^w_{\Om} =}
+ 2 m^w V_\oplus (\ctyw \cos \et + \ctzw \sin \et) $ &
$0$ \\
$F^w_{\Om} = - 2 V_\oplus \al \axw
+ 2 m^w V_\oplus \ctxw \sin^2 \ch$ &
$0$ \\
$F^{\prime w}_{\Om} = - 2 V_\oplus \al \axw
- 2 m^w V_\oplus \ctxw $ &
$0$ \\
$E_{0} = - \half \szz \cos^2 \ch
- \quar \left( \sxx + \syy \right) \sin^2 \ch$ & \\
$ \pt{E_{0} =}
- \fr32 \stt$ &  
$0$ \\
$E_{\om} = -\half \sxz \sin 2 \ch$ &  
$\ps$ \\
$F_{\om} = -\half \syz \sin 2 \ch $ & 
$\ps$ \\
$E_{2 \om} = -\quar (\sxx - \syy) \sin^2 \ch$ & 
$2 \ps $ \\
$F_{2 \om} =  -\quar \sxy \sin^2 \ch$ &
$2 \ps$ \\
$E_{\om + \Om} = \quar V_\oplus \stx \sin \et \sin 2 \ch$ &
$\ps $ \\
$F_{\om + \Om} = \quar V_\oplus \big[ \sty \sin \et $ & \\
$ \pt{F_{\om + \Om} = \quar V_\oplus \big[}
- \stz (1 - \cos \et) \big] \sin 2 \ch $ & 
$\ps$ \\
$E_{\om - \Om} = \quar V_\oplus \stx \sin \et \sin 2 \ch$ &
$\ps$ \\
$F_{\om - \Om} = \quar V_\oplus \big[ \sty \sin \eta $ & \\
$ \pt{F_{\om - \Om} = \quar V_\oplus \big[}
+ \stz (1 + \cos \et) \big] \sin 2 \ch $ & 
$\ps$ \\
$E_{2 \om + \Om} = \quar V_\oplus \sty (1 - \cos \et) \sin^2 \ch$ &
$2 \ps$ \\
$F_{2 \om + \Om} = - \quar V_\oplus \stx (1 - \cos \et) \sin^2 \ch$ &
$2 \ps$ \\
$E_{2 \om - \Om} = - \quar V_\oplus \sty (1 + \cos \et) \sin^2 \ch$ &
$2 \ps $ \\
$F_{2 \om - \Om} = \quar V_\oplus \stx (1 + \cos \et) \sin^2 \ch$  &
$2 \ps$ \\
$E_{\Om} = V_\oplus \big[\sty \cos \et (\half \sin^2 \ch + 3) $ & \\
$ \pt{E_{\Om} = V_\oplus \big[}
+ \stz \sin \et (\cos^2 \ch +3) \big] $ &
$0$ \\
$F_{\Om} = - V_\oplus \stx (\half \sin^2 \ch + 3)$ &
$0$ \\
&\\
\hline
\hline
\end{tabular}
\end{center}

Matter interferometers,
which permit quantum-mechanical laboratory measurements
of the motion of falling matter,
have attained impressive sensitivities
to gravitational acceleration
\cite{pcc}
and to rotational accelerations
via the Sagnac effect 
\cite{glm}.
In the context of gravitational Lorentz violation,
matter interferometry has been used to measure 
combinations of the coefficients 
$\sb_\mn$ and $\kfb^\al_{\pt{\al} \mu \al \nu}$
\cite{gravexpt2}
based on the gravimeter analysis of effects
from the pure-gravity sector of the SME
\cite{qbakpn}.
Here,
we extend the latter analysis
to include effects from the coefficients $\afbw_\mu$, $\cbw_\mn$
and generalize it to other interferometer configurations.

The basic idea of a matter interferometer
is to place the matter in a superposition 
of spatially separated quantum states,
which may acquire a measurable relative phase.
In the gravitational tests considered here,
the behavior of the interferometer
is close to the classical limit,
and a convenient way to perform the analysis 
is to proceed semiclassically
via path integration along the classical motion 
\cite{pcc,sct}.
The phase difference between the final states 
can then be viewed as a sum of three contributions:
the phase difference acquired from the momentum transfers
used to control the beams,
the phase difference accumulated from the classical action 
along the different paths,
and in some configurations a phase difference coming from 
a final separation of the states.
It turns out that the dominant effect for acceleration sensing
is the phase difference acquired through momentum transfers.
Since leading-order Lorentz-violating motional effects
appear as modified accelerations,
the phase difference from the momentum transfers
is the relevant contribution 
in the present context.

For definiteness,
suppose the interferometer paths
trace a parallelogram.
This includes the limiting case 
of temporal path separation,
where the parallelogram has zero area.
Other shapes could also be analyzed
using the equation of motion \rf{labmotion}.
In the Lorentz-invariant case,
the standard result for the phase shift
due to the Earth's gravitational field is
$\De \ph = k_\zhat g \ta^2$,
where $\vec{k}$
is the magnitude of the momentum transfer in the beam splitter
and $\ta$ is the time of flight between impulses.
With Lorentz violation present,
we find the phase shift $\De\ph$
takes a similar form
but with the Newton gravitational acceleration
replaced by the accelerations in Eq.\ \rf{labmotion},
giving
\beq
\Delta \ph = k_\jhat {\rm a}_\jhat \ta^2.
\label{phaseshift}
\eeq
The signal frequencies associated with Lorentz violation
can be identified by substitution 
of the expressions \rf{xacc}-\rf{zacc}
for the acceleration components ${\rm a}_\jhat$.
Note that Lorentz-violating effects on the atomic energy levels 
could generate additional contributions to the phase difference
but are already tightly constrained in other experiments
and so can typically be neglected.
Note also that possible Lorentz-violating effects 
varying with the particle spins,
which are described explicitly by the relativistic hamiltonian 
of Sec.\ \ref{relativham},
are disregarded here as outside our present scope.
A comprehensive investigation of their
implications for matter interferometry may be of interest
\cite{abl}.

Several atom interferometers currently or recently operating 
are relevant to free-fall gravimeter searches for Lorentz violation.
An impressive sensitivity of about $1 \times 10^{-10}g$ 
to the vertical acceleration
was achieved by Peters, Chung, and Chu 
\cite{pcc}.
In another apparatus,
a differential-acceleration sensitivity
of $3 \times 10^{-9} g/\sqrt{\rm Hz}$ has been demonstrated 
\cite{mffsk}.
An interferometer designed for experiments in space
is expected to achieve 
sensitivity of about $3\times 10^{-9}g$
in ground operations
\cite{ykkm}.
Initial sensitivities to accelerations in each direction
of about $6\times 10^{-7}g$
after 10 minutes of averaging 
have been attained
in a device using highly parabolic trajectories
\cite{canuel}.
Recent estimates suggest that future measurements
of vertical acceleration could achieve sensitivities
at the level of about $10^{-15}g$ 
\cite{dghk}.

Given this information and the phase shift \rf{phaseshift},
we can use Tables \laba, \labb, and \labc\
to obtain crude estimates for attainable sensitivities
to coefficients for Lorentz violation
in existing or near-future atom interferometers.
With present capabilities,
sensitivities at the level of parts in $10^{5}$
could in principle be obtained 
to combinations of the coefficients $\al \afbw_J$, $J=X,Y,Z$
and of several currently unconstrained
components of the coefficients
$\cbw_{\Si \Xi}$,
including $\cbn_{(TJ)}$ for the neutron.
The relevant signals are associated
with the Earth's boost as it revolves about the Sun,
so they exhibit an annual periodicity. 
The next generation of atom interferometers 
could in principle improve this sensitivity 
to parts in $10^{10}$.
The boost of the laboratory due to the Earth's rotation
provides sensitivities that have sidereal periodicities
instead but that are weaker by a factor of about 100.
Note that the boost suppressions could in principle
be avoided for certain coefficients,
including presently unbounded combinations involving $\cbn_{ZZ}$,
by the use of an interferometer sensitive to the accelerations
${\rm a}_\xhat$, ${\rm a}_\yhat$
that is placed on a rotating turntable.
Note also that individual sensitivities to neutron coefficients
can in principle be extracted by performing atom interferometry
with different neutral atoms
having distinct proton-to-neutron ratios.
Another possibility with weaker existing sensitivity 
includes neutron interferometry 
\cite{ninterf},
which could provide independent and clean bounds 
on neutron coefficients.

\subsection{Force-comparison gravimeter tests}
\label{fcgravimeters}

Another class of gravimeter tests
is based on the idea of countering the gravitational force
with an appropriate electromagnetic force.
Force-comparison gravimeter tests
can be performed with 
gravimeters based on systems of springs and masses
\cite{mf}
and with superconducting gravimeters 
\cite{mf,wg,ss}.
At present,
the latter devices have sensitivities
competitive with those of existing atom interferometers.
Certain experiments studying short-range gravity 
may also offer relevant sensitivities
\cite{shortrange}.

The signals for gravitational Lorentz violation 
in a given force-comparison gravimeter 
can be extracted from the \pno3 lagrangian $L^{(3)}_{a,c,s}$ 
discussed in Sec.\ \ref{labtheory}.
Since macroscopic bodies are involved,
the analysis must include an assessment
of their composition.
Note also that conventional intuition 
from Newton's second law can be misleading 
because the effective inertial masses 
depend on the coefficients $\cbw_\mn$,
as discussed following Eq.\ \rf{inertialm}.

Superconducting gravimeters
have already been proposed as suitable devices
for measuring the gravity-sector coefficients $\sb_\mn$ 
for Lorentz violation 
\cite{qbakpn}.
Here,
we extend this discussion
to include effects from the coefficients 
$\afbw_\mu$ and $\cbw_\mn$.
The analysis proceeds directly 
from the \pno3 lagrangian $L^{(3)}_{a,c,s}$ 
by noting that the device 
is designed to maintain $\ddot{x}_\jhat=0$.
The applied force required to hold this constraint
can be taken as the relevant observable
and can be written
\beq
F_{\zhat^\prime} = F_\zhat \cos \ze +  F_\xhat \sin \ze,
\eeq
where $\ze$ is the deviation angle defined in Eq.\ \rf{zetadef}.
To maintain consistent counting of small effects,
we restrict terms independent of velocity 
to first order in $\ze$
and terms containing a power of velocity 
to zeroth order in $\ze$.

We find that the relevant contributions
to the force $F_{\zhat^\prime}$
can be decomposed by frequency as 
\bea
F_{\zhat^\prime} &=&
\mt g(1 - \ze \tan \ch - \frac{3}{2} \ze^2) 
- \mt g 
\nonumber \\
& &
\hskip -25pt
\times
\sum_{n \neq 0,w} \Bigg[ 
\left(\fr{N^w}{\mt} G_n^w 
+ \fr{N^w_\oplus}{\ms} E_n^{\prime w} 
+ \frac{1}{3} G_n \right) 
\cos (\om_n T + \ps_n) 
\nonumber \\
& & 
+ \left(\fr{N^w}{\mt} H_n^w +
 \fr{N^w_\oplus}{\ms} F_n^{\prime w} 
+ \frac{1}{3} H_n \right) \sin (\om_n T + \ps_n)  
\Bigg],
\nonumber\\
\label{Fz}
\eea
where constant effects
that are unobservable in superconducting gravimeters 
are neglected.
In this expression,
the amplitudes $G^w_n$, $H^w_n$ 
and their phases are given in Table \labe,
while $E_n^{\prime w} $ and $F_n^{\prime w}$
are listed in Table \labc.
The remaining amplitudes $G_n$ and $H_n$ can be expressed 
in terms of amplitudes given in 
Tables \laba\ and \labc\ as
\bea
G_n &=& A_n \ze + E_n, 
\qquad 
H_n = B_n \ze + F_n .
\eea

The frequency decomposition \rf{Fz}
can be examined to extract 
crude estimates of attainable sensitivities to Lorentz violation.
In this way,
we estimate that
the presently unbounded coefficients $\al \afbw_J$ and $\ctjw$
could be measured at the level of parts in $10^7$
using existing data from superconducting gravimeters 
\cite{ss}.
Improved sensitivities are likely to be attainable
in a dedicated experiment of this type.

\begin{center}
\begin{tabular}{lc}
\multicolumn{2}{c} 
{Table \labe.\ Amplitudes for the force $F_{\zhat^\prime}$.} \\
\hline
\hline
Amplitude & Phase\\ 
\hline
$G^w_{\om} = 2 m^w \ze \cxzw$ & \\
$ \pt{G^w_{\om} =}
- \fr45 V_L \al \afbw_Y \sin \ch 
- 2 m^w V_L \ctyw \sin \ch$ &  
$\ps$ \\
$H^w_{\om} = 2 m^w \ze \cyzw $ & \\
$ \pt{H^w_{\om} =}
+ \fr{4}{5} V_L \al \afbw_X \sin \ch
+ 2 m^w V_L \ctxw \sin \ch$ & 
$\ps$ \\
$G^w_{2 \om} = m^w \ze (\cxxw - \cyyw)$ & 
$2 \ps $ \\
$H^w_{2 \om} =  2 m^w \ze \cxyw $ &
$2 \ps$ \\
$G^w_{\Om} = 2 V_\oplus \al (\ayw \cos \et + \azw \sin \et)$ & \\
$ \pt{G^w_{\Om} =}
+ 2 m^w V_\oplus \big[\ctyw \cos \et \sin^2 \ch $ & \\
$ \pt{G^w_{\Om} = + 2 m^w V_\oplus \big[}
+ 2 \ctzw \sin \et \cos^2 \ch \big] $ &
$0$ \\
$H^w_{\Om} = 2 V_\oplus \al \axw
+ 2 m^w V_\oplus \ctxw \sin^2 \ch$ &
$0$ \\
\hline
\hline
\end{tabular}
\end{center}

For tests of short-range gravity
and certain other applications,
it is useful to consider the standard case of
two point masses $m_1$ and $m_2$
at coordinate locations $\vec x_1$ and $\vec x_2$.
With this setup,
the modified Newton potential $V$ at \pno2 
in the laboratory frame 
can be obtained from $L^{(3)}_{a,c,s}$.
We find
\bea
V &=& - \fr {\G m_1 m_2}{|\vec x_1 - \vec x_2|} 
\bigl[
1+ \fr{2 \al}{m_1} \afbx1_{\hat t}
+ \fr{2 \al}{m_2} \afbx2_{\hat t}
\nonumber\\
&&
\hskip 60pt
+ \cbx1_{\hat t \hat t} 
+ \cbx2_{\hat t \hat t} 
+ \frac 12 \hat x^{\hat j} \hat x^{\hat k} \sb^{\hat j\hat k}
\bigr],
\qquad
\label{modpotV}
\eea
where $\hat x = (\vec x_1 - \vec x_2) /|\vec x_1 - \vec x_2|$.
This modified potential exhibits 
the usual inverse-distance dependence,
and it generalizes Eq.\ (137) of Ref.\ \cite{qbakpn}.
The corresponding modified Newton force 
typically has a component perpendicular
to the unit vector $\hat x$,
while obtaining the accelerations
requires determining also the effective inertial masses.
As usual,
any motion of the masses relative to the Sun-centered frame
implies time dependence of the laboratory-frame coefficients.
In principle, 
the above modified Newton potential could be used
in conjunction with integration or finite-element methods
to determine the effects of the coefficients 
$\afbw_\mu$, $\cbw_\mn$, $\sb_\mn$
on the behavior of two interacting bodies.

\subsection{Free-fall WEP tests}
\label{epfreefall}

In this subsection,
we consider WEP tests
in which signals for Lorentz violation
can be sought by monitoring the relative motion
of two freely falling bodies
of different composition.
Typical free-fall WEP tests are sensitive 
to motion along the direction
of the net acceleration $\acc_{\zhat^{\prime}}$.
This acceleration is the combination
\beq
\acc_{\zhat^\prime} = \acc_\zhat \cos \ze + \acc_\xhat \sin \ze
\eeq
of the component accelerations \rf{xacc} and \rf{zacc},
weighted by the deviation angle $\ze$ given in Eq.\ \rf{zetadef}.
In what follows,
terms containing both a boost factor and a factor of $\ze$
are treated as higher order and negligible,
as in the previous subsection.

The relevant observable for free-fall WEP tests
is the relative position
$\Delta \zhat^\prime$ 
of two test bodies 1 and 2 in a given drop.
It can be written as
\beq
\Delta \zhat^\prime = 
\left[ (v_o)_{\zhat^\prime}^1 - (v_o)_{\zhat^\prime}^2 \right]\that 
+ \half \left( \acc_{\zhat^\prime}^1 
- \acc_{\zhat^\prime}^2 \right) \that^2.
\label{relpos}
\eeq
This relative position varies with the canonical time $T$.
Decomposing by frequency yields the expression
\bea
\Delta \zhat^\prime &=& 
\sum_{n,w} \left( 
\fr{N^w_1}{m_1} - \fr{N^w_2}{m_2} \right) T^2
\nonumber \\ 
& & 
\times [I^w_n \cos (\om_n T + \ps_n) 
+ J^w_n \sin (\om_n T + \ps_n)].
\nonumber \\
\label{deacc}
\eea
The amplitudes $I^w_n$ and $J^w_n$
can be expressed as
\bea
I^w_n & = & A^w_n \ze + E^w_n, 
\qquad
J^w_n = B^w_n \ze + F^w_n ,
\label{fcepamp}
\eea
where $A^w_n$, $B^w_n$ are listed
in Table \laba\
and $E^w_n$, $F^w_n$ are given 
in Table \labc,
along with the associated phases $\ps_n$.
In Eq.\ \rf{deacc},
the quantities $N^w_1$ and $N^w_2$
are the numbers of particles of type $w$
appearing in the test bodies 1 and 2,
respectively,
while $m_1$ and $m_2$
are the corresponding conventional masses.

The frequency decomposition \rf{deacc} of the signal \rf{relpos}
can be used to provide rough estimates
of attainable sensitivities to Lorentz violation
in existing or near-future free-fall WEP tests.
We combine values for the fractional acceleration sensitivity 
${\De \acc}/{\acc}$ 
discussed in 
Refs.\ \cite{dghk,kknm,nmf,fdhw,poem,great,hdcm}
with the result \rf{deacc}
to compile some estimates in Table \labf. 
In this table,
the first row lists
the fractional acceleration sensitivity,
while each of the other rows
concerns a particular combination of coefficients.
For brevity,
in the first column we adopt the notations
\bea
\afb_{Y+Z} &=& \afb_Y \cos \et + \afb_Z \sin \et,
\nonumber \\
\cqbx{w} &=& \cxxw + \cyyw - 2 \czzw,
\label{YplusZ}
\eea
along with those introduced 
in Eqs.\ \rf{afbpen} and \rf{cbpen}.
We follow common procedure in the literature
\cite{tables}
by taking 
$\cbw_{TT}$,
$\cbw_{XX}-\cbw_{YY}$,
and $\cqbx{w}$
as the relevant independent combinations
of the traceless coefficients $\cbw_{\Si \Xi}$.

Each column in Table \labf\ 
lists estimated attainable sensitivities
on the moduli of various quantities 
in specified types of free-fall WEP test,
expressed to the nearest order of magnitude.
Values listed with neither brackets nor braces
are limits based on published data
that are implied by our present analysis.
Values shown in brackets
are our estimate of sensitivities 
that could in principle be obtained
from a suitable reanalysis of existing data.
Values shown in braces
represent our estimate of sensitivities
attainable using data from future tests.

The second column of the table
concerns free-fall WEP tests 
using falling corner cubes 
\cite{kknm,nmf}.
In the second entry of this column,
we present a single bound 
on the time-independent portion of the signal
implied by existing data.
The remainder of this column
lists crude estimates of sensitivities that could be attained 
through sidereal and annual analysis of the same data.

The third and fourth columns of the table 
list sensitivities
from free-fall WEP tests 
using atom interferometry.
In the second entry of the third column,
we present a single bound 
extracted from existing data
\cite{fdhw}.
The fourth column concerns proposals
for future tests with atom interferometers
\cite{dghk},
based on the idea that the relative vertical acceleration
of two different atoms may be measured
using a simultaneous dual-species fountain 
\cite{marion}.

The remaining columns of the table
concern other proposed free-fall WEP tests.
Crude estimates are provided of the sensitivities 
that might be achieved 
in the Principle of Equivalence Measurement (POEM)
\cite{poem},
via balloon drops 
in the General Relativity Accuracy Test (GReAT)
\cite{great},
and using the Bremen drop tower \cite{hdcm}.

In the table,
the estimates for the coefficients listed
in the second and third rows 
and for $\cqbx{n}$ in the penultimate row
all arise from the time-independent component of the data.
A nonzero signal for any of these measurements
would therefore be challenging 
to distinguish from other potential sources of WEP violation.
Note that obtaining the independent sensitivities 
in the third and penultimate rows
requires combining data taken
in free-fall WEP tests performed at different colatitudes $\ch$.

Independent sensitivities can also be achieved
via other techniques.
One possibility is positronium interferometry
\cite{positronium},
which via comparison with ordinary matter
could yield a bound on different linear combinations
of $\afbw_T$ and $\cbw_{TT}$.
Also,
some independent measurements can be extracted
by combining results from free-fall WEP tests
with those from the force-comparison WEP tests
discussed below.

The next generation of the POEM experiment 
\cite{poem}
is the proposed Sounding Rocket POEM (SR-POEM)
\cite{srpoem},
which is a WEP test 
designed to measure to $10^{-16}$ 
the relative acceleration of freely falling test bodies 
on a sounding rocket during certain phases of its flight.
Although not terrestrial,
this experiment can also be analyzed using the methods 
presented here.
A competitive sensitivity is anticipated 
for measurements of the combination
$\al \afbx{\pen}_T -  m^p \cbx{\pen}_{TT}/3$ 
of isotropic coefficients.
Obtaining sufficient data to resolve the periodic changes 
necessary for sensitivity to other coefficient combinations
would be challenging.

\onecolumngrid
\begin{center}
\begin{tabular}{@{\extracolsep{5pt}}lcccccccc}
\multicolumn{6}{c} 
{Table \labf.\ Sensitivities for free-fall WEP tests.} \\
\hline
\hline
 & Falling & \multicolumn{2}{c} {Atom}& Tossed & Balloon & Drop \\
Coefficient & corner cube & \multicolumn{2}{c} {interferometry}& masses & drop & tower \\ 
combination&\cite{kknm,nmf}& \cite{fdhw} & \cite{dghk}         & \cite{poem}   & \cite{great}  & \cite{hdcm} \\ 
\hline
$\frac{\De \acc}{\acc}$ 
           & $10^{-10}$           &$10^{-7}$    & \{$10^{-17}$\}            &\{$10^{-14}$\}   &\{$10^{-15}$\}   &\{$10^{-12}$\}\\
$\al \afbx{\pen}_T - \fr13 m^p \cbx{\pen}_{TT}$ &&&&&& \\            
$ \pt{aaa}+ (\half \cos^2 \ch - \fr16) m^n \cqbx{n}$
           & $10^{-8}$ GeV         &$10^{-5}$ GeV & \{$10^{-15}$ GeV\}         &\{$10^{-12}$ GeV\}&\{$10^{-13}$ GeV\}&\{$10^{-10}$ GeV\} \\
$\al \afbx{\pen}_T - \fr13 m^p \cbx{\pen}_{TT}$             
           & $\dots$              &$\dots$      & \{$10^{-15}$ GeV\}         &\{$10^{-12}$ GeV\}&\{$10^{-13}$ GeV\}&\{$10^{-10}$ GeV\} \\
$\al \afbx{\pen}_X$   
           & [$10^{-4}$ GeV]       &$\dots$      &\{$10^{-11}$ GeV\}          &\{$10^{-8}$ GeV\}&\{$10^{-9}$ GeV\} &\{$10^{-6} $ GeV\} \\
$\al \afbx{\pen}_{Y+Z}$
           & [$10^{-4}$ GeV]       &$\dots$      & \{$10^{-11}$ GeV\}         &\{$10^{-8}$ GeV\}&\{$10^{-9}$ GeV\} &\{$10^{-6} $ GeV\} \\
$\al \afbx{\pen}_Y $
           & [$10^{-2}$ GeV]       &$\dots$      & \{$10^{-9}$ GeV\}          &\{$10^{-6}$ GeV\} &\{$10^{-7}$ GeV\} &\{$10^{-4}$ GeV\} \\
$\al \afbx{\pen}_Z $
           & [$10^{-2}$ GeV]       &$\dots$      & \{$10^{-9}$ GeV\}          &\{$10^{-6}$ GeV\} &\{$10^{-7}$ GeV\} &\{$10^{-4}$ GeV\} \\
$\cqbx{n}$ 
           & $\dots$              &$\dots$      & \{$10^{-15}$\}            &\{$10^{-12}$\}   &\{$10^{-13}$\}   &\{$10^{-10}$\} \\
$\cbn_{(TJ)}$
           & $[10^{-4}]$          &$\dots$      & \{$10^{-11}$\}            &\{$10^{-8}$\}   &\{$10^{-9}$\}    &\{$10^{-6}$\}  \\
\hline
\end{tabular}
\end{center}
\vskip 5pt
\twocolumngrid

\subsection{Force-comparison WEP tests}
\label{tpend}

Typical force-comparison WEP tests 
can be viewed as comparing the motion of two or more bodies
joined through electromagnetic forces 
with that predicted by an equation 
of the modified form \rf{labeom}.
The predicted motion depends
on the details of the configuration,
so a unified analysis for all force-comparison WEP tests
is impractical.
Here,
we consider as an illustration 
a sensitive existing force-comparison WEP test 
based on a torsion pendulum
\cite{scwga,shaghss}.
Exceptional sensitivity to Lorentz violation 
can be achieved using a torsion pendulum 
with a spin-weighted bob
\cite{haccss,hccass,hnlbk},
but here we treat instead a bob with a dipolar composition.
We remark in passing that another interesting option 
for force-comparison WEP tests
is the use of superconducting gravimeters 
to compare gravitational forces
on parts of the Earth having different compositions
\cite{ss2},
although present sensitivities to coefficients for Lorentz violation
are likely to be somewhat weaker.

For the torsion pendulum,
a simple model of the bob is a dumbbell 
viewed as a rod with test bodies 1 and 2 placed on each end.
The two test bodies are composed of different materials,
and the bob is suspended by a torsion fiber
attached at the midpoint between them.
The resulting pendulum is 
typically rotated in the laboratory
to improve the modulation of the signal.
The relevant observable in such tests
is the twist angle $\th(T)$ of the torsion fiber.
This angle can be calculated from the Newton second law
as modified by the presence of Lorentz violation.

At zeroth order in Lorentz violation,
the pendulum hangs at an angle $\ze$ 
from the local vertical in the laboratory
given by Eq.\ \rf{zetadef}.
Lorentz-violating corrections to $\ze$ exist,
but these make no contribution to the signal
to the order at which we work.
This angle represents the equilibrium position 
for the swing mode of the pendulum.
Lorentz-violating modifications to this position
could drive small excitations of the swing mode,
but experiments tuned to the torsion mode
are typically comparatively insensitive to other modes 
\cite{shaghss}.

The orientation of the bob 
about the axis perpendicular to both the torsion fiber
and to the dumbbell dipole moment
can also be considered.
This is the equilibrium position for the wobble mode
of the pendulum.
For simplicity,
we assume here that the bob is suspended 
at a point $P$ equidistant between the centers of mass 
of the test bodies 
and that the test bodies are constructed 
to ensure the dumbbell is perpendicular to the torsion fiber.
In the absence of Lorentz violation,
this implies equality of the two masses $m_1$ and $m_2$.
However,
in the presence of Lorentz violation,
$m_1$ and $m_2$ differ 
at leading order in the coefficients for Lorentz violation.
As the pendulum rotates,
this difference could shift the dumbbell orientation 
away from its equilibrium point and generate small excitations
of the wobble mode about $P$.
Again, 
experiments are comparatively insensitive to this mode. 

\onecolumngrid
\begin{center}
\begin{tabular}{lc}
\multicolumn{2}{c} 
{Table \labg.\ Amplitudes for torsion-pendulum tests.} \\
\hline
\hline
Amplitude & Phase $\al_n$\\ 
\hline
$K_{\om_e} = 
- \half \left( m^w \cxxw + m^w \cyyw - 2 \al \afbw_T \right) 
\left( 1 + \frac{\om^2 R_\oplus}{g} \sin^2 \ch \right) \sin 2 \ch $ & 
$\ph$ \\
$K_{\om_e + \Om} = 
- \half  V_\oplus \al ( \ayw \cos \et + \azw \sin \et ) \sin 2 \ch
+ \half m^w V_\oplus \ctyw \cos \et \sin 2 \ch $ & 
$\ph$ \\
$L_{\om_e + \Om} =
- \half V_\oplus \al \axw \sin 2 \ch
+ \half m^w V_\oplus \ctxw \sin 2 \ch $ &
$\ph$ \\
$K_{\om_e - \Om} =
- \half V_\oplus \al (\ayw \cos \et + \azw \sin \et)\sin 2 \ch
+ \half m^w V_\oplus \ctyw \cos \et \sin 2 \ch $ &
$\ph$ \\
$L_{\om_e - \Om} =
\half V_\oplus \al \axw \sin 2 \ch
- \half m^w V_\oplus \ctxw \sin 2 \ch $ &
$\ph$ \\
$K_{\om_e + \om} =
m^w \cxzw \left( 1 - \frac{\om^2 R_\oplus}{g} \cos^2 \ch \right) \sin^2 \ch
- \frac{g V_L \al}{5 \om^2 R_\oplus} \afbw_Y (1 + \cos \ch)$ &
$ \ph + \ps $ \\
$L_{\om_e + \om} =
- m^w \cyzw \left( 1 - \frac{\om^2 R_\oplus}{g} \cos^2 \ch \right) \sin^2 \ch
- \frac{g V_L \al}{5 \om^2 R_\oplus} \afbw_X (1 + \cos \ch) $ &
$ \ph + \ps $ \\
$K_{\om_e - \om} =
m^w \cxzw \left( 1 - \frac{\om^2 R_\oplus}{g} \cos^2 \ch \right) \sin^2 \ch
+ \frac{g V_L \al}{5 \om^2 R_\oplus} \afbw_Y (1 - \cos \ch) $ &
$ \ph - \ps $ \\
$L_{\om_e - \om} =
m^w \cyzw \left( 1 - \frac{\om^2 R_\oplus}{g} \cos^2 \ch \right) \sin^2 \ch
+ \frac{g V_L \al}{5 \om^2 R_\oplus} \afbw_X (\cos \ch -1) $ &
$ \ph - \ps $ \\
$K_{\om_e + 2 \om} =
- \half m^w (\cxxw - \cyyw) \left( \sin \ch + \sin \ch \cos \ch + \frac{\om^2 R_\oplus}{g} \sin^3 \ch \cos \ch \right) $ &
$ \ph + 2 \ps $ \\
$L_{\om_e + 2 \om} =
m^w \cxyw \left( \sin \ch + \sin \ch \cos \ch + \frac{\om^2 R_\oplus}{g} \sin^3 \ch \cos \ch \right) $ &
$ \ph + 2 \ps $ \\
$K_{\om_e - 2 \om} =
\half m^w (\cxxw - \cyyw) \left( \sin \ch - \sin \ch \cos \ch - \frac{\om^2 R_\oplus}{g} \sin^3 \ch \cos \ch \right) $ &
$ \ph - 2 \ps $ \\
$L_{\om_e - 2 \om} =
m^w \cxyw \left( \sin \ch - \sin \ch \cos \ch - \frac{\om^2 R_\oplus}{g} \sin^3 \ch \cos \ch \right) $ &
$ \ph - 2 \ps $ \\
$K_{\om_e + \om + \Om} =
- \half m^w V_\oplus \ctxw \sin \et \sin^2 \ch $ &
$ \ph + \ps $ \\
$L_{\om_e + \om + \Om} =
\half m^w V_\oplus \big[ \ctyw \sin \et - \ctzw (1-\cos \eta) \big] \sin^2 \ch $ &
$\ph + \ps$ \\
$K_{\om_e + \om - \Om} =
- \half m^w V_\oplus \ctxw \sin \et \sin^2 \ch $ &
$ \ph + \ps $ \\
$L_{\om_e + \om - \Om} =
\half m^w V_\oplus \big[ \ctyw \sin \et + \ctzw (1 + \cos \eta) \big] \sin^2 \ch $ &
$\ph + \ps$ \\
$K_{\om_e - \om + \Om} =
- \half m^w V_\oplus \ctxw \sin \et \sin^2 \ch $ &
$ \ph - \ps $ \\
$L_{\om_e - \om + \Om} =
- \half m^w V_\oplus \big[ \ctyw \sin \et + \ctzw (1 + \cos \eta) \big] \sin^2 \ch $ &
$\ph - \ps$ \\
$K_{\om_e - \om - \Om} =
- \half m^w V_\oplus \ctxw \sin \et \sin^2 \ch $ &
$ \ph - \ps $ \\
$L_{\om_e - \om - \Om} =
- \half m^w V_\oplus \big[ \ctyw \sin \et - \ctzw (1-\cos \eta) \big] \sin^2 \ch $ &
$\ph - \ps$ \\
$K_{\om_e + 2 \om + \Om} =
\half m^w V_\oplus \ctyw (1- \cos \et) \sin \ch (1 + \cos \ch) $ &
$\ph + 2 \ps$ \\
$L_{\om_e + 2 \om + \Om} =
\half m^w V_\oplus \ctxw (1- \cos \et) \sin \ch (1 + \cos \ch) $ &
$\ph + 2 \ps$ \\
$K_{\om_e + 2 \om - \Om} =
- \half m^w V_\oplus \ctyw (1 + \cos \et) \sin \ch (1 + \cos \ch) $ &
$\ph + 2 \ps$ \\
$L_{\om_e + 2 \om - \Om} =
- \half m^w V_\oplus \ctxw (1 + \cos \et) \sin \ch (1 + \cos \ch) $ &
$\ph + 2 \ps$ \\
$K_{\om_e - 2 \om + \Om} =
- \half m^w V_\oplus \ctyw (1 + \cos \et) \sin \ch ( \cos \ch - 1) $ &
$ \ph - 2 \ps $ \\
$L_{\om_e - 2 \om + \Om} =
\half m^w V_\oplus \ctxw (1 + \cos \et) \sin \ch ( \cos \ch - 1) $ &
$ \ph - 2 \ps $ \\
$K_{\om_e - 2 \om - \Om} =
\half m^w V_\oplus \ctyw (1- \cos \et) \sin \ch (\cos \ch -1) $ &
$\ph - 2 \ps$ \\
$L_{\om_e - 2 \om - \Om} =
\half m^w V_\oplus \ctxw (1- \cos \et) \sin \ch (1 - \cos \ch) $ &
$\ph - 2 \ps$ \\
\hline
\hline
\end{tabular}
\end{center}
\twocolumngrid

To analyze the torsion mode,
it is convenient to express 
the relevant contributions to the difference $m_1 - m_2$
of the test-body masses
in a form displaying the dependence on particle species.
This form can be obtained from Eq.\ \rf{lvforcez},
giving
\beq
m_1 - m_2 = - \sum_{w} \left( N^w_1 - N^w_2 \right)
\left( 2 \al \afbw_T + m^w \cbw_{TT} \right).
\eeq
At leading order in Lorentz violation,
the oscillations of the system
are determined by the second-order differential equation
\beq
I \fr{d^2 \th}{dT^2} + 2 \ga I \fr{d \th}{d T} + \ka \th = \ta,
\label{torsionaleom}
\eeq
where $\ga$ is the torsional damping constant
and $\ka$ is the torsional spring constant.
The moment of inertia $I$ can be taken as
\beq
I = (m_1 + m_2) r_0^2 + I^\prime ,
\eeq
where $r_0$ is the distance from $P$ to the test bodies
and $I^\prime$ is the moment of inertia
of the remaining matter comprising the dumbbell.
The torque $\ta$ 
includes Lorentz-violating effects and
is determined by the forces on the test bodies
calculated from the \pno3 lagrangian $L^{(3)}_{a,c,s}$.

The damping term in Eq.\ \rf{torsionaleom}
ensures that free oscillations vanish in the steady state.
The time dependence of the steady-state solution
is therefore determined by the rotations
of the pendulum relative to the Sun-centered frame.
Neglecting possible torques other than those 
implied by $L^{(3)}_{a,c,s}$,
the steady-state solution 
can be written in frequency-decomposed form as 
\bea
\th (T) &=& \sum_{n,w}
\fr{\left( N^w_1 - N^w_2 \right) \om^2 R_\oplus r_0}
{I \sqrt{(\om_0^2 - \om_n^2)^2 + 4 \ga^2 \om_n^2}}
\nonumber \\
& &
\hskip 10pt
\times \big[K_n \sin(\om_n T + \be_n + \al_n) 
\nonumber \\
& & 
\hskip 20pt
+ L_n \cos(\om_n T + \be_n + \al_n) \big],
\label{pendeom}
\eea
where
$\be_n = 2 \ga \om_n/(\om_0^2 - \om_n^2)$
and where $\al_n$ is a phase fixing the relationship 
between the time coordinate in the turntable frame
and the Sun-centered time $T$.
The amplitudes $K_n$, $L_n$ and the phase $\al_n$
are given in Table \labg.
With the exception of the first row in the table,
these signals for Lorentz violation
are distinguished from other potential sources of WEP violation
by their characteristic time dependence.

In the above analysis,
the assumption of a steady-state solution implies 
the pendulum motion is governed by leading-order Lorentz violation,
while the torque $\ta$ is taken as
the only relevant source of Lorentz violation.
Note that Lorentz-violating contributions 
to the moment of inertia $I$ can be neglected here 
because they enter only at higher order.
These are a manifestation 
of the angular-momentum nonconservation
that accompanies Lorentz violation,
and they are analogous to the 
Lorentz-violating contributions  
to the effective inertial mass 
in the modified Newton second law \rf{labeom}.
In principle,
the rotation of the apparatus in the laboratory
introduces similar effects 
proportional to $\cb_{JK}$ and $\om_e^2 r_0$.
These may be comparable in magnitude
to effects listed in Table \labg\
that are suppressed by $\om^2 R_\oplus V_\oplus$,
but they offer no additional advantage 
in terms of sensitivity and so are disregarded here.

The above analysis can be used to extract
constraints on Lorentz violation
from the results of the torsion-pendulum WEP tests
reported in Refs.\ \cite{scwga,shaghss}.
The attained sensitivity to the differential acceleration 
of Be and Ti test bodies 
at the level of $10^{-15}$ ms$^{-2}$
\cite{scwga}
is the experimental basis for our limit \rf{aeffprl},
which extends an earlier bound 
\cite{akjt}
to include the coefficients $\eb_\mu$.
In the remainder of this subsection,
we revisit this issue 
to incorporate the slightly weaker constraints
from torsion-pendulum experiments using 
Al, Be, Cu, and Si test bodies 
\cite{shaghss},
and we consider implications of nonzero $\cbw_\mn$.

First, suppose $\cbw_\mn=0$.
Inclusion of data from tests with different materials
permits the extraction of some independent sensitivities
to neutron coefficients
and to combinations of electron and proton coefficients.
This treatment relies on differences in binding energy
between the materials involved,
so the signal sensitivity of $10^{-15}$ ms$^{-2}$ relative to
$\om^2R_\oplus \simeq 3\times 10^{-3}$ ms$^{-2}$
is suppressed both by the typical material-dependence factor
of $(N^w_1 - N^w_2)\simeq 10^{-2}$ 
appearing in Eq.\ \rf{pendeom}
and by another order of magnitude
from the binding-energy difference. 
By combining available Be-Ti and Al-Be data
\cite{shaghss}
we obtain the estimated bounds
\bea
|\al \afbx{\epcombo}_T| &\lsim& 10^{-10} {\rm ~GeV},
\nonumber\\
|\al \afbn_T | &\lsim& 10^{-10} {\rm ~GeV},
\label{viabinding}
\eea
valid for $\cbw_\mn=0$.

If instead nonzero coefficients $\cbw_\mn$ are present,
then we obtain the estimated bound
\beq
|\al \afbx{\pen}_T - \frac13 m^p \cbx{\pen}_{TT} 
- \frac16 m^n \cqbx{n}| 
\lsim 10^{-11} {\rm ~GeV}.
\label{nobinding}
\eeq
The contributions due to the spatial neutron coefficient $\cqbx{n}$
cannot be disentangled from those due to the temporal components
at this order in the analysis.
However,
this separation becomes feasible
when the result \rf{nobinding}
is combined with the limit achieved via free-fall WEP tests
given in row 2 of Table \labf. 
We thereby obtain the constraints 
\bea
|\al \afbx{\pen}_T - \frac13 m^p \cbx{\pen}_{TT} | 
&\lsim& 10^{-8} {\rm ~GeV},
\nonumber \\
|\cqbx{n}| &\lsim& 10^{-8}.
\label{fftorsionp}
\eea
As discussed following Eq.\ \rf{cbpen},
the possibility of $\cbw_\mn$-type Lorentz-violating effects
in the binding energy impedes 
its direct use in extracting independent sensitivities to
$|\al \afbx{\epcombo}_T - \frac13 m^p \cbx{\epcombo}_{TT} |$
and $|\al \afbx{n}_T - \frac13 m^p \cbx{n}_{TT} |$.

In addition to the constraints
\rf{viabinding}, \rf{nobinding}, and \rf{fftorsionp},
other new bounds could be placed 
on the moduli of certain coefficients for Lorentz violation
by reanalyzing the time dependence of the data 
obtained in the experiments of Refs.\ \cite{scwga,shaghss}
using the result \rf{pendeom}.
Crude estimates of these sensitivities 
are given in Table \labh.
These are obtained 
disregarding binding-energy considerations
but making the strong assumption
that all relevant frequencies
in Table \labg\ can be studied in the data.
Allowing for binding-energy effects
could yield independent sensitivities
to the neutron coefficients
and to a combination of proton and electron coefficients,
both reduced by roughly a factor of 10.

\begin{center}
\begin{tabular}{cc}
\multicolumn{2}{c} 
{Table \labh. Sensitivities for torsion-pendulum tests.} \\
\hline
\hline
Coefficient & Sensitivity\\ 
\hline
$\cb_{(TJ)}^n$                           & $[10^{-7}]$ \\
$ \al \afbx{\pen}_X$                    & $[10^{-8}$ GeV] \\
$ \al \afbx{\pen}_{Y+Z}$                & $[10^{-7}$ GeV] \\
$ \al \afbx{\pen}_Y $                   & $[10^{-8}$ GeV] \\
$ \al \afbx{\pen}_Z $                   & $[10^{-7}$ GeV] \\
\hline
\hline
\end{tabular}
\end{center}

\section{Satellite-based WEP Tests}
\label{space}

Space-based platforms offer certain advantages
in tests of gravity 
\cite{spacegrav} 
and searches for Lorentz violation 
\cite{spaceexpt}.
The long free-fall times that may be attainable
on a drag-free spacecraft make satellite-based WEP tests 
particularly attractive.
Several proposals are in an advanced stage of development, 
including
the Micro-Satellite \`a tra\^in\'ee Compens\'ee
pour l'Observation du Principe d'Equivalence
(MicroSCOPE)
\cite{muscope}, 
the Satellite Test of the Equivalence Principle (STEP) 
\cite{step},
and the Galileo Galilei (GG) mission 
\cite{gg}.
A WEP reach similar to that of STEP
has also been suggested for the 
Grand Unification and Gravity Explorer
(GaUGE) mission
\cite{gauge}.

The basic idea underlying these missions 
is to monitor the relative motion
of test bodies made of different materials
as they orbit the Earth in a satellite.
In the presence of nonzero coefficients for Lorentz violation
$\afbw_\mu$ and $\cbw_\mn$,
the orbits of the test bodies become material dependent.
In this section,
we determine the resulting apparent WEP violations
and then obtain crude estimates of the 
sensitivities to $\afbw_\mu$ and $\cbw_\mn$
attainable in MicroSCOPE, STEP, and GG.

\subsection{Theory}
\label{Theory}

The basic observable for a satellite-based WEP test
is the differential local acceleration
between the test bodies.
The typical design goal is to achieve excellent sensitivity 
to one or two components of this acceleration.
For present purposes,
we can idealize the situation as a pair of test bodies 
aboard a satellite traveling in a circular orbit.
In what follows,
we allow for the possibility that the test bodies 
are also rotating about an axis perpendicular
both to the direction of motion of the satellite
and to the direction of acceleration sensitivity.

Some notation relevant for our analysis of satellite-based WEP tests
is summarized in Table \spacea. 
Paralleling the analysis of terrestrial experiments
in Sec.\ \ref{labtheory},
it is convenient to introduce 
an Earth-centered frame with coordinates 
$x^\muearth = (\tearth,\xearth,\yearth,\zearth)$,
chosen so that $\tearth = T$
and so that the spatial components match 
those of the Sun-centered frame
at leading post-newtonian order.
The Earth-centered coordinates
can be related to the Sun-centered ones
as discussed in Sec.\ \ref{Frames}.
The angles $\xi_1$, $\xi_2$ in the table are defined
relative to the basis vectors of the Earth-centered frame.
The notation for properties of the test masses 1 and 2 
follows that of Sec.\ \ref{Sensitivities}.

\begin{center}
\begin{tabular}{ll}
\multicolumn{2}{c} 
{Table \spacea. Notation for satellite-based WEP tests.} \\
\hline
\hline
Quantity & Definition\\ 
\hline
$R_{\oplus}$            & mean Earth radius \\
$V_{\oplus}$    & mean Earth orbital speed \\
$r^J$            & Earth-satellite separation\\
$\om_s$          & satellite orbital frequency\\
$\om_r$          & satellite rotational frequency\\
$\xi_1$          & inclination of satellite orbit\\
$\xi_2$          & longitude of satellite-orbit node\\
$\th_1$          & phase fixing satellite location at $T=0$\\
$\th_2$          & phase fixing satellite orientation at $T=0$\\
\hline
\hline
\end{tabular}
\end{center}

Establishing the signal arising 
from nonzero coefficients for Lorentz violation
requires the transformation from the Sun-centered frame 
to a frame comoving with the satellite.
The satellite frame serves as the equivalent
of the laboratory frame for terrestrial searches.
We denote coordinates in the satellite frame
by $x^{\hat{\mu}}$. 

Since the satellite orbit is inclined 
relative to the Earth-centered frame,
it is also useful to introduce
an intermediate frame aligned with the satellite orbit
and hence rotated with respect to the Earth-centered frame.
The intermediate coordinates are denoted by $x^{\mu^\prime}$.
The rotation transformation 
from $x^{j^\prime}$ to $x^\jearth$
can be written as the matrix
\beq
R_1^{\tilde{j} k^\prime} = \left( \begin{array}{ccc}
\cos \xi_2 & - \cos \xi_1 \sin \xi_2 & \sin \xi_1 \sin \xi_2 \\
\sin \xi_2 & \cos \xi_1 \cos \xi_2   & - \sin \xi_1 \cos \xi_2 \\
0          & \sin \xi_1              & \cos \xi_1 \\
\end{array} \right)
\eeq
using the angles $\xi_1$ and $\xi_2$
defined in Table \spacea.

The connection between the satellite coordinates 
and the Earth-centered coordinates can be written 
\beq
x^{\tilde{j}} = 
R_1^{\tilde{j} k^\prime}
( R_2^{k^\prime \lhat} x^\lhat + x_s^{k^\prime} ).
\label{satellitetoearth}
\eeq
Here,
$x_s^{k^\prime}$ is the world line of the satellite
in the intermediate coordinate system.
This world line can be parametrized as
\beq
x_s^{k^\prime} = \left( 
r \cos (\om_s T + \th_1), r \sin (\om_s T + \th_1 ), 0 
\right),
\eeq
where $r$ is the magnitude of the Earth-satellite separation.
The satellite therefore orbits in the $x^\prime$-$y^\prime$ plane. 
Also,
in Eq.\ \rf{satellitetoearth}
the rotation $R_2^{k^\prime \lhat}$ of the satellite 
is given by the matrix
\beq
R_2^{k^\prime \lhat} = \left( \begin{array}{ccc}
\cos (\om_r T + \th_2) & - \sin (\om_r T + \th_2) & 0 \\
\sin (\om_r T + \th_2) & \cos(\om_r T + \th_2) & 0 \\
0 & 0 & 1\\
\end{array} \right).
\eeq
The axis of the satellite rotation 
is therefore along $\zhat$.

\vskip -20pt
\onecolumngrid
\begin{center}
\begin{tabular}{lc}
\multicolumn{2}{c} 
{Table \spaceb. Amplitudes for satellite-based WEP tests.} \\
\hline
\hline
Amplitude & phase\\ 
\hline
&\\
$P_{\om_r} = m^w r \om_s \Big[ \ctyw \sin \xi_1 +\ctxw \cos \xi_1  \Big] 
+ \fr{\om R_\oplus^2 \al \cos \xi_2}{5 r} \Big[\axw \cos \xi_1 + \ayw \sin \xi_1 \Big]$ & $\th_2$\\
$Q_{\om_r} = m^w r \om_s  \Big[ \ctxw \sin \xi_1 \cos \xi_2 -\ctyw \cos \xi_1 
\cos \xi_2 -\ctzw \sin \xi_2  \Big] $ & \\
$\pt{Q_{\om_r} =}
+ \fr{\om R_\oplus^2 \al}{5 r} \Big[
\axw \sin \xi_1 - \ayw \cos \xi_1 \Big]$
& $\th_2$ \\
$P_{\om_r + \om_s} = 2 m^w \Big[ 
\cos \xi_2 \cos 2 \xi_1 \cxyw 
+\sin \xi_2 \sin \xi_1 \cyzw$ 
& \\
\pt{$P_{\om_r + \om_s} = 2 m^w \Big[$} 
$+ \half \sin 2 \xi_1 \cos \xi_2 (\cyyw - \cxxw)
+\sin \xi_2 \cos \xi_1 \cxzw \Big] $ & $\th_1 + \th_2$ \\
$Q_{\om_s + \om_r} = m^w \Big[ ( \cos^2 \xi_2 \cos^2 \xi_1 - \sin^2 \xi_1 + \half \sin^2 \xi_2) (\cxxw - \cyyw) 
+ \half \sin^2 \xi_2 ( \cxxw + \cyyw - 2 \czzw)$ & \\
$\pt{Q_{\om_s + \om_r} = m^w \Big]}
- \cos \xi_1 \sin 2 \xi_2 \cyzw + \sin \xi_1 \sin 2 \xi_2 \cxzw 
+ \sin 2 \xi_1 (1 + \cos^2 \xi_2)\cxyw \Big]$ & $\th_1 + \th_2$ \\
$Q_{\om_s-\om_r} = m^w \Big[
\left( \cos^2 \xi_1 \sin^2 \xi_2 + \half \cos^2 \xi_2 + \half \right) \left( \cxxw - \cyyw \right)$
& \\
\pt{$Q_{\om_s-\om_r} = m^w \Big[$}
$- \half \sin^2 \xi_2 \left( \cxxw + \cyyw - 2 \czzw \right)
+ 2 \cyyw$ & \\
$\pt{Q_{\om_s-\om_r} = m^w \Big[}
+ \sin 2 \xi_1 \left( 1 - \cos^2 \xi_2 \right) \cxyw
- \sin \xi_1 \sin 2 \xi_2 \cxzw 
+ \cos \xi_1 \sin 2 \xi_2 \cyzw \Big]
- 2 \alpha \afbw_T $ & $\th_1 - \th_2$\\
$P_{2\om_s-\om_r} = - m^w r \om_s \Big[ \ctxw \cos \xi_1 
+\ctyw \sin \xi_1  \Big]
- \fr{3 \om R_\oplus^2 \al \cos \xi_2}{5 r}
\Big[ \axw \cos \xi_1 + \ayw \sin \xi_1 \Big]$ & $2 \th_1 - \th_2$ \\
$Q_{2\om_s-\om_r} = m^w r \om_s \Big[ \ctyw \cos \xi_1 \cos \xi_2 
-\ctxw \sin \xi_1 \cos \xi_2 
+\ctzw \sin \xi_2 
 \Big]$ & \\
$ \pt{Q_{2\om_s-\om_r} =}
- \fr{3 \om R_\oplus^2 \al}{5 r} \Big[
\axw \sin \xi_1 - \ayw \cos \xi_1 \Big]$
& $2 \th_1 - \th_2$ \\
$P_{\Om + \om_s + \om_r} = m^w V_\oplus \Big[ 
\left( \cos^2 \xi_1  - \sin^2 \xi_1 \cos^2 \xi_2 
- \cos \et \cos \xi_2 \cos 2 \xi_1 
- \sin \et \sin \xi_2 \cos \xi_1 \right) \ctxw$ & \\
$ \pt{P_{\Om + \om_s + \om_r} = m^w V_\oplus \Big[ }
+ \sin \xi_1 \sin \xi_2 \left( \cos \xi_2 - \cos \et \right) \ctzw$ & \\
$ \pt{P_{\Om + \om_s + \om_r} = m^w V_\oplus \big[}
+ \left(\cos \xi_1 
+ \cos \xi_1 \cos^2 \xi_2 
- \sin \et \sin \xi_2
- 2 \cos \et \cos \xi_1 \cos \xi_2 \right) \sin \xi_1 \ctyw  \Big]$ & $\th_1 + \th_2$ \\
$Q_{\Om + \om_s + \om_r} = m^w V_\oplus \Big[
\left(2 \cos \xi_1 \cos \xi_2 
- \sin \et \sin \xi_2 \cos \xi_2 
- \cos \et \cos \xi_1 (1 + \cos^2 \xi_2) \right) \sin \xi_1 \ctxw$ & \\
$ \pt{Q_{\Om + \om_s + \om_r} = m^w V_\oplus \Big[ }
- \left(\cos 2 \xi_1 \cos \xi_2 
- \sin \et \cos \xi_1 \sin \xi_2 \cos \xi_2
+ \cos \et ( 1 - \cos^2 \xi_1 \sin^2 \xi_2)
\right)\ctyw $ & \\
$ \pt{Q_{\Om + \om_s + \om_r} = m^w V_\oplus \Big[ }
- \left( \cos \xi_1 
- \sin \et \sin \xi_2
- \cos \et \cos \xi_1 \right) \sin \xi_2 \ctzw \Big]$ & $\th_1 + \th_2$ \\
$P_{\Om + \om_s - \om_r} = m^w V_\oplus \Big[
 \left( 1 - \sin^2 \xi_1 \sin^2 \xi_2 \right) \ctxw 
+ \half \sin 2 \xi_1 \sin^2 \xi_2 \ctyw 
- \half \sin \xi_1 \sin 2 \xi_2 \ctzw \Big]$
& \\
\pt{$P_{\Om + \om_s - \om_r} = m^w V_\oplus \Big[$}
$ -\alpha V_\oplus \axw$ & $\th_1 - \th_2$ \\
$Q_{\Om + \om_s - \om_r} = 
- m^w V_\oplus \Big[\half \left( \cos \et \sin 2 \xi_1 \sin^2 \xi_2 - \sin \et \sin \xi_1 \sin 2 \xi_2 \right) \ctxw$ & \\
$ \pt{Q_{\Om + \om_s - \om_r} = -m^2 V_\oplus}
+ \left( \half \sin \et \cos \xi_1 \sin 2 \xi_2 
+ (1 - \sin^2 \xi_2 \cos^2 \xi_1) \cos \et \right) \ctyw$ & \\
$ \pt{Q_{\Om + \om_s - \om_r} = -m^2 V_\oplus}
+ \left( \sin \et \sin^2 \xi_2 
+ \half \cos \et \cos \xi_1 \sin 2 \xi_2 \right) \ctzw  \Big]
+ \al V_\oplus \big[ \azw \sin \et + \ayw \cos \et \Big]$ & $\th_1 - \th_2$ \\
$P_{\Om - \om_s + \om_r} = m^w V_\oplus \Big[
(1 - \sin^2 \xi_1 \sin^2 \xi_2) \ctxw
+ \half \sin 2 \xi_1 \sin^2 \xi_2 \ctyw 
- \half \sin \xi_1 \sin 2 \xi_2 \ctzw \Big]$
& \\
\pt{ $P_{\Om - \om_s + \om_r} = m^w V_\oplus \Big[$}
$- \al V_\oplus \axw$ & $- \th_1 + \th_2$\\
$Q_{\Om - \om_s + \om_r} = m^w V_\oplus \Big[ 
\half \left( \sin \et \sin \xi_1 \sin 2 \xi_2 - \cos \et \sin2 \xi_1 \sin^2 \xi_2 \right) \ctxw$ & \\
$\pt{Q_{\Om + \om_s - \om_r} = m^w V_\oplus} 
- \left(\half \sin \et \cos \xi_1 \sin 2 \xi_2 + \cos \et (1 - \cos^2 \xi_1 \sin^2 \xi_2) \right) \ctyw $ & \\
$\pt{Q_{\Om + \om_s - \om_r} = m^w V_\oplus} 
- \left( \sin \et \sin^2 \xi_2 + \half \cos \et \cos \xi_1 \sin 2 \xi_2 \right) \ctzw \Big]
+ \alpha V_\oplus \Big[ \azw \sin \et + \ayw \cos \et \Big]$ & $- \th_1 + \th_2$ \\
$P_{\Om - \om_s - \om_r} = m^w V_\oplus \Big[
\left(\cos^2 \xi_1 - \sin^2 \xi_1 \cos^2 \xi_2 
+ \sin \et \cos \xi_1 \sin \xi_2 
+ \cos \et \cos 2 \xi_1 \cos \xi_2 \right) \ctxw$ & \\
$\pt{P_{\Om - \om_s - \om_r} = m^w V_\oplus}
+ \left( \half \sin 2 \xi_1 (1 + \cos^2 \xi_2)
+ \sin \et \sin \xi_1 \sin \xi_2 
+ \cos \et \sin 2 \xi_1 \cos \xi_2 \right) \ctyw$ & \\
$\pt{P_{\Om - \om_s - \om_r} = m^w V_\oplus}
+ \left( \half \sin 2 \xi_2 + \cos \et \sin \xi_2 \right) \sin \xi_1 \ctzw \Big]$ & $-\th_1 - \th_2$ \\
$Q_{\Om - \om_s - \om_r} = m^w V_\oplus \Big[ 
- \left( \sin 2 \xi_1 \cos \xi_2 
+ \half \sin \et \sin \xi_1 \sin 2 \xi_2 + \half \cos \et \sin \xi_1 (1 + \cos^2 \xi_2) \right) \ctxw $ & \\
$ \pt{P_{\Om - \om_s - \om_r} = m^w V_\oplus \Big[ }
+ \left( \cos 2 \xi_1 \cos \xi_2 
+ \half \sin \et \cos \xi_1 \sin \xi_2 - \cos \et (\sin^2 \xi_1 - \cos^2 \xi_1 \cos^2 \xi_2)\right)\ctyw
$ & \\
$ \pt{P_{\Om - \om_s - \om_r} = m^w V_\oplus \Big[ }
+ \left( \cos \xi_1 \sin \xi_2 + \sin \et \sin^2 \xi_2 + \half \cos \et \cos \xi_1 \sin 2 \xi_2 \right) \ctzw \Big]$ & $-\th_1 - \th_2$ \\
&\\
\hline
\hline
\end{tabular}
\end{center}
\twocolumngrid

For our purposes,
it suffices to obtain explicitly 
the local differential acceleration $\De \acc^\xhat$ 
of the test bodies in the $\xhat$ direction.
We have 
\beq
\De \acc^\xhat \equiv 
\fr {d^2 \De \xhat}{d\that^2} 
= \De \acc^\xhat_{\rm tidal} + \De \acc^\xhat_{\rm LV} + \ldots.
\label{spaceacc}
\eeq
The first term on the right-hand side of this expression
is the conventional Newton tidal term.
It takes the form
\bea
\De \acc^\xhat_{\rm tidal} &=& 
- \Big( 
\frac{3}{2} \om_s^2 
\cos ( 2 \om_r T - 2 \om_s T + \th_2 -\th_1 ) 
\nonumber \\
& & 
\hskip 15pt
+ \om_r^2 + \half \om_s^2
\Big)  \De \xhat.
\eea
The second term in Eq.\ \rf{spaceacc} 
contains Lorentz-violating contributions
to the differential acceleration.
It can be written
\bea
\De \acc^\xhat_{\rm LV} &=& 
r \om_s^2 \sum_{w,n} 
\left( \fr{N_1^w}{m_1} - \fr{N_2^w}{m_2} \right)
\label{lvspaceacc} \\
& & 
\nonumber
\times
\Big( P_n \sin (\om_n T +\al_n) 
+ Q_n \cos (\om_n T + \al_n) \Big).
\eea
The amplitudes $P_m$, $Q_m$
and the corresponding phases
are provided in Table \spaceb.
Finally,
the ellipsis in Eq.\ (\ref{spaceacc})
represents higher-order general-relativistic corrections
and Lorentz-violating effects
at the same post-newtonian order
as $\De \acc^\xhat_{\rm tidal}$.
The latter are typically of lesser interest.
If desired,
the differential acceleration $\De \acc^\yhat$ 
along $\yhat$ 
can be obtained by performing the transformation
$\om_r T \rightarrow \om_r T - \pi /2$
on Eq.\ \rf{spaceacc}.

\subsection{MicroSCOPE and STEP}
\label{MicroSCOPE and STEP}

Within our idealized scenario,
MicroSCOPE 
\cite{muscope}
and STEP
\cite{step}
can be analyzed in parallel.
Each apparatus consists of a pair of cylindrical test bodies 
made of different material
but having a common symmetry axis.
The test bodies are free to move along this axis.
In satellite coordinates,
this direction lies along $\xhat$ 
and is perpendicular 
both to the direction of motion of the satellite
and to the axis of the satellite rotation.

One prosaic origin of relative motion of the test bodies 
along the $\xhat$ direction
could be the influence of tidal forces
on a misalignment of the two centers of mass,
which would lead to the acceleration 
$\De \acc^\xhat_{\rm tidal}$
in Eq.\ \rf{spaceacc}.
This can be separated 
from the acceleration due to WEP violations
stemming from Lorentz-invariant sources,  
which enters with the characteristic frequency 
$\om_s - \om_r$.
Here,
we are interested in 
a WEP-violating acceleration $\De \acc^\xhat_{\rm LV}$
arising from the coefficients $\afbw_\mu$ and $\cbw_\mn$
for Lorentz violation.
This can be distinguished from both the above effects
through careful separation of the frequencies 
associated with the amplitudes in Table \spaceb,
except for the amplitude $Q_{\om_s - \om_r}$.

The sensitivity goals of MicroSCOPE and STEP
are $\De \acc / r \om_s^2 < 10^{-15}$
and $\De \acc / r \om_s^2 < 10^{-18}$,
respectively.
These sensitivities and the results in Table \spaceb\
can be used to obtain rough estimates
of the reach of these experiments
for studies of Lorentz violation.
For this purpose,
we take the quantity
$N_1^w/m_1 - N_2^w/m_2$
appearing in Eq.\ \rf{lvspaceacc}
to be of order $10^{-2}$ GeV$^{-1}$,
which is the best available value
with the Pt-Ir, Be, and Nb test bodies 
presently proposed for STEP.
Note that the bounds scale linearly with this difference,
so a careful choice of test-body material
can maximize sensitivity to Lorentz violation.
Moreover,
combining results for different test materials
can yield additional independent sensitivities.
Note also that
the experimental reach may vary with the choice of orbit.
For definiteness,
we suppose the sines and cosines of $\xi_1$ and $\xi_2$
are of order one.

Our crude estimates for attainable sensitivities 
to the moduli of $\afbw_\mu$ and $\cbw_\mn$
for MicroSCOPE and STEP
are presented in Table \spacec.
In each row, 
the listed sensitivities are obtained under the assumption 
that all coefficients vanish 
except those appearing in the first entry.
The key factor underlying the difference in reach
for the various coefficient combinations
is the boost entering the relevant amplitude 
in Table \spaceb.
Amplitudes containing $V_\oplus$
are suppressed by roughly $10^{-4}$,
while those containing $r \om_s$
are suppressed by about $10^{-5}$.
As before,
the braces indicate the estimated sensitivities
involve data from future tests.

\subsection{Galileo Galilei}
\label{Galileo Galilei}

Certain design features of GG 
\cite{gg}
differ from those of MicroSCOPE and STEP
in ways that are significant 
for studies of Lorentz violation.
Although GG also uses coaxial cylindrical test bodies,
it is sensitive to accelerations
in the plane perpendicular to the axis of the cylinders.
Also,
the cylinders are rotated about their axis
at a comparatively high frequency of about 2 Hz.

In applying the generic analysis 
of Sec.\ \ref{Theory} to GG,
it is convenient to take the cylinder axes 
to lie along $\zhat$.
The experiment is then sensitive to accelerations
in the $\xhat$-$\yhat$ plane.
The differential acceleration $\De \acc^\xhat$ 
along $\xhat$ 
is given in Eq.\ \rf{spaceacc},
while $\De \acc^\yhat$ 
can be obtained by adjusting the phase $\th_2$.

The sensitivity goal
of GG is $\De \acc / r \om_s^2 < 10^{-17}$.
In Table \spacec,
we present rough estimates
of the corresponding reach 
for measurements of the coefficients $\afbw_\mu$ and $\cbw_\mn$
for Lorentz violation,
obtained using the result \rf{lvspaceacc}.
The values for GG in the table are based on the same
assumptions as those discussed above for MicroSCOPE and STEP. 
This includes the material-dependent factor,
with the proposed materials for the GG test bodies being Be and Cu.
The boost factors leading to the varying sensitivities
for GG listed in the table
are also of the same order of magnitude
as for the other satellite experiments.

We remark in passing that the comparatively high rotation rate
for the GG cylinders could introduce
additional Lorentz-violating effects.
Typically,
the presence of nonzero $\cbw_\mn$
introduces modifications to the effective moment of inertia
of a body.
This can affect the dynamical balance of the system,
which can lead to observable signals.
For example,
potential effects of this type on the timing of pulsar signals
have been used to constrain some combinations of $\cbn_\mn$
\cite{bapulsar}.
In the present context,
the observable signals could include 
a material-dependent Lorentz-violating wobble
varying at the satellite frequency
and at the Earth's orbital frequency.
It is conceivable that these Lorentz-violating effects
could be detected by the GG apparatus
that senses the test-body location.
Notice that the signals would be independent of gravity.
They may be detectable 
using sophisticated terrestrial dynamical-balancing equipment,
perhaps including that used in the  
Galileo Galilei on the Ground (GGG) experiment
\cite{ggg}.
The investigation of these effects
represents an interesting open question
for future work.

\onecolumngrid
\begin{center}
\begin{tabular}{cccc}
\multicolumn{4}{c} 
{Table \spacec. Sensitivities for satellite-based WEP tests.} \\
\hline
\hline
Coefficient & MicroSCOPE & GG & STEP \\ 
\hline
$\al \afbx{\pen}_T - \fr13 m^p \cbx{\pen}_{TT}$ & \{$10^{-13}$ GeV\} & \{$10^{-15}$ GeV\} & \{$10^{-16}$ GeV\} \\
$\al \afbx{\pen}_X$                              & \{$10^{-9}$ GeV\}  & \{$10^{-11}$ GeV\} & \{$10^{-12}$ GeV\} \\
$\al \afbx{\pen}_{Y+Z}$                          & \{$10^{-9}$ GeV\}  & \{$10^{-11}$ GeV\} & \{$10^{-12}$ GeV\} \\
$\al \afbx{\pen}_Y$                              & \{$10^{-7}$ GeV\}  & \{$10^{-9}$ GeV\}  & \{$10^{-10}$ GeV\} \\
$\al \afbx{\pen}_Z$                              & \{$10^{-7}$ GeV\}  & \{$10^{-9}$ GeV\}  & \{$10^{-10}$ GeV\} \\
$\cqbx{n}$               & \{$10^{-13}$\} & \{$10^{-15}$\} & \{$10^{-16}$\} \\
$\cbn_{(TJ)}$                                    & \{$10^{-9}$\}  & \{$10^{-11}$\} & \{$10^{-12}$\} \\
\hline
\hline
\end{tabular}
\end{center}
\twocolumngrid

\section{Exotic gravitational tests}
\label{exotic}

In this section,
we offer a few remarks 
about some gravitational searches for Lorentz violation 
using material test bodies other than 
neutral bulk matter, neutral atoms, or neutrons.
These more exotic searches typically present 
unique experimental challenges,
but they could provide access 
to combinations of coefficients for Lorentz violation
that are awkward or impossible to isolate and measure
in other searches discussed in this paper.
Here,
we briefly consider 
tests with electrons and ions,
studies with antihydrogen,
and experiments using particles
from the second and third generation of the SM.

\subsection{Tests with electrons and ions}
\label{charged}

Measurements of the gravitational acceleration 
of charged matter remain of definite theoretical interest 
because the WEP and other foundational aspects 
of gravity are comparatively poorly tested
in this regime.
In this subsection,
we consider possible signals 
from studies of charged electrons or ions.
Given the experimental challenges of these tests
and their limited attainable sensitivities,
we restrict attention here to effects from $\afbw_\mu$,
setting other coefficients to zero for simplicity.

In the context of searches for Lorentz violation,
gravitational tests with charged matter
offer unique access to the coefficients $\afbw_\mu$.
For example,
measurements of this kind can disentangle 
coefficients for Lorentz violation 
in the proton and electron sectors. 
They can also detect certain countershaded effects 
that are otherwise invisible.
In particular,
some models have coefficients $\afbw_\mu$ 
proportional to electric charge,
which would evade detection
in searches with neutral test bodies 
\cite{akjt}.
This possibility is a natural consequence 
for theories in which the photon modes are interpreted 
as NG bosons from spontaneous Lorentz breaking
and in which $\afbw_\mu$ remains physically observable,
such as nonminimally coupled bumblebee electrodynamics 
\cite{lvng}.

One candidate technique to measure gravitational effects 
from the coefficients $\afbw_\mu$
is charged-particle interferometry.
Electron interferometry has been used
to measure the Sagnac effect at the 30\% level 
\cite{fhmn},
while ion interferometry is under investigation
as a practical tool for sensitive tests
of Coulomb's law 
\cite{ncd}.
In the present context,
electron or ion interferometry
offers an interesting alternative prospect
to the free-fall tests with neutral matter
discussed in Sec.\ \ref{labtests}.
For a given geometry,
the observed phase shift
can be determined using the methods
of Sec.\ \ref{ffgravimeter}.
In the limit of interest here,
the vertical acceleration $\acc_\zhat$ 
of the electron or ion T 
in the gravitational field of the Earth S
is given at \pno2 by
\beq
\acc_\zhat = 
- g 
- \fr{2 g \al}{\mt} \abt_\that 
- \fr{2 g \al}{\ms} \abs_\that .
\eeq
As before,
the \pno3 version of this acceleration 
can be frequency decomposed relative to the Sun-centered frame,
with the corresponding amplitudes 
depending on the coefficients $\afbw_\mu$
as given in Table \labc.

In principle,
a charged-particle interferometer
can be used for free-fall gravimeter tests
of the type discussed in Sec.\ \ref{ffgravimeter} 
or for free-fall WEP tests
as in Sec.\ \ref{epfreefall}.  
A free-fall gravimeter test is insensitive to $\afbw_T$
and has only boost-suppressed signals from $\afbw_J$,
so a substantial improvement over the existing reach 
of charged-matter interferometers
would be required to achieve a sensitivity
compatible with perturbative consistency.
In contrast,
a free-fall WEP test is directly sensitive to $\afbw_T$ 
but requires a simultaneous measurement
with two test bodies.
One option along these lines could be 
a direct comparison with neutral matter 
via a falling corner cube or an atom interferometer.

Another approach to gravitational tests with charged matter
is to study the motion of charged particles
in a vertical metallic drift tube.
This setup is accompanied by 
gravitationally induced electric forces
caused by the sagging of the tube
\cite{drifttube},
along with a variety of challenging systematics.
An experiment of this type with electrons
\cite{fwwf}
confirmed that the gravitational forces 
on the electrons in the tube 
and on the electrons within the metal
are comparable to about 10\%.
An analogous experiment involving cold antiprotons
\cite{tgmn}
was designed to achieve a sensitivity
of $0.1\%$ to the gravitational acceleration
\cite{ps200}.
These measurements are all experimentally challenging,
and their interpretation is theoretically subtle
\cite{drom}.

In the present context
of gravitational Lorentz violation 
involving the coefficients $\afbw_\mu$,
intuition for the theoretical implications 
of ballistic tests of this type
can be gained by considering the idealized case
and working at \pno2.
We suppose a test particle T
of charge $\qt$
moves along the symmetry axis 
of a vertical cylindrical metallic drift tube 
with body comprised of a lattice of ions of type $I$
and conduction electrons $e$.
Disregarding applied fields,
stray fields, and various systematics,
the overall conventional force on the particle T
is the sum of the direct gravitational force on T 
from the Earth S
and the net force on T from the electromagnetic field 
arising from the gravitationally induced sagging of the tube.
The presence of nonzero coefficients $\afbw_\mu$
introduces corrections to both these forces.
At \pno2,
the gravitational force on T is given by 
\bea
(F_{\rm grav})_\zhat &=& 
- \mt g 
- 2 g \al \abt_\that 
- 2 g \al \fr{\mt}{\ms} \abs_\that ,
\nonumber\\
\label{grzforce}
\eea
while the vertical component of the force on T 
from the gravitationally induced electric field is
\bea
(F_{\rm em})_\zhat &=& 
\fr {\qt}{e} 
\left( m^e g 
+ 2 g \al \afbe_\that + 2 g \al\fr{m^e}{\ms} \abs_\that
\right)
\nonumber\\
&&
+ \ga \fr {\qt}{e} 
\left( m^I g
+ 2 g \al \afbx{I}_\that + 2 g \al\fr{m^I}{\ms} \abs_\that
\right).
\nonumber\\
\label{emzforce}
\eea
In these expressions,
$m^e$ and $m^I$ are the masses
of an electron and an ion in the tube lattice,
respectively, 
while 
$e$ and $q^I$ are the corresponding charges. 
The factor $\ga$ is a constant,
set by the properties of the metal lattice.
In Eq.\ \rf{emzforce},
the first three terms arise from the
sagging of the electrons in the tube walls,
while the last three 
are proportional to the dilation derivative 
of the work function for the metal 
and originate in the longitudinal compression of the lattice.
These expressions reduce to standard ones
when the coefficients $\afbw_\mu$ vanish.

Although the expressions 
\rf{grzforce} and \rf{emzforce}
hold in an idealized situation,
they suffice to demonstrate in principle
that experiments of this type
are sensitive to nonzero coefficients $\afbw_\mu$,
even when these coefficients 
are undetectable with neutral matter.
This is also true if the particle T is an electron,
when the sum of the forces
$(F_{\rm grav})_\zhat$ and $(F_{\rm em})_\zhat$ 
leaves only the last three terms in  Eq.\ \rf{emzforce}.
In practice,
however,
the reported reach of drift-tube experiments 
to date is insufficient to achieve useful sensitivity
in gravimeter tests.
A WEP test relating a drift-tube setup 
to an independent gravimeter may be of more interest. 
For a given experiment,
specific sensitivities
can be estimated using the analyses 
presented in Sec.\ \ref{labtests}.

A third methodology 
for investigating gravitational Lorentz violation 
from nonzero coefficients $\afbw_\mu$
could conceivably be to adopt as the gravimeter
a device that wholly confines charged particles.
For example,
a single charged particle can be trapped
for long periods using a Penning trap
\cite{lbgg}.
Measuring gravitational effects in this way is ambitious,
as can be appreciated from the size of the quantity 
$|m^e g/e| \simeq 6\times 10^{-12}$ V/m.
Nonetheless,
the feasibility of
gravitational measurements with trapped antiprotons
at a sensitivity of about 1\% has been suggested,
using a gravity-induced shift of radial orbits
\cite{llmth}.
This would also lead to sensitivity 
to the coefficients $\afbw_\mu$
via an analysis similar to those discussed above.

\subsection{Tests with antimatter}
\label{antimatter}

The study of antimatter offers another realm
in which to search for Lorentz and CPT violation.
Antihydrogen has been detected
\cite{ps210,e862}
and produced in copious amounts 
\cite{athenaatrap},
while prospects for studies of trapped cold antihydrogen
are excellent 
\cite{alphaatrap2}.
Antihydrogen spectroscopy could yield special sensitivity 
to nongravitational SME coefficients for Lorentz and CPT violation
\cite{bkrhhbar},
and the experiment for  
Atomic Spectroscopy And Collisions Using Slow Antiprotons
(ASACUSA)
expects to achieve sensitivities of parts in $10^{-7}$ 
to the predicted shifts in hyperfine transitions
\cite{asacusa}.
 
To study the interaction of gravity and antimatter,
various ideas for measuring the gravitational acceleration
of antihydrogen have been advanced.
Among them are methods involving 
trapped antihydrogen 
\cite{gblmstp},
antihydrogen interferometry 
\cite{tjpage},
antihydrogen free fall from an antiion trap
\cite{jwth},
and tests in space
\cite{weax}.
One approved project,
the Antimatter Experiment: Gravity, Interferometry, Spectroscopy
(AEGIS)
\cite{aegis},
has an interoferometric design
with an initial sensitivity goal of 1\%
to the gravitational acceleration of antihydrogen.

In the context of gravitational Lorentz and CPT violation,
these experiments offer the prospect of special sensitivities
to the coefficients $\afbw_\mu$ and $\cbw_\mn$.
The key point is that a CPT transformation
has the net effect of reversing the sign of $\afbw_\mu$ 
while leaving $\cbw_\mn$ unchanged.
As a result,
experiments with antihydrogen 
could in principle observe distinctive and novel behaviors.
Moreover,
when compared with similar measurements on hydrogen, 
the results would offer 
the opportunity for clean separation of effects. 
For instance,
free-fall WEP tests comparing hydrogen and antihydrogen
could yield independent sensitivity to $\cbx{e+p}_{TT}$.
In general,
the theoretical treatment 
of prospective free-fall gravimeter or WEP tests with antihydrogen
follows the same path as described in Sec.\ \ref{labtests},
except with the sign of $\afbw_\mu$ reversed throughout. 

The literature contains numerous attempts
to place indirect limits on the possibility
of unconventional antimatter-gravity interactions,
many of which are reviewed and critiqued
in Ref.\ \cite{mntg}.
In the present context,
the SME offers a general field-theoretic approach
that can elucidate aspects of this issue
and provide new insights about possible limitations on effects.
We next present an explicit toy model
that evades some previous indirect limits 
on large unconventional effects in antihydrogen.

For simplicity,
we choose to work within the isotropic limit of the SME.
In any specified inertial frame $O$,
a subset of Lorentz-violating operators
in the SME Lagrange density preserves rotational symmetry.
Setting the coefficients of all other operators to zero
produces an interesting limiting case.
The frame $O$ then becomes a preferred frame,
since the rotation invariance is broken
in any frame $O^\prime$ boosted with respect to $O$.
Physical effects of Lorentz violation
are then isotropic in $O$ but not in $O^\prime$.
The frame $O$ could in principle be identified as the
rest frame $U$ of the cosmic microwave background (CMB),
the Sun-centered frame $S$,
or any other desired choice.  
Isotropic models of this type are sometimes called
`fried-chicken' models
because of their popularity and simplicity.

In Minkowski spacetime,
toy isotropic models can be used to show
that Lorentz- and CPT-violating effects
could in principle be substantially larger 
in antihydrogen than hydrogen.
One example is the isotropic `invisible' model (IIM) 
\cite{akifc},
which is defined in the CMB frame $U$
and yields effects challenging to see
in searches with ordinary matter.
Denoting coordinates in $U$ 
by $(T^\prime, X^\prime, Y^\prime, Z^\prime)$,
the IIM assumes
the only nonzero coefficients for Lorentz violation
are $(b^p)_{T^\prime}$ and isotropic $(d^p)_{\Xi^\prime\Xi^\prime}$
obeying the simple condition 
\beq
(b^p)_{T^\prime} = k m^p (d^p)_{T^\prime T^\prime}
\eeq
for a suitable choice of constant $k$.
In the Sun-centered frame $S$,
this one-parameter model generates nonzero coefficients
$(b^p)_J$ and $(d^p)_{JT}$.
The dominant signals in terrestrial experiments with hydrogen
appear in the hyperfine structure
and involve the combination
$(b^p)_J - m^p (d^p)_{JT}$,
which vanishes for suitable $k$.
These experiments can therefore detect
only effects suppressed by at least one power
of the boost of the Earth around the Sun,
which is about $10^{-4}$
and requires an experiment sensitive to annual modulations.
In contrast,
the dominant effects in experiments with antihydrogen
involve the combination
$(b^p)_J + m^p (d^p)_{JT}$,
which produces unsuppressed signals in the hyperfine structure.
The IIM thus provides a toy field-theoretic scenario 
in which observable effects in antihydrogen
are at least 10,000 times greater than those in hydrogen
or other nonrelativistic neutral matter.

The IIM involves spin-dependent operators
for Lorentz and CPT violation in Minkowski spacetime.
In this work,
the focus is on the gravitational couplings
of spin-independent operators
with coefficients $\afbw_\mu$ and $\cbw_\mn$.
At the end of Sec.\ \ref{Theory},
we remark on the difficulty 
of observing with matter
any signals depending on the combination 
$\al \abt_T - \mt \cbt_{TT}/3$
of isotropic coefficients.
Here,
we consider some implications for antimatter gravity
of a specific toy model,
the isotropic `parachute' model (IPM),
in which unobserved combinations of this type 
provide the dominant source of Lorentz-violating effects
and could yield significant \it a priori \rm differences 
in the gravitational accelerations of hydrogen and antihydrogen.

To construct the IPM,
consider the Lagrange density of the SME  
in the Sun-centered frame $S$,
with nonzero coefficients
restricted to $\afbw_T$ and isotropic $\cbw_{\Si\Xi}$.
Following the derivation in the early sections of this work,
we can extract the \pno3 effective classical lagrangian 
for a test particle T moving in the gravitational field
of a source S.
This can be written in the suggestive form
\beq
L_{\rm IPM} = \half \mt_i v^2 + \fr{\G \mt_g \ms_g}{r},
\eeq
where $\mt_i$ is the effective inertial mass of T,
while $\mt_g$ and $\ms_g$ 
are the effective gravitational masses 
of T and S, respectively.
All these effective masses are defined in terms of 
the coefficients $\afbw_T$, $\cbw_{TT}$ for Lorentz violation
and the body masses $m^{\rm B}$ of Eq.\ \rf{bodymass}.
We find
\bea
m^{\rm B}_i &=& 
m^{\rm B} + \sum_w \frac53 (N^w+N^{\bar{w}}) m^w \cbw_{TT}, 
\nonumber \\
m^{\rm B}_g &=& 
m^{\rm B} + \sum_w \Big( (N^w+N^{\bar{w}}) m^w \cbw_{TT}
\nonumber \\ 
& & \pt{m + \sum \Big(}
+ 2 \al (N^w-N^{\bar{w}}) \afbw_T \Big),
\label{friedmasses}
\eea
where B is either T or S.
These expressions adopt the notation
$N^w$ and $N^{\bar{w}}$
for the number of particles and antiparticles of type $w$,
respectively,
while as before $m^w$ is the mass of a particle of type $w$.
Note that for a given body 
the passive and active gravitational masses are identical,
reflecting the preservation of Newton's third law
in the model.

For electrons, protons, and neutrons,
the IPM is defined by the three conditions
\beq
\al \afbw_T = \frac 13 m^w \cbw_{TT},
\label{lfccondition}
\eeq
where $w$ ranges over $e$, $p$, $n$.
Since there are three independent conditions 
on six real parameters,
this produces a three-parameter IPM. 
The condition \rf{lfccondition}
ensures that for a matter body B
the effective inertial and gravitational masses are equal,
\beq
m^{\rm B}_i = m^{\rm B}_g \qquad ({\rm matter}),
\eeq
and hence no Lorentz-violating effects 
appear in gravitational tests to \pno3
using ordinary matter.
However,
for an antimatter test body T this condition fails,
\beq
m^{\rm T}_i \neq m^{\rm T}_g \qquad ({\rm antimatter}),
\eeq
so observable signals arise in comparisons 
between the gravitational responses of matter and antimatter
or between different types of antimatter.
Ensuring the validity of perturbation theory 
requires that the coefficients $\al \afbw_T = m^w \cbw_{TT}/3$
are perturbatively small relative to $m^w$. 
With theoretically conceivable values 
perhaps even as large as $0.5 m^w$,
the gravitational accelerations of hydrogen and antihydrogen 
might differ at the 50\% level.

Rather than a serious effort at a realistic theory,
the IPM is constructed as a simplistic playground 
within which to explore field-theoretic limitations
on unconventional properties of antimatter and antihydrogen.
In the next few paragraphs we treat it as such, 
briefly addressing some concerns
about unconventional signals in this context.

One issue is whether energy remains conserved
when matter and antimatter have different
gravitational interactions
\cite{pm}.
For the analysis of the SME in the present work,
this issue is moot because 
an explicit conserved energy-momentum tensor exists.
As an illustration,
consider the gedanken experiment 
in which a particle-antiparticle pair
is lowered in a gravitational field,
converted to a photon pair,
raised to the original location,
and finally reconverted to the particle-antiparticle pair.
In generic scenarios
the particle, antiparticle, and photons
each provide different contributions to the energy
and so problems can arise.
However,
in the IPM these complications are avoided.
The photons make no contribution
because they are conventional,
partly via the coordinate choice \rf{coordchoice}.
The particle and antiparticle do contribute to the energy 
via the coefficient $\afbw_T$,
but the two contributions cancel.
Contributions involving the coefficient $\cbw_{TT}$ 
exist and combine during the lowering procedure,
but the definition \rf{cenergy} of the conserved energy 
also contains $\cbw_{TT}$ and so the net change remains zero
at the end of the experiment.
The resolution of this and other illustrative scenarios
is less transparent when more nonzero coefficients
for Lorentz violation are present,
but the existence of a conserved energy-momentum tensor
ensures that no contradictions arise. 

Another attempt to argue against the possibility
of an anomalous antimatter response to gravity
is based on the large binding-energy content of 
baryons, atoms, and bulk matter
\cite{lis2}.
For hydrogen and antihydrogen,
a modern version of the argument could proceed by first noting
that the quarks in hydrogen contain only about 10\% of the mass
with most of the remainder contained 
in the gluon and sea binding,
and then concluding that 
since the binding forces are comparable 
for hydrogen and antihydrogen
their gravitational response cannot differ 
by more than about 10\%.
This type of reasoning implicitly assumes 
that the gravitational response of a body is determined
by its mass and hence also by its binding energy.
However,
as shown generically in Sec.\ \ref{Theory},
the coefficient $\afbw_T$ in the IPM 
leads to a correction to the gravitational force
that is independent of mass and can vary with flavor.
Indeed,
the binding forces are largely conventional in the IPM,
and the gravitational responses of hydrogen and antihydrogen
are primarily determined by the flavor content
of the valence particles.
It is even conceivable in principle 
that a large gravity effect could be associated
purely with the positron,
as occurs in the IPM when only $\afbe_T$ is nonzero
and satisfies the condition \rf{lfccondition}.
A careful treatment of this issue in the IPM
would require consideration of radiative effects
involving $\afbw_T$, $\cbw_{TT}$,
and other SME coefficients for Lorentz violation 
\cite{ck,renorm},
perhaps imposing the condition \rf{lfccondition} 
only after renormalization.
In any case,
the essential points illustrated with the IPM remain valid:
the gravitational response of a body
can be independent of mass,
can vary with flavor,
and can differ between particles and antiparticles.

The gravitational response of antimatter
could in principle also be restricted
by the results of experiments studying kaons 
\cite{mlg}
and other neutral-meson systems,
which are natural interferometers
mixing strong-interaction particle and antiparticle eigenstates 
via weak-interaction effects.
When analyzed in the context of the SME in Minkowski spacetime,
neutral-meson mixing places tight constraints 
on certain differences of the coefficients $\afbw_\mu$
for $w$ ranging over several quark flavors
\cite{mesons,akmesons}.
However,
these constraints have no dominant implications
for leptons or for baryons,
which involve three valence quarks rather than
a quark and an antiquark as in mesons.
Moreover,
the neutral-meson constraints 
necessarily involve valence $s$, $c$, and $b$ quarks,
which are largely irrelevant for protons and neutrons.
In the presence of gravitational interactions,
the same line of reasoning holds,
with the flavor dependence of Lorentz and CPT violation
leading to the conclusion
that the IPM evades restrictions from meson oscillations. 

We can also use the IPM 
to illustrate a type of constraint
on more realistic model building
arising from the extensive searches
for Lorentz and CPT violation in Minkowski spacetime.
The key point is that
the mixing of Lorentz-violating operators
under rotations and boosts
can imply indirect limits on some coefficients.
In the IPM,
for example,
the coefficient $\afbw_T$ is unobservable 
in Minkowski spacetime,
as discussed in Sec.\ \ref{Observability},
but certain nongravitational experiments 
could in principle obtain boost-suppressed sensitivity 
to $\cbw_{TT}$ for some $w$
via measurements of the coefficients $\cbw_{JK}$.
As one illustration,
a measurement with a Cs-Rb double fountain clock
over a total of five weeks in the spring and fall of 2005
achieved a sensitivity of parts in $10^{25}$ 
on some combinations of the coefficients $\cbx{p}_{JK}$
\cite{wcbc}.
This suggests that continuing an experiment of this type
over a longer period could attain parts in $10^{17}$ 
on the coefficient $\cbx{p}_{TT}$
by analyzing the data allowing for the Earth's orbital boost
$V_\oplus \simeq 10^{-4}$.
Similarly,
a careful analysis of multiple searches 
for Lorentz violation involving the electron sector
could be used to measure $\cbx{e}_{TT}$ 
at the level of parts in $10^{15}$
\cite{ba-ce}.
Although these types of nongravitational studies
remain to be performed,
they could in principle place experimental limits 
on the magnitude of the anomalous gravitational response 
of antihydrogen in the IPM
and possibly also in more realistic models.
We remark in passing that these kinds of constraints 
nonetheless leave considerable room 
for realistic model building,
in particular when operators of arbitrary dimension
are incorporated in the framework
\cite{km-nr2}.

\subsection{Tests with matter beyond the first generation}
\label{hgm}

Most studies of fermion-gravity couplings to date
have involved particles from the first generation of the SM.
However,
the SME coefficients for Lorentz and CPT violation 
can differ between sectors,
so investigations of higher-generation matter-gravity couplings
are of independent interest.
Since fermion masses and hence fermion-gravity couplings
typically increase with the generation,
it is conceivable that an unconventional gravitational coupling 
may be more readily identified
in gravitational tests with higher-generation matter. 
Comparatively few results exist 
for the coefficients $\afbw_\mu$ and $\cbw_\mn$
for particles $w$ beyond the first generation 
\cite{tables},
so there is considerable room 
for measurements of effects involving gravity couplings.

The comparatively long lifetime of the muon 
makes it an interesting candidate 
for gravitational tests of Lorentz violation
with second-generation particles.
Several muon coefficients for Lorentz and CPT violation 
have already been measured
\cite{muexpt},
but the sensitivities are largely limited
to spin-dependent effects.
Measurements of Lorentz-violating gravitational couplings 
of the muon could be achieved via muonium interferometry,
with an estimated initial reach of 10\% 
\cite{kk}.
Interferometry with muonic hydrogen
may also be possible 
\cite{bl}.
In principle,
these experiments could yield first measurements 
of some components of the coefficients 
$\afbm_\mu$ and $\cbm_\mn$ in the muon sector.  
In particular,
free-fall WEP tests using muonium interferometry
to search for Lorentz and CPT violation
offer the prospect of direct sensitivity to the coefficients 
$\afbm_T$ and $\cbm_{TT}$.
In contrast,
performing free-fall gravimetric tests with muonium interferometry 
is unlikely to be useful in the near future 
because the dominant signals appear at annual frequencies 
and are suppressed by the boost $V_\oplus$.

Consider for definiteness
a free-fall WEP experiment comparing 
the gravitational acceleration
of muonium with that of neutral matter $N$.
Muonium is a bound system containing an antimuon
and an electron,
so its spin-independent Lorentz-violating gravitational properties 
are determined by the coefficients 
$-\afbx{\mu}_\mu$, $\cbx{\mu}_\mn$, 
$\afbx{e}_\mu$ and $\cbx{e}_\mn$.
Following the line of reasoning in Sec.\ \ref{Sensitivities},
we find that the dominant observable combination 
of coefficients for CPT-odd effects 
in a free-fall WEP experiment is 
\beq
\afbx{\mu+e-N}_\mu = 
-\afbx{\mu}_\mu + \afbx{e}_\mu - \fr{m^\mu + m^e}{m^N}\afbx{N}_\mu,
\eeq
where $m^N$ is the mass of $N$ 
and $\afbx{N}_\mu$ is its effective coefficient 
for Lorentz and CPT violation.
Assuming $N$ is composed of first-generation particles,
the existing constraints on coefficients
\cite{tables}
imply that for most models it is a good approximation
to neglect all but the first term on the right-hand side
of this equation.
For CPT-even effects,
the relevant observable combination of coefficients is
\bea
\cbx{\mu+e-N}_\mn &=& 
\cbx{\mu}_\mn + \fr{m^e}{m^\mu}\cbx{e}_\mn 
- \fr{m^\mu + m^e}{m^\mu} \cbx{N}_\mn .
\nonumber \\
\eea
Again,
only the first term is likely to be significant
in practice.
Similar expressions hold for muonic hydrogen,
with the replacements 
$e \to p$ for the superscripts
and $\afbx{\mu}_\mu \to -\afbx{\mu}_\mu$
for the muon coefficient for Lorentz violation.

Searches for Lorentz-violating gravitational couplings 
of other second- and third-generation particles
could also be countenanced.
The typically short lifetimes of these particles
can in principle be overcome by boosting,
so accelerator experiments are likely to provide
the best laboratory prospects.
Studying the gravitational infall 
of particles of extraterrestrial origin
in the context of free-fall WEP searches  
might be a source of additional constraints. 
 
The physical mixing of uncharged particles 
of different flavors $w$
offers an interesting alternative method
to achieve sensitivity to the coefficients $\afbx{w}_\mu$.
Examples already yielding SME constraints 
on Lorentz and CPT violation
include the interferometric oscillations 
of neutral mesons
\cite{mesons,akmesons}
and of neutrinos
\cite{neutrinos}.
Particle mixing implies nondiagonal terms
in the propagator matrix,
so field redefinitions of the type \rf{redefa}
cannot be used to remove the coefficients 
$\afbx{w}_\mu$
from the theory.
Differences between the coefficients $\afbx{w}_\mu$ 
then become observable even in Minkowski spacetime,
offering sensitivity to effects 
that would otherwise be undetectable.
For instance,
the recent observation of anomalous CP-violating effects
in $B$-meson oscillations
\cite{d0}
could originate in one or more nonzero coefficients 
$\afbx{w}_T$ 
for Lorentz and CPT violation in the quark sector,
since these control CP-odd but T-even operators
that contribute to the effective hamiltonian
for the mixing
\cite{akmesons}.
Spin-independent CPT-odd Lorentz violation 
involving coefficients such as $\abw_\mu$
could also underlie
the observed baryon asymmetry in the Universe 
\cite{bnsyn}.

In terms of the perturbative counting scheme 
of Sec.\ \ref{Perturbation scheme},
the existing SME studies using
neutral-meson and neutrino oscillations lie at O(1,0).
Incorporating leading-order gravitational couplings
along the lines in this paper
would introduce O(1,1) oscillation effects,
including species-dependent modifications 
of the meson or neutrino trajectories
with characteristic time dependences
similar to the WEP-violating effects
discussed in Sec.\ \ref{epfreefall}.
Possible O(1,0) contributions to the oscillations 
can be distinguished from O(1,1) ones
via the dependences on energy, baseline, flavor, and time.
The advent of neutrino-oscillation experiments
with long and very long baselines of order 100-1000 km
and corresponding changes in gravitational potential
along the beams
may offer particularly interesting options 
for free-fall WEP tests of Lorentz and CPT violation
of this type.
A detailed consideration of these possibilities 
would be a worthwhile subject for future investigation.

\section{Solar-system tests}
\label{Solar-system tests}

Studies of the motion of bodies within the solar system
provide an important source 
of information about gravitational couplings to matter.
In this section,
we investigate the effects of 
nonzero coefficients $\afb_\mu$ and $\cb_\mn$
for Lorentz violation
in two solar-system contexts:
lunar and satellite laser ranging,
and perihelion precession.
The analysis here neglects effects that act merely 
to scale the mass of the gravitational source.
These are unobservable using solar-system observations alone,
but they may be detectable in combined measurements
using photon tests.
This latter issue is revisited in Sec.\ \ref{photon}.

\subsection{Lunar and satellite laser ranging}
\label{llr}

Lunar and satellite laser ranging provides a sensitive test 
of gravitational physics.
The relevant orbital perturbations 
to the motion of a satellite orbiting the Earth
that arise from nonzero Lorentz violation
in the pure-gravity sector of the minimal SME
have been established 
\cite{qbakpn}
and used to constrain some of the coefficients $\sb_\mn$
\cite{gravexpt1}.
Here,
we seek to extend these results
to include dominant effects
from nonzero coefficients $\afb_\mu$ and $\cb_\mn$.

Where possible in this subsection,
we follow the conventions of Ref.\ \cite{qbakpn}.
A summary of our notation is given in Table \llra.
The flavor dependence of the matter-gravity couplings
leads to composition-dependent factors
in some of the equations to follow.
To simplify these expressions,
it is useful to define the eight combinations 
\bea
\nwr1 &=&  
N^w_1 + N^w_2, 
\nonumber \\
\nwr2 &=& 
N^w_1 - N^w_2,
\nonumber \\
\nwr3 &=& 
M \left( \fr{N^w_1}{m_1} + \fr{N^w_2}{m_2} \right), 
\nonumber \\
\nwr4 &=& 
M \left( \fr{N^w_2}{m_2} - \fr{N^w_1}{m_1} \right),
\nonumber \\
\nwr5 &=& 
\fr{1}{M}(m_1 N^w_2 + m_2 N^w_1),
\nonumber \\
\nwr6 &=& 
\fr{1}{M} \left( m_1 N^w_2 - m_2 N^w_1 \right),
\nonumber \\
\nwr7 &=&  
\fr{m_2}{m_1} N^w_1 + \fr{m_1}{m_2} N^w_2,
\nonumber \\
\nwr8 &=& 
\fr{1}{M} \left( \fr{m^2_2}{m_1} N^w_1 
- \fr{m^2_1}{m_2} N^w_2 \right).
\label{compdepfact}
\eea

\onecolumngrid
\begin{center}
\begin{tabular}{ll}
\multicolumn{2}{c} 
{Table \llra. Notation for laser-ranging tests.} \\
\hline
\hline
Quantity & Definition\\ 
\hline
$m_1$                 & satellite mass\\
$N^w_1$               & number of particles of species $w$ in the satellite\\
$m_2$                 & Earth mass\\
$N^w_2$               & number of particles of species $w$ in the Earth\\
$M=m_1 + m_2$         & total Earth-satellite mass\\
$\de m = m_2 - m_1$   & Earth-satellite mass difference\\
$m_n$                 & mass of the $n$th perturbing body\\
$M_{\odot}$           & Sun mass \\
$N^w_{\odot}$         & number of particles of species $w$ in the Sun\\
$r^J_1$               & satellite position \\ 
$r^J_2$               & Earth position \\
$r^J = r^J_1 - r^J_2 = (x,y,z)$ 
                      & Earth-satellite separation,
                        of magnitude $r = |\vec r_1 - \vec r_2|$\\
$R^J = (m_1 r^J_1 + m_2 r^J_2)/M$ 
\qquad\qquad 
               & position of Newton center of mass for Earth-satellite system \\
$\Om_\oplus = \sqrt{\G M_\odot/R^3}$ & mean Earth orbital frequency \\
$v^J = v^J_1 - v^J_2 = dr^J/dT$ 
                      & relative Earth-satellite velocity \\ 
$V^J = (m_1 v^J_1 + m_2 v^J_2)/M$ 
\qquad\qquad 
                      & 
velocity of Newton center of mass for Earth-satellite system \\
\hline
\hline
\end{tabular}
\end{center}
\twocolumngrid
The primary observable in laser-ranging tests
is the coordinate acceleration 
$\acc^J_{\rm ES}$
of the relative Earth-satellite separation.
Working in the Sun-centered frame,
we can obtain this acceleration
from the equation of motion \rf{oogeo}.
The relevant contributions 
to the coefficient and metric fluctuations
from $\afb_\mu$ and $\cb_\mn$
can be found in Sec.\ \ref{fluct},
while those from $\sb_\mn$
are given in Ref.\ \cite{qbakpn}.

Incorporating perturbative effects of other bodies
including the Sun,
the coordinate acceleration can be written
\beq
\acc^J_{\rm ES} \equiv \fr {d^2 r^J}{dT^2}
= \acc^J_{\rm N} + \acc^J_{\rm T} + \acc^J_{\rm Q} 
+ \acc^J_{\rm LV} + \ldots .
\label{acces}
\eeq
The first three terms in this expression involve effects
independent of Lorentz violation.
They represent the acceleration due to 
the Newton gravitational field 
of the Earth-satellite system,
the Newton tidal quadrupole term,
and the quadrupole moment of the Earth,
respectively.
Their explicit form is given 
in Ref.\ \cite{qbakpn}.
The leading Lorentz-violating contributions 
to the acceleration
are represented by the fourth term $\acc^J_{\rm LV}$.
This term can itself be split into four pieces,
\beq
\acc^J_{\rm LV} = 
\acc^J_{\afbnp,\cb,{\rm ES}} 
+ \acc^J_{\afbnp,\cb,{\rm tidal}}
+ \acc^J_{\sb,{\rm ES}} 
+ \acc^J_{\sb,{\rm tidal}} .
\label{lvacc}
\eeq
The first two terms are the ones of interest
in the present work
and are discussed below.
The last two depend on the coefficient $\sb_\mn$,
with $\acc^J_{\sb,{\rm ES}}$ 
arising from the Earth-satellite system 
and $\acc^J_{\sb,{\rm tidal}}$
involving perturbations due to other bodies.
The explicit form of these two quantities
is provided in Ref.\ \cite{qbakpn}.

The term $\acc^J_{\afbnp,\cb,{\rm ES}}$ 
in Eq.\ \rf{lvacc}
provides the Lorentz-violating acceleration
of the Earth-satellite system 
from the matter-gravity couplings $\afb_\mu$ and $\cb_\mn$.
It takes the form
\begin{widetext}
\bea
\acc^J_{\afbnp,\cb,{\rm ES}} &=& \fr{\G}{r^3} \sum_{w} \Big[
- 2 \nwr3 \al \afbw_T r^J
- \nwr1  m^w \cttw r^J
+ 2 \nwr7 m^w \et^{JK} \cbw_{(KL)} r^L
\nonumber \\
& & 
\pt{\fr{G}{r^3} \sum_{w} \Big[}
- 2 \nwr3 \al \akw V^K r^J
- 2 \nwr2 \al \akw \et^{JK} v_{L} r^L
+ 2 \nwr2 \al \akw v^{K} r^J
+ 2 \nwr7 m^w \cbw_{(TK)} V^J r^K
\nonumber \\
& & 
\pt{\fr{G}{r^3} \sum_{w} \Big[}
- 2 \nwr1 m^w \cbw_{(TK)} V^K r^J
+ 2 \nwr6 m^w \cbw_{(TK)} v^K r^J
+ 2 \nwr7 m^w \et^{JK} \cbw_{(TK)} V_L r^L
\nonumber \\
& & 
\pt{\fr{G}{r^3} \sum_{w} \Big[}
- 2 ( \nwr6 - 2 \nwr8) m^w \et^{JK} \cbw_{(TK)} v_L r^L  
+ 2 \nwr8 m^w \cbw_{(TK)} v^J r^K \Big].
\label{lves}
\eea
In principle,
$\acc^J_{\afbnp,\cb,{\rm ES}}$ 
also acquires contributions proportional to $R_\oplus \om$,
but these are neglected here 
because they are typically suppressed
compared to effects proportional to $V^J$ and $v^j$.

In Eq.\ \rf{lvacc},
the term $\acc^J_{\afbnp,\cb,{\rm tidal}}$
contains the Lorentz-violating tidal acceleration
involving $\afb_\mu$ and $\cb_\mn$,
which arises from perturbing bodies.
When the satellite is taken as the Moon,
the dominant tidal contributions are due to the Sun
and can be written
\bea
\acc^J_{\afbnp,\cb,{\rm tidal}} &=& \Om^2_\oplus \sum_{w} 
\Big{\{}
\Big[\fr{N^w_\odot}{m_\odot} 
\left( 2 \al \afbw_T + m^w \cttw \right)
- \fr{2}{M} \nwr1 \al \afbw_T \Big] 
\left( 3 r^L \hat{R}_L \hat{R}^J - r^J \right)
\nonumber \\
& & 
\pt{\Om^2_\oplus \sum_{w} \Big{\{}}
+ \fr{2}{M} \nwr4 \al \afbw_T R^J
- 2 \fr{m^w}{M} \nwr4 \et^{JK} \cbw_{KL} R^L
- 2 \fr{m^w}{M} \nwr7 \et^{JK} \cbw_{(KL)}
\left( 3 r^M \hat{R}_M \hat{R}^L - r^L \right)
\nonumber \\
& & 
\pt{\Om^2_\oplus \sum_{w} \Big{\{}}
- 2 \fr{m^w}{M} \Big[ 
2 \nwr4 \et^{JK} V_L \cbw_{(TK)} 
- 2 \nwr7 \et^{JK} v_L \cbw_{(TK)} 
+  \nwr4  V^J \cbw_{(TL)} 
-  \nwr7  v^J \cbw_{(TL)}\Big] R^L
\nonumber \\
& & 
\pt{\Om^2_\oplus \sum_{w} \Big{\{}}
+ 4 \fr{N^w_\odot}{m_\odot} \al \afbw_K \et^{J[K} v^{L]} R_L
- 4 \fr{N^w_\odot}{m_\odot} \al \afbw_K 
\left(\fr{\de m}{M} \et^{J[K} v^{L]} 
+ \et^{J[K} V^{L]} \right) 
\left( 3 r^M \hat{R}_M \hat{R}_L - r_L \right)
\nonumber \\
& & 
\pt{\Om^2_\oplus \sum_{w} \Big{\{}}
- 2 \fr{m^w}{M} \Big[ 
2 \nwr7 \et^{JK} V_L \cbw_{(TK)} 
+ 2 \nwr8 \et^{JK} v_L \cbw_{(TK)} 
\nonumber \\
& & 
\pt{\Om^2_\oplus \sum_{w} \Big{\{}- \fr{m^w}{M} \Big[}
+  \nwr7  V^J \cbw_{(TL)} 
+  \nwr8  v^J \cbw_{(TL)}\Big] 
\left( 3 r^M \hat{R}_M \hat{R}^L - r^L \right)
\Big{\}}.
\label{lvtide}
\eea
\end{widetext}
If instead the satellite is artificial,
then there are tidal effects from both the Sun and the Moon.
However,
these are suppressed relative to 
the Earth-satellite acceleration \rf{lves}.

The Lorentz-violating coordinate accelerations
given by Eqs.\ \rf{lves} and \rf{lvtide}
exhibit some interesting features.
The first two terms in Eq.\ \rf{lves} 
and the first term in Eq.\ \rf{lvtide}
are composition-dependent scalings
of the corresponding Newton accelerations. 
These terms are therefore detectable 
only by comparison to results obtained
using satellites of different compositions.
Also,
unlike the contributions $\acc^J_{\sb,{\rm tidal}}$
obtained in Ref.\ \cite{qbakpn},
here the tidal acceleration \rf{lvtide} from the Sun
on the Moon-Earth system involves nontrivial WEP violations
because the Moon and the Earth fall differently towards the Sun
when the coefficients $\afbw_\mu$ and $\cbw_\mn$ are nonzero.
It is also interesting to note that
the tidal acceleration \rf{lvtide}
contains contributions at \pno2 that are independent of $r^J$
and hence are enhanced at this order 
relative to other contributions by a factor of $R/r$.
This too is a consequence
of the WEP violations arising from $\afb_\mu$.
Similar terms appear at \pno3 as well.

A typical experiment measures the time of flight
for laser photons to travel from the Earth
to a reflector on the satellite and back.
To analyze the results,
the laser-ranging data can be fitted
by incorporating Eq.\ \rf{acces}
and other conventional perturbing effects
into an appropriate modeling code.
An alternative approach is to perform 
an analytical perturbative expansion 
along the lines of the one performed
for the $\sb_\mn$ contributions 
in Ref.\ \cite{qbakpn}
and then match to the data.
This latter method is adopted in 
Ref.\ \cite{gravexpt1}
to constrain combinations of the coefficients $\sb_\mn$.

In the present context, 
we can obtain crude estimates
of sensitivities to $\afbw_\mu$ and $\cbw_\mn$
attainable in lunar laser ranging 
via either of these procedures,
by using term-by-term comparison of the accelerations
\rf{lves} and \rf{lvtide}
to the accelerations
$\acc^J_{\sb,{\rm ES}}$ and $\acc^J_{\sb,{\rm tidal}}$
obtained for the coefficient $\sb_\mn$
in Ref.\ \cite{qbakpn}.
With the precision already achieved in lunar laser ranging 
\cite{llrexpt},
we thereby find estimated sensitivities
at parts in $10^{10}$ to combinations of $\cbw_{(JK)}$
and $\sb_{JK}$,
and parts in $10^6$ 
to various combinations of $\afbw_J$, $\cbw_{(TJ)}$
and $\sb_{TJ}$.
Actual measurements at roughly these levels
can be expected to result from a reanalysis
of existing data.
A significant further improvement 
is likely to be possible using data from the 
Apache Point Observatory Lunar Laser-Ranging Operation (APOLLO)
\cite{apollo}.
Assuming that millimeter ranging is achieved as expected
and disregarding probable substantially improved statistics,
we anticipate competitive estimated 
sensitivities of $10^{-7}$ GeV on various combinations of 
$\al \afbx{w}_X$ and $\al \afbx{w}_{Y+Z}$,
and a sensitivity of $10^{-7}$ on
$\cbx{n}_{(TJ)}$,
where the notation of Eq.\ \rf{YplusZ} is used
and the dependence on the coefficients $\sb_\mn$
has been omitted for simplicity.

Ranging to artificial satellites
with orbit orientations different from that of the Moon
can yield sensitivity 
to additional independent linear combinations
of $\afbw_\mu$ and $\cbw_\mn$.
Typically,
the reach of satellite ranging is expected
to be about an order of magnitude less 
than lunar laser ranging.
Other possibilities for gravitational tests of Lorentz violation 
include ranging to objects orbiting bodies other than the Earth.
For example,
the time variation $\G^{-1}d{\G}/dt$ 
of the Newton gravitational constant 
has been constrained by 
ranging data to the Viking landers on Mars,
to the Mariner 9 spacecraft orbiting Mars,
and to other bodies including the Moon
\cite{gdotexpt}.
These studies primarily seek secular changes 
in the gravitational force.
Although secular changes in coupling constants
can result from Lorentz violation
\cite{klp},
the signals of interest in the present context are periodic.
Reanalysis of existing data 
to seek periodic effects in $\G^{-1}d{\G}/dt$ 
would yield sensitivities to Lorentz violation
estimated to be somewhat less than lunar laser ranging
but involving different combinations 
of $\afbw_\mu$ and $\cbw_\mn$.

We conclude this subsection with some comments 
about the coordinate location $R^J$ of the center of mass
of an Earth-satellite system.
Boost invariance normally ensures this location is fixed,
but the presence of Lorentz violation
means it can be time dependent,
although the effect may be unobservable via laser ranging.
Neglecting the effects of other bodies
and working at \pno3,
the Lorentz-violating contributions
to the equation of motion for the center of mass
of the Earth-satellite system can be written as the sum 
\beq
\ddot{R}^J \supset
\ddot{R}^J_{\afbnp} + \ddot{R}^J_{\cb} + \ddot{R}^J_{\sb}
\label{Rddot}
\eeq
of contributions from 
$\afbw_\mu$, $\cbw_\mn$, and $\sb_\mu$.
Explicitly,
we find 
\beq
\ddot{R}^J_{\afbnp} =
\sum_w \fr{2 \G \nwr5 \et^{JK} \al \afbw_K v_L r^L}{r^3},
\eeq
which contains only \pno3 effects
involving the internal motion of the system.
The second term in Eq.\ \rf{Rddot} is
\bea
\ddot{R}^J_{\cb} &=& \sum_w \fr{2 \G m_1 m_2 m^w}{M^2 r^3} 
\nonumber \\
& & 
\times \Big[ \half \nwr2 \cttw r^J
+ \nwr2 \et^{JK} \cbw_{KL} r^L 
\nonumber \\
& & 
\hskip 20pt
+ \et^{JK} \cbw_{(TK)}(\nwr2 V_L + \nwr5 v_L) r^L
\nonumber \\
& &
\hskip 20pt
+ \cbw_{(TK)} (\nwr2 V^K + \nwr5 v^K) r^J
\nonumber \\
& &
\hskip 20pt
+ \cbw_{(TK)} (\nwr2 V^J + \nwr5 v^J) r^K \Big].
\eea
The first two terms are at \pno2
and reflect the modification of the effective Newton inertial mass
in the presence of nonzero $\cb_\mn$,
while the remaining terms are at \pno3.
The ones proportional to $V^J$
arise as a result of the system boost in the Sun-centered frame,
and those proportional to $v^J$ 
are due to the internal motion of the system.
The last term in Eq.\ \rf{Rddot} is
\bea
\ddot{R}^J_{\sb} &=& \fr{\G m_1 m_2}{M r^3} \Big[
3 \et^{JK} \sb_{TK} v_L r^L
- \sb_{TK} v^J r^K
\nonumber \\
& & 
\hskip 50pt
- \sb_{TK} v^K r^J
+ 3 \sb_{TK}r^J v_L \hat{r}^K \hat{r}^L \Big],
\nonumber \\
\eea
which again consists only of \pno3 effects 
proportional to the internal motion of the system.
Note that all these contributions introduce an oscillatory motion
for the center of mass,
and their presence is required by momentum conservation.

\subsection{Perihelion precession}
\label{pprecession}

The presence of nonzero coefficients 
$\afb_\mu$ and $\cb_\mn$
for Lorentz violation
leads to corrections to the motion 
of a test body in a gravitational field.
These corrections can be calculated
from the equation of motion \rf{oogeo}
and from the expressions 
for the coefficient and metric fluctuations
given in Sec.\ \ref{fluct}.
In this subsection,
we determine the effect 
of nonzero $\afb_\mu$ and $\cb_\mn$
on the perihelion precession for planetary orbits. 
We follow the treatment of Ref.\ \cite{qbakpn},
which obtains the perihelion shift arising 
from nonzero SME coefficients $\sb_\mn$.
Our notation matches that of Table \llra\ 
and Eq.\ \rf{compdepfact}
in Sec.\ \ref{llr},
with the labels 1 and 2 representing
the planet and Sun,
respectively.

The derivation of the perihelion precession used here 
relies on the method of osculating elements
\cite{vab},
in which the instantaneous motion of the planet is treated
as part of an ellipse.
The ellipse is characterized using
the standard Kepler orbital elements,
and the motion of the planet
is described by specifying them as a function of time. 
The relevant orbital elements in the present case are 
the angle $\om$ between the line of ascending nodes
and the semimajor axis of the ellipse,
the longitude $\Om$ of the ascending node,
and the inclination $i$ with respect to the ecliptic.
These are specified in the reference coordinate system,
which can be taken as the Sun-centered frame
for the planetary orbits considered here.
More generally,
the reference frame is related 
to the Sun-centered frame by a rotation
and possibly a boost,
as discussed in Sec.\ V E 5 of Ref.\ \cite{qbakpn}.
The physical quantity relevant for the perihelion precession
is the change per period $\De \tilde{\om}$
of the perihelion angle $\tilde{\om}$ with respect to the equinox.
In terms of the basic orbital elements,
$\tilde{\om}$ can be expressed as
\beq
\tilde{\om} = \om + \Om \cos i.
\label{omt}
\eeq
For the cases of interest here,
the angle $i$ can be assumed small.

The secular changes in the orbital elements
arising from $\afb_\mu$ and $\cb_\mn$
can be obtained by considering the relative acceleration
of the planet and the Sun,
which has the form
\bea
\fr{d^2 r^j}{dt^2} 
&=& 
- \fr{\G}{r^3} 
\sum_{w} \Big[ 
M
+ 2 \nwr3 \al \afbw_0 r^j 
+ \nwr1  m^w \cbw_{00} r^j 
\nonumber \\
& & 
- 2 \et^{jk}  \nwr7  m^w \cbw_{(kl)} r^l 
+ 2 \nwr2 \al \afbw_k \et^{jk} v_{l} r^l 
\nonumber \\
& & 
- 2 \nwr2 \al \afbw_k v^{k} r^j 
- 2 \nwr6 m^w \cbw_{(0k)} v^k r^j 
\nonumber \\
& & 
+ 2 \et^{jk} ( \nwr6 - 2 \nwr8) m^w \cbw_{(0k)} v_l r^l  
\nonumber \\
& & 
- 2 \nwr8 m^w \cbw_{(0k)} v^j r^k \Big].
\label{ppacc}
\eea
The unperturbed ellipse is given as the solution $\vec r_0$
of the Kepler-type equation
\beq
\fr{d^2 r_0^j}{dt^2} = 
- \fr{\G}{r^3} \sum_{w} 
\Big[ M
+2 \nwr3  \al \afbw_0 
+ \nwr1 m^w \cbw_{00} \Big] r^j.
\label{keplereq}
\eeq
This shows that the frequency $n$ and semimajor axis $a$ 
of the unperturbed elliptic motion
are related according to 
\beq
n^2 a^3 = \G \sum_{w} 
\Big[ M
+ 2 \nwr3 \al \afbw_0 
+ \nwr1  m^w \cbw_{00} \Big].
\eeq
Note that the right-hand side of this expression
depends on the composition of the planet and the Sun.

The orientation of the orbit
can be specified using three unit vectors
$\vec{k}$, $\vec{P}$, and $\vec{Q}$.
The first is chosen perpendicular to the orbit,
the second points from the focus to the perihelion,
and the third completes the orthonormal set.
Their explicit form in terms of orbital elements
is given in Eq.\ (116) of Ref.\ \cite{qbakpn}.
In terms of this basis set,
the unperturbed elliptical orbit
can be expressed as 
\beq
\vec r_0 
= \fr{a (1-e^2)}{1+e \cos f} (\vec{P} \cos f + \vec{Q} \sin f),
\label{ur}
\eeq
where $e$ is the eccentricity
and $f$ is the true anomaly.

The perturbing acceleration $\acc^{\prime j}$
consists of the terms in Eq.\ \rf{ppacc}
that are absent from Eq.\ \rf{keplereq},
\beq
\acc^{\prime j} = 
\fr{d^2 r^j}{dt^2} -\fr{d^2 r_0^j}{dt^2}.
\label{pertaccel}
\eeq
The time dependence of the orbital elements
can be extracted from this equation
via the method of osculating elements.
The general procedure is to insert 
the unperturbed solution \rf{ur} for $\vec{r}$ 
into the expression \rf{pertaccel} for $\acc^{\prime j}$,
to project the result as desired,
and to integrate over the true anomaly.

To obtain the perihelion precession,
the final results for the orbital elements $\om$ and $\Om$
must be combined according to Eq.\ \rf{omt}.
After some calculation,
we obtain the expression 
\bea
\De \tilde{\om} &=& 2 \pi \sum_{w}
\bigg[ \fr{(e^2 - 2 \ep)}{M e^4} 
\nwr7  m^w
(\cbw_{QQ} - \cbw_{PP})
\nonumber \\
&&
\hskip 30pt
- \fr{2 n a (e^2 - \ep)}{e^3 M \sqrt{1-e^2}}
\big[(\nwr6 - 2 \nwr8) m^w \cbw_{(0Q)}
\nonumber \\
&& 
\hskip 100pt
+ \nwr2 \al \afbw_Q \big] \bigg]
\label{Deomtw}
\eea
for the shift in the perihelion per orbit.
Here,
the subscripts $P$ and $Q$ 
on the coefficients for Lorentz violation
indicate projections along the directions 
$\vec{P}$ and $\vec{Q}$,
respectively.
The quantity $\ep$ is the eccentricity function,
defined by $\ep = 1 - \sqrt{1-e^2}$. 

The result \rf{Deomtw}
reveals that the perihelion precession
depends on the orbit orientation
through the projections of the 
coefficients $\afb_\mu$ and $\cb_\mn$
along the directions $\vec{P}$ and $\vec{Q}$.
Also,
the factors scaling the coefficients
in Eq.\ \rf{Deomtw}
vary with the composition of the orbiting body.
This means that the orbits of different planets or, 
more generally,
different satellites
are affected by different linear combinations
of coefficients for Lorentz violation.
It is therefore valuable to consider data
from multiple systems
so that independent measurements can be obtained.

To illustrate the sensitivities that can be achieved,
we consider explicitly 
the perihelion precessions of Mercury and of the Earth.
Substituting the relevant orbital data 
for the two planets into Eq.\ \rf{Deomtw} in turn,
taking the planetary mass 
as small compared to the solar mass $m_\odot$,
and incorporating the results for the coefficients $\sb_\mn$
obtained in Eq.\ (190) of Ref.\ \cite{qbakpn},
we find the overall perihelion shifts
$\dot{\tilde{\om}}_{\mercury}$ of Mercury
and $\dot{\tilde{\om}}_{\oplus}$ of the Earth
are given in units of arc seconds per century C 
by the expressions
\bea
\dot{\tilde{\om}}_{\mercury} &\approx& 
\fr{7 \times 10^7{}^{\prime\prime}}{\rm C} \sb_{\mercury} 
\nonumber \\
&&
\hskip -20pt
+\fr{1 \times 10^8 {}^{\prime\prime}} {\rm C}
\sum_w \Big( 
3 \times 10^{-3} \fr{N^w_{\odot}}{m_\odot} \afbw_{\mercury}
-\fr{N^w_{\mercury} m^w}{m_{\mercury}} \cbw_{\mercury}\Big),
\nonumber \\
\dot{\tilde{\om}}_{\oplus} &\approx& 
\fr{2 \times 10^7{}^{\prime\prime}}{\rm C} \sb_{\oplus} 
\nonumber \\
&&
\hskip -20pt
+\fr{4 \times 10^7 {}^{\prime\prime}}{\rm C}
\sum_w \Big( 3 \times 10^{-2} \fr{N^w_{\odot}}{m_\odot} \afbw_\oplus
-\fr{N^w_\oplus m^w}{m_\oplus} \cbw_\oplus \Big).
\nonumber\\
\label{ppshift}
\eea
The combinations of coefficients for Lorentz violation
appearing in these equations are defined as
\bea
\nonumber
\afbw_{\mercury} &=& 
\al \afbw_Q, 
\nonumber \\
\afbw_{\oplus} &=& 
\al \afbw_Q, 
\nonumber \\
\cbw_{\mercury} & \approx & 
[\cbw_{QQ} - \cbw_{PP}] - 6 \times 10^{-3} \cbw_{(0Q)},
\nonumber \\
\cbw_{\oplus} & \approx & 
[\cbw_{QQ} - \cbw_{PP}] - 5 \times 10^{-2} \cbw_{(0Q)},
\nonumber\\
\sb_{\mercury} & \approx & 
(\sb_{PP}-\sb_{QQ})- 6 \times 10^{-3} \sb_{(0Q)},
\nonumber\\
\sb_{\oplus} & \approx & 
(\sb_{PP}-\sb_{QQ}) - 5 \times 10^{-2} \sb_{(0Q)}.
\qquad
\label{lvshift}
\eea
Note that the subscripts $P$, $Q$ here
represent projections that differ for Mercury and the Earth. 

The chemical composition of the Sun 
is believed to be over 70\% hydrogen 
and about 27\% helium by mass
\cite{dba}.
The factors in Eq.\ \rf{ppshift}
that depend on the solar composition
can therefore be estimated as
${N^e_\odot}/{m_\odot} = 
{N^p_\odot}/{m_\odot} \simeq 0.9$ GeV$^{-1}$
and
${N^n_\odot}/{m_\odot} \simeq 0.1$ GeV$^{-1}$.
As can be seen from Eq.\ \rf{lvshift},
these factors suffice for placing approximate bounds
on the coefficients $\afbw_\mu$
from knowledge of the perihelion precessions. 
The composition of Mercury is believed to be
about 70\% iron and about 30\% rocky material 
\cite{dba},
so the analogous ratios for Mercury
are roughly 
${N^e_{\mercury}}/{m_{\mercury}} = 
{N^p_{\mercury}}/{m_{\mercury}} \simeq 0.4$ GeV$^{-1}$
and
${N^n_{\mercury}}/{m_{\mercury}} \simeq 0.6$ GeV$^{-1}$.
For the Earth,
using Ref.\ \cite{earthcomp}
and following the discussion of
Sec.\ \ref{Test and source bodies},
we find 
${N^e_\oplus}/{m_\oplus} = 
{N^p_\oplus}/{m_\oplus} \approx
{N^n_\oplus}/{m_\oplus} \simeq 0.5$ GeV$^{-1}$.
However,
for the approximate bounds obtained below on 
the coefficients $\cb_\mn$,
it suffices that the composition-dependent factors
for the planets are of order $10^{-1}$ GeV$^{-1}$.

We are now in a position to place constraints
on some combinations of the coefficients 
$\afbw_\mu$, $\cb_\mn$, and $\sb_\mn$
by adopting the established error bars
in the existing data for perihelion shifts.
These error bars are 
$0.043^{\prime \prime}$ C$^{-1}$ for Mercury 
and $0.4^{\prime \prime}$ C$^{-1}$ for the Earth
\cite{cmw,cmw2}.
Taking the error bars to be upper bounds
on the perihelion shifts in Eq.\ \rf{ppshift},
we obtain the order-of-magnitude constraints
\bea
|\sb_{\mercury}
+10^{-3}[\afbe_{\mercury} + \afbp_{\mercury}] 
+ 10^{-4} \afbn_{\mercury} 
\hskip 50pt
\nonumber\\
- 10^{-4} \cbe_{\mercury} 
- 10^{-1} \cbp_{\mercury}
- 10^{-1} \cbn_{\mercury} |
\lsim 10^{-9} {\rm ~GeV},
\nonumber\\
| \sb_{\oplus}
+ 10^{-2}[\afbe_{\oplus} + \afbp_{\oplus}] 
+ 10^{-3} \afbn_{\oplus} 
\hskip 50pt
\nonumber\\
- 10^{-4} \cbe_{\oplus} 
- 10^{-1} \cbp_{\oplus}
- 10^{-1} \cbn_{\oplus} |
\lsim 10^{-8} {\rm ~GeV}.
\nonumber\\ 
\label{acsperi}
\eea
Assuming a model 
with nonzero coefficients $\afbw_\mu$ only, 
this yields the approximate constraints
\bea
|\afbe_{\mercury} + \afbp_{\mercury} 
+ 0.1 \afbn_{\mercury} | \lsim 10^{-6} {\rm ~GeV},
\nonumber\\
|\afbe_{\oplus} + \afbp_{\oplus} 
+ 0.1 \afbn_{\oplus} | \lsim 10^{-6} {\rm ~GeV}.
\label{aperi}
\eea

Similarly,
assuming a model 
with nonzero coefficients $\cbw_\mn$ only
and making use of 
existing limits on $\cb_\mn$ for protons and electrons
\cite{tables},
we obtain the approximate constraints  
\bea
| \cbn_{\mercury} | \lsim 10^{-8},
\qquad
| \cbn_{\oplus} | \lsim 10^{-7}.
\label{cperi}
\eea
A careful reanalysis of the existing data
for multiple bodies in the solar system
could yield sharper sensitivities.

The result \rf{aperi} represents first constraints
on the spatial coefficients $\al\afbw_J$.
A sense of the maximal attained sensitivity
to the nine components in $\al\afbw_J$
can be obtained by taking each component in turn 
to be the only nonzero one.
Extracting these sensitivities 
requires the explicit form of the vectors 
$\vec Q_{\mercury}$ and $\vec Q_\oplus$.
The relevant orbital elements 
in heliocentric coordinates are
$\om_{\mercury} \simeq 29^\circ$, 
$\Om_{\mercury} \simeq 48^\circ$,
$i_{\mercury} \simeq 7^\circ$
and 
$\om_\oplus \simeq 103^\circ$, 
$\Om_\oplus = 0^\circ$
$i_\oplus = 0^\circ$.
Converting to the Sun-centered frame
using a counterclockwise rotation 
by $\et \simeq 23.5^\circ$ about the $X$ axis 
yields 
\bea
\vec Q_{\mercury} & \simeq &
-0.97 {\bf e}_X + 0.15 {\bf e}_Y + 0.18 {\bf e}_Z ,
\nonumber\\
\vec Q_\oplus & \simeq &
-0.97 {\bf e}_X - 0.21 {\bf e}_Y - 0.10 {\bf e}_Z.
\eea
Taking each component $\al\afbw_J$ 
as the only nonzero coefficient in turn
yields the order-of-magnitude sensitivities 
\bea
| \al \afbe_X | 
& \lsim & 10^{-6} {\rm ~ GeV},
\nonumber\\
| \al \afbe_Y |, 
| \al \afbe_Z | 
& \lsim & 10^{-5} {\rm ~ GeV},
\nonumber\\
| \al \afbp_X | 
& \lsim & 10^{-6} {\rm ~ GeV},
\nonumber\\
| \al \afbp_Y |,
| \al \afbp_Z | 
& \lsim & 10^{-5} {\rm ~ GeV}, 
\nonumber\\
| \al \afbn_X | 
& \lsim & 10^{-5} {\rm ~ GeV}, 
\nonumber\\
| \al \afbn_Y |,
| \al \afbn_Z | 
& \lsim & 10^{-4} {\rm ~ GeV}. 
\eea
These results are the maximal sensitivities 
achieved to date on the coefficients $\al\afbw_J$.

\section{Photon tests}
\label{photon}

In this penultimate section,
we consider searches for gravitational Lorentz violation
involving the trajectories of photons.
With the coordinate choice
\rf{coordchoice} adopted in this work,
photons follow null geodesics.
The signals of interest therefore arise
from the modifications to the metric,
which are associated with 
Lorentz-violating matter-gravity couplings of the source body 
and in certain cases also 
of the clocks and rods used for measurements.

Photon tests for Lorentz violation
involving the coefficients $\sb_\mn$
in the pure-gravity sector of the SME
have been studied in  
Refs.\ \cite{qbakpn,qbtd}.
Here,
this analysis is extended to include
the matter-sector coefficients $\afbw_\mu$ and $\cbw_\mn$. 
The treatment and notation of Ref.\ \cite{qbtd}
is adopted where possible.
Some quantities relevant for the analysis 
are listed in Table \photona.

\begin{center}
\begin{tabular}{ll}
\multicolumn{2}{c} 
{Table \photona. Notation for photon tests.} \\
\hline
\hline
Quantity & Definition\\ 
\hline
$x^\mu_E = (t_E, \vec{r}_E)$         
& coordinates of event $E$\\
$r_E$                                          
& magnitude of $\vec{r}_E$\\
$x^\mu_P = (t_P, \vec{r}_P)$         
& coordinates of event $P$\\
$r_P$                                          
& magnitude of $\vec{r}_P$\\
$\vec{R} = \vec{r}_P - \vec{r}_E$                   
& zeroth-order light trajectory\\
$\hat{R} = \vec{R}/R$                               
& unit vector along $\vec R$\\
$R = |\vec{R}|$                                     
& magnitude of $\vec{R}$ \\
$b^j = r^j_P 
- \hat{R}^j \vec{r_P} \cdot \hat{R}$                
\pt{xx}
& impact-parameter vector\\
$b$ 
& magnitude of $\vec b$\\
$\ms$                                               
& mass of source body\\
$l_P = \vec{r_P} \cdot \hat{R}$                     
& $\la$ at $P$\\
$ - l_E = \vec{r_E} \cdot \hat{R}$                  
& $\la$ at $E$\\
$\ta_E$                                        
& proper time of $E$ \\
$\ta_P$                                        
& proper time of $P$\\
$u^\mu_E = {d x^\mu_E}/{d \ta_E}$  
\pt{xx}
& 4-velocity of $E$ \\
$u^\mu_P = {d x^\mu_P}/{d \ta_P}$  
& 4-velocity of $P$ \\
$\vec{v}= {d \vec{r}_E}/{dt}$                     
& 3-velocity of $E$\\
$\vec{w}= {d \vec{r}_P}/{dt}$                     
& 3-velocity of $P$\\
$\nu_E$                                        
& frequency at $E$ \\
$\nu_P$                                        
& frequency at $P$\\
\hline
\hline
\end{tabular}
\end{center}

In what follows,
we consider various effects on a light signal as it travels
from an emission event $E$ 
to a spacetime point $P$ located near a massive body.
The light path can be specified 
parametrically as $x^\mu = x^\mu(\la)$,
where $\la$ is the path parameter.
The wave 4-vector $p^\mu$ of the ray tangent to the path is 
\beq
p^\mu = \fr{dx^\mu}{d \la},
\eeq
and it obeys the conditions
\bea
\fr{dx^\mu}{d \la} &=& 
- \Ga^\mu_{\pt{\mu} \al \be} p^\al p^\be,
\nonumber \\
p^\mu p^\nu g_\mn &=& 0.
\eea
The wave 4-vector can be linearized as
\beq
p^\mu = \pb^\mu + \dep^\mu,
\eeq
where the first term is the zeroth-order wave vector
and the second term contains gravitational corrections.
Our interest here lies in the O(1,1) contributions to $\dep^\mu$.
The basic procedure is to insert the modifications 
\rf{metricc} and \rf{metrica} 
of the metric arising from matter-sector effects 
into the general expressions obtained in Ref.\ \cite{qbtd}.
We consider in turn Lorentz-violating contributions 
to the Shapiro time delay,
to the gravitational Doppler shift,
to the gravitational redshift,
and to the null redshift,
and we compare the results to the effective mass 
of a gravitational source as measured by orbital tests.
We also offer some comments
about the implications of the results 
for various experiments.

\subsection{Shapiro time delay}
\label{timedelay}

In this subsection,
we obtain the Lorentz-violating modifications 
to the Shapiro time delay of a light signal
as it passes from a source to a detector
in the presence of a massive body such as the Sun.
The one-way time delay $t_P - t_E$ can be determined
by integrating $\de p^\mu$ along the path
and applying the null condition,
\beq
t_P - t_E = 
R + \half \int^{l_P}_{-l_E} h_\mn \pb^\mu \pb^\nu d \la.
\eeq
Inserting the Lorentz-violating metric modifications 
\rf{metricc} and \rf{metrica} 
and integrating, 
we find the delay can be written in the form 
\bea
t_P - t_E &=& 
R + (t_P - t_E)_{\rm GR}
\nonumber\\
&&
+ (t_P - t_E)_{\afbnp,\cb} 
+ (t_P - t_E)_{\sb}.
\label{generaltd}
\eea
Here, 
$R$ is the zeroth-order time difference.
The second term is the standard GR contribution,
which at O(0,1) and \pno2 takes the form 
\beq
(t_P - t_E)_{\rm GR} = 
2\G\ms \ln\left(\fr{r_E + r_P + R}{r_E + r_P - R} \right).
\eeq
The third term of Eq.\ \rf{generaltd}
consists of contributions 
from Lorentz-violating matter-gravity couplings
associated with the source body S.
At O(1,1) and \pno2,
these contributions are
\bea
&&
\hskip -20pt
(t_P - t_E)_{\afbnp,\cb} = 
\nonumber \\
&&
\hskip 20pt
2\G\ms 
\Big( \fr{\al}{\ms} \abs_0 + \fr{\al}{\ms} \abs_j \hat{R}^j
+ \cbs_{00} \Big) 
\nonumber \\
&& 
\hskip 50 pt
\times \ln\left(\fr{r_E + r_P + R}{r_E + r_P - R} \right)
\nonumber \\
&& 
\hskip 20pt
- \G \al ( \abs_0 + \abs_j \hat{R}^j) 
\left(\fr{l_E}{r_E} + \fr{l_P}{r_P} \right)
\nonumber \\
&& 
\hskip 20pt
- \G \al \abs_k b^k \left(\fr{r_E - r_P}{r_E r_P} \right).
\label{mattertd}
\eea
The final term in Eq.\ \rf{generaltd}
arises from gravitational Lorentz violation
involving the coefficient $\sb_\mn$
and is given in Ref.\ \cite{qbtd}.

In typical time-delay measurements,
an observer emits a light signal at $E$
that is reflected at the spacetime point $P$
and subsequently detected by the observer at $E^\prime$. 
The round-trip coordinate travel time $\De t$,
which is related to the measured proper time $\De \ta_E$ 
by the factor $d\ta_E/dt$,
can be written to O(1,1) and \pno2 as 
\beq
\De t = 
2 R (1 + v^2 - \vec{v} \cdot \hat{R}) 
+ (\De t)_{\rm GR} + (\De t)_{\afbnp,\cb} + (\De t)_{\sb}.
\qquad
\label{rttime}
\eeq
The zeroth-order term in this expression
incorporates Lorentz-violating corrections 
to the trajectory of the emitter,
which here can depend on particle species.
These can in principle be determined 
by modeling the relevant orbits
along the lines of the treatment
in Sec.\ \ref{Solar-system tests}
and Ref.\ \cite{qbakpn}.
The second term in Eq.\ \rf{rttime}
contains the leading GR corrections,
\beq
\De t_{\rm GR} =
4\G\ms \ln\left(\fr{r_E + r_P + R}{r_E + r_P - R} \right).
\label{rtgr}
\eeq
The third term in Eq.\ \rf{rttime}
contains the leading contributions 
from nonzero $\afb_\mu$ and $\cb_\mn$,
\bea
\De t_{\afbnp,\cb} &=& 
4\G\ms \left( \fr{\al}{\ms} \abs_0 + \cbs_{00} \right) 
\nonumber \\
&&
\hskip 30pt 
\times \ln\left(\fr{r_E + r_P + R}{r_E + r_P - R} \right)
\nonumber \\
&& 
- 2 \G \al \abs_0 \left(\fr{l_E}{r_E} + \fr{l_P}{r_P} \right).
\label{rtac}
\eea
The last term of Eq.\ \rf{rttime}
contains corrections involving the coefficient $\sb_\mn$ 
and is given in Ref.\ \cite{qbtd}.
Note that contributions from the coefficients $\afb_j$ 
and $\sb_{0j}$ cancel in the round-trip expression,
a result that can be traced to the parity-odd nature
of the corresponding Lorentz-violating operators.
Note also that 
the time-delay signal changes over two relevant time scales,
the conjunction time $b/v$
and the typically longer orbital time $r/v$,
which enables separation of the zeroth-order 
and gravitational effects.

The dominant Lorentz-violating corrections to $\De t$
are proportional to the logarithm in Eq.\ \rf{rtac}.
The primary effect of the Lorentz-violating matter-gravity couplings 
is therefore to scale the factor of $\G\ms$
in the usual GR time delay \rf{rtgr}.
The scaling can be interpreted
as an effective value $(\G M)_{\rm TD}$
for the source body relevant for time-delay tests,
\beq
(\G M)_{\rm TD} 
= \G \ms \left(1 + \fr{\al}{\ms} \abs_0 
+ \cbs_{00} + \sb_{00} \right).
\label{mtd}
\eeq
This scaling is unobservable in time-delay tests alone.
However,
we show in what follows that other tests can yield
different effective values of $\G\ms$,
so suitable comparisons can reveal
signals for Lorentz violation.
This prospect is considered in Sec.\ \ref{photonexpt} below.

\subsection{Gravitational Doppler shift}
\label{doppler}

When light passes near a massive body,
it suffers a frequency shift as well as a time delay.
In this and the subsequent subsections,
we consider the corrections to the frequency shift
due to the matter-sector coefficients $\afb_\mu$ and $\cb_\mn$.

The relevant quantity is the ratio of frequencies 
observed at the two events $E$ and $P$,
\beq
\fr{\np}{\ne} = \fr{(u^\mu p_\mu)_P}{(u^\nu p_\mu)_E}.
\eeq
At \pno3,
this can be written as
\beq
\fr{\np}{\ne} = 
\sqrt{\fr{1-v^2}{1-w^2}}
\left(
\fr{1 - \vec{w} \cdot \hat{R}} {1 - \vec{v} \cdot \hat{R}}
\right)
\left[ 1 + \left( \fr{\np}{\ne} \right)_g \right].
\eeq
Here,
the term labeled $g$ 
contains gravitational effects involving 
both the Doppler shift and the redshift,
\beq
\left( \fr{\np}{\ne} \right)_g = 
\left( \fr{\np}{\ne} \right)_{\rm DS} 
+ \left( \fr{\np}{\ne} \right)_{\rm RS}. 
\label{gfreq}
\eeq
This subsection treats the gravitational Doppler shift,
while the redshift effects are discussed 
in the next subsection.
 
Corrections to the gravitational Doppler shift 
$({\np}/{\ne})_{\rm DS}$
depending on the coefficients $\afb_\mu$, $\cb_\mn$, and $\sb_\mn$
can be obtained by inserting 
into Eq.\ (31) of Ref.\ \cite{qbtd}
the modifications to the metric 
from Eqs.\ \rf{metricc} and \rf{metrica},
along with those due to $\sb_\mn$ given in Ref.\ \cite{qbakpn}.
Near conjunction,
we find that the dominant effects take the form
\bea
\left(\fr{\np}{\ne}\right)_{\rm DS} & \approx & 
\fr{4 \G \ms}{b} \Big( 1 + \fr{\al}{\ms} \abs_0 
+ \fr{\al}{\ms} \abs_j \hat{R}^j
\nonumber \\
& & 
\hskip 40pt
+ \cbs_{00} \Big) 
\fr{db}{dt} 
+ \left(\fr{\np}{\ne}\right)_{{\rm DS},\sb},
\qquad
\eea
where the last term contains the contributions 
from $\sb_\mn$ found in Ref.\ \cite{qbtd}.

Typical searches measure the round-trip frequency shift,
\bea
\left(\fr{\de \nu}{\nu}\right)_{\rm DS} &=& 
\fr{8 \G \ms}{b} 
\left( 1 + \fr{\al}{\ms} \abs_0 + \cbs_{00} \right)
\fr{db}{dt} 
\nonumber\\
&&
+ \left(\fr{\de \nu}{\nu}\right)_{{\rm DS},\sb} .
\eea
Note that the effects from parity-odd operators again cancel.
The coefficients $\afb_0$, $\cb_{00}$, and $\sb_{00}$
associated with isotropic Lorentz violation 
in the chosen inertial frame
act to scale the factor $\G \ms$
in the usual expression for the gravitational Doppler shift,
leading to an effective value $(\G M)_{\rm DS}$ 
given by
\beq
(\G M)_{\rm DS} = 
\G \ms \left(1 + \fr{\al}{\ms} \abs_0 
+ \cbs_{00} + \sb_{00} \right).
\label{md}
\eeq
The scaling \rf{md} is unobservable in Doppler-shift tests alone.
This result for $(\G M)_{\rm DS}$ 
is identical in form to that
of the time-delay value $(\G M)_{\rm TD}$ 
in Eq.\ \rf{mtd}.

\subsection{Gravitational redshift}
\label{redshift}

The Lorentz-violating contributions
to the term $(\np/\ne)_{\rm RS}$ in Eq.\ \rf{gfreq}
for the gravitational redshift 
can be viewed as subdominant 
to the time delay or Doppler shift
because they occur at the slow time scale.
However,
in dedicated redshift measurements,
the Lorentz-violating gravitational redshift 
can appear as the dominant effect.
In this subsection,
we discuss Lorentz-violating modifications
to the usual gravitational redshift
and effects in null-redshift tests.

To place in context the results in this subsection,
we note that clocks can be used to perform 
three distinct types of gravitational tests
that are often convolved in the literature
under the term `redshift tests.' 
The first type,
which measures the traditional gravitational redshift,
involves two clocks held at different gravitational potentials
whose frequency is compared 
using light or some other signal passing between them. 
This type of test is discussed in Sec.\ \ref{rsstandard} below.
The second type of test is called a null-redshift test,
and it involves monitoring the frequencies of
two clocks of different composition
as they move together through the gravitational potential.
This is discussed in Sec.\ \ref{rsnull}.
The third kind of test
involves synchronizing two clocks 
and then moving one of them around a closed path 
in the gravitational potential.
The signal in this case 
is the accumulated phase difference between the clocks.
An example of this `twin-paradox' redshift test
is the free-fall gravimeter measurement with interferometers
discussed in Sec.\ \ref{epfreefall}.
These three kinds of tests produce related signals in GR.
However,
they can yield distinct sensitivities 
in a more general context such as the SME,
as is demonstrated in what follows.

\subsubsection{Modified redshift}
\label{rsstandard}

The term $(\np/\ne)_{\rm RS}$ in Eq.\ \rf{gfreq}
for the gravitational redshift 
can be understood as the product 
\beq
\left(\fr{\np}{\ne}\right)_{\rm RS} =  
\left( \fr{dt}{d\ta_P} \right)
\left( \fr{d\ta_E}{dt}\right), 
\eeq
of the factors relating proper and coordinate times 
for the clocks at the two points $E$ and $P$.
Each factor is determined by the dispersion relation 
for the corresponding clock,
which depends on coefficients for Lorentz violation 
via its material composition
and on the Lorentz violation 
associated with the gravitational field.

For simplicity in what follows,
we assume the sending and receiving clocks are identical.
This eliminates the need to consider O(1,0) effects,
which have been sought in numerous
clock-comparison experiments 
performed with both clocks
at the same gravitational potential
\cite{tables}.
To the order at which we work,
the redshift can then be expanded as  
\beq
\left(\fr{\np}{\ne}\right)_{\rm RS} = 
\left(\fr{\np}{\ne}\right)^{(0,1)}_{\rm RS}
+ \left(\fr{\np}{\ne}\right)^{(1,1)\rm S}_{\rm RS} 
+ \left(\fr{\np}{\ne}\right)^{(1,1)\rm T}_{\rm RS},
\label{gred}
\eeq
where the term at O(0,1) is the conventional redshift,
the term at O(1,1) labeled by S contains
Lorentz-violating corrections from the gravitational source,
and the last term labeled by T involves O(1,1) contributions
from the clocks.
For our present purposes,
it suffices to work at \pno2.

For an ideal clock,
the Lorentz-violating contributions 
to the first two terms in Eq.\ \rf{gred} can be calculated
by inserting into the usual redshift equation
the modifications \rf{metricc} and \rf{metrica} to the metric
from the coefficients $\afb_\mu$ and $\cb_\mn$,
along with the corrections from Ref.\ \cite{qbakpn}
involving the coefficients $\sb_\mn$.
This gives
\bea
\left(\fr{\np}{\ne}\right)^{(0,1)}_{\rm RS}
+ \left(\fr{\np}{\ne}\right)^{(1,1)\rm S}_{\rm RS} && 
\nonumber\\
&&
\hskip -50pt
=\sqrt{\fr{1-(h^{(0,1)}_{00})_E - (h^{(1,1)}_{00})_E}
{1-(h^{(0,1)}_{00})_P-(h^{(1,1)}_{00})_P}}.
\qquad 
\eea
Expanding to \pno2 and keeping leading-order 
terms in Lorentz violation,
we obtain the conventional \pno2 result,
\beq
\nonumber
\left(\fr{\np}{\ne}\right)^{(0,1)}_{\rm RS}
= \G \ms 
\left( \fr{r_e - r_p}{r_e r_p} \right) ,
\eeq
together with the correction
\bea
\left(\fr{\np}{\ne}\right)^{(1,1)\rm S}_{\rm RS} &=& 
\G \ms 
\Big( \fr{2 \al}{\ms} \abs_0 + \cbs_{00} \Big)
\left( \fr{r_e - r_p}{r_e r_p} \right) 
\nonumber\\ 
&&
+ \left(\fr{\np}{\ne}\right)_{\rm RS,\sb}.
\eea
The last term contains the contributions from $\sb_\mn$
given in Ref.\ \cite{qbtd}.

For the remaining term in Eq.\ \rf{gred},
the situation is more complicated
because the clock frequency must be calculated directly
and typically depends on the structure and composition
of the clock.
Moreover,
although our interest is at O(1,1),
all three of the perturbative contributions
O(1,0), O(0,1), O(1,1) must be treated 
due to the appearance of cross terms in the calculation.
For convenience,
we can express the last term in Eq.\ \rf{gred} in the form
\bea
\left(\fr{\np}{\ne}\right)^{(1,1)\rm T}_{\rm RS} &=& 
\G \ms \cskc \left( \fr{r_e - r_p}{r_e r_p} \right) ,
\label{nupnueT}
\eea
where $\cskc$ is a function 
of the coefficients for Lorentz violation 
associated with the clock.
If the clock's ticking rate is set by its inertial properties,
as is the case for most atomic clocks, 
then $\cskc$ can be expected to depend
on the coefficients $\cbw_\mn$.
If the clock's ticking rate depends intrinsically
on the local gravitational acceleration,
as occurs for a pendulum clock, 
then $\cskc$ can be expected to depend
on the coefficients $\afbw_\mu$.
In general,
the value of $\cskc$ can depend
on both sets of coefficients, 
\beq
\cskc = \cskc \big( \afbw_\mu, \cbw_\mn \big).
\eeq
The key point is that different clocks have different $\cskc$
according to the details of their construction
and flavor content.

Combining the above results,
we see that the dominant Lorentz-violating effects
for the gravitational redshift
can be represented as an effective value 
$(\G M)_{\rm RS}$ implementing a scaling of $\G \ms$,
in parallel with the results for the time delay and
the gravitational Doppler shift.
We obtain
\bea
(\G M)_{\rm RS} &=& 
\G\ms \Big(1 + \fr{2 \al}{\ms} \abs_0 
+ \cbs_{00} + \frac 53 \sb_{00}
\nonumber\\
&&
\hskip 30pt
+ \cskc \Big).
\label{meffrs}
\eea
This represents an unobservable scaling
in any particular redshift test,
but comparing redshift tests performed with different clocks 
could yield access to differences in $\cskc$.
Moreover,
the result for $(\G M)_{\rm RS}$ 
differs from both the
time-delay value $(\G M)_{\rm TD}$
in Eq.\ \rf{mtd}
and the Doppler-shift value $(\G M)_{\rm DS}$ 
in Eq.\ \rf{md},
so comparing results from different tests
could yield independent sensitivities to $\abs_0$
that are inaccessible in other searches with ordinary matter.
This prospect is considered in Sec.\ \ref{photonexpt}.

We remark in passing that for certain special models
the observable redshift effects in $(\G M)_{\rm RS}$ 
may be hidden in WEP tests.
A simple example is provided by the isotropic parachute model 
discussed in Sec.\ \ref{antimatter}.
By virtue of Eq.\ \rf{lfccondition},
the effective inertial and gravitational masses in this model 
are equal for a test body made of ordinary matter,
so no signals are observable in WEP tests.
However,
the presence of nonzero $\cbw_\mn$
implies a nonzero rescaling of $(\G M)_{\rm RS}$,
which is observable by comparing to  
$(\G M)_{\rm TD}$ or $(\G M)_{\rm DS}$.
Signals from this model could also arise
in the null-redshift tests 
discussed in Sec.\ \ref{rsnull} below.

We conclude this subsection
with an illustrative calculation of $\cskc$ 
for a simplified clock 
based on transitions between the Bohr levels of hydrogen,
for which we determine 
$({\np}/{\ne})^{(1,1)\rm T}_{\rm RS}$
and $\cskc \equiv \csk_{\rm H,Bohr}$
assuming both the clocks and the gravitational source 
are at rest.
This calculation is straightforward due to 
the spherical symmetry and the zero velocity,
and also because a simple match exists
between the zeroth-order hamiltonian $h^{(0,0)}$ 
and the kinetic contributions to 
the sum $h^{(1,0)}+h^{(0,1)}+h^{(1,1)}$
of the perturbative corrections
presented in Sec.\ \ref{quantummech}.
By matching these expressions,
we find that 
the kinetic portion of the hamiltonian 
in the presence of gravity and Lorentz violation
can be obtained from the zeroth-order one 
by the following simple replacements
for the proton and electron mass:
\bea
\fr{1}{m^p} & \rightarrow & 
\fr{1}{m^p}
\left(1 - \frac32 h_{00} 
+ \frac53 \cbp_{00} + \frac{13}{6} \cbp_{00} h_{00}\right),
\nonumber\\
\fr{1}{m^e} & \rightarrow & 
\fr{1}{m^e}
\left(1 - \frac32 h_{00} 
+ \frac53 \cbe_{00} + \frac{13}{6} \cbe_{00} h_{00}\right).
\nonumber\\
\eea
Also,
the source term in the Maxwell equations
is corrected by the vierbein determinant $e$,
and the result can be obtained by 
a simple replacement for the proton charge, 
\beq
q^p \rightarrow \fr{q^p}{e} \approx q^p ( 1 - h_{00} ).
\eeq
It follows that the calculation of interest
can be directly performed
by implementing the above replacements 
in the standard result for the Bohr energy levels.
This yields 
\bea
E &\rightarrow &
E\Big(1 - \half h_{00} 
\nonumber\\
&&
\hskip 10pt
+ \fr{1}{m^p + m^e} [m^p \cbe_{00} + m^e \cbp_{00}]
(\frac 53 - \half h_{00}) \Big).
\nonumber\\
\eea
The modification \rf{nupnueT} to the gravitational redshift
is therefore given by
\bea
\left(\fr{\np}{\ne}\right)^{(1,1)\rm T}_{\rm RS} &=& 
- \fr{2 \G\ms}{3(m^p + m^e)}(m^p \cbe_{00} + m^e \cbp_{00}) 
\nonumber\\
&&
\times \left(\fr{r_e - r_p}{r_e r_p}\right)
\eea
when the clock transitions are those 
of the Bohr levels of hydrogen.
This implies the result 
\beq
\csk_{\rm H,Bohr} 
= - \fr{2}{3(m^p + m^e)}(m^p \cbe_{00} + m^e \cbp_{00}).
\eeq
The value of $\cskc$ for a realistic clock
can be obtained via calculation
if the hamiltonian describing the clock is known.

\subsubsection{Null redshift}
\label{rsnull}

Another Lorentz-violating signal
can be accessed by
comparing two clocks of different types
as they explore the gravitational potential together.
This type of measurement is called a null-redshift test
\cite{cmw3}.

Consider comparing the frequencies of two clocks $A$ and $B$
having different values 
$\cskc = \csk_A$ and $\cskc = \csk_B$
that are located at a point $P$ 
with gravitational potential $h^P_{00}$.
The frequency ratio is given by
\beq
\left( \fr{\nu_A}{\nu_B} \right)^P = 
\left[1 + \half (\csk_A - \csk_B) h^P_{00} \right]
\left( \fr{\nu^{(0)}_A}{\nu^{(0)}_B} \right),
\label{nuab}
\eeq
where the superscript $(0)$ denotes a frequency 
at a hypothetical zero gravitational potential $h_\mn =0$. 
This frequency ratio depends inseparably 
on the potential $h^P_{00}$ at point $P$
and the ratio in zero potential.

When the same two clocks are moved to a point $Q$
at potential $h^Q_{00}$,
the frequency ratio takes a new value.
If the values $\csk_A$ and $\csk_B$ differ,
then so do the frequency ratios at $P$ and $Q$.
The ratio of frequency ratios then shifts away from 1
and is given by
\beq
\left(\fr{\nu_A}{\nu_B}\right)^P 
\left(\fr{\nu_B}{\nu_A}\right)^Q =
1 - \half (\csk_A - \csk_B) 
(h^Q_{00} - h^P_{00}).
\label{clockshift}
\eeq
The shift is an observable,
and it depends on the difference 
$\De\csk_{AB}= \csk_A-\csk_B$
and also on the potential difference between $P$ and $Q$. 

For a gravitational source with
$h_{00} = 2\G \ms/r$ at \pno2,
we obtain
\beq
\left(\fr{\nu_A}{\nu_B}\right)^P 
\left(\fr{\nu_B}{\nu_A}\right)^Q =
1 - \G \ms \De\csk_{AB} 
\fr{(r_P - r_Q)}{r_P r_Q}.
\label{clockshift2}
\eeq
Unlike the other Lorentz-violating photon effects discussed here,
all of which represent scalings of $\G \ms$,
this result is a qualitative change
from conventional gravity.
It is also strictly a gravitational effect,
vanishing in Minkowski spacetime.

Since the shift varies with spacetime position,
it exhibits features analogous to 
violations of local position invariance,
which have been the subject of numerous studies
\cite{cmw}.
In the present case,
these features arise from the Lorentz-violating 
flavor dependence of the clock material.
Note also that the observable \rf{clockshift2}
contains the same information
as the result of two separate redshift tests
performed with different clocks 
but the same gravitational source.
This can be verified by inspection 
of the effective value $(\G M)_{\rm RS}$ 
in Eq.\ \rf{meffrs}.
Some relevant experiments 
are described in Sec.\ \ref{photonexpt}.

\subsection{Comparison to effective orbital mass}
\label{Orbiting Massive Bodies}

The preceding subsections
reveal that the Lorentz-violating contributions 
to the Shapiro time delay,
the gravitational Doppler shift,
and the gravitational redshift
are all controlled by the effective value of $\G M$
for the gravitational source.
In the context of the solar-system tests 
discussed in Sec.\ \ref{Solar-system tests},
rescalings of $\G M$ also occur
but can be disregarded as unobservable.
Here,
we determine the effective value of $\G M$
relevant to observations of orbiting bodies,
$(\G M)_{\rm OB}$.

For Lorentz violation
involving the coefficients $\afb_\mu$ and $\cb_\mn$,
the secular changes in the orbital elements 
for the trajectory of an orbiting body
are given by Eq.\ \rf{ppacc}.
The analogous result for the coefficients $\sb_\mn$
is given in Eq.\ (162) of Ref.\ \cite{qbakpn}.
Inspecting these equations,
we can deduce the effective reduced mass 
of the source and test bodies
and hence extract the effective value $(\G M)_{\rm OB}$.
Making no additional assumptions 
about the masses of the source and test bodies,
we find
\bea
(\G M)_{\rm OB} &=& 
\G M \Big( 1 + \fr{2 \al}{\ms} \abs_0 
+ \fr{2 \al}{\mt} \abt_0  + \frac 53 \sb_{00}
\nonumber \\
& & 
+ \fr{\ms - \fr23 \mt}{M} \cbs_{00} 
+ \fr{\mt - \fr23 \ms}{M} \cbt_{00} \Big).
\nonumber\\
\eea

To obtain an expression that is more readily comparable
to the effective values of $(\G M)$ 
measured in photon tests,
we note that $\mt\ll\ms$ under typical circumstances.
The above result then reduces to 
\bea
(\G M)_{\rm OB} &=& 
\G \ms \Big( 1 + \fr{2 \al}{\ms} \abs_0 + \cbs_{00} 
+ \frac 53 \sb_{00}
\nonumber \\
& & 
\hskip 35pt
+ \fr{2 \al}{\mt} \abt_0 - \frac 23 \cbt_{00} \Big).
\label{mso}
\eea
This expression for $(\G M)_{\rm OB}$ 
contains a linear combination of coefficients for Lorentz violation
that is independent of the three combinations
$(\G M)_{\rm TD}$, $(\G M)_{\rm DS}$, and $(\G M)_{\rm RS}$
obtained for photon tests.
Some comments about tests with this result
are provided in the next subsection.

\subsection{Experiments}
\label{photonexpt}

The above subsections show that
each type of photon test of Lorentz symmetry 
is sensitive to an effective value of $\G M$
that contains a combination
of coefficients for Lorentz violation.
The time-delay value $(\G M)_{\rm TD}$
is given by Eq.\ \rf{mtd},
and it depends on the coefficients
$\al \abs_0$, $\cbs_{00}$, and $\sb_{00}$.
The gravitational Doppler shift
involves the value $(\G M)_{\rm DS}$ in Eq.\ \rf{md}
and involves the same combination of the three coefficients.
The value $(\G M)_{\rm RS}$ for the gravitational redshift 
is given by Eq.\ \rf{meffrs},
which contains a different combination of coefficients
and varies also with $\cskc$.
All three of these photon tests
yield sensitivities differing from those
in orbital tests,
which involve the value $(\G M)_{\rm OB}$ in Eq.\ \rf{mso}
that depends also on the test-body coefficients
$\al \abt_0$ and $\cbt_{00}$.
Note that no qualitatively new signals
are involved in any of these cases,
since the effects are merely scalings
of established physics.
In contrast,
the null-redshift observable 
given in Eq.\ \rf{clockshift},
which depends on the difference
of clock quantities $\De\csk_{AB}$,
represents a qualitative departure
from conventional gravitational physics.

Comparisons of the time-delay value $(\G M)_{\rm TD}$
or the Doppler-shift value $(\G M)_{\rm DS}$ 
to the redshift value $(\G M)_{\rm RS}$ 
for the same source body 
can be used to obtain sensitivity 
to combinations of coefficients for Lorentz violation.
High-quality data for the time delay 
and the gravitational Doppler shift
have been obtained by tracking the Cassini spacecraft 
\cite{cassini}
in the gravitational field of the Sun.
Proposed missions such as 
the Astrodynamical Space Test of Relativity 
using Optical Devices (ASTROD) 
\cite{astrod},
the Mercury Orbiter Radio-science Experiment (MORE)
\cite{more},
the Search for Anomalous Gravitation using Atomic Sensors (SAGAS) 
\cite{sagas},
and 
the Solar System Odyssey (SSO) 
\cite{odyssey}
have the potential to improve these measurements
using the Sun as the gravitational source,
while the Beyond Einstein Advanced Coherent Optical Network (BEACON)
\cite{beacon} 
could sharpen results 
using the Earth as the gravitational source.
Another relevant recent proposal 
involves the use of very-long-baseline interferometry (VLBI)
\cite{vlbi}
to measure the deflection of radio waves from distant sources
by solar-system objects.
The sensitivity of this measurement to Lorentz violation
is likely to be comparatively weaker but may be offset
by the enhanced access to independent coefficient combinations
offered by multiple measurements
and perhaps by access to anisotropic effects
involving spatial components of $\sb_\mn$.

Redshift tests permit sensitivities 
to effects controlled by $\cskc$.
These can be isolated either by comparing separate redshift tests
performed with different clocks 
in the same gravitational source 
or more directly by null-redshift tests,
in which the signal depends on the difference $\De\csk_{AB}$
between two clocks $A$, $B$
and vanishes in the absence of gravity.
The results of some investigations of local position invariance
can be reinterpreted as measurements of $\De\csk_{AB}$.
For example,
a recent Earth-based test
comparing a hydrogen maser with a Cs fountain 
\cite{ahjp}
obtained a sensitivity that corresponds to the bound
\beq
|\csk_{\rm H} - \csk_{\rm Cs}| < (0.1 \pm 1.4) \times 10^{-6},
\eeq
while another comparing a hydrogen maser
with a cryogenic sapphire oscillator 
\cite{twb}
yields the measurement
\beq
\csk_{\rm H} - \csk_{\rm CSO} = (-2.7 \pm 1.4) \times 10^{-4}.
\eeq 
These results offer a benchmark 
for currently attainable sensitivities to $\cskc$.
The two experiments involve different clocks
and hence likely different sensitivities 
to the coefficients $\cbw_\mn$.
Calculating the specific constraints on $\cbw_\mn$
and possibly other coefficients for Lorentz violation 
from these and other tests is an interesting open project.
Note that Earth-based searches of this type
typically take advantage 
of the annual and diurnal variations  
in the gravitational potential of the Sun
as experienced in the laboratory.
In searches using the annual variation,
it is challenging and perhaps impossible
to disentangle gravitational effects of nonzero $\cskc$ 
from other Lorentz-violating effects in Minkowski spacetime.
However,
diurnal searches can distinguish the two types of effects
because the Minkowski-spacetime signals
occur at the sidereal frequency instead. 
Note also that other clock-comparison tests
normally viewed as sensitive to SME coefficients
in Minkowski spacetime may also have sensitivity to $\cskc$. 
One intriguing possibility is that suitable choices of clocks
could separate effects from $\cbe_{TT}$ and $\cbp_{TT}$,
which would then lead to 
independent sensitivities to $\afbe_T$ and $\afbp_T$,
a result otherwise challenging to achieve.

Satellites carrying two different clocks offer interesting prospects  
for improved null-redshift searches for Lorentz violation.
Since the attainable sensitivities improve 
with the gravitational potential difference
according to Eq.\ \rf{clockshift},
it is desirable to acquire elliptical orbits.
The Space-Time Asymmetry Research (STAR) program 
\cite{star}
presently under development
proposes to compare two different clocks 
on a satellite traveling in an elliptical orbit.
This mission could improve sensitivities to $\cskc$
by an order of magnitude or more
relative to ground-based tests.
Improved sensitivities may also be possible
by comparing clocks aboard the proposed SAGAS spacecraft.
Note also that experiments in highly elliptical orbits
can be expected to have increased sensitivity
to anisotropic effects on the redshift produced by $\sb_{JK}$.

Provided effects due to $\cskc$ are excluded,
either through independent experiments 
or by using a clock with $\cskc =0$,
then the dependence of $(\G M)_{\rm RS}$ on $\abs_T$
implies that measurements of the gravitational redshift 
can be compared with other photon tests
performed with the same gravity source
to obtain independent sensitivities to $\afbw_T$.
The Gravity Probe A (GPA) mission
\cite{rfcv},
which used the Earth as the gravity source,
confirmed the conventional gravitational redshift
to parts in $10^4$.
This result could eventually be combined 
with proposed time-delay or Doppler-shift measurements 
of the BEACON type to yield sensitivity 
to the coefficient $\abs_T$ for the Earth.
Improved tests of the gravitational redshift  
are also proposed for 
the Atomic Clock Ensemble in Space (ACES)
\cite{aces}, 
SAGAS, and STAR missions.
With the Sun as the gravity source instead,
the Galileo space probe
obtained sensitivity to deviations for the gravitational redshift 
at the level of parts in $10^2$
\cite{galileo}.
Given knowledge of $\cskc$ for the Galileo clock,
this result could be combined with the Cassini results
to yield sensitivity 
to the coefficient $\abs_T$ for the Sun.
If feasible,
a redshift test performed directly with Cassini
would be of interest in this respect. 
Other gravitational sources could also be used.
For example,
the gravitational redshift was measured to parts in $10^{2}$ 
using Saturn as the source 
during the flyby of Voyager
\cite{saturnrs}.
Time-delay or Doppler-shift data could therefore permit sensitivity 
to the coefficient $\afb_T$ for Saturn.

Combinations of photon tests with measurements
of the effective orbital mass are also of interest.
In the limit of zero matter-sector Lorentz violation,
the result \rf{mso} for $(\G M)_{\rm OB}$ 
has been combined with 
Eq.\ \rf{mtd} for $(\G M)_{\rm TD}$
to extract sensitivity to $\sb_{TT}$
\cite{qbtd}.
However,
with nonzero coefficients $\afb_\mu$ and $\cb_\mn$,
the effective value $(\G M)_{\rm OB}$ 
involves properties of the test body
as well as the source. 
Note that these appear in the familiar combination 
$\al \abt_T - \mt \cbt_{TT}/3$
discussed in Secs.\ \ref{labtests} and \ref{antimatter}.
The WEP tests considered in this work
constrain the degree to which this combination can differ
between neutrons and neutral combinations of electrons and protons,
but only through the indirect arguments involving binding energy 
described in Sec.\ \ref{Sensitivities}.
In contrast,
comparing $\G M$ factors
for measurements with orbiting bodies and for photon tests 
offers the opportunity to obtain direct sensitivity
to $\al \afbn_T - \mt \cbn_{TT}/3$
and $\al \afbx{e+p}_T - \mt \cbx{e+p}_{TT}/3$.
Comparisons with sensitive gravimeters
may also be of interest in this respect.

We remark in passing that proposed sensitive experiments 
to measure gravitational light bending,
including
the Laser Astrometric Test of Relativity (LATOR)
\cite{lator}
and 
the Space Interferometry Mission (SIM)
\cite{sim}, 
are likely to have signals affected by Lorentz violation.
The attainable sensitivities can be expected
to be similar to those discussed above,
but the analysis of this possibility 
lies beyond our present scope.

\section{Summary}
\label{summary}

This work studies the gravitational couplings of matter
in the presence of Lorentz violation.
The framework for the investigation is 
the fermion sector of the gravitationally coupled minimal SME
in a post-newtonian expansion.
Our primary goal is to develop a suitable methodology
for searches for Lorentz and CPT violation
that exploit the couplings of matter to gravity,
incorporating in particular effects 
that are challenging or impossible to detect
in Minkowski spacetime.

Section \ref{theory} presents 
the basic formalism for the work.
The action for the gravity-matter system is given
in Sec.\ \ref{Action},
and the linearization procedure is outlined
in Sec.\ \ref{Linearization}.
Some types of Lorentz violation are unobservable in principle.
This issue is discussed in Sec.\ \ref{Observability},
which also fixes the coordinate choice \rf{coordchoice}
used in this work.
The metric and coefficient fields for Lorentz violation
can fluctuate about their background values,
and the corresponding interactions
must be incorporated in analyses of experiments.
In Sec.\ \ref{Perturbation scheme},
we develop general perturbative techniques to analyze
these fluctuations.
Two notions of perturbative order are introduced.
One is denoted O($m,n$)
and tracks the orders in Lorentz violation and in gravity,
while the other is denoted \pno{$p$}
and tracks the post-newtonian order. 
The goal of this work is to investigate
dominant terms involving Lorentz violation in gravity,
which are at O(1,1).

Section \ref{quantumthry} studies
the quantum theory of the gravity-matter system.
Starting from the field-theoretic action,
we construct the relativistic quantum mechanics 
in the presence of gravitational fluctuations
and Lorentz violation.
Formulating the quantum theory for matter 
in the presence of gravitational fluctuations
is a standard challenge.
In Sec.\ \ref{Time dependence},
we present a solution to this problem
via a field redefinition,
which yields a hamiltonian that is hermitian with respect to
the usual scalar product for wave functions.
We then use this procedure in Sec.\ \ref{relativham}
to extract the explicit form of the relativistic hamiltonian
involving all coefficients for Lorentz violation
in the minimal QED extension. 
The result forms the appropriate starting point
for general investigations of Lorentz and CPT violation
in matter-gravity couplings.
To maintain a reasonable scope in this work,
we subsequently specialize our focus to the study
of spin-independent Lorentz-violating effects,
which are governed by the coefficient fields $\af_\mu$, \c\
and the metric fluctuation $h_\mn$.
The nonrelativistic quantum hamiltonian 
for this case is obtained in Sec.\ \ref{quantummech}
using the standard Foldy-Wouthuysen procedure.

Measurements of gravity-matter couplings
typically are performed at the classical level.
Section \ref{classicaltheory}
constructs the classical theory
associated with the quantum-mechanical dynamics
of matter involving nonzero $\af_\mu$, \c, and $h_\mn$.
The behavior of test and source bodies 
in the presence of Lorentz violation
is the subject of Sec.\ \ref{action}.
Working from the action for a point particle,
we provide expressions for the mass and for 
the effective coefficients for Lorentz violation
for a test or source body,
along with the effective action \rf{actioneq}
describing the dynamics of the body.
These results enable the derivation in Sec.\ \ref{eom}
of the modified Einstein equation
and the equation \rf{oogeo}
for the trajectory of a test particle.
To apply this equation in practice
requires knowledge of the coefficient and metric fluctuations.
In Sec.\ \ref{fluct},
we develop a systematic methodology 
for calculating this information in perturbation theory
and obtain general expressions 
for the coefficient and metric fluctuations
to O(1,1)
in terms of various gravitational potentials
and the background coefficient values $\afb_\mu$ and $\cb_\mn$.

To illustrate the application of the general formalism,
we consider in Sec.\ \ref{bumblebee}
a specific class of bumblebee models,
which are theories with a vector field 
driving spontaneous Lorentz breaking.
The action for the bumblebee field $B_\mu$
is given in Sec.\ \ref{Bumblebee model},
where a match at the field-theoretic level 
to the general formalism of earlier sections is made 
and the coefficient fields $\af_\mu$ and \c\ 
are identified in terms of $B_\mu$ and the metric.
In Sec.\ \ref{Solving the Model},
we explicitly solve the model 
at the relevant order in perturbation theory,
extract the modified Einstein equation,
and derive the equation for the trajectory of a test particle.
The results are shown to match those
obtained using the general formalism developed 
in the earlier sections.

The largest portion of the paper
is devoted to a discussion of experiments and observations
that can achieve sensitivity to 
the coefficients $\afb_\mu$ and $\cb_\mn$.
Section \ref{experiment}
presents some general material 
broadly applicable to searches for Lorentz violation.
Various choices of reference frame
and their relationship to the canonical Sun-centered frame
are discussed in Sec.\ \ref{Frames}.
Attainable sensitivities 
to the coefficients $\afb_\mu$ and $\cb_\mn$
in any measurement procedure
are constrained by certain generic features.
Section \ref{Sensitivities}
considers some of these,
including the role of binding energy 
in impeding or aiding the analysis
of WEP tests for signals of Lorentz violation.

A major class of searches for Lorentz violation
involves laboratory tests with ordinary neutral bulk matter,
neutral atoms, and neutrons.
Section \ref{labtests} treats this topic.
The \pno3 lagrangian describing the dynamics of a test body
moving near the surface of the Earth
in the presence of Lorentz violation
is considered in Sec.\ \ref{labtheory}.
Expressions are given in an Earth-centered frame
and the transformation to the laboratory frame
is outlined.
The resulting description 
of laboratory signals for gravitational Lorentz violation
includes effects 
from the matter-sector coefficients $\afb_\mu$ and $\cb_\mn$
and ones from the gravity-sector coefficients $\sb_\mn$
obtained in Ref.\ \cite{qbakpn}.
It reveals that the gravitational force acquires tiny corrections
both along and perpendicular to the usual free-fall trajectory
near the surface of the Earth,
while the effective inertial mass of a test body
becomes a direction-dependent quantity.
These effects can be sought in numerous laboratory experiments.
Since the standard relationship between force and acceleration
is modified,
it is useful to distinguish tests 
measuring gravitational acceleration 
from ones comparing forces.
In Sec.\ \ref{ffgravimeter},
we consider free-fall gravimeter tests
such as falling corner cubes and atom interferometry.
Force-comparison gravimeter tests
using equipment such as superconducting gravimeters
are studied in Sec.\ \ref{fcgravimeters}.
An important potential signal
for gravitational Lorentz violation
arises from the flavor dependence of the effects,
which implies signals in WEP tests.
A variety of free-fall WEP tests
is considered in Sec.\ \ref{epfreefall},
while force-comparison WEP tests
with a torsion pendulum are treated in Sec.\ \ref{tpend}.
For all the tests considered,
the possible signals for Lorentz violation
are decomposed according to their time dependence,
and estimates of the attainable sensitivities are obtained.

\onecolumngrid									
\begin{center}									
\begin{tabular}{lcccc}									
\multicolumn{5}{c}									
{Table \summarya.\									
Summary of actual and attainable sensitivities in past or present tests.} \\									
\hline									
\hline									
Coefficient     	&	Gravimeter	&	       Free-fall       	&	Force-comparison	&	Solar	\\
combinations    	&		&	WEP	&	WEP	&	system	\\
\hline
&&&&\\
$\al \afbx{\epcombo}_X$ 	&	       [$10^{-7}$ GeV] 	&	       [$10^{-3}$ GeV] 	&	       [$10^{-7}$ GeV] 	&	       $\dots$ 	\\
$\al \afbx{\epcombo}_{Y+Z}$     	&	       [$10^{-7}$ GeV] 	&	       [$10^{-3}$ GeV] 	&	       [$10^{-6}$ GeV] 	&	       $\dots$ 	\\
$\al \afbx{\epcombo}_Y$ 	&	       [$10^{-5}$ GeV] 	&	       $\dots$ 	&	       [$10^{-7}$ GeV] 	&	       $\dots$ 	\\
$\al \afbx{\epcombo}_Z$ 	&	       [$10^{-5}$ GeV] 	&	       $\dots$ 	&	       [$10^{-6}$ GeV] 	&	       $\dots$ 	\\
$\al \afbx{\epcombo}_T$ 	&	       $\dots$ 	&	       $10^{-7}$ GeV$^\dag$   	&	       $10^{-10}$ GeV$^\dag$   	&	       $\dots$ 	\\
$\al \afbx{n}_X$        	&	       [$10^{-7}$ GeV] 	&	       [$10^{-3}$ GeV] 	&	       [$10^{-7}$ GeV] 	&	       $\dots$ 	\\
$\al \afbx{n}_{Y+Z}$    	&	       [$10^{-7}$ GeV] 	&	       [$10^{-3}$ GeV] 	&	       [$10^{-6}$ GeV] 	&	       $\dots$ 	\\
$\al \afbx{n}_Y$        	&	       [$10^{-5}$ GeV] 	&	       $\dots$ 	&	       [$10^{-7}$ GeV] 	&	       $\dots$ 	\\
$\al \afbx{n}_Z$        	&	       [$10^{-5}$ GeV] 	&	       $\dots$ 	&	       [$10^{-6}$ GeV] 	&	       $\dots$ 	\\
$\al \afbx{n}_T $       	&	       $\dots$ 	&	       $10^{-7}$ GeV$^\dag$   	&	       $10^{-10}$ GeV$^\dag$   	&	       $\dots$ 	\\
$\al \afbx{\pen}_X$     	&	       [$10^{-7}$ GeV] 	&	       [$10^{-4}$ GeV] 	&	       [$10^{-8}$ GeV] 	&	       [$10^{-6}$ GeV] 	\\
$\al \afbx{\pen}_{Y+Z}$ 	&	       [$10^{-7}$ GeV] 	&	       [$10^{-4}$ GeV] 	&	       [$10^{-7}$ GeV] 	&	       [$10^{-6}$ GeV] 	\\
$\al \afbx{\pen}_Y$     	&	       [$10^{-5}$ GeV] 	&	       $\dots$ 	&	       [$10^{-8}$ GeV] 	&	       $\dots$ 	\\
$\al \afbx{\pen}_Z$     	&	       [$10^{-5}$ GeV] 	&	       $\dots$ 	&	       [$10^{-7}$ GeV] 	&	       $\dots$ 	\\
$\al \afbx{\pen}_T - \fr13 m^p \cbx{\pen}_{TT}$ 	&	       $\dots$ 	&	       $10^{-8}$ GeV$^\ddag$  	&	       $10^{-8}$ GeV$^\ddag$   	&	       $\dots$ 	\\
$\al \afbx{\pen}_T - \fr13 m^p \cbx{\pen}_{TT}$ 	&	               	&	               	&	               	&	               	\\
$ \pt{a} + (\half \cos^2 \ch - \fr16) m^n \cbx{n}_Q$    	&	       $\dots$ 	&	       $10^{-8}$ GeV   	&	       $\dots$ 	&	       $\dots$ 	\\
$\al \afbx{\pen}_T - \fr13 m^p \cbx{\pen}_{TT} $        	&	       	&	       	&	       	&	       	\\
$ \pt{a} - \fr16 m^n \cbx{n}_Q$ 	&	       $\dots$ 	&	       $\dots$ 	&	       $10^{-11}$ GeV  	&	       $\dots$ 	\\
$\afbe_{\mercury} + \afbp_{\mercury} + 0.1 \afbn_{\mercury} $   	&	       $\dots$ 	&	       $\dots$ 	&	       $\dots$ 	&	       $10^{-6}$ GeV$^\dag$   	\\
$\afbe_{\oplus} + \afbp_{\oplus} + 0.1 \afbn_{\oplus}$  	&	       $\dots$ 	&	       $\dots$ 	&	       $\dots$ 	&	       $10^{-6}$ GeV$^\dag$   	\\
$\cbx{n}_{(TJ)}$        	&	       $[10^{-7}]$     	&	       $[10^{-4}]$     	&	       $[10^{-7}]$     	&	       $[10^{-6}]$     	\\
$\cbx{n}_Q$       	&	       $\dots$ 	&	       $10^{-8}{}^\ddag$    	&	       $10^{-8}{}^\ddag$       	&	       $\dots$ 	\\
$\cbn_{\mercury}$       	&	       $\dots$ 	&	       $\dots$ 	&	       $\dots$ 	&	       $10^{-8}{}^\dag$       	\\
$\cbn_{\oplus}$ 	&	       $\dots$ 	&	       $\dots$ 	&	       $\dots$ 	&	       $10^{-7}{}^\dag$        	\\
&&&&\\
\hline									
\end{tabular}									
\end{center}									
\twocolumngrid									
Section \ref{space} considers satellite-based WEP tests,
which offer interesting prospects for improved sensitivities
to Lorentz violation.
In this context, 
the signal for Lorentz violation
is an anomalous time variation 
of the relative local acceleration
between two test bodies of differing composition
located on the satellite.
We derive the frequency decomposition of the signal
for Lorentz violation,
and we consider idealized scenarios
for several proposed satellite-based WEP tests.
Based on the design reach of the missions,
we estimate the sensitivities that could be achieved
to various combinations
of the matter-sector coefficients $\afb_\mu$ and $\cb_\mn$.

Studies of the gravitational couplings 
of charged particles, antimatter,
and second- and third-generation particles
present distinct experimental challenges
but can yield sensitivities to Lorentz and CPT violation 
that are otherwise difficult or impossible to achieve.
Section \ref{exotic} addresses some 
of these possibilities,
including charged-particle interferometry,
ballistic tests with charged particles,
gravitational experiments with antihydrogen,
and signals in muonium free fall.
For antihydrogen experiments,
simple toy models are introduced 
to illustrate aspects of their discovery potential
and to address attempts to place indirect limits
on possible effects.

Traditional tests of gravity couplings to matter
include observations of the motion of bodies
within the solar system.
Section \ref{Solar-system tests}
contains a discussion of the signals accessible 
via lunar and satellite laser ranging
and via measurements of the precession of the perihelion
of orbiting bodies.
A reanalysis of existing data from lunar laser ranging 
could yield interesting sensitivities
to some combinations 
of the matter-sector coefficients $\afb_\mu$ and $\cb_\mn$.
We use the established advance of the perihelion
for Mercury and for the Earth to obtain constraints
on combinations of $\afb_\mu$, $\cb_\mn$, and $\sb_\mn$.

The interaction of photons with gravity
offers a different arena 
in which to seek Lorentz and CPT violation.
Section \ref{photon} is devoted to this topic.
We consider signals arising 
in measurements of the photon time delay,
studies of the gravitational Doppler and redshifts,
and comparisons of the behaviors of photons and massive bodies. 
A variety of existing and proposed experiments on spacecraft
offer interesting prospects for these measurements.
 
Tables \summarya\ and \summaryb\ 
collect estimated sensitivities
to the matter-sector coefficients $\afbw_\mu$ and $\cbw_\mn$
obtained from many of the measurements discussed in this work.
These tables disregard possible effects 
from the pure-gravity coefficients $\sb_\mn$
that could in principle be relevant to solar-system tests.
Table \summarya\ concerns existing data,
while Table \summaryb\ 
tabulates future prospects. 
One result omitted from these tables
is the generalization \rf{aeffprl}
of the constraint obtained in Ref.\ \cite{akjt}
using data from force-comparison WEP tests
with a torsion pendulum.
In Sec.\ \ref{tpend},
multiple data sets are combined
to separate this constraint into 
the two limits \rf{viabinding},
and both of these are included in Table \summarya\ instead.

\onecolumngrid									
\vskip 10 pt

\begin{center}									
\begin{tabular}{lccccc}									
\multicolumn{5}{c}									
{Table \summaryb.\									
Summary of attainable sensitivities in future tests.} \\									
\hline									
\hline									
Coefficient	&	Free-fall	&	Free-fall	&	Satellite	&	Solar	\\
combinations	&	gravimeter	&	WEP	&	WEP 	&	system	\\
\hline	
&&&&\\
$\al \afbx{\epcombo}_X$ 	&	       $10^{-10}$ GeV   	&	       $10^{-10}$ GeV  	&	       $10^{-11}$ GeV  	&	       $\dots$ 	\\
$\al \afbx{\epcombo}_{Y+Z}$     	&	       $10^{-10}$ GeV   	&	       $10^{-10}$ GeV  	&	       $10^{-11}$ GeV  	&	       $\dots$ 	\\
$\al \afbx{\epcombo}_Y$ 	&	       $10^{-8}$ GeV   	&	       $10^{-8}$ GeV   	&	       $10^{-9}$ GeV   	&	       $\dots$ 	\\
$\al \afbx{\epcombo}_Z$ 	&	       $10^{-8}$ GeV   	&	       $10^{-8}$ GeV   	&	       $10^{-9}$ GeV   	&	       $\dots$ 	\\
$\al \afbx{\epcombo}_T$ 	&	       $\dots$ 	&	       $10^{-14}$ GeV$^\dag$   	&	       $10^{-15}$ GeV$^\dag$   	&	       $\dots$ 	\\
$\al \afbx{n}_X$        	&	       $10^{-10}$ GeV   	&	       $10^{-10}$ GeV  	&	       $10^{-11}$ GeV  	&	       $\dots$ 	\\
$\al \afbx{n}_{Y+Z}$    	&	       $10^{-10}$ GeV   	&	       $10^{-10}$ GeV  	&	       $10^{-11}$ GeV  	&	       $\dots$ 	\\
$\al \afbx{n}_Y$        	&	       $10^{-8}$ GeV   	&	       $10^{-8}$ GeV   	&	       $10^{-9}$ GeV   	&	       $\dots$ 	\\
$\al \afbx{n}_Z$        	&	       $10^{-8}$ GeV   	&	       $10^{-8}$ GeV   	&	       $10^{-9}$ GeV   	&	       $\dots$ 	\\
$\al \afbx{n}_T $       	&	       $\dots$ 	&	       $10^{-14}$ GeV$^\dag$   	&	       $10^{-15}$ GeV$^\dag$   	&	       $\dots$ 	\\
$\al \afbx{\pen}_X$     	&	       $10^{-10}$ GeV  	&	       $10^{-11}$ GeV  	&	       $10^{-12}$ GeV  	&	       $10^{-7}$ GeV   	\\
$\al \afbx{\pen}_{Y+Z}$ 	&	       $10^{-10}$ GeV  	&	       $10^{-11}$ GeV  	&	       $10^{-12}$ GeV  	&	       $10^{-7}$ GeV   	\\
$\al \afbx{\pen}_Y$     	&	       $10^{-8}$ GeV   	&	       $10^{-9}$ GeV   	&	       $10^{-10}$ GeV  	&	       $\dots$ 	\\
$\al \afbx{\pen}_Z$     	&	       $10^{-8}$ GeV   	&	       $10^{-9}$ GeV   	&	       $10^{-10}$ GeV  	&	       $\dots$ 	\\
$\al \afbx{\pen}_T $    	&	               	&	               	&	               	&	               	\\
$\pt{+} - \fr13 m^p \cbx{\pen}_{TT}$    	&	       $\dots$ 	&	       $10^{-15}$ GeV  	&	       $10^{-16}$ GeV  	&	       $\dots$ 	\\
$\cbx{n}_{(TJ)}$        	&	       $10^{-10}$      	&	       $10^{-11}$      	&	       $10^{-12}$      	&	       $10^{-7}$       	\\
$\cbx{n}_Q$     	&	       $\dots$ 	&	       $10^{-15}$      	&	       $10^{-16}$      	&	       $\dots$ 	\\
&&&&\\
\hline									
\end{tabular}									
\end{center}									
\twocolumngrid									
 
The formalism and the analytical results 
for gravitational signals of Lorentz violation
presented in this work apply 
to the nonzero matter-sector coefficients 
$\afbw_\mu$, $\cbw_\mn$
and in some cases also to the gravity-sector coefficients $\sb_\mn$.
Comparatively little is known about 
the coefficients $\afbw_\mu$, 
and scenarios exist in which they could be countershaded,
having large values while still escaping notice in searches to date
\cite{akjt}.
However,
nongravitational measurements have already yielded 
impressive sensitivities to various components of $\cbw_\mn$ 
\cite{tables}.
The estimated attainable sensitivities to $\cbw_\mn$
derived in this work are therefore primarily restricted to 
components of $\cbx{n}_\mn$,
for which existing constraints are weaker.
Tables \summarya\ and \summaryb\
reflect these facts,
containing mostly entries for  
combinations of the coefficients $\afbw_\mu$
along with some results for $\cbw_\mn$. 

Table \summarya\ summarizes 
actual sensitivities or estimated attainable ones 
using data from past or present measurements.
The table is based on the calculations presented in this work
and includes only sensitivities below parts in $10^2$.
Each entry in the first column of this table
represents a linear combination of coefficients 
that is accessible in principle 
via existing searches.  
Each of the other four columns 
contains our estimates for sensitivities
that could be achieved in the listed class of tests,
expressed to the nearest order of magnitude.
Values in these four columns
that are shown without brackets 
represent order-of-magnitude sensitivities 
implied by our present analysis
to the modulus of the coefficient combination displayed.
Values appearing in brackets in the table 
represent our estimate of sensitivities 
that could in principle be obtained
from a suitable reanalysis of existing data.
An obelisk ($\dag$) following a value 
indicates a limit attainable 
under the assumption that 
either $\afbw_\mu$ or $\cbw_\mn$
is negligibly small or vanishes.
A diesis ($\ddag$) indicates a sensitivity 
that is attained by combining data
from two different classes of experiments,
and this sensitivity is placed in each 
of the two corresponding columns in the table.

Table \summaryb\ contains future attainable sensitivities 
to the moduli of various combinations
of the matter-sector coefficients $\afbw_\mu$ and $\cbw_\mn$,
as estimated in previous sections of this work.
The structure of this table
is similar to that of Table \summarya.
The listed entries are based on the best design reach
and are given to the nearest order of magnitude.
For each class of search,
we assume enough measurements have been performed
to achieve the maximum number of independent sensitivities.
The reader is cautioned that 
for certain coefficients a single measurement cannot attain 
the indicated sensitivity,
but instead only a linear combination of coefficients
with multipliers controlled by 
composition and orientation factors.
Note that elsewhere in this work
the convention is to display 
values of future sensitivities in braces,
but this convention is suppressed in Table \summaryb\ 
because all entries are of this type.
Note also that further improvements 
in theoretical techniques and experimental design
in all types of searches,
including ones not listed in Table \summaryb\
such as exotic gravitational tests 
or photon tests, 
are expected to yield additional interesting prospects
for future attainable sensitivities.
 
Taken together,
Tables \summarya\ and \summaryb\
reveal excellent prospects for using matter-gravity couplings 
to seek effects of Lorentz violation.
The opportunities for measuring 
the countershaded coefficients $\afbw_\mu$
at sensitive levels are of particular interest
in this context,
as these coefficients typically cannot be detected 
in nongravitational searches.
Indeed,
the spatial components of $\afbw_\mu$
remain essentially unconstrained to date.
The tests proposed here can be performed with existing 
or near-future technology,
and they offer a promising new arena
for searches for signals from the Planck scale.

\vskip 20pt

\pt{x}

\section*{Acknowledgments}

This work was supported in part
by the Department of Energy
under grant DE-FG02-91ER40661
and by the Indiana University Center for Spacetime Symmetries.


\begin{thebibliography}{99}

\bibitem{ksp}
V.A.\ Kosteleck\'y and S.\ Samuel,
Phys.\ Rev.\ D {\bf 39}, 683 (1989);
V.A.\ Kosteleck\'y and R.\ Potting,
Nucl.\ Phys.\ B {\bf 359}, 545 (1991).

\bibitem{tables}
{\it Data Tables for Lorentz and CPT Violation,}
V.A.\ Kosteleck\'y and N.\ Russell,
2010 edition,
Rev.\ Mod.\ Phys., in press
[arXiv:0801.0287v3].

\bibitem{akgrav}
V.A.\ Kosteleck\'y,
Phys.\ Rev.\ D {\bf 69}, 105009 (2004).

\bibitem{gravexpt1}
J.B.R.\ Battat, J.F.\ Chandler, and C.W.\ Stubbs, 
Phys.\ Rev.\ Lett.\ {\bf 99}, 241103 (2007).

\bibitem{gravexpt2}
K.-Y.\ Chung, S-w.\ Chiow, S.\ Herrmann, S.\ Chu, and H.\ M\"uller,
Phys.\ Rev.\ D {\bf 80}, 016002 (2009);
H.\ M\"uller, S.-w.\ Chiow, S.\ Herrmann, S.\ Chu, and K.-Y.\ Chung,
Phys.\ Rev.\ Lett.\ 100, 031101 (2008).

\bibitem{gravexpt3}
W.M.\ Jensen, S.M.\ Lewis, and J.C.\ Long, 
in V.A.\ Kosteleck\'y, ed.,
{\it CPT and Lorentz Symmetry IV}, 
World Scientific, Singapore, 2008;
J.M.\ Overduin, 
\it ibid.\ \rm

\bibitem{qbakpn}
Q.G.\ Bailey and V.A.\ Kosteleck\'y,
Phys.\ Rev.\ D {\bf 74}, 045001 (2006).

\bibitem{cmw}
C.M.\ Will,
{\it Theory and Experiment in Gravitational Physics},
Cambridge University Press, Cambridge, 1993.

\bibitem{kp}
V.A.\ Kosteleck\'y and R.\ Potting,
Phys.\ Rev.\ D {\bf 51}, 3923 (1995).

\bibitem{owg}
O.W.\ Greenberg,
Phys.\ Rev.\ Lett.\ {\bf 89}, 231602 (2002).

\bibitem{akjt}
V.A.\ Kosteleck\'y and J.D.\ Tasson,
Phys.\ Rev.\ Lett.\ {\bf 102}, 010402 (2009).

\bibitem{gravexpt4}
L.\ Carbone, H.\ Panjwani, C.C.\ Speake, T.J.\ Quinn
and C.J.\ Collins,
in T.\ Damour, R.T.\ Jantzen, and R.\ Ruffini, eds.,
{\it Proceedings of the Twelfth Marcel Grossmann
Meeting on General Relativity},
World Scientific, Singapore, 2010.

\bibitem{ng}
Y.\ Nambu,
Phys.\ Rev.\ Lett.\ {\bf 4}, 380 (1960);
J.\ Goldstone,
Nuov.\ Cim.\ {\bf 19}, 154 (1961);
J.\ Goldstone, A.\ Salam, and S.\ Weinberg,
Phys.\ Rev.\ {\bf 127}, 965 (1962).

\bibitem{lvng}
R.\ Bluhm and V.A.\ Kosteleck\'y,
Phys.\ Rev.\ D {\bf 71} 065008 (2005).

\bibitem{lvmm}
R.\ Bluhm \etal, 
Phys.\ Rev.\ D {\bf 77}, 065020 (2008).

\bibitem{uk}
R.\ Utiyama,
Phys.\ Rev.\ {\bf 101}, 1597 (1956);
T.W.B.\ Kibble,
J.\ Math.\ Phys.\ {\bf 2}, 212 (1961).

\bibitem{krt}
V.A.\ Kosteleck\'y, N.\ Russell, and J.D.\ Tasson,
Phys.\ Rev.\ Lett.\ {\bf 100}, 111102 (2008).

\bibitem{ck}
D.\ Colladay and V.A.\ Kosteleck\'y,
Phys.\ Rev.\ D {\bf 55}, 6760 (1997);
Phys.\ Rev.\ D {\bf 58}, 116002 (1998).

\bibitem{cm}
D.\ Colladay and P.\ McDonald,
J.\ Math.\ Phys.\ {\bf 43}, 3554 (2002).

\bibitem{mbak}
M.S.\ Berger and V.A.\ Kosteleck\'y,
Phys.\ Rev.\ D {\bf 65}, 091701(R) (2002).

\bibitem{km}
V.A.\ Kosteleck\'y and M.\ Mewes,
Phys.\ Rev.\ D {\bf 66}, 056005 (2002).

\bibitem{qbak04}
Q.G.\ Bailey and V.A.\ Kosteleck\'y,
Phys.\ Rev.\ D {\bf 70}, 076006 (2004).

\bibitem{ba-f}
B.\ Altschul,
J.\ Phys.\ A {\bf 39} 13757 (2006).

\bibitem{rl-b}
R.\ Lehnert,
Phys.\ Rev.\ D {\bf 75}, 041301 (2007).

\bibitem{km-nr}
V.A.\ Kosteleck\'y and M.\ Mewes,
Phys.\ Rev.\ D {\bf 80}, 015020 (2009);
Ap.\ J.\ Lett.\ {\bf 689}, L1 (2008).

\bibitem{adp}
C.\ Armendariz-Picon, A.\ Diez-Tejedor, and R.\ Penco,
JHEP {\bf 1010}, 079 (2010).

\bibitem{cardinal}
V.A.\ Kosteleck\'y and R.\ Potting, 
Gen.\ Rel.\ Grav.\ {\bf 37}, 1675 (2005);
Phys.\ Rev.\ D {\bf 79}, 065018 (2009).

\bibitem{ctw}
S.M.\ Carroll, H.\ Tam, and I.K.\ Wehus,
Phys.\ Rev.\ D {\bf 80}, 025020 (2009).

\bibitem{ahclt}
N.\ Arkani-Hamed, H.-C.\ Cheng, M.\ Luty, and J.\ Thaler,
JHEP {\bf 0507}, 029 (2005).

\bibitem{abk}
B.\ Altschul \etal,
Phys.\ Rev.\ D {\bf 81}, 065028 (2010).

\bibitem{photons}
Y.\ Nambu,
Prog.\ Theor.\ Phys.\ Suppl.\ Extra 190 (1968);
J.D.\ Bjorken,
Ann.\ Phys.\ {\bf 24}, 174 (1963);
P.G.O.\ Freund,
Acta Phys.\ Austriaca {\bf 14}, 445 (1961);
W.\ Heisenberg,
Rev.\ Mod.\ Phys.\ {\bf 29}, 269 (1957);
P.A.M.\ Dirac,
Proc.\ R.\ Soc.\ Lon.\ {\bf A209}, 291, (1951).

\bibitem{gravitons}
Z.\ Berezhiani and O.V.\ Kancheli,
arXiv:0808.3181;
P.\ Kraus and E.T.\ Tomboulis
Phys.\ Rev.\ D {\bf 66}, 045015 (2002);
D.\ Atkatz,
Phys.\ Rev.\ D {\bf 17}, 1972 (1978);
H.C.\ Ohanian,
Phys.\ Rev.\ {\bf 184}, 1305 (1969);
P.R.\ Phillips, 
Phys.\ Rev.\ {\bf 146}, 966 (1966).

\bibitem{scwga}
S.\ Schlamminger, K.-Y.\ Choi, T.A.\ Wagner,
J.H.\ Gundlach, and E.G.\ Adelberger,
Phys.\ Rev.\ Lett.\ {\bf 100}, 041101 (2008).

\bibitem{is}
I.L.\ Shapiro, 
Phys.\ Rep.\ 357, 113 (2002).

\bibitem{haccss}
B.R.\ Heckel, E.G.\ Adelberger, C.E.\ Cramer,
T.S.\ Cook, S.\ Schlamminger, and U.\ Schmidt,
Phys.\ Rev.\ D {\bf 78}, 092006 (2008).

\bibitem{bkr}
R.\ Bluhm \etal,
Phys.\ Rev.\ D {\bf 57}, 3932 (1998).

\bibitem{kl}
V.A.\ Kosteleck\'y and C.D.\ Lane,
J.\ Math.\ Phys.\ {\bf 40}, 6245 (1999).

\bibitem{kleh}
V.A.\ Kosteleck\'y and R.\ Lehnert,
Phys.\ Rev.\ D {\bf 63}, 065008 (2001).

\bibitem{lp}
L.\ Parker,
Phys.\ Rev.\ D {\bf 22}, 1922 (1980).

\bibitem{xhlp}
X.\ Huang and L.\ Parker,
Phys.\ Rev.\ D {\bf 79}, 024020 (2009).

\bibitem{hccass}
B.R.\ Heckel, C.E.\ Cramer, T.S.\ Cook, 
E.G.\ Adelberger, S.\ Schlamminger, and U.\ Schmidt,
Phys.\ Rev.\ Lett.\ {\bf 97} 021603 (2006).

\bibitem{fw}
L.L.\ Foldy and S.A.\ Wouthuysen,
Phys.\ Rev.\ {\bf 78}, 29 (1950).

\bibitem{gos}
B.\ Goncalves, Y.N.\ Obukhov, and I.L.\ Shapiro,
Phys.\ Rev.\ D {\bf 80}, 125034 (2009).

\bibitem{aknr}
V.A.\ Kosteleck\'y and N.\ Russell,
Phys.\ Lett.\ B {\bf 693}, 443 (2010).

\bibitem{kl2}
V.A.\ Kosteleck\'y and C.D.\ Lane,
Phys.\ Rev.\ D {\bf 60}, 116010 (1999).

\bibitem{earthcomp}
C.J.\ All\`egre, J.-P.\ Poirier, E.\ Humler, and A.W.\ Hofmann,
Earth Planet.\ Sci.\ Lett.\ {\bf 134}, 515 (1995).

\bibitem{lis}
L.I.\ Schiff,
Am.\ J.\ Phys.\ {\bf 28}, 340 (1960).

\bibitem{aad}
The breaking of general coordinate invariance is studied in
M.M.\ Anber, U.\ Aydemir, and J.F.\ Donoghue,
Phys.\ Rev.\ D {\bf 81}, 084059 (2010). 

\bibitem{ksbb}
V.A.\ Kosteleck\'y and S.\ Samuel,
Phys.\ Rev.\ D {\bf 40}, 1886 (1989);
Phys.\ Rev.\ Lett.\ {\bf 63}, 224 (1989).

\bibitem{bumblebee}
M.D.\ Seifert,
Phys.\ Rev.\ D {\bf 81}, 065010 (2010);
Phys.\ Rev.\ D {\bf 79}, 124012 (2009); 
J.\ Alfaro and L.F.\ Urrutia, 
Phys.\ Rev.\ D {\bf 81}, 025007 (2010);
J.L.\ Chkareuli, C.D. Froggatt, and H.B.\ Nielsen,
Nucl.\ Phys.\ B {\bf 821}, 65 (2009);
C.\ Armendariz-Picon and A.\ Diez-Tejedor,
JCAP {\bf 0912}, 018 (2009).

\bibitem{stability}
S.M.\ Carroll, T.R.\ Dulaney, M.I.\ Gresham, and H.\ Tam,
Phys.\ Rev.\ D {\bf 79}, 065011 (2009);
R.\ Bluhm, N.L.\ Gagne, R.\ Potting, and A.\ Vrublevskis,
Phys.\ Rev.\ D {\bf 77}, 125007 (2008);
M.D.\ Seifert,
Phys.\ Rev.\ D {\bf 76}, 064002 (2007).

\bibitem{gl}
G.\ Leibbrandt, 
Rev.\ Mod.\ Phys.\ {\bf 59}, 1067 (1987).

\bibitem{mesons}
KTeV Collaboration,
H.\ Nguyen, 
in V.A.\ Kosteleck\'y, ed.,
{\it CPT and Lorentz Symmetry II}, 
World Scientific, Singapore, 2002
[hep-ex/0112046];
A.\ Di Domenico,
KLOE Collaboration,
J.\ Phys.\ Conf.\ Ser.\ {\bf 171}, 012008 (2009);
FOCUS Collaboration,
J.M.\ Link \etal,
Phys.\ Lett.\ B {\bf 556}, 7 (2003);
BaBar Collaboration,
B.\ Aubert
\etal,
Phys.\ Rev.\ Lett.\ {\bf 100}, 131802 (2008);
hep-ex/0607103.

\bibitem{akmesons}
V.A.\ Kosteleck\'y,
Phys.\ Rev.\ Lett.\ {\bf 80}, 1818 (1998);
Phys.\ Rev.\ D {\bf 61}, 016002 (1999);
Phys.\ Rev.\ D {\bf 64}, 076001 (2001);
V.A.\ Kosteleck\'y and R.\ Van Kooten,
Phys.\ Rev.\ D {\bf 82}, 101702 (R) (2010).

\bibitem{neutrinos}
MINOS Collaboration,
P.\ Adamson \etal,
Phys.\ Rev.\ Lett.\ {\bf 101}, 151601 (2008);
LSND Collaboration,
L.B.\ Auerbach \etal, 
Phys.\ Rev.\ D {\bf 72}, 076004 (2005);
M.D.\ Messier (SK), 
in V.A.\ Kosteleck\'y, ed.,
{\it CPT and Lorentz Symmetry II}, 
World Scientific, Singapore, 2005;
V.A.\ Kosteleck\'y and M.\ Mewes,
Phys.\ Rev.\ D {\bf 69}, 016005 (2004);
Phys.\ Rev.\ D {\bf 70}, 031902 (2004);
Phys.\ Rev.\ D {\bf 70}, 076002 (2004);
T.\ Katori \etal, 
Phys.\ Rev.\ D {\bf 74}, 105009 (2006);
V.\ Barger, D.\ Marfatia, and  K.\ Whisnant,
Phys.\ Lett.\ B {\bf 653}, 267 (2007);
J.S.\ D\'\i az \etal,
Phys.\ Rev.\ D {\bf 80}, 076007 (2009).

\bibitem{qbgem}
Q.G.\ Bailey,
Phys.\ Rev.\ D {\bf 82}, 065012 (2010).

\bibitem{mf}
I.\ Marson and J.E.\ Faller,
J.\ Phys.\ E {\bf 19}, 22 (1986).

\bibitem{pcc}
A.\ Peters, K.Y.\ Chung, and S.\ Chu,
Nature {\bf 400}, 849 (1999);
Metrologia {\bf 38}, 25 (2001).

\bibitem{glm}
T.L.\ Gustavson, A.\ Landragin, and M.A.\ Kasevich,
Class.\ Quantum.\ Grav.\ {\bf 17}, 2385 (2000);
T.L.\ Gustavson, P.\ Bouyer, and M.A.\ Kasevich,
Phys.\ Rev.\ Lett.\ {\bf 78}, 2046 (1997).

\bibitem{sct}
P.\ Story and C.\ Cohen-Tannoudji,
J.\ Phys.\ II France {\bf 4}, 1999 (1994).

\bibitem{abl}
Special forms of spin-dependent Lorentz violation
in matter interferometers have been discussed in 
J.\ Audretsch, U.\ Bleyer, and C.\ L\"ammerzahl,
Phys.\ Rev.\ A {\bf 47}, 4632 (1993).

\bibitem{mffsk}
J.M.\ McGuirk, G.T.\ Foster, J.B.\ Fixler, M.J.\ Snadden,
and M.A.\ Kasevich,
Phys.\ Rev.\ A {\bf 65}, 033608 (2002).

\bibitem{ykkm}
N.\ Yu, J.M.\ Kohel, J.R.\ Kellogg, and L.\ Maleki,
Appl.\ Phys.\ B {\bf 84}, 647 (2006).

\bibitem{canuel}
B.\ Canuel \etal,
Phys.\ Rev.\ Lett.\ {\bf 97}, 010402 (2006).

\bibitem{dghk}
S.\ Dimopoulos, P.W.\ Graham, J.M.\ Hogan, and M.A.\ Kasevich, 
Phys.\ Rev.\ Lett.\ {\bf 98}, 111102 (2007);
Phys.\ Rev.\ D {\bf 78}, 042003 (2008).

\bibitem{ninterf}
H.\ Kaiser \etal,
Physica B {\bf 385-386}, 1384 (2006).

\bibitem{wg}
R.J.\ Warburton and J.M.\ Goodkind,
Astrophys.\ J.\ {\bf 208}, 881 (1976).

\bibitem{ss}
S.\ Shiomi,
arXiv:0902.4081.

\bibitem{shortrange}
A.A.\ Geraci, S.J.\ Smullin, D.M.\ Weld,
J.\ Chiaverini, and A.\ Kapitulnik,
Phys.\ Rev.\ D {\bf 78}, 022002 (2008);
R.S.\ Decca, D.\ L\'opez, E.\ Fischbach, 
G.L.\ Klimchitskaya, D.E.\ Krause, and V.M.\ Mostepanenko,
Phys.\ Rev.\ D {\bf 75}, 077101 (2007);
D.J.\ Kapner, T.S.\ Cook, E.G.\ Adelberger,
J.H.\ Gundlach, B.R.\ Heckel, C.D.\ Hoyle, and H.E.\ Swanson,
Phys.\ Rev.\ Lett.\ {\bf 98}, 021101 (2007);
J.C.\ Long, H.W.\ Chan, A.B.\ Churnside,
E.A.\ Gulbis, M.C.M.\ Varney, and J.C.\ Price,
Nature {\bf 421}, 922 (2003);
T.J.\ Quinn, C.C.\ Speake, S.J.\ Richman,
R.S.\ Davis, and A.\ Picard,
Phys.\ Rev.\ Lett.\ {\bf 87}, 111101 (2001).

\bibitem{kknm}
K.\ Kuroda and N.\ Mio,
Phys.\ Rev.\ D {\bf 42}, 3903 (1990).

\bibitem{nmf}
T.M.\ Niebauer, M.P.\ McHugh, and J.E.\ Faller,
Phys.\ Rev.\ Lett.\ {\bf 59}, 609 (1987).

\bibitem{fdhw}
S.\ Fray, C.A.\ Diez, T.W.\ H\"ansch, and M.\ Weitz,
Phys.\ Rev.\ Lett.\ {\bf 93}, 240404 (2004).

\bibitem{poem}
R.D.\ Reasenberg, 
in V.A.\ Kosteleck\'y, ed.,
{\it CPT and Lorentz Symmetry II}, 
World Scientific, Singapore, 2005.

\bibitem{great}
V.\ Iafolla, S.\ Nozzoli, E.C.\ Lorenzini, I.I.\ Shapiro,
and V.\ Milyukov,
Class.\ Quantum Grav.\ {\bf 17}, 2327 (2000).

\bibitem{hdcm}
H.\ Dittus and C.\ Mehls,
Class.\ Quantum Grav.\ {\bf 18}, 2417 (2001).

\bibitem{marion}
H.\ Marion \etal,
Phys.\ Rev.\ Lett.\ {\bf 90}, 150801 (2003).

\bibitem{positronium}
M.K.\ Oberthaler,
Nucl.\ Instr.\ and Meth.\ B {\bf 192}, 129 (2002).

\bibitem{srpoem}
R.D.\ Reasenberg and J.D.\ Phillips,
Class.\ Q.\ Grav.\ {\bf 27}, 095005 (2010).

\bibitem{shaghss}
Y.\ Su, B.R.\ Heckel, E.G.\ Adelberger, J.H.\ Gundlach, 
M.\ Harris, G.L.\ Smith, and H.E.\ Swanson,
Phys.\ Rev.\ D {\bf 50}, 3614 (1994).

\bibitem{hnlbk}
L.-S.\ Hou, W.-T.\ Ni, and Y.-C.M.\ Li,
Phys.\ Rev.\ Lett.\ {\bf 90}, 201101 (2003);
R.\ Bluhm and V.A.\ Kosteleck\'y,
Phys.\ Rev.\ Lett.\ {\bf 84}, 1381 (2000).

\bibitem{ss2}
S.\ Shiomi,
Phys.\ Rev.\ D {\bf 78}, 042001 (2008).

\bibitem{spacegrav}
For reviews of space-based tests of relativity see,
for example
C.\ L\"ammerzahl, 
C.W.F.\ Everitt, 
and F.W.\ Hehl, eds.,
{\it Gyros, Clocks, Interferometers \ldots :
Testing Relativistic Gravity in Space},
Springer, Berlin, 2001.

\bibitem{spaceexpt}
R.\ Bluhm \etal,
Phys.\ Rev.\ Lett.\ {\bf 88}, 090801 (2002);
Phys.\ Rev.\ D {\bf 68}, 125008 (2003).

\bibitem{muscope}
P.\ Touboul, M.\ Rodrigues, G.\ M\'etris, and B.\ Tatry,
Comptes Rendus de l'Acad\'emie des Sciences, Series IV,
{\bf 2}, 1271 (2001).

\bibitem{step}
T.J.\ Sumner \etal,
Adv.\ Space Res.\ {\bf 39}, 254 (2007).

\bibitem{gg}
A.M.\ Nobili \etal,
Exp.\ Astron.\ {\bf 23}, 689 (2009).

\bibitem{gauge}
G.\ Amelino-Camelia \etal,
Exp.\ Astron.\ {\bf 23}, 549 (2008).

\bibitem{bapulsar}
B.\ Altschul,
Phys.\ Rev.\ D {\bf 75}, 023001 (2007).

\bibitem{ggg}
G.L.\ Comandi \etal,
Rev.\ Sci.\ Instrum.\ {\bf 77}, 034501 (2006).

\bibitem{fhmn}
F.\ Hasselbach and M.\ Nicklaus,
Phys.\ Rev.\ A {\bf 48}, 143 (1993);
R.\ Neutze and F.\ Hasselbach,
Phys.\ Rev.\ A {\bf 58}, 557 (1998).

\bibitem{ncd}
B.\ Neyenhuis, D.\ Christensen, and D.S.\ Durfee,
Phys.\ Rev.\ Lett.\ {\bf 99}, 200401 (2007).

\bibitem{drifttube}
L.I.\ Schiff and M.V.\ Barnhill, 
Phys.\ Rev.\ {\bf 151}, 1067 (1966);
A.J.\ Dessler, F.C.\ Michel,
H.E.\ Rorschach, and G.T.\ Trammell,
Phys.\ Rev.\ {\bf 168}, 737 (1968);
C.\ Herring,
Phys.\ Rev.\ {\bf 171}, 1361 (1968);
L.I.\ Schiff,
Phys.\ Rev.\ B {\bf 1}, 4649 (1970).

\bibitem{fwwf}
F.S.\ Witteborn and W.M.\ Fairbank,
Phys.\ Rev.\ Lett.\ {\bf 19}, 1049 (1967);
Rev.\ Sci.\ Instrum.\ {\bf 48}, 1 (1977).

\bibitem{tgmn}
T.\ Goldman and M.M.\ Nieto,
Phys.\ Lett.\ B {\bf 112}, 437 (1982).

\bibitem{ps200}
M.H.\ Holzscheiter \etal,
Nucl.\ Phys.\ A {\bf 558}, 709c (1993).

\bibitem{drom}
T.W.\ Darling, F.\ Rossi, G.I.\ Opat, and G.F.\ Moorhead,
Rev.\ Mod.\ Phys.\ {\bf 64}, 237 (1992).

\bibitem{lbgg}
L.S.\ Brown and G.\ Gabrielse,
Rev.\ Mod.\ Phys.\ {\bf 58}, 233 (1986).

\bibitem{llmth}
V.\ Lagomarsino, V.\ Lia, G.\ Manuzio, and G.\ Testera, 
Phys.\ Rev.\ A {\bf 50}, 977 (1994);
V.\ Lagomarsino, G.\ Manuzio, G.\ Testera, 
and M.H.\ Holszscheiter,
Hyperfine Int.\ {\bf 100}, 153 (1996).

\bibitem{ps210}
PS210 Collaboration,
G.\ Baur \etal,
Phys.\ Lett.\ B {\bf 368}, 251 (1996).

\bibitem{e862}
E862 Collaboration,
G.\ Blanford, D.C.\ Christian, K.\ Gollwitzer,
M.\ Mandelkern, C.T.\ Munger, J.\ Schultz,
and G.\ Zioulas,
Phys.\ Rev.\ Lett.\ {\bf 80}, 3037 (1998).

\bibitem{athenaatrap}
ATHENA Collaboration,
M.\ Amoretti \etal,
Nature {\bf 419}, 456 (2002);
ATRAP Collaboration,
G.\ Gabrielse \etal,
Phys.\ Rev.\ Lett.\ {\bf 89}, 213401 (2002).

\bibitem{alphaatrap2}
ALPHA Collaboration,
G.\ Andresen \etal,
Phys.\ Rev.\ Lett.\ {\bf 98}, 023402 (2007);
ATRAP Collaboration,
G.\ Gabrielse \etal,
Phys.\ Rev.\ Lett.\ {\bf 98}, 113002 (2007).

\bibitem{bkrhhbar}
R.\ Bluhm \etal,
Phys.\ Rev.\ Lett.\ {\bf 82}, 2254 (1999).

\bibitem{asacusa}
B.\ Juh\'asz and E.\ Widmann,
Hyperfine Int.\ {\bf 193}, 305 (2009).

\bibitem{gblmstp}
G.\ Gabrielse,
Hyperfine Int.\ {\bf 44}, 349 (1988);
N.\ Beverini, V.\ Lagomarsino, G.\ Manuzio, F.\ Scuri,
and G.\ Torelli,
Hyperfine Int.\ {\bf 44}, 357 (1989);
R.\ Poggiani,
Hyperfine Int.\ {\bf 76}, 371 (1993).

\bibitem{tjpage}
T.J.\ Phillips,
Hyperfine Int.\ {\bf 109}, 357 (1997);
AGE Collaboration,
A.D.\ Cronin \etal,
\it Letter of Intent: 
Antimatter Gravity Experiment (AGE) at Fermilab, \rm
February 2009.

\bibitem{jwth}
J.\ Walz and T.W.\ H\"ansch,
Gen.\ Rel.\ Grav.\ {\bf 36}, 561 (2004);
P.\ P\'erez, L.\ Liszkay, B.\ Mansouli\'e, J.M.\ Rey,
A.\ Mohri, Y.\ Yamazaki, N.\ Kuroda, and H.A.\ Torii,
\it Letter of Intent to the CERN-SPSC, \rm
November 2007.

\bibitem{weax}
F.M.\ Huber, E.W.\ Messerschmid, and G.A.\ Smith,
Class.\ Quantum Grav.\ {\bf 18}, 2457 (2001).

\bibitem{aegis}
AEGIS Collaboration,
A.\ Kellerbauer \etal,
Nucl.\ Instr.\ Meth.\ B {\bf 266}, 351 (2008).

\bibitem{mntg}
M.M.\ Nieto and T.\ Goldman,
Phys.\ Rep.\ {\bf 205}, 221 (1991).

\bibitem{akifc}
V.A.\ Kosteleck\'y, unpublished (2003).

\bibitem{pm}
P.\ Morrison,
Am.\ J.\ Phys.\ {\bf 26}, 358 (1958).

\bibitem{lis2}
L.I.\ Schiff,
Phys.\ Rev.\ Lett.\ {\bf 1}, 254 (1958),
Proc.\ Natl.\ Acad.\ Sci.\ {\bf 45}, 69 (1959).

\bibitem{renorm}
R.\ Jackiw and V.A.\ Kosteleck\'y,
Phys.\ Rev.\ Lett.\ {\bf 82}, 3572 (1999);
M.\ P\'erez-Victoria, JHEP {\bf 0104}, 032 (2001);
V.A.\ Kosteleck\'y, C.D.\ Lane, and A.G.M.\ Pickering,
Phys.\ Rev.\ D {\bf 65}, 056006 (2002);
V.A.\ Kosteleck\'y and A.G.M.\ Pickering,
Phys.\ Rev.\ Lett.\ {\bf 91}, 031801 (2003);
B.\ Altschul, 
Phys.\ Rev.\ D {\bf 69}, 125009 (2004);
Phys.\ Rev.\ D {\bf 70}, 101701 (2004);
B.\ Altschul and V.A.\ Kosteleck\'y, 
Phys.\ Lett.\ B {\bf 628}, 106 (2005); 
H.\ Belich, T.\ Costa-Soares, M.M.\ Ferreira, and J.A.\ Helayel-Neto,
Eur.\ Phys.\ J.\ C {\bf 42}, 127 (2005);
T.\ Mariz, J.R.\ Nascimento, E.\ Passos, 
R.F.\ Ribeiro, and F.A.\ Brito,
JHEP {\bf 0510}, 019 (2005);
G.\ de Berredo-Peixoto and I.L.\ Shapiro,
Phys.\ Lett.\ B {\bf 642}, 153 (2006);
P.\ Arias, H.\ Falomir, J.\ Gamboa, F.\ M\'endez,
and F.A.\ Schaposnik,
Phys.\ Rev.\ D {\bf 76}, 025019 (2007);
D.\ Colladay and P.\ McDonald,
Phys.\ Rev.\ D {\bf 75}, 105002 (2007);
Phys.\ Rev.\ D {\bf 77}, 085006 (2008);
Phys.\ Rev.\ D {\bf 79}, 125019 (2009);
M.\ Gomes, T.\ Mariz, J.R.\ Nascimento,
E.\ Passos, A.Yu.\ Petrov, and A.J.\ da Silva,
Phys.\ Rev.\ D {\bf 78}, 025029 (2008);
D.\ Anselmi, 
Ann.\ Phys.\ {\bf 324}, 874 (2009);
Ann.\ Phys.\ {\bf 324}, 1058 (2009).

\bibitem{mlg}
M.L.\ Good,
Phys.\ Rev.\ {\bf 121}, 311 (1961).

\bibitem{wcbc}
P.\ Wolf, F.\ Chapelet, S.\ Bize, and A.\ Clairon,
Phys.\ Rev.\ Lett.\ {\bf 96}, 060801 (2006).

\bibitem{ba-ce}
B.\ Altschul, 
Phys.\ Rev.\ D {\bf 82}, 016002 (2010).

\bibitem{km-nr2}
V.A.\ Kosteleck\'y and M.\ Mewes,
in preparation.

\bibitem{muexpt}
G.W.\ Bennett \etal,
Muon $g$--2 Collaboration,
Phys.\ Rev.\ Lett.\ {\bf 100}, 091602 (2008);
B.\ Altschul,
Astropart.\ Phys.\ {\bf 28}, 380 (2007);
V.W.\ Hughes \etal,
Phys.\ Rev.\ Lett.\ {\bf 87}, 111804 (2001);
R.\ Bluhm \etal,
Phys.\ Rev.\ Lett.\ {\bf 84}, 1098 (2000).

\bibitem{kk}
K.\ Kirch,
arXiv:physics/0702143.

\bibitem{bl}
B.\ Lesche,
Gen.\ Rel.\ Grav.\ {\bf 21}, 623 (1989).

\bibitem{d0}
D0 Collaboration,
V.M.\ Abazov \etal,
Phys.\ Rev.\ D {\bf 82}, 032001 (2010).

\bibitem{bnsyn}
O.\ Bertolami \etal,
Phys.\ Lett.\ B {\bf 395}, 178 (1997);
G.\ Lambiase,
Phys.\ Rev.\ D {\bf 72}, 087702 (2005);
J.M.\ Carmona, J.L.\ Cort\'es, A.\ Das,
J.\ Gamboa, and F.\ M\'endez,
Mod.\ Phys.\ Lett.\ {\bf A21}, 883 (2006);
S.M.\ Carroll and J.\ Shu,
Phys.\ Rev.\ D {\bf 73}, 103515 (2006).

\bibitem{llrexpt}
J.G.\ Williams, S.G.\ Turyshev, and H.D. Boggs, 
Phys.\ Rev.\ Lett.\ {\bf 93}, 261101 (2004).

\bibitem{apollo}
T.W.\ Murphy \etal, 
Pub.\ Astron.\ Soc.\ Pac.\ {\bf 120}, 20 (2008).

\bibitem{gdotexpt}
J.\ M\"uller and L.\ Biskupek,
Class.\ Quantum Grav.\ {\bf 24}, 4533 (2007);
R.W.\ Hellings \etal,
Phys.\ Rev.\ Lett.\ {\bf 51}, 1609 (1983).

\bibitem{klp}
V.A.\ Kosteleck\'y, R.\ Lehnert, and  M.J.\ Perry, 
Phys.\ Rev.\ D {\bf 68}, 123511 (2003).

\bibitem{vab}
V.A.\ Brumberg,
{\it Essential Relativistic Celestial Mechanics,}
Adam Hilger, Bristol, 1991.

\bibitem{dba}
P.\ Moore,
{\it The Data Book of Astronomy},
Institute of Physics Publishing, Bristol, 2000.

\bibitem{cmw2}
C.M.\ Will,
Living Rev.\ Relativity {\bf 4}, 4 (2001)
[gr-qc/0510072].

\bibitem{qbtd}
Q.G.\ Bailey,
Phys.\ Rev.\ D {\bf 80}, 044004 (2009).

\bibitem{cmw3}
C.M.\ Will,
Phys.\ Rev.\ D {\bf 10}, 2330 (1974);
J.P.\ Turneaure, C.M.\ Will, B.F.\ Farrell, 
E.M.\ Mattison, and R.F.C.\ Vessot,
Phys.\ Rev.\ D {\bf 27 }, 1705 (1983).

\bibitem{cassini}
B.\ Bertotti, L.\ Iess, and P.\ Tortora,
Nature {\bf 425}, 374 (2003).

\bibitem{astrod}
T.\ Appourchaux \etal, 
Exp.\ Astron.\ {\bf 23}, 491 (2009).

\bibitem{more}
L.\ Iess and S.\ Asmar,
Int.\ J.\ Mod.\ Phys.\ D {\bf 16}, 2117 (2007).

\bibitem{sagas}
P.\ Wolf \etal,
Exp.\ Astron.\ {\bf 23}, 651 (2008);
S.\ Reynaud, C.\ Salomon, and P.\ Wolf,
Space Sci.\ Rev.\ {\bf 148}, 233 (2009).

\bibitem{odyssey}
B.\ Christophe \etal, 
Exper.\ Astron.\ {\bf 23}, 529 (2008).

\bibitem{beacon}
S.G.\ Turyshev, B.\ Lane, M.\ Shao, and A.\ Girerd, 
Int.\ J.\ Mod.\ Phys.\ D {\bf 18}, 1025 (2009).

\bibitem{vlbi}
S.B.\ Lambert and C.\ Le Poncin-Lafitte,
Astron.\ Astrophys.\ {\bf 499}, 331 (2009).

\bibitem{ahjp}
N.\ Ashby, T.P.\ Heavner, S.R.\ Jefferts, T.E.\ Parker,
A.G.\ Radnaev, and Y.O.\ Dudin,
Phys.\ Rev.\ Lett.\ {\bf 98}, 070802 (2007).

\bibitem{twb}
M.E.\ Tobar, P.\ Wolf, S.\ Bize,
G.\ Santarelli, and V.\ Flambaum,
Phys.\ Rev.\ D {\bf 81}, 022003 (2010).

\bibitem{star}
R.\ Byer,
{\it Space-Time Asymmetry Research},
Stanford University proposal,
January 2008.

\bibitem{rfcv}
R.F.C.\ Vessot \etal,
Phys.\ Rev.\ Lett.\ {\bf 45}, 2081 (1980).

\bibitem{aces}
L.\ Cacciapuoti and C.\ Salomon,
Eur.\ Phys.\ J.\ Spec.\ Top.\ {\bf 172}, 57 (2009).

\bibitem{galileo}
T.P.\ Krisher, D.D.\ Morabito, and J.D.\ Anderson,
Phys.\ Rev.\ Lett.\ {\bf 70}, 2213 (1993).

\bibitem{saturnrs}
T.P.\ Krisher, J.D.\ Anderson, and J.K.\ Campbell,
Phys.\ Rev.\ Lett.\ {\bf 64}, 1322 (1990).

\bibitem{lator}
S.G. Turyshev and M.\ Shao,
Int.\ J.\ Mod.\ Phys.\ D {\bf 16}, 2191 (2007).

\bibitem{sim}
S.C.\ Unwin \etal,
Pub.\ Astron.\ Soc.\ Pacific {\bf 120}, 38 (2008).

\end{thebibliography}
\end{document}